\documentclass[a4paper,11pt,subfigure]{book}

\usepackage[all]{xy}
\usepackage{mathrsfs}
\usepackage{graphicx}  
\usepackage{rotating}
\usepackage{bm}        
\usepackage[english]{babel}
\usepackage{amsfonts,amsmath,amssymb,mathrsfs}
\usepackage[T1]{fontenc}
\usepackage{multirow}

\usepackage{latexsym}

\newcommand{\be}{\begin{equation}}
\newcommand{\ee}{\end{equation}}
\newcommand{\bea}{\begin{eqnarray}}

\newcommand{\eea}{\end{eqnarray}}

\begin{document}
\frontmatter

\begin{titlepage}

\begin{figure} [h!] 
\centering
\includegraphics[width=4cm]{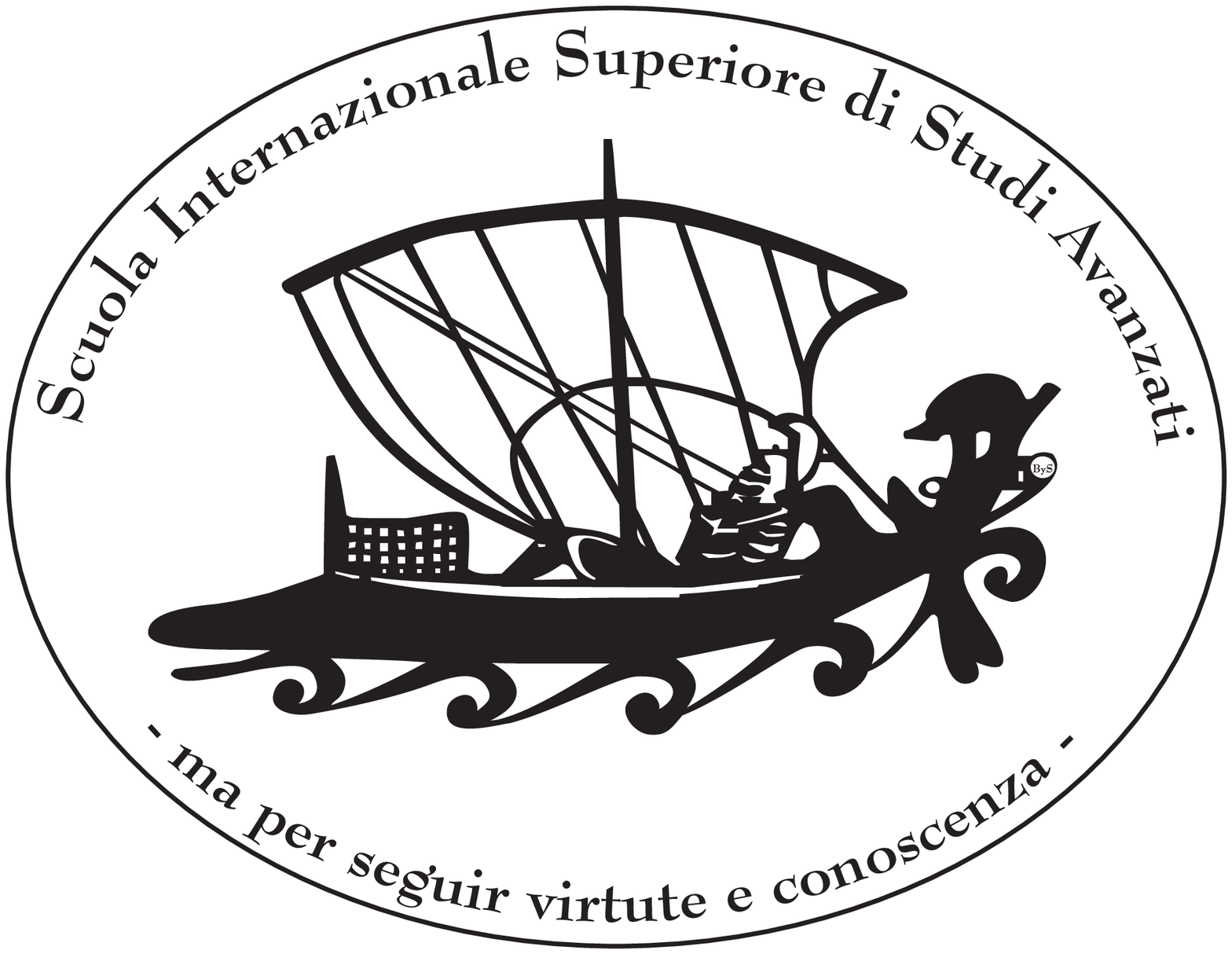}
\end{figure} 

\begin{centering}
\vspace*{0.4cm}
{\Large \bf The distribution of the dark matter in galaxies as the imprint of its Nature}

\vspace*{2cm}
\large
Thesis submitted for the degree of\\
``Doctor Philosophi\ae''

\vspace*{1cm}
September 2008
\vspace*{3cm}

\begin{tabular*}{400pt}{l @{\extracolsep\fill} r}

CANDIDATE & SUPERVISOR\\
&\\
Christine Frigerio Martins & Paolo Salucci
\end{tabular*}

\vspace*{4cm}
International School for Advanced Studies\\
Via Beirut 2-4, 34014 Trieste, Italy.\\
E-mail: {\sl martins@sissa.it} \\
\end{centering}
\end{titlepage}

\cleardoublepage

\begin{figure} [h!] 
\centering
\hskip -0.4cm
\vskip 3cm
\includegraphics[width=10cm]{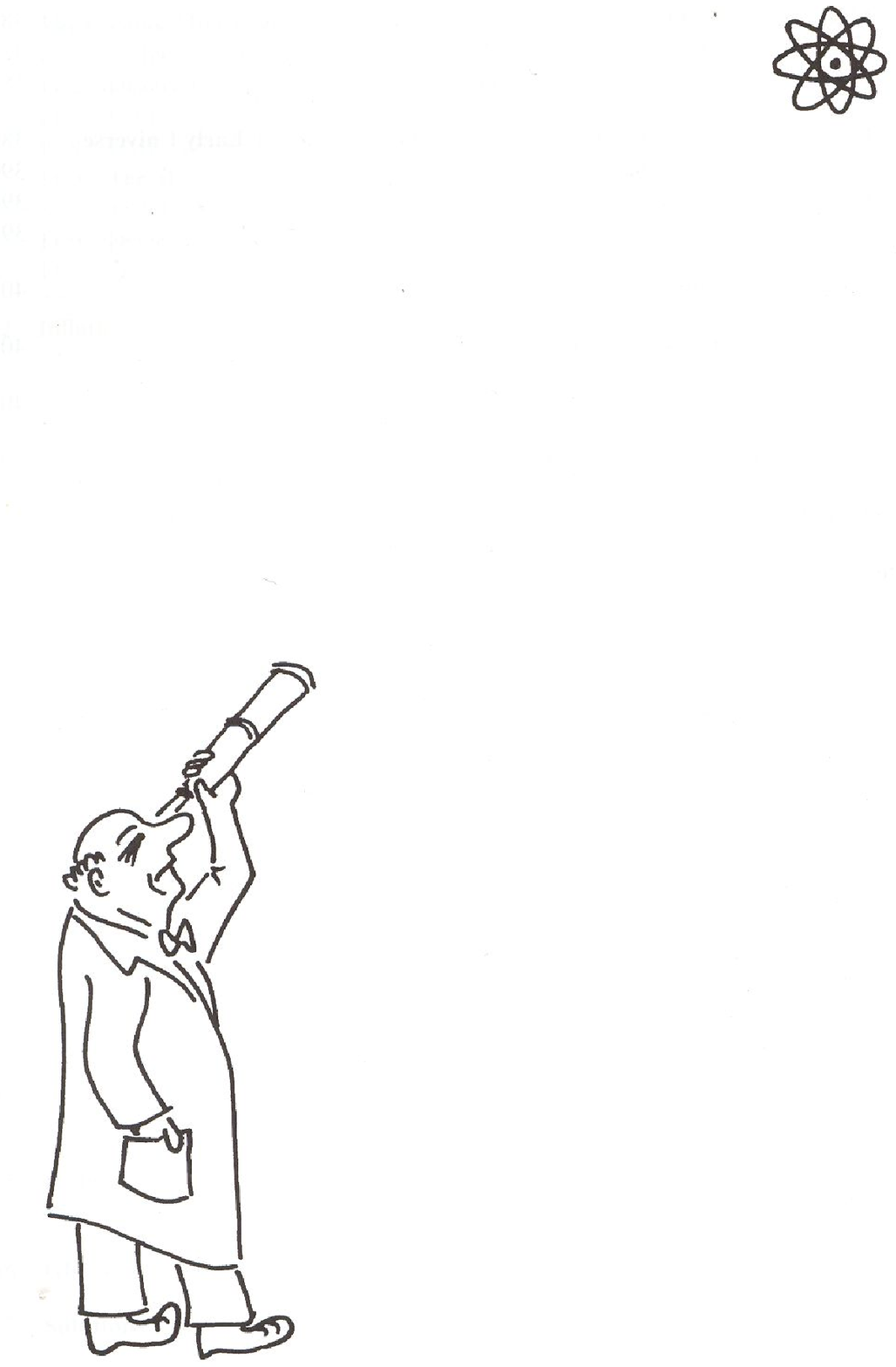}
\vskip -2cm
\end{figure}

\begin{quote}
\em
\raggedleft

``Dark Matter drives me crazy.''

{\bf Uli Klein}
\end{quote}

\newpage

\chapter*{Abstract}

The standard framework within which cosmological measurements are confronted and interpreted nowadays, called $\Lambda$ Cold Dark Matter, presents a Universe dominated by unknown forms of energy and matter.

My Thesis is devoted to investigate the distribution of dark matter in galaxies and addresses the fact that the local universe-the small objects that orbit galaxies and the galaxy cores-turns out to be a marvelous laboratory for examining the nature of dark matter and the fundamental physics involved in the structure formation and evolution.

I develop tests, based on mass modeling of rotation curves, for the validation of dark matter  models on galactic scales.
These tests have been applied in analyzing the phenomenology of the cusp vs core controversy, and the phenomenon of non-Keplerian rotation curves as modification of the laws of gravity.
I further investigate the properties and scaling laws of dark matter halos.

My conclusion is that galactic observations provide strong imprints on the nature of dark matter. 

\chapter*{Papers}

The research presented in this Thesis was mainly conducted in SISSA-International School for Advanced Studies from October 2004. This Thesis is the result of the author's own work, as well as the outcome of scientific collaborations stated below.

The content of this Thesis is based on the following research papers published in refereed Journals:

\begin{itemize}
\small{
\item
{\bf ``Analysis of Rotation Curves in the Framework of the Gravitational Supression Model''}
\\{}C.~Frigerio Martins \& P.~Salucci,
\\{}Phys.~Rev.~Lett.~{\bf 98} 151301 (2007)
[arXiv:hep-ph/0611028]}

\item
{\bf ``Analysis of rotation curves in the framework of $R^n$ graviy''}
\\{} C.~Frigerio Martins \& P.~Salucci,
\\{} Mon.~Not.~Roy.~Astron.~Soc.~{\bf 381} 1103-1108 (2007)
[arXiv:astro-ph/0703243]

and on the following works to be submitted:

\item
{\bf ``Dwarf spheroidal galaxy kinematics and  spirals scaling laws''}
\\{} P.~Salucci, M.~I.~Wilkinson, G.~Gilmore, E.~Grebel, A.~Koch, R.~Wyse, C.~Frigerio Martins, G.~Gentile

\item
{\bf ``A constant Dark Matter Halo Surface density in Galaxies''}
\\{} F.~Donato, G.~Gentile, P.~Salucci, C.~Frigerio Martins, M.~I.~Wilkinson, G.~Gilmore, E.~Grebel, A.~Koch, R.~Wyse

\end{itemize}

\chapter{Preface}

Cosmological observations provide compelling evidence that about 95\% of the content of the Universe resides in two unknown forms of energy that we call dark matter (DM) and dark energy: the first residing in bound objects as non-luminous matter, the latter in the form of a zero-point energy that pervades the whole Universe~\cite{wmap}.
The DM is thought to be composed of cold, neutral, weakly interacting particles, beyond those existing in the Standard Model of Particle Physics, and not yet detected in accelerators  or in dedicated direct and indirect searches.
In the standard $\Lambda$ Cold Dark Matter ($\Lambda$CDM) scenario  primordial density fluctuations are generated during an inflationary period and become the seeds of the bottom-up structure formation model.
This scenario  successfully describes the accelerated expansion of the Universe, the observed temperature fluctuations in the cosmic microwave background radiation, the large scale matter distribution, and the main aspects of the formation and the evolution of virialized cosmological objects~\cite{ostriker}.

Despite these important achievements, at galactic scales  of about 10 kpc, where today most of the mass is located, the $\Lambda$CDM model meets with severe difficulties in explaining the observed distribution of the invisible matter around the luminous one.
In fact, on the one hand, N-body simulations performed in this scenario, unambiguously predict that every halo in the Universe hosting and surrounding a galaxy, must have a very specific density profile.
This features a well pronounced central cusp, obeying to the well known 
Navarro, Frenk \& White (NFW) profile~\cite{nfw96}.
On the observational side instead, high-resolution rotation curves (RCs) show that the actual distribution of DM is much shallower than the above, and it presents a density profile with  a nearly constant density core~\cite{sb00,47salucci03,gentile04,donato04,blok05,deNaray06,3741,spano08} that is  well represented by a Burkert profile~\cite{sb00}.

The cusp vs core controversy, together with other present main failures of the the standard collisionless particle paradigm, such as the issue on the number of sub-halos~\cite{moore99}, has far-reaching consequences in the researches on the nature of DM and unveil the elusive knowledge on it.

My aim is to show how a systematic comparison of cosmological and particle physics models with galactic observations provides strong bounds on the properties of DM.
The outline of this Thesis is then as follows. 
In the Introduction, the basis of the $\Lambda$CDM scenario is summarized, presenting fundamental physical ingredients and its important predictions. 
In Chapter 2 a picture of the fundamental properties of DM as non-relativistic particles is given, as well as their devoted searches.
There are in fact hints that the phenomenon of the missing mass is linked to a new high energy phenomenology not included in the standard model of particle physics and foreseeing the existence of new elementary particles with a mass roughly above the hundred GeV scale.  

Chapter 3 is devoted to a discussion on the mass modeling of RCs as major tools for investigating the distribution and nature of DM in galaxies.
A deep understanding of the  mass models is a fundamental ingredient for a future discovery of the DM particles or alternatives theories, in what it provides the link between the microphysics phenomenology and the observations.
Moreover this chapter gives an exhaustive discussion on the current status of the cusp vs core controversy with extensive references to relevant literature.

In Chapter 4 basic ideas behind the most popular alternatives to DM, MOND and f(R) theories of gravity, are summarized.
In Chapter 5 I develop a test for analyzing the models which aims at solving the cusp vs core controversy by resorting to the best available galaxy kinematics.
I apply this test to an example for such models.
In Chapter 6  instead I develop a test for analyzing the models which aims at solving the phenomenon of the RCs by resorting to modifications of the laws of gravity.
I focus on f(R) theories of gravity.  
This work represents a step forward on  the issue in what for the first time a complete analysis with a devised RC sample has been  performed.

I have also further investigated the properties of DM halos. 
Kinematic observations of the dwarf spheroidal (dSphs) satellites of the Milky Way are revealing hints about the structure of DM halos.
I investigate whether the extrapolation of the scaling relations of brighter galaxies to the low end of the galaxy luminosity regime is consistent with the observed internal kinematics of dSphs.
In Chapters 7 and 8 I discuss the implication of such relations for the comprehension  of the nature of DM.
Finally I conclude in Chapter 9.

A number of people have contributed in this Thesis in various ways.
First of all I would like to thank my supervisor Prof. Paolo Salucci whose great enthusiasm, knowledge and experience in the work are contagious: Thank you for your availability for daily  important discussions, advices, support and magic power.
You are the best.
And the RCs are definitely not flat.

I would like to thank CAPES for my Brazilian fellowship as well as the High Energy and Astrophysical SISSA faculty members for the unique opportunity to study in SISSA and participating in such a vivid scientific atmosphere.
My warmly thanks to Profs. Marco Fabbrichesi, Petcov and Bilenky.

So nice having met {\it i carissimi} Stefano, Francesco, Christoph, Max, Irina, Lucia, Filippo, Lucia del bar, Luca! Thank you for your help, support and friendship.

I warmly hold my dear friends from {\it saudosas}  Birigui and Londrina, for their presence, faith and kindness.
Thank you Prof. Helayel. 

Special thanks to my Sicilian and his lovely family.
Beautiful days in Trieste with Pietro.

And my hearty thanks to my brother, my mother and my father.

\vspace{1cm}
{\bf Trieste, 03/09/2008}  \hspace{2.5cm} {\bf Christiane Frigerio Martins}

\newpage

\begin{tabular*}{344pt}{l @{\extracolsep\fill} r}
&\\
&\\
&\\
&\\
&\\
&\\
&\\
&\\
&\\
& {\it ``Pois h\'{a} menos peixinhos a nadar no mar} \\
& {\it do que os beijinhos que eu darei na sua boca...''} \\
&\\
& {\it Para meu sicilian\'{i}ssimo futuro esposo Pietro}
\end{tabular*}

\tableofcontents
\mainmatter
\chapter{Introduction}
More than eighty years ago E. Hubble established the expansion of the Universe with his pioneering observations of galaxies.
Since then galaxies have been fundamental tools for understanding the structure and evolution of our Universe.
Today they are crucial laboratories where microphysics phenomena, up to now not detected  by particle physics experiments, emerge with unprecedented clarity.
In particular the great improvement in quality and quantity of the measurements of galaxy kinematics spanning a large range in luminosity, has provided precise tests for evaluating theories both of cosmological and particle physics relevance.

The study of the micro and macro cosmo today produced the $\Lambda$CDM scenario (not yet a theory!) which allows the study of the formation and evolution of cosmic structure from first principles, and embraces cosmological theories (Big Bang and Inflation), particle physics models (the standard model and extensions) and astrophysical models and observations.   

The fact that we need a mysterious new form of matter having a dominant role in structure formation and evolution represents for the first time a demonstration from the cosmological side that the standard model of particle physics needs a deep extension.

In this Chapter I first give a brief introduction to modern cosmology with reference to the latest precision measurements of its most important parameters. 
The first paragraph provides the cosmological basis for the $\Lambda$CDM paradigm.
Then I introduce the theory of structure formation and the 
growth of perturbations in the primordial Universe.
I finally describe the particle physics basis aspects relevant for this paradigm.

\section[Large Scale Structure of the Universe]{Observations of the Large Scale Structure of the Universe}

Within the current $\Lambda$CDM paradigm of structure formation and evolution, cosmology provides the initial and boundary conditions that together with astrophysical models  allow to make definite predictions about the visible Universe.
The systematic comparison between these predictions and the astrophysical observations are fundamental tests of any cosmological model.

Our modern theory of the universe, started with the work of Einstein and Friedman in the 1920s, is based on the Einstein's theory of space-time developed few years before.
It starts from the  assumption of homogeneous and isotropic universe at large scales, described by the Friedmann-Lema\^{i}tre-Robertson-Walker (FLRW) metric:
\begin{equation}
ds^2=dt^2-R^2(t)\left\{ \frac{dr^2}{1-k\,r^2}+r^2\,d\theta^2+r^2\,sin^2\theta\, d\phi^2\right\},
\end{equation}
where $(t,r,\theta,\phi)$ are co-moving coordinates, $R(t)$ is the cosmic scale factor, and $k$ is a curvature parameter which can be chosen to be $+1,0$ or $-1$ for positive, flat or negative curvature respectively. 
For a test particle moving freely in such a metric the geodesic equation reduces to:
\begin{equation}
 \frac{1}{\left| \vec{p} \right|} \cdot \frac{d\left| \vec{p} \right|}{dt}= -\frac{1}{R}\cdot\frac{dR}{dt}, 
\end{equation}
where $\vec{p}$ is the particle momentum.
This equation shows that the relativistic momentum is red-shifted by an amount $z\equiv\frac{R(t_1)}{R(t_0)}-1$ as the scale factor expands.

In the Big Bang model the scale factor evolves over time and its evolution is related to the energy density by the two Friedmann's equations:
\begin{equation}
\frac{\dot{R}}{R}+\frac{k}{R}=\frac{8\pi G}{3}\rho , 
\end{equation}
and 
\begin{equation}
\frac{\ddot{R}}{R}=-\frac{4\pi G}{3}\left(\rho+3p\right) , 
\end{equation}
where $G$ is the gravitational constant and $\rho$ and $p$ are the energy density and pressure of the universe.
In a Newtonian interpretation the first equation is the energy balance in a central force problem, while the second one is the analogous of the Newton law $\vec{F}=m\vec{a}$. 
It is possible to define the critical density of the universe as:
\begin{equation}
\rho_c=\frac{3H^2}{8\pi G} , 
\end{equation}
which corresponds to the density of a flat Universe.

The energy density and pressure in general receive contributions from several kind of sources like photons, baryons, DM and several others:
\begin{equation}
\rho = \rho_{\gamma}+\rho_{b}+\rho_{DM}+ ...  \qquad p = p_{\gamma}+p_{b}+p_{DM}+ ... \;.
\end{equation}
From general thermodynamic reasoning however all these sources respect a general relation between density and pressure: $p=w\rho$, where $w=1/3$ is valid for an ultra-relativistic fluid  (radiation), $w\simeq0$ is valid for non relativistic species (matter) and $w=-1$ is valid for vacuum energy.
This implies that the energy density of radiation scales with the expansion of the universe as $\rho_{\gamma}\propto R^{-4}$, while for the non relativistic components and for vacuum energy we have respectively: $\rho_{CDM}\propto R^{-3}$ and $\rho_{\lambda}\propto R^0$. 
It follows that the early Universe was dominated by the radiation energy density while at later stage it became matter dominated, with estimates of the redshift of the transition epoch  of  $z\simeq 10^4$ when it was about $ t\simeq 5 \cdot 10^4 \mbox{ years} $ old.   
The time of radiation-matter equality is of fundamental importance for the understanding of the formations of the structure of the Universe as it represents the moment when the primordial density fluctuations start to have a significant growth.
In the last stages vacuum energy however dominates.

From the 1970s the FLRW cosmology is rooted in three observational evidences: the expansion of the Universe, discovered in the 1930s by E.Hubble 
observing the recession of galaxies as a function of their distance. 
The second evidence is the Primordial Nucleosynthesis, pioneered in the 1940s by G.Gamow.
Finally the Cosmic Microwave Background (CMB), which is the fossil radiation of the primordial universe discovered in the 1960s by A.A.Penzias and R.W.Wilson 
and today has a temperature of $T_0 \simeq 2.7 \mbox{ K}$.

The Hubble law is described by the equation $ V = H_0 d $, where $V$ is the recession velocity, $d$ is the galaxy distance and $H_0 \simeq 71\mbox{km }\mbox{ s}^{-1}\mbox{ Mpc}^{-1}$ is the Hubble constant.   
The Hubble constant is linked to the scale factor $R$ by a Taylor expansion:
\begin{equation}
\frac{R(t)}{R(t_0)}=1+H_0\left(t-t_0 \right)-\frac{1}{2}q_0 H^2_0\left(t-t_0 \right)^2 +...  
\end{equation}
where $q_0$ is the so-called deceleration parameter linked to the second derivative of the scale factor.

\begin{figure}[t!]
\centering
\includegraphics[width=8cm]{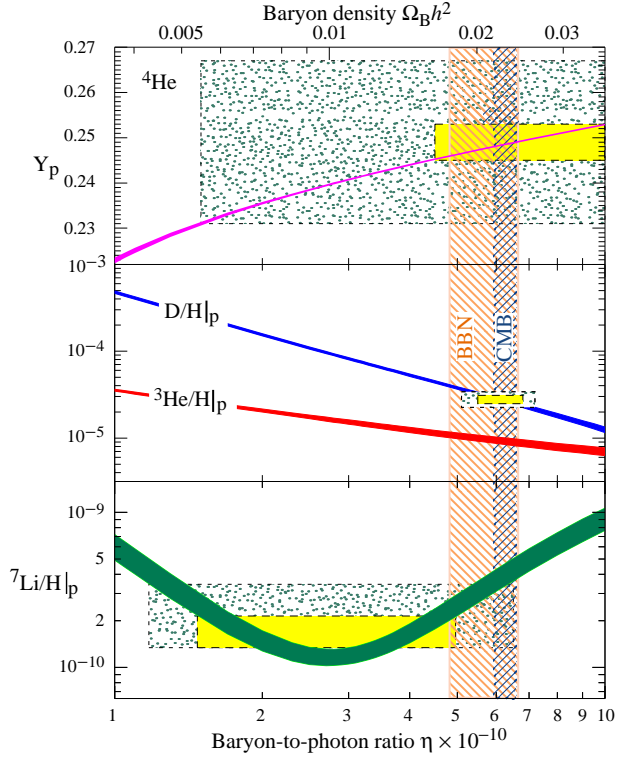}
\caption{Solid lines: predictions of light elements abundances from BBN. Shaded areas: best fit for the fundamental baryon-to-photon parameter. Rectangles with dashed contours: measurements of light elements abundances \cite{pdg}.}
\label{fig:bbn07}
\end{figure}

The Primordial Nucleosynthesis explains the relative abundances of light elements (Hydrogen, Deuterium, Helium-3, Helium-4 and Lithium-7, see Fig. \ref{fig:bbn07}) produced during the first 20 minutes of the Universe.
The prediction depends on one free parameter: the baryon-to-photon ration $\eta=273\cdot10^{-10}\Omega_b h^2$ \cite{olive}, where $\Omega_b$ is the ratio of the baryon density to the critical density and $h$ defined such that $H_0 = 100 h \mbox{km }\mbox{ s}^{-1}\mbox{ Mpc}^{-1} $. 
Measurements give $4.7 \cdot 10^{-10} \le \eta \le 6.5 \cdot 10^{-10}$ \cite{olive}, giving a precise measurement of the baryonic content of the Universe.
Moreover these measurements yield an Helium-4 mass fraction $Y_p\simeq 0.25$.
This mass fraction is of great importance in developing the mass models of gaseous disks in spirals.

Despite the success of the FLRW cosmological model, the Universe is clearly neither homogeneous nor uniform on scales smaller than about $100 \mbox{ Mpc}$.
The modern trend in cosmology, both theoretical and observational, is to try to understand the formation and evolution of the inhomogeneities of cosmological relevance.
What follows is a brief description of the most important observations on the very large scales.

After the discovery by G.Smooth and J.Mather with the COBE mission of fluctuations in the CMB temperature of the order $\delta T / T \simeq 10^{-5}$ (after the subtraction of the variation due to earth motion of order $\delta T / T \simeq 10^{-3}$), these measurements played a major role in the development of the cosmological model and of the $\Lambda$CDM paradigm of structure formation.
The importance relies on the fact that the fluctuations in the CMB temperature reflects the fluctuations of the matter density at the time when the primordial plasma became neutral, at a temperature $T \simeq 3000 \mbox{ K}$ and redshift $z \simeq 1100$:
\begin{equation}
\frac{\delta \rho}{\rho}\simeq const \times \frac{\delta T}{T}, 
\end{equation}
where the constant depends of the kind of matter considered.
Up to the recombination epoch the temperature fluctuations evolve under the influence of sound waves propagating in the hot plasma.

\begin{figure}[h!]
\centering
\includegraphics[width=11cm]{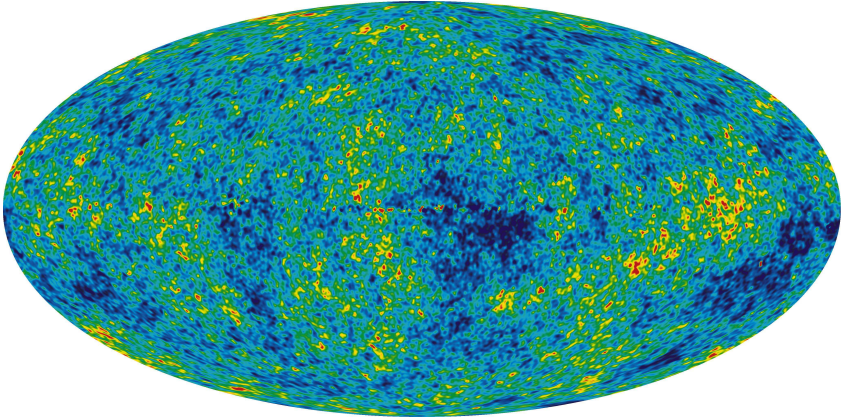}
\caption{The full sky 5-years WMAP image of the CMB temperature fluctuations after dipole subtraction and foreground reduction \cite{wmap5}.} 
\label{fig:WMAP}
\end{figure}

\begin{figure}[h!]
\centering
\includegraphics[width=10cm]{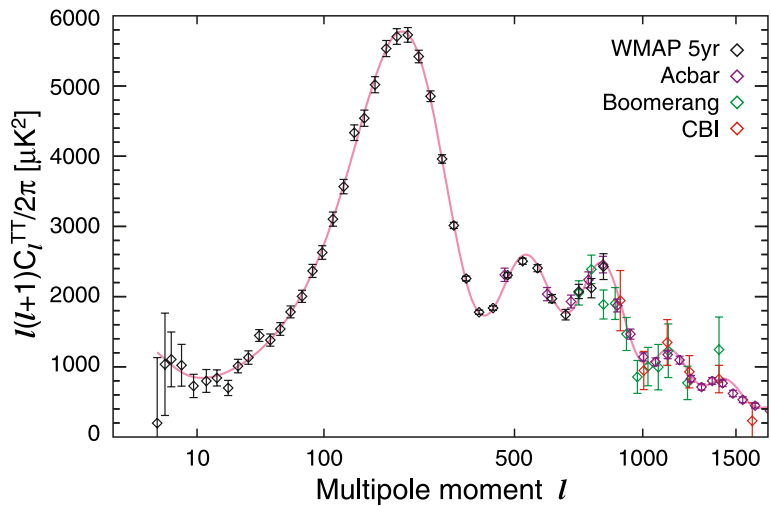}
\caption{The WMAP 5-year power spectrum along with recent results from the
ACBAR, Boomerang, and CBI experiments \cite{wmap5}.
The red curve is the best-fit $\Lambda$CDM model to the WMAP data.}
\label{fig:WMAPspec}
\end{figure}

Fig. \ref{fig:WMAP} shows the more recent CMB measurements from WMAP.
From this figure fluctuations with an angular size of about half a degree are clearly visible.
A quantitative analysis performed on the multipole decomposition of the 2-point correlation function (see Fig. \ref{fig:WMAPspec}) shows that multipoles with $l < 10^2$ corresponds sound waves with periods bigger than the age of the Universe at decoupling.
The multipoles with $10^2 < l < 10^3$ show clearly the oscillations of the sound waves with period short enough to undergo at least one oscillation before the decoupling.
The position of the first peak in this region is sensitive to the flatness of the Universe while the ratio of the height of the even peaks with respect to the odd ones gives a measurement of the ratio between DM and Baryon content of the Universe.
Multipoles with $l > 10^3$ are suppressed due to the fact that the recombination did not happened instantaneously, but the last scattering surface had a finite thickness.  

\begin{figure}[t!]
\centering
\includegraphics[width=9cm]{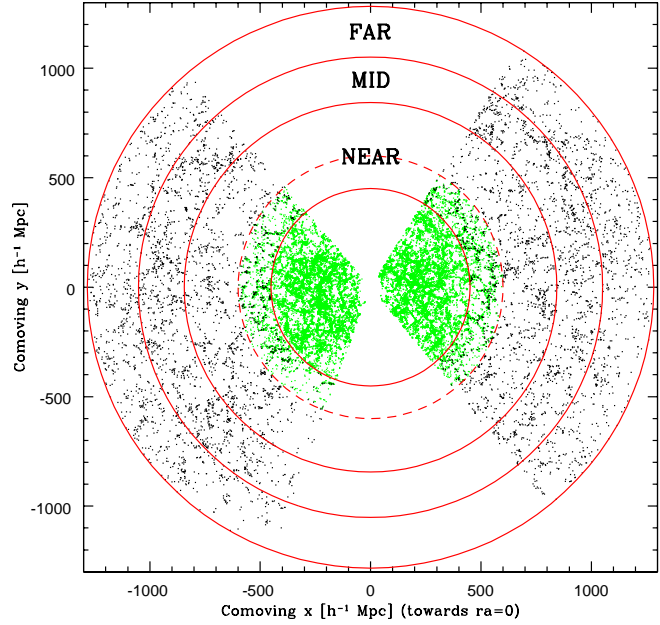}
\caption{Two dimensional distributions of galaxies within $1.25^\circ$ of the Equatorial plane \cite{tegmark06}.}
\label{fig:SDSS}
\end{figure}

\begin{figure}[t!]
\centering
\includegraphics[width=10cm]{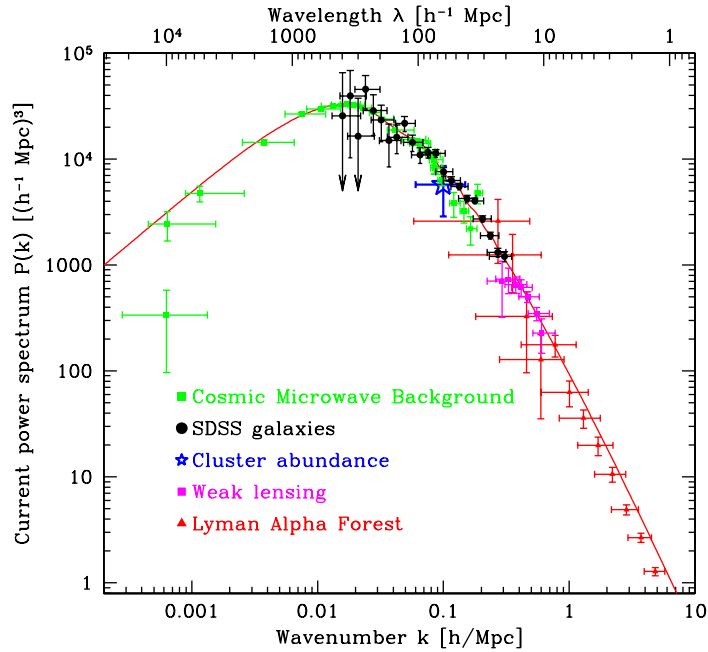}
\caption{Comparison of current power spectrum with observations from CMB, galaxy surveys, cluster, lensing and Ly$\alpha$ forest \cite{tegmark06}.}
\label{fig:all4power}
\end{figure}

The second pillar of the modern observational cosmology relies on the measurements of the galaxy distribution over large portions of the visible Universe (see Fig. \ref{fig:SDSS}).
One of the main challenges of any theory of structure formation and evolution is to explain how the tiny fluctuations in the baryon density measured by the CMB evolved under the influence of gravity up to the stage visible today within the known age of the Universe (this is precisely one of the main success of the $\Lambda$CDM paradigm). 
From the Fourier analysis of the two-point galaxy correlation function (see Fig \ref{fig:all4power}) it has been possible recently to measure the imprint of the primordial sound waves (in this context called Baryon Acoustic Oscillations, BAO) in the visible Universe.
The primordial fluctuations start to oscillate due to the interplay of the pressure of the hot plasma and the attraction of gravity as soon as their size is below the horizon of a given epoch.
At the time of the baryon-photon decoupling the plasma becomes neutral and pressure drops arresting the oscillations and leaving only gravity as dominant force.
The imprints of the primordial oscillations however is still visible in the large scale matter distribution: galaxies in fact are encountered more often in the large overdense regions than in the depleted ones.

The third fundamental observation of modern cosmology is measurement of the distance of Type Ia supernovae which allowed the discovery of the accelerated expansion of the Universe.
 
\begin{figure}[t!]
\centering
\includegraphics[width=10cm]{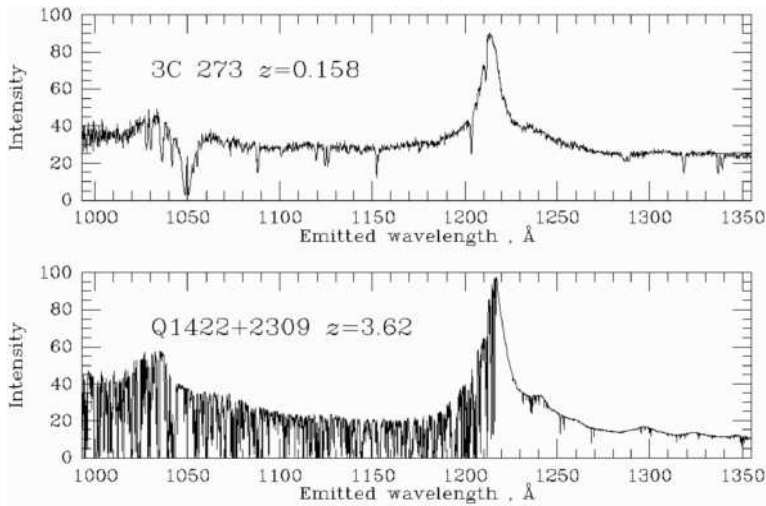}
\caption{Spectra of two Quasars. Top: near one. Bottom: a distant one featuring Lyman alpha absorption.}
\label{fig:Lya-forest-60.eps}
\end{figure}

Other important measurements include the matter power spectrum as measured from Lyman Alpha absorbers and the cosmic shear (cosmological weak lensing). 
The Lyman Alpha forest in particular is the sum of absorption lines arising from the neutral hydrogen Lyman Alpha transitions and is visible in the spectra of distant objects (see \ref{fig:Lya-forest-60.eps}).
These absorption lines are due to clouds of neutral hydrogen which the emitted light encounters while traveling to earth.
Their amplitude and position depend on the matter density as a function of the redshift and hence is a good probe of the matter power spectrum.

The combination of the CMB, the BAO and the SN Ia data are well fitted by the $\Lambda$CDM  cosmological model, whose free parameters together with the best-fit values are shown in table \ref{tab:lcdm}.
The same data imposes important constraints on models with  extended sets of free parameters.
The emerging global picture is a universe with an energy density dominated today by the vacuum energy (for about 72\%).
Important contributions arise also from CDM (about 22\%) and Baryons (about 4.6\%), while for Neutrinos an upper limit of $\Omega_\nu \lesssim 0.026$ is obtained. 

\section{Structure Formation}

The paradigm for structure formation finds its roots in the 
pioneering work by Peebles (see e.g. \cite{peebles70}),
who developed the hierarchical clustering theory.
In this scenario, structure builds up through the aggregation of
nonlinear objects into larger and larger units.
In the current CDM model, the build-up of
structures is governed by the dark dissipationless component, that evolves under gravity from
an initially gaussian distribution of primordial perturbations; small
fluctuations first, and then larger and larger ones, become nonlinear and
collapse when self-gravity dominates their dynamics, to form virialized, gravitationally
bound systems.
As larger perturbations collapse, the smaller objects embedded
in them cluster to form more complex patterns. 
In the meanwhile, the DM provides the potential wells within which the
gas cools and forms galaxies under dissipative collapse.

The development of the proper description of the origin of the structures  (e.g. \cite{kolb}) needs two basic inputs: the initial values of the matter density fluctuations and a proper model for their evolution.

The fluctuations are described by introducing the density contrast:
\begin{equation}
\delta(\vec{x})\equiv \frac{\delta \rho(\vec{x})}{\overline{\rho}}= \frac{\rho(\vec{x})-\overline{\rho} }{\overline{\rho}},
\end{equation}
where $\rho(\vec{x})$ is the local matter density and $\overline{\rho}$ its average, or alternatively, by the Fourier coefficients of the density contrast defined by:
\begin{equation}
\delta_k \equiv V^{-1}\int_{Vol}\delta(\vec{x})exp(i\vec{k} \cdot  \vec{x})d^3 x,
\end{equation}
where $V$ is an appropriate normalization volume, and $k$ is its comoving wavenumber.
Accordingly the physical wavenumber is $k_{phys}=k/R(t)$ and the physical wavelength is then $\lambda_{phys}=R(t)\lambda=2 \pi R(t)/k$.
The density perturbations are also characterized by the mass within a sphere of radius $\lambda/2$ given by:
\begin{equation}
M\equiv \frac{\pi}{6}\lambda^3_{phys}\rho_{m}\simeq 1.5 \times 10^{11} M_\odot (\Omega_m h^2) \lambda^3_{\mbox{Mpc}},
\end{equation}
implying that a galactic mass perturbation corresponds to a scale of a Mpc.

\begin{table}
\centering
\begin{tabular}{|c|c|c|}
\hline
parameter & WMAP + BAO + SN & short description \\
\hline
$100 \Omega_b h^2$ & $2.273 \pm 0.059$ & Hundred times the baryon density \\
$\Omega_c h^2$ & $0.1143 \pm 0.0034$& Cold Dark Matter density \\
$\Omega_\Lambda$ & $0.721 \pm 0.015 $ & Dark Energy density \\
$n_s$ & $0.960^{+0.014}_{-0.013}$ & Scalar Spectral index$^a$\\
$\tau$ & $0.084 \pm 0.016 $  & Reionization optical depth \\
$\Delta^2_\mathcal{R}(k_0^e)$ & $(2.457^{+0.092}_{-0.093}) \times 10^{-9} $ & Amplitude of curvature perturbations$^a$ \\
\hline
$\sigma_8$ & $0.817 \pm 0.026$ & Galaxy fluctuation amplitude \\
$H_0$ & $70.1 \pm 1.3 \mbox{Km/s/Mpc}$ & Hubble constant  \\
$z_{\mbox{reion.}}$ & $10.4 \pm 1.4$ & Redshift of reionization epoch  \\
$t_0$ & $ 13.73 \pm 0.12 \mbox{ Gyr}$ & Age of the universe  \\
$\Omega_b$ & $ 0.0462 \pm 0.015$ & -  \\
$\Omega_c$ & $ 0.233 \pm 0.013 $ & - \\
$\Omega_m h^2$ & $0.1369  \pm 0.0037 $ & Matter density  \\
\hline
\multicolumn{3}{l}{$a)$ estimated at $k_0=0.002/\mbox{Mpc}$}
\end{tabular}
\caption{Summary of the cosmological parameters of the $\Lambda$CDM model and corresponding to 68\% intervals from \cite{wmap5}.}
\label{tab:lcdm}
\end{table}

The primordial fluctuations are generated randomly according to a distribution which is considered as a power law spectrum: $\delta_k\simeq A V k^n$, where A is its characteristic amplitude.
It is useful to introduce the root mean squared density fluctuation as:
\begin{equation}
\frac{\delta \rho}{\rho}=\langle \delta(\vec{x})\delta(\vec{x})\rangle^{1/2},
\end{equation}
which, taking the Fourier transform reduces to: 
\begin{equation}
\left(\frac{\delta \rho}{\rho}\right)^2 = V^{-1}\int_0^\infty \frac{k^3|\delta_k|^2}{2\pi^2} \frac{d k}{k}.
\end{equation}
It is customary to define $P(k)\equiv |\delta_k|^2$ as the power spectrum.

Fluctuations are normally divided in two classes: curvature (or adiabatic) and isocurvature (or isothermal).
The former are authentic fluctuations in the matter density while the latter are fluctuations in the matter composition (e.g. variation in the fraction of baryons) which results in variations in the local equation of state.
The difference between the two types however is relevant only on scales larger than the horizon as on smaller scales microphysics process can transform isothermal in adiabatic fluctuations (and viceversa).
In the following only curvature fluctuations will be considered.

To start the study of the linear description of the perturbation evolutions the simple case of fluctuations in a non expanding universe will be considered first.
This analysis allows the introduction of a fundamental quantity called Jeans Length.
This simple analysis presents however some inconsistencies which can be eliminated in a more complex and rigorous model.

In Eulerian coordinates of a non expanding  Universe the equations describing matter and momentum conservation and the Poisson equations take respectively the following forms: 
\begin{align}
\frac{\partial \rho_1}{\partial t}+  \rho_0 \vec{\nabla} \cdot \vec{v}_1 =0, \\
\frac{\partial \vec{v}_1}{\partial t}+v^2_s\frac{\vec{\nabla}\rho_1}{\rho_0}+\vec{\nabla}\phi=0,\\
\nabla^2 \phi_1=4\pi G \rho_1,
\end{align}
where $\rho$ is the matter density, $p$ and $\vec{v}$ its local pressure and  velocity respectively, and gravitational potential.
The subscript $0$ indicates the homogeneous case (i.e. $\rho_0=const$, $p_0=const$, $\phi_0=const$, $\vec{v}_0=0$) and the subscript $1$ the small perturbations (i.e. $\rho=\rho_0+\rho_1$, $p=p_0+p_1$, $\phi=\phi_0+\phi_1$, $\vec{v}=\vec{v}_0+\vec{v}_1$).
$v_s\equiv\left(\frac{\partial p}{\partial \rho}\right)\simeq \frac{p_1}{\rho_1}$ is the sound speed (in adiabatic conditions).

The equations of the perturbations can be combined in a second order differential equation of the form:
\begin{equation}
\frac{\partial^2 \rho_1}{\partial t^2}-v^2_s \nabla^2 \rho_1=4\pi G \rho_0 \rho_1.
\end{equation}
Assuming solutions of the form 
$
\rho_1 (\vec{r},t)=A e^{\left(-i\vec{k}\cdot\vec{r}+i\omega t\right)}\rho_0,
$
the dispersion relation is obtained: 
$
\omega^2=v^2_s k^2-4\pi G \rho_0
$, 
with $k\equiv |\vec{k}|$.

Defining the critical Jeans wavenumber as:
\begin{equation}
k_j=\left(\frac{4 \pi G \rho_0}{v^2_s}\right)^{1/2},
\end{equation}
it is clear that solutions with wavenumber less than $k_j$ are unstable (either exponentially growing or decaying) while solutions with bigger wavenumbers have oscillatory behavior.

Considering the unperturbed solutions for the matter density, matter velocity and gravitational potential, in an expanding Universe,
a second order differential equation for the Fourier transform of the density contrast is obtained:
\begin{equation}
\ddot{\delta}_k + 2 \frac{\dot{R}}{R}\dot{\delta}_k + \left(\frac{v^2_s k^2}{R^2}- 4 \pi G \rho_0 \right)\delta_k=0.
\end{equation}
In a flat matter-dominated model, the solution of this equation for the unstable ($k<<k_J$) growing ($\delta_{+,k}$) or decaying ($\delta_{-,k}$) mode takes the form:
\begin{equation}
\delta_{+,k}\left(t\right)=\delta_{+,k}\left(t_i\right)\left(\frac{t}{t_i}\right)^{2/3}, \qquad \delta_{-,k}\left(t\right)=\delta_{-,k}\left(t_i\right)\left(\frac{t}{t_i}\right)^{-1} ,
\end{equation}
where $\delta_{+,k}(t_i)$ and $\delta_{-,k}(t_i)$ are the initial values at a chosen reference time $t_i$. 
The exponential evolution obtained in a non expanding Universe becomes a power law evolution in an expanding Universe.
A realistic treatment of the evolution of the perturbations however must consider the dynamics of several fluids, each with a different equation of state.
Moreover the full treatment of the general relativity formalism must be taken into account.

\begin{figure}[t!]
\centering
\vskip -4cm
\includegraphics[width=12.5cm]{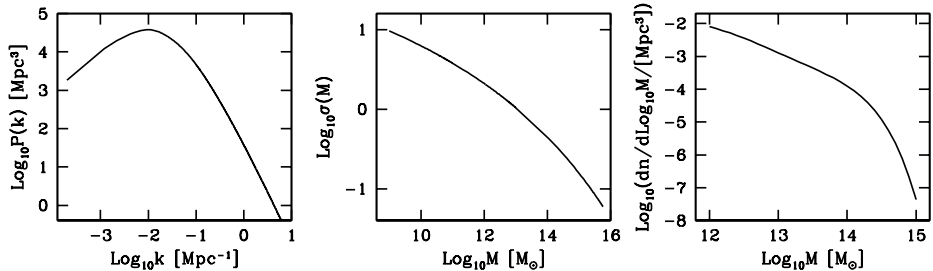}
\vskip -4cm
\caption{From left to right: current linear spectrum, {\it rms} of mass fluctuations and mass function.}
\label{fig:power}
\end{figure}

In the $\Lambda$CDM model the evolution of the linear power spectrum is constructed as:
\begin{equation}
\frac{k^3 P(k,z)}{2\pi^2}= 2.21 \times 10^{-9} \left(\frac{2k^2}{5H^2_0\Omega_{\rm m}}\right)^2  \times {D^2(k,z)} T^2(k)\left(\frac{k}{k_{WMAP}}\right)^{n_s -1},
\label{eq:Delta}
\end{equation}
where $D(k,z)$ and $T(k)$ is the linear growth rate an the matter transfer function (e.g., \cite{bardeen}).
The model with spectral index $n_s\simeq 0.96$ fits the data, indication an almost-free power spectrum\footnote{Inflationary models favour a running spectral index, $n_s(k)=dlnP(k)/dlnk$.}.
Notice that the requirement of hierarchical clustering, that small objects form first, is ensured if $P(k,z)$ is a decreasing function of mass, or correspondingly, an increasing function of the spatial wavenumber $k$.
Using the fitting functions for $D(k,z)$ and $T(k)$ as found in \cite{eisenstein_hu99} I plot in Fig.\ref{fig:power} the current power spectrum. 

The {\it rms} amplitude of mass fluctuations inside a top hat spherically symmetric window of radius $R$ is 
\begin{equation}\label{eq:sigma}
\sigma^2(R,z) = \int^\infty_0 {dk\over k} {k^3 \over 2\pi^2}P(k,z) 
	\left({3\over (kR)^3}(\sin kR - kR\cos kR)\right)^2,
\end{equation}
where $M_R=\frac{4 \pi}{3}\rho_0 R^3$ is the mass enclosed in the window and at the mean density $\rho_0$ of the Universe (see Fig.\ref{fig:power}). 

The mass function can then be defined as
\begin{equation}
f(\sigma,z)\equiv \frac{M}{\rho_0} \frac{dn(M,z)}{d ln \sigma^{-1}},
\end{equation}
where $n(M,z)$ is the abundance of halos with mass less than $M$ at redshift $z$.
In Fig. \ref{fig:power} I plot the mass function at $z=0$ using the formula from the simulations of \cite{jenkins}, $f(M)=0.315 exp (-|ln \sigma^{-1} +0.61|^{3.8})$.

Correcting the linear prediction for the nonlinear dynamics when the density contrast grows above unity (important at small scales) the agreement with the observations is remarkable as shown in Fig. \ref{fig:all4power}. 

The study of hot DM models have been well motivated by the presence of neutrinos in the primordial universe and by the difficulties presented by CDM.
A hot DM species however is characterized by a typical length (called free-streaming length) which is of order of several Mpc.
The hot DM component would dump the fluctuations on scales smaller then the free-streaming, a disfavoured situation. 

\section{Summary of thermal history of the Universe}

A major achievement in the development of cosmology has been to show how the behavior of the Universe on the large scale is dictated in a good part by microphysics (see e.g. \cite{kolb}).
The microphysical laws are incorporated in the study of cosmology and structure formation in a statistical sense by the use the phase space distribution function $f_i(p^\mu,x^\mu)$ of the various species of particles ($i$) considered.

The evolution of the distribution functions is determined by the Boltzmann equation in its general relativistic form:
\begin{equation}
p^\alpha \frac{\partial f}{\partial x^\alpha}-\Gamma^\alpha_{\beta\gamma}p^\beta p^\gamma \frac{\partial f}{\partial x^\alpha}=\mathcal{C}\left[f\right], 
\end{equation}
 where $\Gamma^\alpha_{\beta\gamma}$ is the Christoffel symbol and $\mathcal{C}\left[f\right]$ represent the collision term.

The number density, energy density and pressure of particle specie can be obtained by integrating the distributions (using natural units and the relation $E^2=p^2+m^2$):
\begin{equation}
n(x^\mu)=\frac{g}{(2 \pi )^3}\int f(x^\mu,\overrightarrow{p})d^3p, 
\end{equation}
\begin{equation}
\rho(x^\mu)=\frac{g}{(2 \pi )^3}\int E(\overrightarrow{p})f(x^\mu,\overrightarrow{p})d^3p, 
\end{equation}
and 
\begin{equation}
p(x^\mu)=\frac{g}{(2 \pi )^3}\int \frac{\left| \overrightarrow{p} \right|}{3E}f(x^\mu,\overrightarrow{p})d^3p, 
\end{equation}
where $g$ is the number of internal degrees of freedom of the particle.

In an homogeneous and isotropic universe the distribution function is a function of only energy and time $f(E,t)$, and in the equilibrium condition takes the form:
\begin{equation}
f(E,t)=\frac{1}{e^{\frac{E-\mu}{T}}\pm1}, 
\end{equation}
where the temperature $T$ and the chemical potential $\mu$ are functions of time and the sign is positive for fermions and negative for bosons. 

Particles of specie $\psi$ are kept in thermal equilibrium by interaction processes, which for concreteness can be considered of the form $\psi \psi \leftrightarrow X X$, where $X$ is any kind of final state particle  (the elastic scattering is a simple example of process of this type).
In this case the Boltzmann equation in a FLRW metric reduces to:
\begin{equation}
\frac{dn}{dt}+3H(t)n=-\langle \sigma v \rangle \left( n^2-n^2_{eq}\right)\label{eq:boltz}, 
\end{equation}
where $H(t)\equiv \dot R(t)/R(t)$, $\sigma$ is the cross section of the process considered and $v$ is the velocity of the particle $\psi$, the average been taken over the particle distribution.
It is clear that the particle is kept in thermal equilibrium as far as the term $\langle \sigma v \rangle$ is much bigger that the expansion rate $H(t)$, otherwise it decouples from the thermal bath. 

As in the condition of thermal equilibrium the density and pressure of a non relativistic species (i.e. $T \ll m$, in appropriate units) is much smaller than that of a relativistic one, to a good approximation the two quantities take the form:
\begin{equation}
\rho_R=\frac{\pi^2}{30}g_\ast T^4, 
\end{equation}
and
\begin{equation}
p_R=\rho_R/3=\frac{\pi^2}{90}g_\ast T^4, 
\end{equation}
where $g_\ast$ is the total number of relativistic degrees of freedom:
\begin{equation}
g_\ast=\sum_{i=bosons}g_i\left(\frac{T_i}{T}\right)^4+\frac{7}{8}\sum_{i=fermions}g_i\left(\frac{T_i}{T}\right)^4 \label{eq:gstar}, 
\end{equation}
$g_i$ being the  number of relativistic degrees of freedom of each particle specie and $T_i$ being its temperature (allowing for deviations from the photon temperature).

For temperature $T\gtrsim 300 \mbox{GeV}$ all particles of the Standard Model should be relativistic and a value $g_\ast \sim 100 $ is obtained.
For temperatures $T \simeq 100 \mbox{MeV}$ among the known particles only the electrons, positrons, neutrinos (the tree flavour of them)  and photons remain relativistic and $g_\ast$ drops to about $10$.
As the temperature falls below $0.5 \mbox{MeV}$ however also the electrons and positrons slow down and a value $g_\ast \simeq 3$ is obtained. 

During the radiation-dominated epoch, the use of $\rho_R$ and $p_R$ in the Friedmann's equations yield the following useful relations:
\begin{equation}
H=1.66 g^{1/2}_\ast \frac{T^2}{(8 \pi G)^{1/2}}, 
\end{equation}
and
\begin{equation}
t=0.301 g^{-1/2}_\ast \frac{(8 \pi G)^{1/2}}{T^2}\sim \left( \frac{T}{ \mbox{MeV}}\right)^{-2}\mbox{sec}.  
\end{equation}

The evolution of the temperature with the scale length $R$ can be understood in terms of the conservation of entropy $S$.
For this purpose it useful to define the entropy density as $s=\frac{\rho + p}{T}$.
It follows that whenever $g_\ast$ is constant the result $T \propto R^{-1}$ is obtained.
The number of effective relativistic degrees of freedom for the entropy is defined as:
\begin{equation}
g_{\ast S}=\sum_{i=bosons}g_i\left(\frac{T_i}{T}\right)^3+\frac{7}{8}\sum_{i=fermions}g_i\left(\frac{T_i}{T}\right)^3, 
\end{equation}
with notation similar to Eq. \ref{eq:gstar}.

In brief the thermal history of the primordial plasma is the following: 
in the first phase the space was filled by an almost homogeneous plasma of elementary particles at thermal equilibrium and  at very high temperatures ($10^{-44}\mbox{ s}$, more than $10^{19}\mbox{ GeV}$ characteristic energy).
As the scale factor increases the temperature drops and the plasma undergoes several phase transitions, most notably the inflationary and GUT (Grand Unified Theory, $10^{-38}\mbox{ s}$, $10^{16}\mbox{ GeV}$) ones, then the Electroweak phase transition ($10^{-10}\mbox{ s}$, $10^{2}\mbox{ GeV}$) and  the QCD one (Quantum Chromo Dynamics, $10^{-4}\mbox{ s}$, $10^{-1}\mbox{ GeV}$).  
Among the relics of these phase transitions there are the primordial density fluctuations left from the inflationary epoch and the baryon content of the Universe after the QCD transition.
These eras are followed by the nucleosysnthesis era ($1-200\mbox{ s}$, $1-0.1\mbox{ MeV}$), neutrino decoupling  and electron-positron annihilation ($1\mbox{ min}$, $0.5\mbox{ MeV}$).
Much later the matter and radiation have the same density ($ 10^5\mbox{ yrs}$, $1\mbox{ eV}$), and afterwords the electrons become bound to the nuclei to form atoms ($3 \times 10^5\mbox{ yrs}$, $0.3\mbox{ eV}$).
Then the CMB photons decouple from the plasma traveling freely.
From this epoch on starts the formation via gravitational collapse of visible structures.

The abundance of a particle specie at the decoupling can be estimated by properly manipulating Eq.\ref{eq:boltz}.
For this purpose let us define the two variables $Y=n/s$ and $x=m/T$, with $m$ mass of the particle specie considered and $s$ the entropy density.
From the entropy conservation it follows: 
\begin{equation}
\dot n +3 H n = s \dot Y.  
\end{equation}
Moreover during the radiation dominated epoch the relation between time and temperature obtained above reduces to:
\begin{equation}
t=0.301 g^{-1/2}_\ast \frac{(8 \pi G)^{1/2}}{m^2}x^2.  
\end{equation}
 From  Eq.\ref{eq:boltz} then it follows:
\begin{equation}
\frac{dY}{dt}=\frac{-\langle \sigma v \rangle s }{H x} \left( Y^2-Y^2_{eq}\right)\label{eq:relab}. 
\end{equation}

The exact solution of Eq. \ref{eq:relab} depends on the cross section $\sigma$, which in turn depends by the particle physics model adopted.
However introduction the parametrization:
\begin{equation}
\langle \sigma v \rangle \approx a+6b/x, 
\end{equation}
valid for non relativistic species, the relic density expressed in terms of the critical density assumes a simple form:
\begin{align}
\Omega_{CDM} h^2 & \approx \frac{1.07 \times 10^9 \mbox{ GeV}^{-1}}{(8 \pi G)^{1/2}}\frac{x_F}{\sqrt{g_\ast}}\frac{1}{a+6b/x_F}\\
& \approx \frac{3 \times 10^{-27}\mbox{ cm}^3\mbox{s}^{-1}}{\langle \sigma v \rangle},
\end{align}
where $x_F=m/T_F$ is the $x$ parameter evaluated at freeze-out temperature.
For a particle with  a given mass, the annihilation cross section has an upper bound imposed by the unitarity of the S matrix: $\langle \sigma v \rangle \sim 1 / m^2$.
This limit can be transformed in an upper limit for the DM particle mass by taking the DM abundances from the recent WMAP measurements: $m \lesssim 120 \mbox{ TeV}$.
For more precise estimation of the relic abundances see \cite{masiero} and references therein.

\chapter{Dark Matter Particles}

As discussed in the previous chapter the $\Lambda$CDM paradigm needs a component which behaves like a non-relativistic pressureless dark component.
This component may well be represented by particle candidates which extend the standard model of particle physics at a scale above hundred GeV. 
It is remarkable that such modifications are expected also from a pure theoretical reasoning giving good synergy between astrophysical observations and particle physics.

This chapter discusses the fundamental properties of the hypothetical new particles giving rise to DM.
The possibility of direct or indirect detection is also discussed.
Specific models are hence presented, highlighting the most appealing candidates.

\section{Fundamental properties}

Astrophysical and cosmological measurements provide elements that DM particles were already present in a non-relativistic state in the early Universe.
These observations clearly put constraints on the life-time of the candidate to be $\tau \gtrsim 4.3 \times 10^{17} \mbox{ s}$.
Moreover these particles should interact with the already known particles \emph{at most} weakly, hence excluding charged particles (which would not be dark, if not in very specific models excluded however by experiments) or particles with color quantum numbers (see \cite{masiero} and references therein for a review).

It has been proposed that DM may be subject of self interaction.
This interaction would help in solving the cusp vs core controversy (discussed extensively later in this Thesis) for values of the cross section per unit mass $0.3 \lesssim \sigma/m \lesssim 10^4 \mbox{ cm}^2\mbox{g}^{-1}$.
The recent observation of the merger of two clusters (the so called \emph{Bullet cluster}) however firmly constrains the cross section to $ \sigma/m \lesssim 1 \mbox{ cm}^2\mbox{g}^{-1}$.
Other weaker observations further constrain the allowed self-interaction, making of it a disfavoured hypothesis. 
  
Also the self-annihilation has been proposed as a mechanism to reconcile the cusp vs core controversy, this mechanism however is excluded by both astrophysics (a self-annihilation would produce density cores of same radius for different galaxies, in contradiction with observations) and particle physics measurements.

Another important constrain on the DM properties comes from the models of stellar evolution.
If DM was significantly produced in the interior of sun-like stars (due to the high temperature condition), it would change the energy loss rate of the sun core modifying all the stellar evolution mechanism.
This observation provides strong bounds on CDM candidates based on light particles such as the axion.
Similarly also the BBN measurements provide important constraints on the light candidates properties ($m \lesssim 1 \mbox{ MeV}$) as well as on the decay rates of some heavier particles in some specific models.

Although severe constraints exist on the DM properties, one of the requirements of any realistic model is the correct prediction of its abundance $\Omega_{DM}$.
The various model can be divided in two classes: the one with thermal production (which advocate WIMP candidates) and the other non-thermal models (whose prototype is the axion).
The thermal models are currently more developed and better constrained.
They however require some kind of weak interaction which can be tested in on going experiments or observations.

\section{Direct Searches}

The direct searches are focused in detecting the DM direct interaction with ground based detectors.
The two possible interactions are either with electrons or with nuclei, of the two however only the second have an acceptable sensitivity while the huge background due to natural radioactivity (mainly beta decays) makes the first unfeasible.
Moreover only DM particles with mass sufficiently high can generate a nuclear recoil with detectable energy transfer, making of the  WIMPs the only acceptable candidates for this kind of searches.

The interaction rate on the detector depends on three quantities: the DM flux,  the DM-nucleus cross section and the detector mass.
Assuming a local density of $\rho_{DM} \sim 0.3 \mbox{ GeV cm}^{-3}$ and a mean velocity of the same order of one of the sun around the galactic center ($\overline{v}\sim 220 \mbox{km s}^{-1}$) the expected flux is $\Phi \sim 10^7 (\mbox{GeV}/m_{DM}) \mbox{cm}^{-2}\mbox{s}^{-1}$, where $m_{DM}$ is the DM particle mass expressed in GeV.
Clearly the big astrophysical  uncertainty in the determination of the local density affects directly the detection rate prediction.
Moreover the actual calculation involves not only the mean particle velocity but the full distribution of  velocities, making the prediction even more uncertain.
The revolution of the Earth around the Sun however modulates the mean velocity according to:
\begin{equation}
\overline{v}(t)= 220\mbox{ km/s }\left\{1.05+0.07 cos\left[ 2\pi (t-t_m)\right]/1 \mbox{ year}  \right\}\label{eq:amod},
\end{equation}
where $t_m$ is approximately the begin of June.
This modulation offers an important handle for detecting the DM signal.

Even bigger uncertainties arise on the cross section side.
Not only the cross section depends on the the Particle Physics model under study, but also large uncertainties arise from the theoretical description of the chosen target.
In general two kinds of cross sections are studied: the Spin Independent (SI) and the Spin Dependent (SD).
An important characteristic of the SI coupling is that it is coherently enhanced in nuclei according to:
\begin{equation}
\sigma^{SI}_N \simeq A^2\left( \frac{M_{red}(M_N,M_\chi)}{M_{red}(p,M_\chi)} \right)^2 \sigma^{SI}_p,
\end{equation}
where $A$ is the atomic number, $M_{red}(M_N,M_\chi)$ and  $M_{red}(p,M_\chi)$ denote the reduced mass of the WIMP-Nucleus or WIMP-Proton systems respectively.
The SD coupling normally does not have a similar enhancement, making experiments with heavy nuclei far more sensitive to SI interaction in most of the cases (although particle physics models with SI coupling suppression or SD enhancement exist).

The detection strategy is based in detecting one or more of the following effects generated by DM-nuclei interaction:
\begin{itemize}
\item Ionization: electrons liberated by the atom in primary or secondary interactions 
\item Scintillation: photons emitted by the de-excitation of excited atoms
\item Heat: phonons generated by the displacement of the nucleus with respect to the crystalline structure of the detector 
\end{itemize}
Typically the experiments are sensitive to energies deposited in the detector above the $\mbox{keV}$ magnitude.  
In the last decade several experiments have been run, most often with null results.
What follows is a brief description of the most relevant ones.

\textbf{CDMS - Cryogenic Dark Matter Search} \cite{cdms}:  this experiment, now running deep underground in the Soudan facility, employs crystals of Silicon or Germanium kept at temperatures as low as $10 \mbox{ mK}$.
The detectors, known as ZIP detector and featuring the state of the art thin film superconducting technology, aims at detecting both the phonons and the ionization signals.
The combination of the two signals allows a precise constrain on the background, especially induced by neutrons.  

\textbf{Edelweiss - Experience pour DEtecter Les Wimps} \cite{edelweiss}: as for the CDMS experiment, the Edelweiss detection technique is based on the coincidence of heat and charge detection.
In this case however the heat is measured by very sensitive thermometric sensor glued on the Germanium crystals. 

\textbf{WARP - Wimp ARgon Programme} \cite{warp}: this experiment, located at the Gran Sasso facility, searches for nuclear recoils in liquid Argon with deposited energy in the range $10-100 \mbox{ keV}$ by means of both ionization and scintillation.  
The advantage of this technique over the Silicon or Germanium detectors relies in the capability of the Argon based detectors to be more easily scalable to higher fiductial masses, increasing the sensitivity.

\textbf{XENON Dark Matter Programme} \cite{xenon}: as in the WARP case, this experiment aims at measuring both the charge and the light signal.
Although the technical detection details are different, also the XENON experiment has the advantage of being relatively easy to scale to high fiductial masses.
This experiment has recently published one of the most stringent limits on WIMP particles.

\textbf{CRESST - Cryogenic Rare Event Search with Superconducting Thermometers} \cite{cresst}: in this case the detection is based on the combination of scintillation and phonon detection.
As the active targets are crystals of $\mbox{CaWO}_4$, the SI interaction is enhanced due to the high mass number of tungsten. 

\textbf{DM-TPC - Dark Matter Time Projection Chamber} \cite{dmtpc}: this is a novel detection scheme based on a low pressure gaseous detector.
The experiment should be able to measure a small track of the recoiled nucleus (which should travel few mm), making possible the measurement of the direction of arrival of the WIMP particle and hence providing a powerful tool for the study of the annual modulation of the signal.

\begin{figure}[t]
\centering
\includegraphics[width=12.7cm]{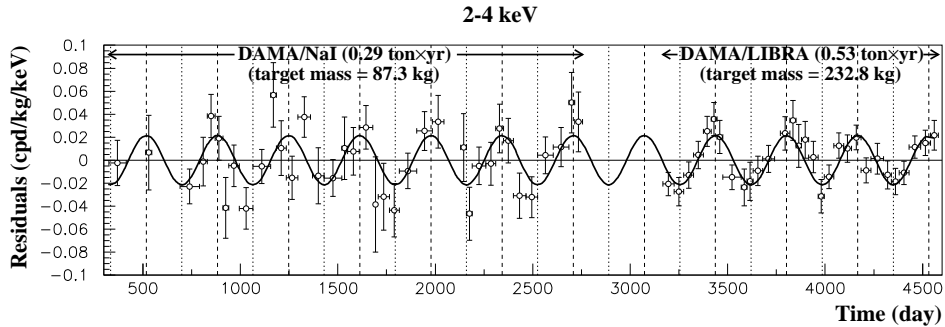}
\caption{Annual modulation of the DAMA signal \cite{dama_exp}: evidence of direct DM Detection?}\label{fig:DAMA}
\end{figure}

\textbf{DAMA} \cite{dama}: this experiment measures the scintillation in NaI crystals.
This highly controversial experiment is the only one having reported a signal detection.
As the experiment is based on only one detection technique, the background suppression and the control of other systematic effects are more difficult, however the advantage is in the capability of lowering the detection energy threshold.
Fig. \ref{fig:DAMA} shows the annual modulation of the detected signal as a function of time for recoil energies between $2$ and $4 \mbox{ keV}$.
It is interesting to note that the amplitude, period and phase of the modulation is actually compatible with Eq. \ref{eq:amod}.
It is puzzling however that other experiments with similar or better sensitivity did not find any signal; it is true however that the comparison between different experiments is somehow model dependent.
The DAMA result, surprising and controversial, will be carefully checked by future experiments and certainly the signal detected by just one group is not sufficient to claim for a discovery. 

\begin{figure}[h!]
\centering
\includegraphics[width=11.5cm]{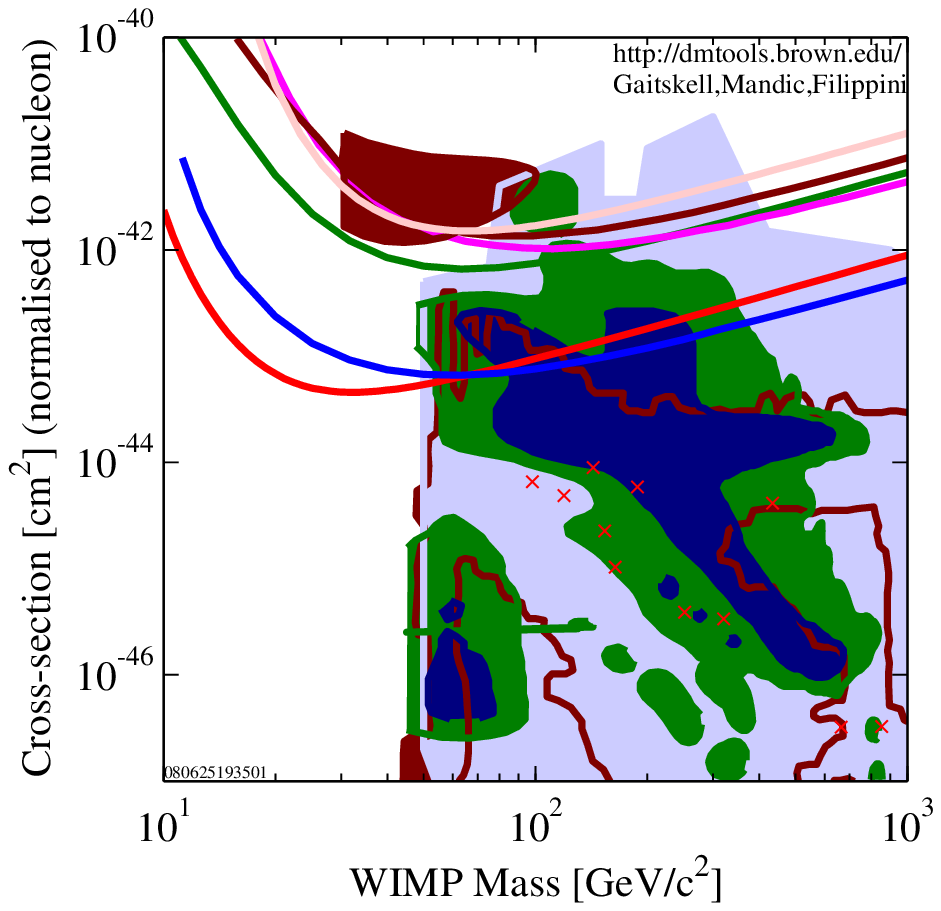}
\includegraphics[width=11.9cm]{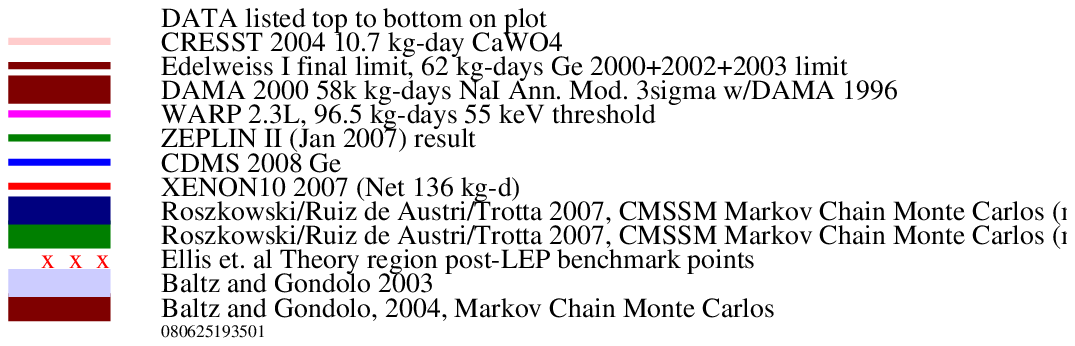}
\caption{Search of SI DM interaction: limits, DAMA signal and theoretical predictions.}\label{fig:DMSlim}
\end{figure}

The results of the most sensitive experiments, together with the DAMA signal and some theoretical predictions are shown in Fig. \ref{fig:DMSlim}.
Clearly most of the models predict a cross section several order of magnitude below the current experimental sensitivity, however future improvements in the fiductial mass of the experiments together with longer time exposure will hopefully improve the situation.
The range of masses and cross sections allowed by the DAMA signal is model-dependent, however the reconciliation with the other experiments is non-trivial.

\section{Indirect Searches}

The Indirect searches of DM are based on astrophysical observations of the products of DM self-annihilation or decay.
Given the known long lifetime of the DM the signal for decay products is suppressed for heavy candidates (due to the combination of low number densities and long lifetime) leaving only the self-annihilation as most sensitive possible source of a signal.

In the case of searches via gamma ray observation, the expected flux in a detector on Earth is given by:
\begin{equation}
\frac{d \Phi_\gamma}{d E_\gamma}\left(E_\gamma,\Delta \psi\right)=\frac{\langle \sigma v\rangle_{ann}}{4 \pi m_\chi^2}\sum_f \mathcal{B}_f \frac{d N^f_\gamma}{d E_\gamma}\times \frac{1}{2}\int_{\Delta \psi}\frac{d \Omega}{\Delta \psi}\int_{l.o.s.}dl\left(\psi\right)\rho^2\left(\mathrm{r}\right),
\end{equation}
where $E_\gamma$ is the photon energy, $m_\chi$ is the DM particle mass, $\Delta \psi$ is the detector opening angle, $\langle \sigma v\rangle_{ann}$ is the mean annihilation cross section times the relative velocity (of order $10^{-26} \mbox{ cm}^3\mbox{s}^{-1}$  for cold WIMP relics from abundances constraints), $\mathcal{B}_f$ indicates the branching fraction in a given channel $f$,  $\frac{d N^f_\gamma}{d E_\gamma}$ is the photon spectrum for a given annihilation channel which depends on the DM model and can have both continuum and discrete lines contributions, $\rho$ is the DM density and the integrals are along the line of sight and over the detector opening angle.   

The Quadratic dependence on $\rho$ suggest that the preferred targets for indirect searches are the places with higher DM concentrations, like the centre of galaxies or galaxy clusters.
It must be noticed however that the galactic centres are very often sources of strong activities due for example of the presence of a Black Holes or other compact objects enhancing the overall background.
Moreover the large uncertainty on the DM density reflects directly on the flux predictions making the searches extremely difficult (although possible enhancements due to local DM over-densities are possible).

Another possibility is to pursue indirect search by looking at charged particles such as positrons or antiprotons, in this case however the galactic magnetic is such that the direction of arrival of the particle does not reflect the production point and the only observable if an excess of antimatter with respect to the expected background due to ordinary cosmic rays (which also suffer from big uncertainties). 

What follows is a brief partial review of the most important facilities looking for indirect signals:

\textbf{XMM Newton and Chandra} \cite{xmm,chandra}: these are two satellites operated by the European Space Agency (ESA) and by the National Aeronautics and Space Administration (NASA) respectively.
They have both imaging and spectroscopic capabilities in a photon energy window between $0.1 \mbox{ keV}$ and $10 \mbox{ keV}$ approximately.
The searched signal is a narrow line not explicable in terms of weak known physics processes and originating either from DM decay or self annihilation.
Their observations put important limits on the medium mass DM candidates (such as the sterile neutrinos).

\textbf{Integral} \cite{integral}: this ESA observatory operating a window of energy of gamma rays between  $15 \mbox{ keV}$ and $10 \mbox{ MeV}$ approximately is complemented by optical instrumentation.
This observatory may detect a signal of DM as a new narrow line, as an excess of $511 \mbox{ keV}$ photons due to positrons annihilation.
The mission has actually published a claim of possible DM detection discussed below.

\textbf{Compton Gamma Ray Observatory - CGRO} \cite{cgro}: this observatory, together with Hubble and Spitzer, is one of the most important research projects of NASA.
Two instruments on board made important DM searches: COMPTEL (Imaging Compton Telescope) operating in an energy range of  $0.75 \mbox{ MeV}$ and $30 \mbox{ MeV}$ and EGRET (Energetic Gamma Ray Experiment Telescope) operating in the window  $20 \mbox{ MeV } - 30 \mbox{ GeV}$.
The EGRET telescope in particular provided important limits both on the DM properties and possible signals of detection discussed below.

\textbf{AGILE and GLAST} \cite{agile,glast}: these are the two recently lunched gamma ray observatories.
The first operates in the energy windows $30 \mbox{ MeV} - 50 \mbox{ GeV}$ and $10 - 40  \mbox{ keV}$, allowing both gamma and X-ray measurements.
The second one has full coverage of the window $10 \mbox{ keV} - 300 \mbox{ GeV}$ with both large opening angle and excellent sensitivity.
GLAST, with its unprecedented sensitivity, is certainly the best observatory to look for indirect searches of WIMP annihilations.

\textbf{CANGAROO, HESS, MAGIC and VERITAS} \cite{cangaroo,hess,magic,veritas}: these are ground based facilities observing Ultra High Energy gamma rays with energies above $100 \mbox{ GeV}$ approximately.
The detection technique is based on the measurements of the Cherenkov light emitted by electromagnetic showers in the upper atmosphere. 
Constraints (not very rigid yet) on self annihilating WIMP particles in nearby halos have been provided.

\textbf{AMANDA, ICECUBE and ANTARES} \cite{amanda,icecube,antares}: these are High Energy Neutrino observatories.
The indirect search of DM with this detector is based on the assumption that a high density of WIMP particles  would accumulate at the core of the Sun or of the Earth due to a combination of the elastic scattering of the particles with the Sun or Earth material followed by a gravitational capture.
The local high density of DM would enhance the self-annihilation which may proceed through a channel which includes neutrinos in the final state.
This neutrinos would then easily escape from the Sun or Earth core allowing the detection on the earth surface.
No signal has been found so far.

\textbf{PAMELA and AMS} \cite{pamela,ams}: these are two satellites which aims at measuring the spectra and properties of primary  cosmic rays in the $ \mbox{ GeV }$ region.
The presence of an unexplained excess of antimatter (either antiprotons or positrons) in the primary cosmic rays can be interpreted as a signal of self-annihilating DM.
It has to be mentioned that few years ago the HEAT experiment found evidence of an excess of positrons with energies of about $7 \mbox{ GeV}$ and the signal has been confirmed later by AMS-I.
The interpretation of the signal as coming from DM annihilations however is problematic due to the much lower excess predicted by the WIMP models.
However new data from PAMELA should clear the uncertainty soon.

Up today several claims of indirect DM detection has been made, sometimes in conflict with each other or with other measurements.
The most significant are however: the positron excess measured by HEAT, the $511 \mbox{ keV}$ line excess measured by INTEGRAL, the EGRET Diffuse Galactic Spectrum, the EGRET Diffuse Extragalactic Spectrum and the so called \emph{WMAP Haze} (an excess of microwave emission around the center of the Milky Way).

The INTEGRAL signal is many order of magnitude above the expected signal from secondary positrons due to cosmic rays and is approximately spherically symmetric with a full width half maximum of about $6^\circ$.
Astrophysical interpretations of the signal are difficult and several interpretations due to indirect DM detection have been proposed.

\begin{figure}[h!]
\centering
\includegraphics[width=5.2cm]{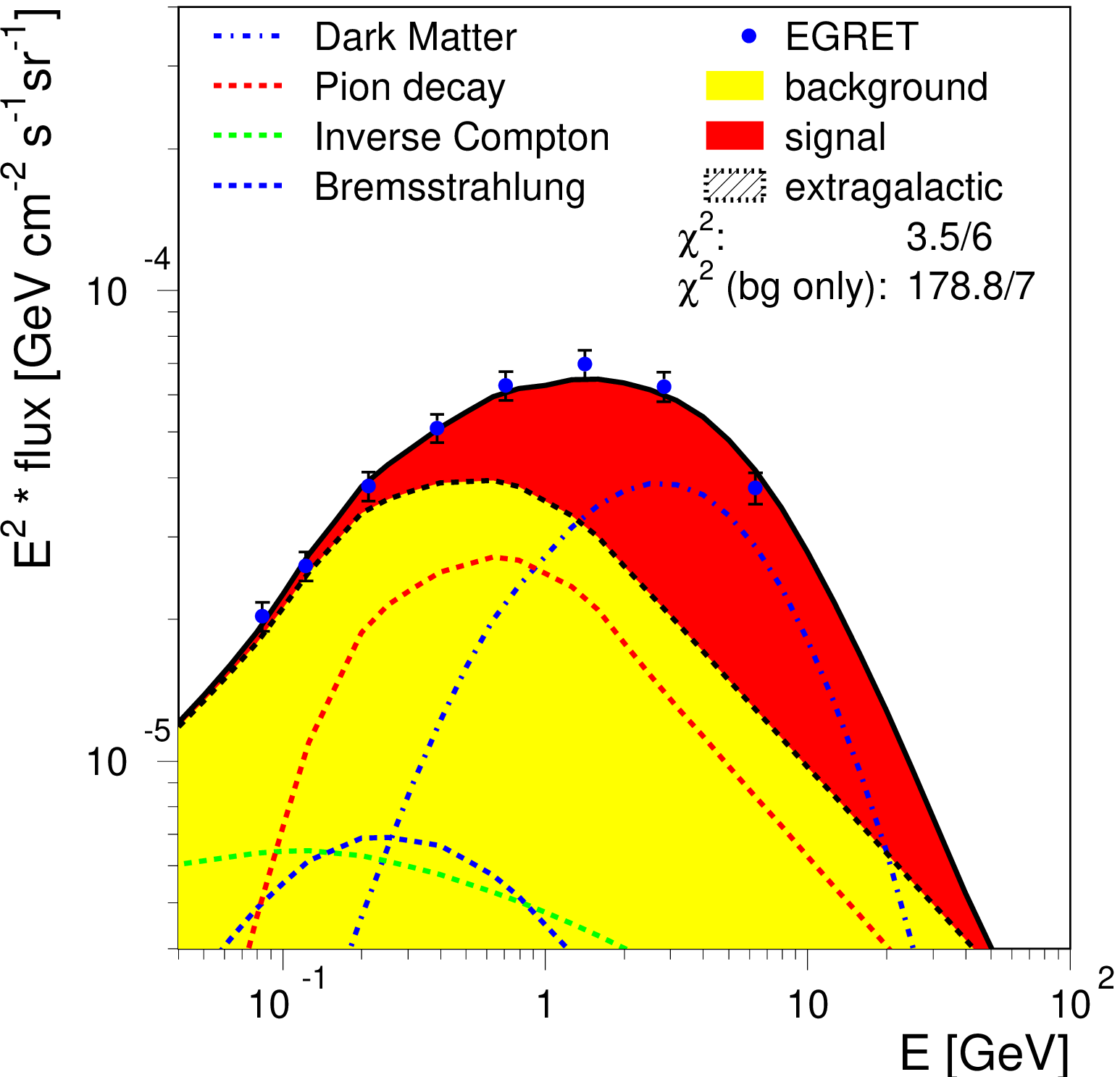}
\includegraphics[width=6.45cm]{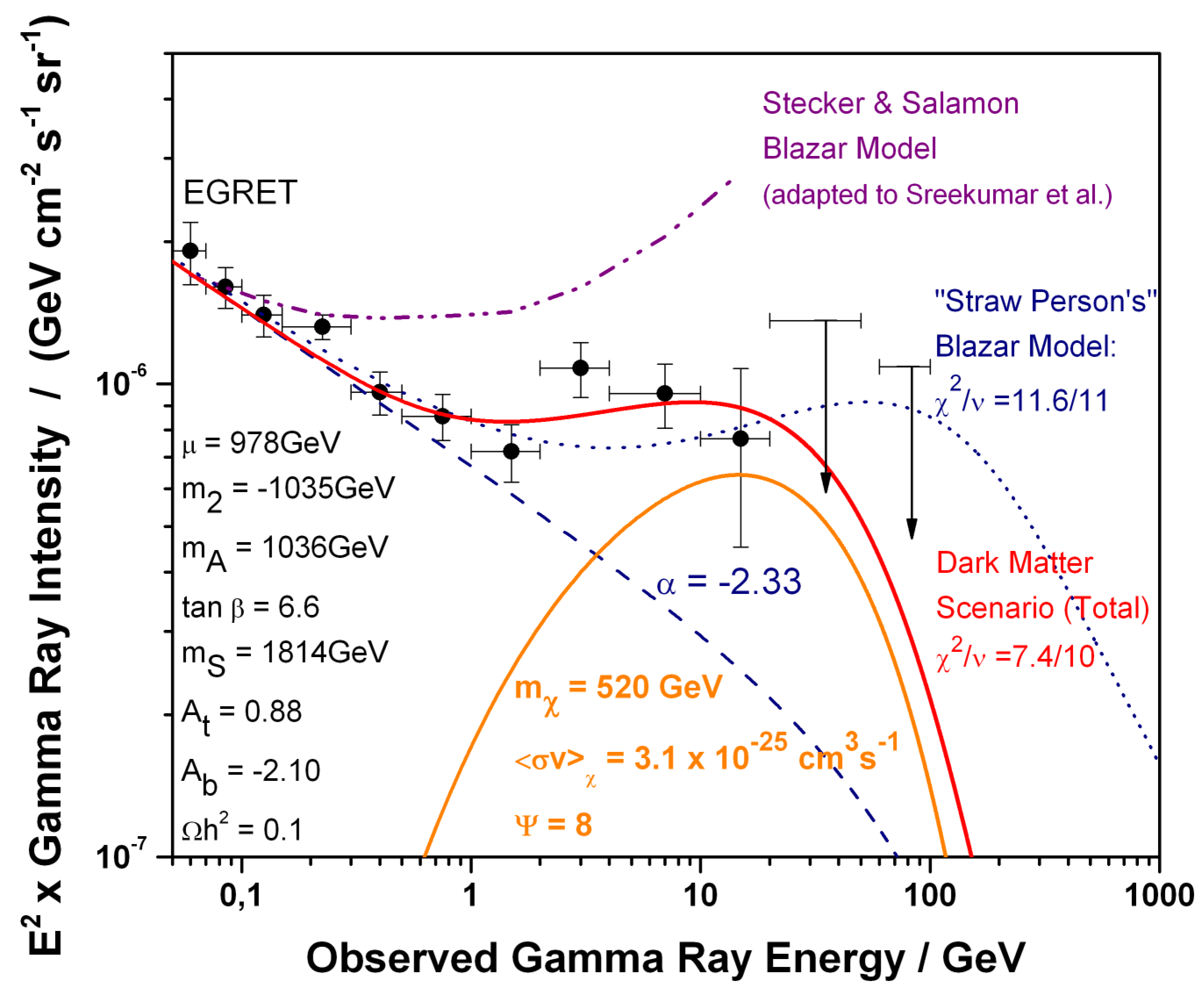}
\caption{On the left: EGRET measurement of the Diffuse Galactic Spectrum on a portion of the sky as evidence of DM (see \cite{hooper07} and references therein). On the right: EGRET measurement of the Diffuse Extragalactic Spectrum as  evidence of DM (with different characteristics from the left panel).}\label{fig:EGRET}
\end{figure}

The EGRET Galactic and Extragalactic Spectra are shown in Fig. \ref{fig:EGRET}.
The Galactic measurements show an excess of photons with energies in the range $1 - 10 \mbox{ GeV}$ approximately.
The interpretation of this signal as DM detection however is questionable due to the mismatch between the knowledge of the halo density distribution and the directional variation of the signal, moreover a large amount of secondary antiprotons would be expected in contrast with observations.
The Extragalactic excess measured by EGRET with energies above $10 \mbox{ GeV}$ can also be interpreted as a DM signal.
In order for this interpretation to be valid however the DM halos have to be very cuspy for most of the galaxies BUT far less cusped for the Milky Way, a rather odd situation indeed (beside being in conflict with other observations). 
 
The last of the above mentioned claims of indirect detection, the WMAP Haze, has been proposed to arise from synchrotron radiation emitted by relativistic positrons or electrons generated by DM annihilations.
If this is correct however an associated prompt gamma ray emission should be within of the recently lunched GLAST experiment.

To conclude this section a final remark is necessary: several claims have already been proposed as indirect DM detection, all of them however competing with other Astrophysical explanations.
It seems clear then that a convincing evidence of indirect detection must come from several complementary measurements (at different energies or with different particles) although all the claims have to be properly studied and possibly eliminated.

\section{The zoo of candidates}

Since the early years following the discovery of DM halos surrounding galaxies particle physicists tried to develop models which included a DM candidate.
Today almost every proposed extension of the Standard Model of Particle Physics (SM in the following) in a way or another includes a candidate.
Most often these candidates are WIMPs produced thermally in the early Universe, and in this case the long lifetime is ensured by including in the model a symmetry which forbids the DM decay.
Other models with non-thermal production however also play an important role in the discussion of the extensions of the SM.

What follows is a brief discussion of the two most attractive candidates:  the supersymmetric models and axions.
The list of other proposed candidates is however very long and includes:  sterile neutrinos, minimal DM models, Little Higgs models, Kaluza-Klein particles, wimpzillas, CHArged Massive Particles (CHAMPs), brane-world DM and many others.

\textbf{Supersymmetric candidates}: In the SM bosons and fermions play two different roles: the bosons act as mediators of fundamental interactions while the fermions are the elementary constituents of matter.
It is natural therefore to ask whether a symmetry exists between these two particle classes providing a sort of unified picture.
This boson-fermion symmetry is called SUper Symmetry (SUSY).

While a review of the SUSY theory is beyond the scope of this Thesis, the discussion here will be concentrated on the most common SUSY DM candidate (the neutralino), referring the reader to some excellent reviews and didactic materials available in the literature (see e.g. \cite{olive08} and references therein).

The benefits of the SUSY models include not only a suitable DM candidates, but help in solving the so called ``hierarchy problem'' (the difference between the electroweak and the Planck scales) and provide a mechanism for the unification of the gauge coupling of the SM at a Grand Unification Scale.

In the Minimal Supersymmetric Standard Model (MSSM) each boson (gluons, $\mbox{W}^\pm$ and $\mbox{B}$) is associated with a fermion (the gluinos, winos and binos), the quarks and leptons are associated to scalars called squarks and sleptons and the Higgs sector is composed by two Higgs doublets associated with spin $1/2$ higgsinos.
Another ingredient of the MSSM is a discrete symmetry called $R$-parity where to each particle is associated a conserved quantum number defined as $R \equiv (-1)^{3B+L+2s}$, where $B$ and $L$ are the baryonic and leptonic number respectively, while $s$ is the particle spin.
Clearly all the SM particles have $R=1$ while the SUSY partners have $R=-1$ and the $R$-parity conservation implies the stability of the lightest SUSY partner.

A consequence of SUSY is that the mass of each particle must be equal to the mass of its super-partner, otherwise SUSY is broken.
Clearly the mass degeneracy predicted by SUSY is not observed in Nature and several SUSY-breaking mechanisms have been proposed.
It is evident that the original SUSY idea led to the formulation of a model with some attractive feature (like grand-unification of the couplings) at the price of an enormous increase of the number of free parameters.
It is common however to try to reduce the number of free parameters by introducing some kind of additional condition like the unification of the gaugino masses at GUT scales or some universality of the couplings.
The most common resulting models are the Constrained MSSM (CMSSM) or the minimal Super Gravity (mSUGRA).

\begin{figure}[t!]
\centering
\includegraphics[width=6cm]{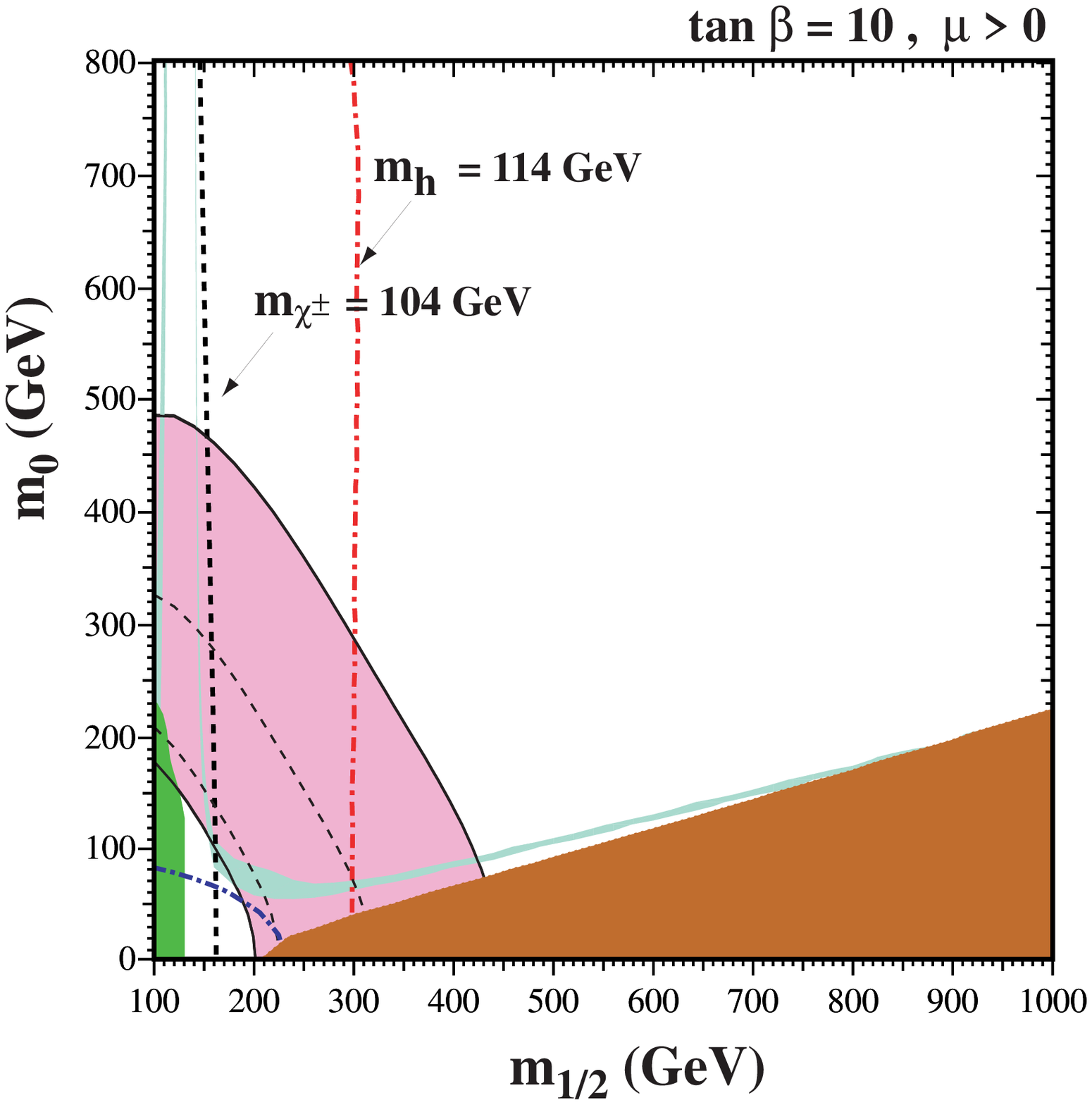}
\includegraphics[width=6cm]{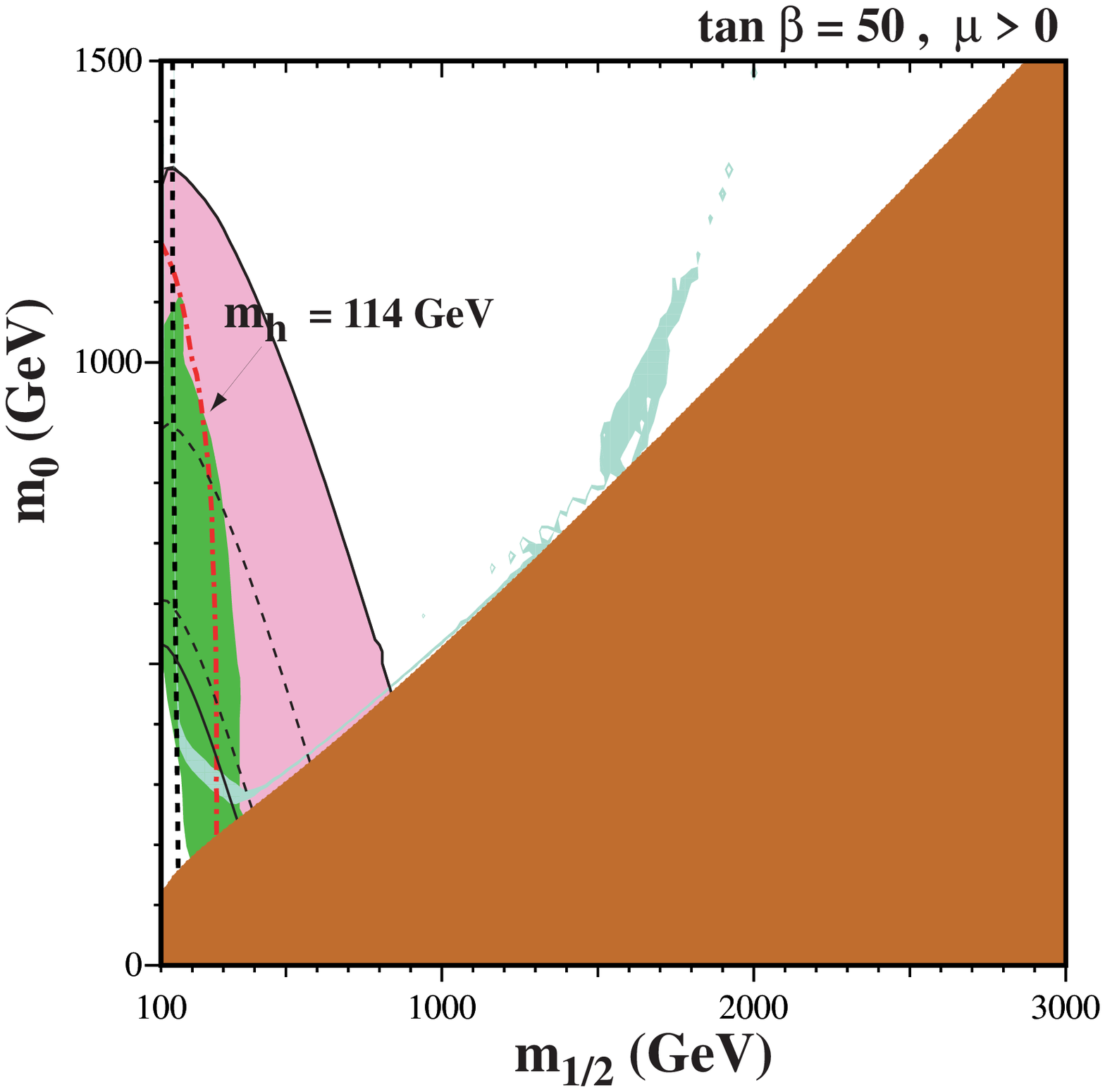}
\caption{On the left: the $(m_{1/2}, m_0)$ planes for $\tan \beta = 10$ and $\mu > 0$, assuming $A_0 = 0, m_t = 175$~GeV and $m_b(m_b)^{\overline {MS}}_{SM} = 4.25$~GeV. 
The near-vertical (red) dot-dashed lines are the contours $m_h = 114$~GeV, and the near-vertical (black) dashed line is the contour $m_{\chi^\pm} = 104$~GeV. 
Also shown by the dot-dashed curve in the lower left is the corner excluded by the LEP bound of $m_{\tilde e} > 99$ GeV. 
The medium (dark green) shaded region is excluded by $b \to s \gamma$, and the light (turquoise) shaded area is the cosmologically preferred region. 
In the dark (brick red) shaded region, the LSP is the charged ${\tilde \tau}_1$.
The region allowed by the E821 measurement of $a_\mu$ at the 2-$\sigma$ level, is shaded (pink) and bounded by solid black lines, with dashed lines indicating the 1-$\sigma$ ranges. 
On the right: $\tan \beta= 50$ \cite{olive08}.}
\label{fig:cmssm}
\end{figure}

As the MSSM (with its variants) received a lot of attention in recent years both from the theoretical and experimental community important constraints exist on the parameter space.
Fig. \ref{fig:cmssm} shows for the CMSSM in the parameter space $(m_0 ; m_{1/2})$ (universal sfermion mass in the vertical axis and universal gaugino mass in the horizontal one) for two definite values of the other CMSSM parameters the region allowed by the WMAP measurements (turquoise) together with bounds from accelerator measurements or other observations.
An immediate observation which can be drawn from these diagrams is that the parameter space compatible with the cosmological bounds is well constrained and an important portion of this region is not compatible with the other measurements.
Clearly all the bounds can be relaxed in a less constrained MSSM, in this case however the huge number of free parameters makes the model less attractive. 

\textbf{Axion}: One of the open problems of the SM is that the gauge theory responsible for the strong interaction foresees the possibility of a strong CP violation (see \cite{raffelt07} and references therein).
This strong CP violation however is not observed in Nature, hence the so called strong CP problem arises.
One of the possibility for its solution is that Nature respects a symmetry, called Peccei-Quinn symmetry, which allows the restoration of the CP conservation in the strong sector.
The PQ symmetry however should be spontaneously broken giving rise to a new Nambu-Goldstone called Axion.

The Axion has a specific property of being coupled to two photons as:
\begin{equation}
{\cal L}_{a\gamma}= -g_{a\gamma}\,{\bf E}\cdot{\bf B}\,a\, ,
\end{equation}
where  ${\bf E}$ and ${\bf B}$ are the electric and magnetic fields respectively, $a$ is the axion field and the coupling constant $g_{a\gamma}$ is related to more fundamental parameters of the theory such as the axion mass: $g_{a\gamma} \propto m_a$.

\begin{figure}[h!]
\centering
\includegraphics[width=8cm]{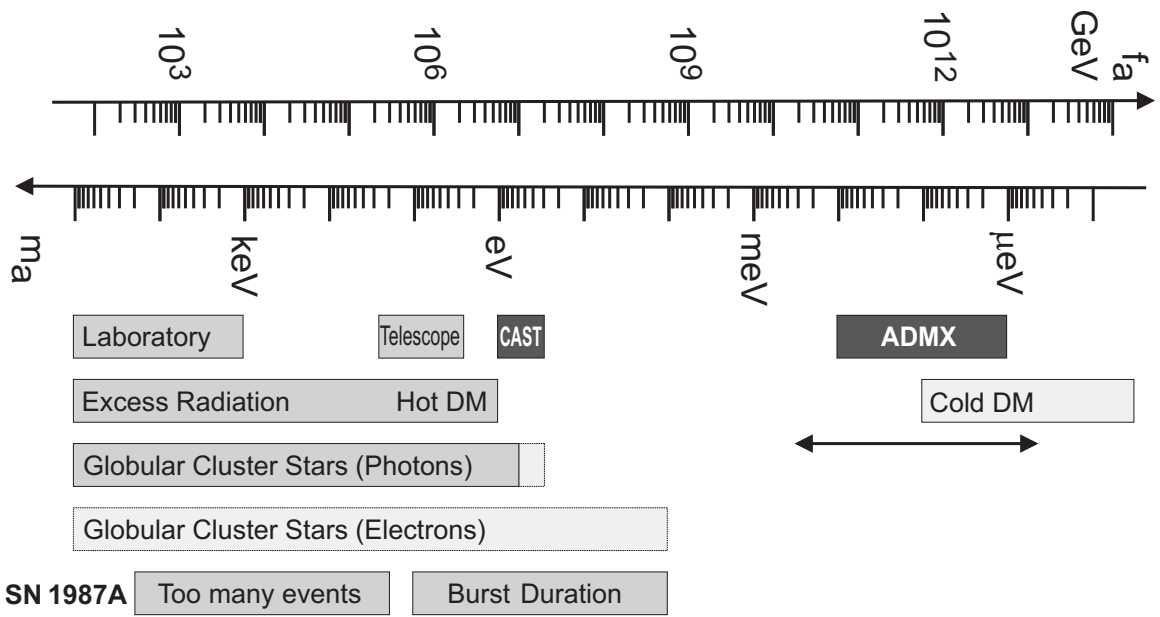}
\caption{Axion limits and foreseen search ranges \cite{raffelt07}.}
\label{fig:axion}
\end{figure}

The coupling plays a fundamental role in the Axion searches.
Axions in fact can transform into photons when propagating in an external magnetic field in a way similar to neutrino oscillations.
As visible in Fig. \ref{fig:axion}, both astrophysical and laboratory measurements impose strong constraints on the properties of a hypothetical axion with mass as low as about $1\mbox{ eV}$. 
From the current DM search point of view the on-going Axion DM eXperiment (ADMX) is searching for a signal in a mass region close to the $\mu\mbox{eV}$ scale.
An axion with a mass around this scale in fact is a theoretically well motivated cold DM candidate.

While axions with a mass above the $\mbox{eV}$ scale would be produced thermally in the early universe (and hence would be a hot DM candidate similar to the neutrinos) for masses lower than the $\mu\mbox{eV}$ the production would be non-thermal and linked to the Peccei-Quinn phase transition by the so-called misalignment mechanism.
The relic axion density can then be calculated according to:
 \begin{equation}
 \Omega_ah^2\approx
 0.7\,\left(\frac{f_a}{10^{12}~{\rm GeV}}\right)^{7/6}\,
 \left(\frac{\Theta_{\rm i}}{\pi}\right)^2\,,
\end{equation}
where $-\pi\leq\Theta_{\rm i}\leq\pi$ is the initial ``misalignment angle'' relative to the CP-conserving position and $f_a\propto g^{-1}_{a\gamma}$ is the Peccei-Quinn scale.

In this case an axion with $m_a\approx  10 \mu\mbox{eV}$ would provide a CDM density in agreement with WMAP measurements, however, this number sets only a crude scale of the expected mass for axion DM, with uncertainties coming both from the particle physics and the cosmological models.
It has to be mentioned that in the non thermal axion production mechanism the effective temperature today is of order $10^{-34}\left( \frac{10^{-5}\mbox{eV}}{m_a}\right)^{2/3}\mbox{K}$: an extraordinary low temperature!

\section{Concluding remarks}

The dedicated searches of the DM particle candidate have seen an important boost in recent years with relevant and costly experiments been planned and executed.
The success of these searches however crucially depends on our capability of predicting the signal expected for each particle model.
This capability in turn relies on our knowledge of the DM distribution in bound objects.
The mass distribution in galaxies is then the theme of the next chapter.   

\chapter{The mass distribution in Spiral Galaxies}

The presence of large amounts of unseen matter in spiral galaxies with a distribution different from that of stars and gas is well established.
The primary observational evidence for the existence of DM, under the assumption of Newtonian gravity, comes from optical and 21 cm RCs of spirals which do not show the expected Keplerian drop-off at large radii but remain increasing, flat or gently decreasing over their entire observed range \cite{faber79,bosma81,rubin80,rubin85}.
The invisible mass component becomes progressively more abundant at outer radii and for the less luminous galaxies \cite{PS88,broeils92}.
The distribution of matter in disk systems has become a benchmark for the present understanding of the process of galaxy formation and, with the help of the available observational tools, crucial questions can be addressed:
\begin{itemize}
\item has the dark matter an universal distribution reflecting its very Nature?
\item how and why the dark-to-luminous mass ratio and other physical quantities vary in objects of different Hubble type?
\item how dark matter affects the fate of the universe?
\end{itemize}

It is well known that numerical simulations performed in the $\Lambda$CDM scenario predict a well-defined density profile for the virialized halos surrounding and hosting the galaxies.
This profile leads to structural properties of galaxies \cite{nfw96} that are in strong disagreement with observations.
Moreover the mechanism of galaxy formation, as currently understood, involves the cooling and the condensation of HI gas inside the gravitational potential well of DM halos.
Part of the condensed gas then transforms into stars which reheat the former by the feedback of  SN explosions.
It is clear then that a mistake in the model of the halo potential has a deep impact on the complex dynamics of stars and gas.
\begin{figure} [h!] 
\centering
\includegraphics[width=6cm]{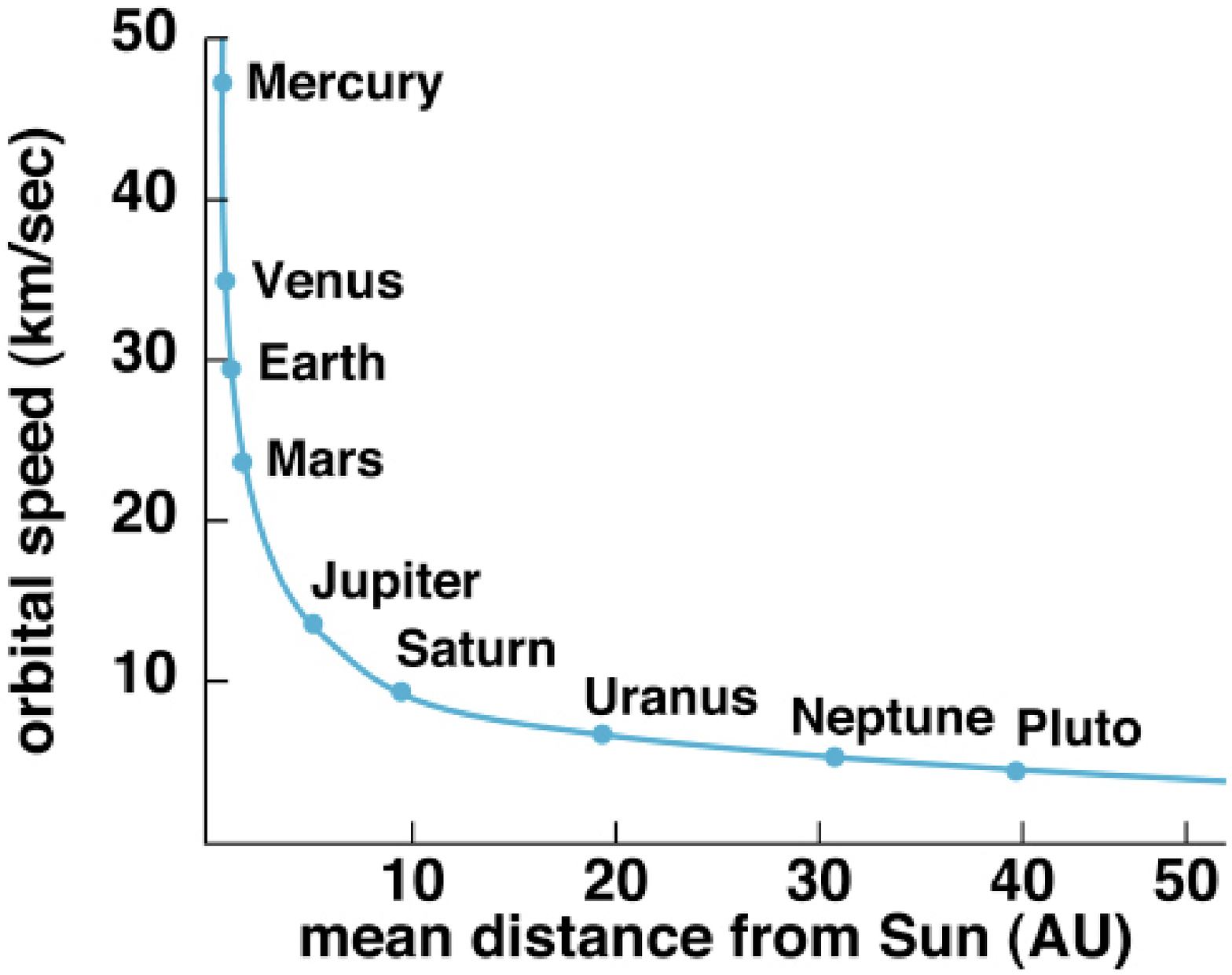}
\includegraphics[width=5.6cm]{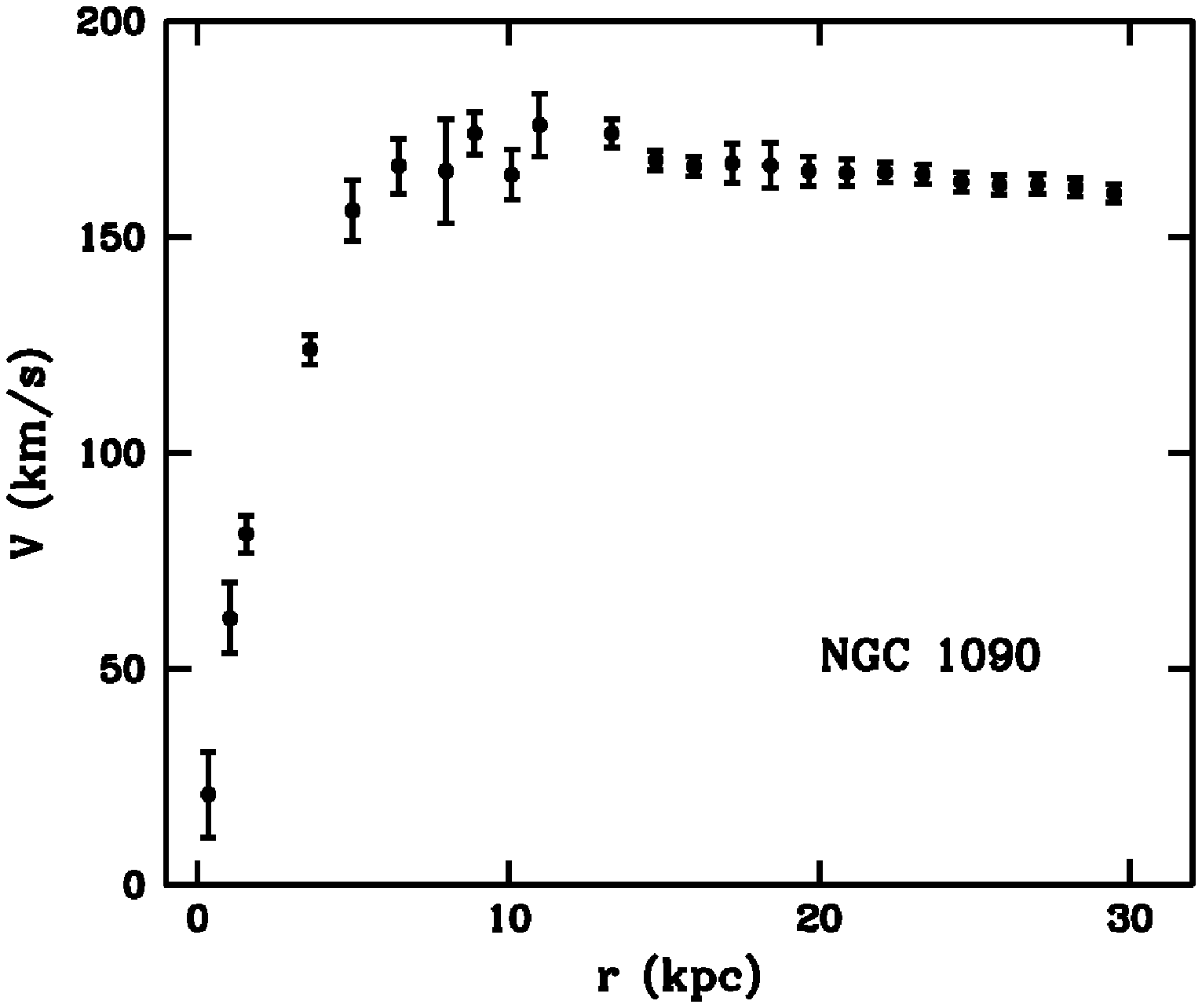}
\vskip -0cm
\caption{On the left orbital velocity of the planets of the solar system as a function from the distance from the Sun ({\it Copyright \copyright Addison Wesley}).
This picture shows the typical Keplerian falloff planetary systems as well as demonstrating that on these scales  the gravitational potential is dominated by the Sun mass with negligible contribution from DM.
On the right circular velocity of the NGC 1090 disk.
It clearly shows  the absence of any Keplerian falloff and hence revealing the presence of DM.}
\label{fig:planets}
\end{figure} 

It is widely accepted that the mass distribution of spiral galaxies, which can be derived from observations, bears the imprint of the Nature and the cosmological history of DM.
Moreover it reflects the interaction (possibly not only gravitational) between dark and luminous matter.

Although the DM presence is widely accepted a strong debate about its spatial distribution is ongoing and very little is known about its Nature.

After a brief historical introduction, RCs are described as main tracers of gravitational field in galaxies. 
Then their most important properties, including their slope as well as scaling relations, are presented.
These properties are well described by the Universal Rotation Curve (URC) paradigm.

The predictions of the halo properties from numerical simulations is the subject of a later paragraph, which is followed by an extensive discussion of the the comparison with the observational properties of the inner distribution of DM.

\section{Historical Introduction}

The use of galaxy kinematic as a tool for studying the mass distribution has a long history (see \cite{sofue01} and references therein) starting in the second decade of the XX century by works of Slipher \cite{slipher14} and Wolf \cite{wolf14}.
In particular it is due to Slipher the discovery that the Andromeda (M 31) galaxy is approaching the Milky Way with a speed of order $100$ Km/s and its disk is rotating around its center with  a steeply rising velocity in the inner region.
Only more than 20 years later however the measurements of Babcock \cite{babcock39} and Oort \cite{oort40} where precise enough to show that the total mass distribution in spirals is not simply proportional to the distribution of light emitted by stars as it was expected.
It has to be mentioned that in 1933 Zwicky, in a failed attempt of analyzing the coma cluster dynamics, made the hypothesis of the existence of a mysterious dark component.
He was not aware however at that time of the importance of its fortuitous statement \cite{zwicky37}.  
The technological advances after the second world war allowed astronomers to routinely detect the RCs of tens of galaxies.
In 1972 Whitehurst \& Roberts \cite{whitehurst72} found an anomalous high velocity of neutral hydrogen gas around M 31 giving a first hint of non-keplerian fall at large radii. 
Only with the work of Rubin \cite{rubin80} however it was clear that "the conclusion is inescapable that non-luminous matter exists beyond the optical galaxy".
This work opened the so called "dark matter problem" in galaxies.
The work of Rubin, based on optical observations, received an important confirmation and extension by the work of Bosma \cite{bosma81} with HI measurements up to larger galactocentric radii.
Clearly at that time the Standard Model of Particle Physics was not complete yet so the nature of DM component was an open question.
However neutrinos provided a viable candidate (but soon clearly understood to be excluded).
The phenomenological analysis of the RCs had an important step further in 1988 when Persic \& Salucci \cite{PS88} found a general trend with larger baryonic to dark mass discrepancy in fainter galaxies and vice-versa.
By that time it was clear that the disks of spiral galaxies are embedded in a much bigger spherical halo whose nature still remained unclear. 
Under the assumption of DM being a cold collisionless particle in 1996 Navarro, Frenk \& White \cite{nfw96} developed a computational model for the formation of the halo and gave a simple parametrization of the halo mass distribution.    
On the phenomenological side instead the study of more than a thousand of galactic RCs by Persic, Salucci \& Sersic \cite{PSS96} revealed that they can be well-represented by a URC, function of the galaxy luminosity.

\section{Rotation Curves as gravitational field tracers}
A RC of a spiral galaxy can be defined as the diagram of the circular velocity as a function of the galactrocentric distance, and is the fundamental probe of the behaviour of the gravitational potential of the system.

The mass distribution in a spiral can be modeled as the sum of three discrete components: a halo of DM, a disk of stars and gas, and a stellar bulge.
The halo and the bulge are assumed to have a spherical distribution while the disk is approximately as infinitesimally thin, with the centre of the three distributions being coincident.
The total gravitational potential $\phi_{tot}$ can then be decomposed as 
\begin{equation}
\phi_{tot}=\phi_{DM}+\phi_{disk,stars}+\phi_{disk,gas}+\phi_{bulge}. \label{eq:3_phi_tot}
\end{equation}
\begin{figure}[h!] 
\centering
\vskip -0cm
\includegraphics[width=6.4cm]{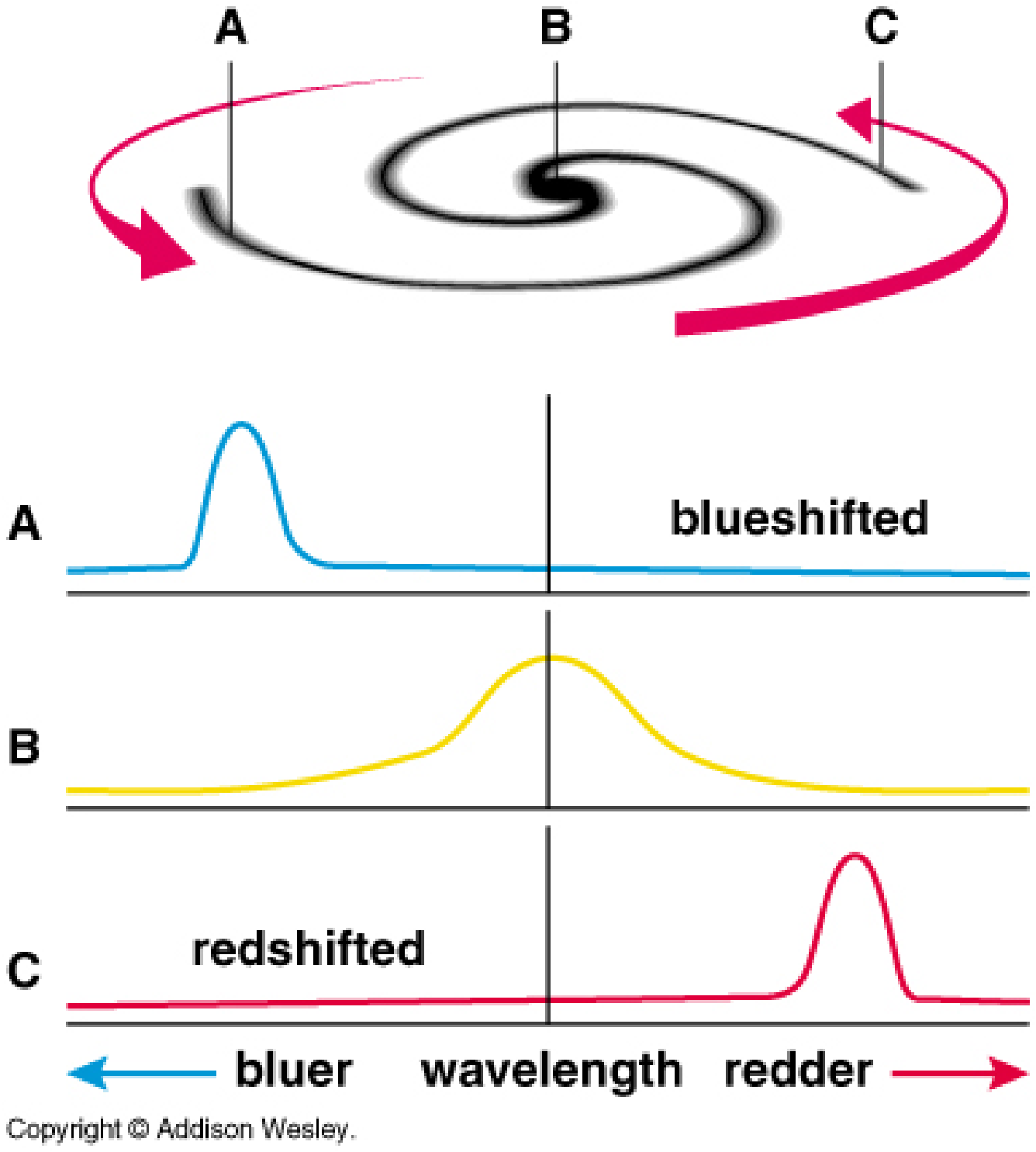}
\includegraphics[width=6.1cm]{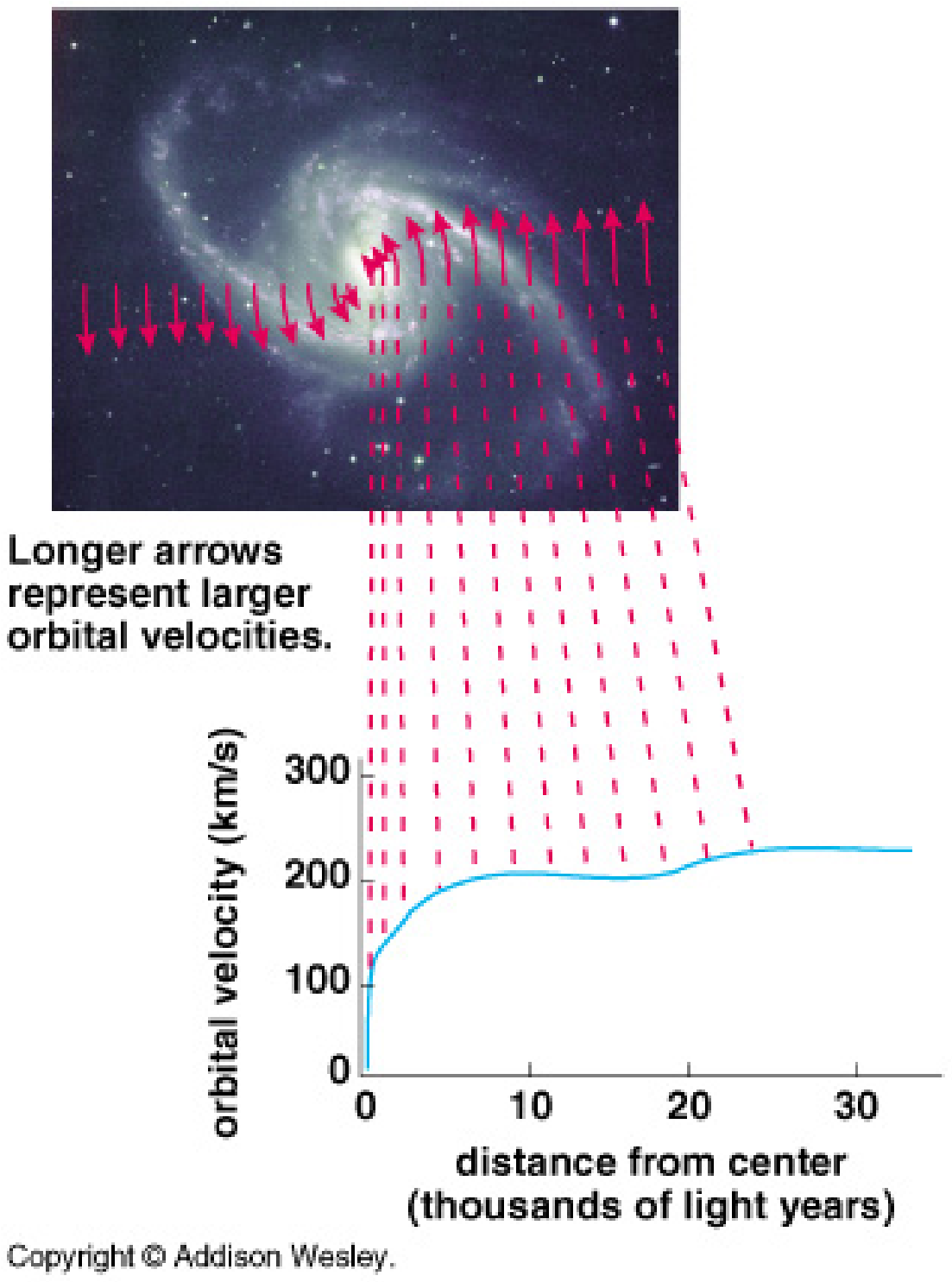}
\vskip -0cm
\caption{Illustration of the measurement of a RC. On the left: points on the major axis equidistant from the centre are red-shifted or blue-shifted by the same amount. On the right: the RC is obtained from the Doppler shift measurements along the major axis.} 
\label{fig:doppler}
\end{figure}
Assuming that the components of the disk have circular orbits with velocity $V_{tot}(r)$ at a radius $r$ the following relations hold:
\begin{equation}
V^2_{tot}(r)=r\frac{d}{dr}\phi_{tot}=V^2_{DM}+V^2_{disk,stars}+V^2_{disk,gas}+V^2_{bulge}, \label{eq:3_vel_tot}
\end{equation}
where we defined $V^2_{DM} \equiv r\:d \phi_{DM}/dr$, and similarly for the disk and bulge.

RCs are obtained by measuring the Doppler shift of absorption or emission lines of disk material (see Fig. \ref{fig:doppler}).
The most commonly used tracers are H$\alpha$ emission lines obtained by optical spectroscopy, that gives the kinematics of the inner part of the galaxy (stars), and neutral hydrogen HI (``21-cm line'') obtained by radio measurements, that extends up to a larger radii.
Radio observations have an angular resolution bigger then optical, but better spectral resolution corresponding to smaller errors in the velocity.

It is possible to estimate the contribution of disk and bulge to the total gravitational potential from the measurements of their mass surface densities.
These in turn yield the contributions $V^2_{disk,stars}$, $V^2_{disk,gas}$ and $V^2_{bulge}$.

The stellar mass distribution is given by its luminosity distribution multiplied by a mass-to-light ratio, which is assumed to be constant within each bulge/disk component. 
Note that from optical measurements it is difficult to disentangle the mass surface density of the disk and of the bulge in the inner region of the galaxy.
For this reason the best mass models are obtained in galaxies with negligible bulges.
Moreover more complex mass models have been tried introducing more components.
However these additional components increase the degeneracy  between free parameters without adding much physical information.

The situation is different for HI measurements where the surface luminosity density distribution $\Sigma_{gas}$ gives a direct measurement of the gas mass.
The halo mass distribution can be either parametrized by a theoretical or an empirical model, or derived from the observed RC inverting Eq. \ref{eq:3_vel_tot} and using appropriate models for bulge and disk.

Optical observations show that very often the stars in the disk follow the exponential Freeman profile \cite{freeman}
\begin{equation}
\Sigma_{D}(r)=\frac{M_{D}}{2 \pi R_{D}^{2}}\: e^{-r/R_{D}},\label{eq:3_sigma_stars}
\end{equation}
where $M_D$ is the disk mass and $R_D$ is the scale length, the latter being measured directly from the observations.
It is useful to define the optical radius (the radius enclosing 83$\%$ of the total light, see \cite{PSS96}), $R_{opt} \equiv 3.2\,R_D$, as the ``size'' of the stellar disk.
In the same way the stars in the bulge very often are distributed according to the S\'{e}rsic  mass density profile (e.g. \cite{aceves06} and references therein), which yield the following surface mass density profile:
\begin{equation}
\Sigma_{b}(r)=\frac{M_{b} \:\alpha^{2n}}{2 \pi R_{e}^{2}\: n \:\Gamma[2n]}\:e^{-\alpha (r/R_{e})^{1/n}}, \label{eq:3_sigma_bulge}
\end{equation}
where $M_b$ is the total projected mass, $r$ is the projected spherical radius, $R_e$ is the effective radius, $n$ is the index of the profile, $\alpha \sim 2n - 0.324$ and $\Gamma[2n]$ is the complete gamma function.
The index $n$ is associated with the curvature and the concentration of the profile; $n=1$ corresponds to an exponential profile, while the classical de Vaucouleurs  profile is obtained for $n=4$. 

From the Poisson equation and using cylindrical coordinates, the potential due to disk material reads
\begin{equation}
\phi_{disk}(r)=-G\int^{\infty}_{0}dr'\:r'\Sigma_{disk}(r')\int^{2\pi}_{0} \frac{d\theta}{|\textbf{r}-\textbf{r'}|}.\label{eq:3_phi_disk}
\end{equation}
$\Sigma_{disk}(r')$ is the surface density distribution of the stars in the disk $\Sigma_{D}(r')$, given by (\ref{eq:3_sigma_stars}), or of the gas $\Sigma_{gas}(r')$, given by an interpolation of the HI data points up to the last measured point.
Having the stars a simple distribution, equation (\ref{eq:3_sigma_stars}) can be integrated in terms of Bessel functions and results in the usual expression \cite{freeman}:
\begin{equation}
V_{disk,stars}^{2}(r)=\frac{G M_{D}}{2R_{D}} x^{2}B\left(\frac{x}{2}\right)\label{eq:3_V_freeman},
\end{equation}
where $x\equiv r/R_{D}$, $G$ is the gravitational constant and the quantity $B=I_{0}K_{0}-I_{1}K_{1}$ is a combination of Bessel functions.

For a spherically symmetric bulge distribution, one has simply $V^{2}_{b}(r)=G m(r)/r$, where $m(r)$ is the mass interior to radius $r$.
Following \cite{kent}, the bulge mass gives 
\begin{equation}
m(r)=\int^{r}_{0}dr'\:2\:\pi\:r'\Sigma_{B}(r') +\int^{\infty}_{r}dr'\:[sin^{-1}(r/r')- r\:(r'^{2}-r^{2})^{-1/2}]\:4\:r'\:\Sigma_{B}(r'). \label{eq:3_mass_bulge}
\end{equation}
\begin{figure} 
\centering
\vskip -0.7cm
\includegraphics[width=12cm]{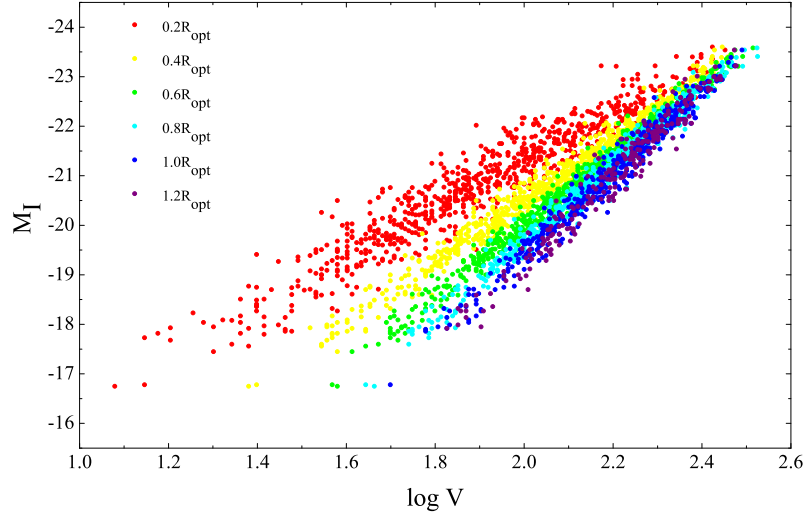}
\vskip -0.7cm
\caption{The Radial TF relation \cite{RTF07}.
Each one of the six relations is indicated with different colours.} 
\label{fig:RTF}
\end{figure}

The above mass model of RCs is valid under the hypothesis of circular motions.
The issue of testing this hypothesis then arises.
Tully \& Fisher \cite{TF77} discovered that the maximal rotational velocity $V_{max}$ of a spiral galaxy, measured by the full width at half-maximum of the neutral hydrogen 21-cm line, correlates with the galaxy luminosity by means of a power law of exponent $a \sim 4$.
This equivalently reads $M=a\: logV_{max}+b$, where $M$ is the absolute magnitude in a certain band and $b$ is a constant.
This relation is a powerful tool to determine the distances of galaxies and to study their dynamics \cite{PS88}.
The rotational velocity reflects the equilibrium configuration of the underlying galaxy gravitational potential. 
In a recent work it has been found a new Tully Fisher relation for spirals holding at different galactocentric radii, called {\em Radial Tully-Fisher relation} \cite{RTF07}: 
\begin{equation}
M_{band} = a_n \log V_n + b_n , \label{eq:RTF}
\end{equation}
where  $V_n \equiv V_{rot}(R_n)$, and $a_n$, $b_n$ are the slope and zero-point of the relations, with $R_n \equiv (n/5)R_{opt}$.
This relation proves that the rotation velocity of spirals is a good measure of their gravitational potential (see Fig. \ref{fig:RTF}).
More specifically, the fact that in any object and at any radius, the rotation velocity can be predicted just by the galaxy luminosity implies that non circular motions are generically negligible.
\begin{figure} [h!] 
\centering
\vskip -0.7cm
\includegraphics[width=13cm]{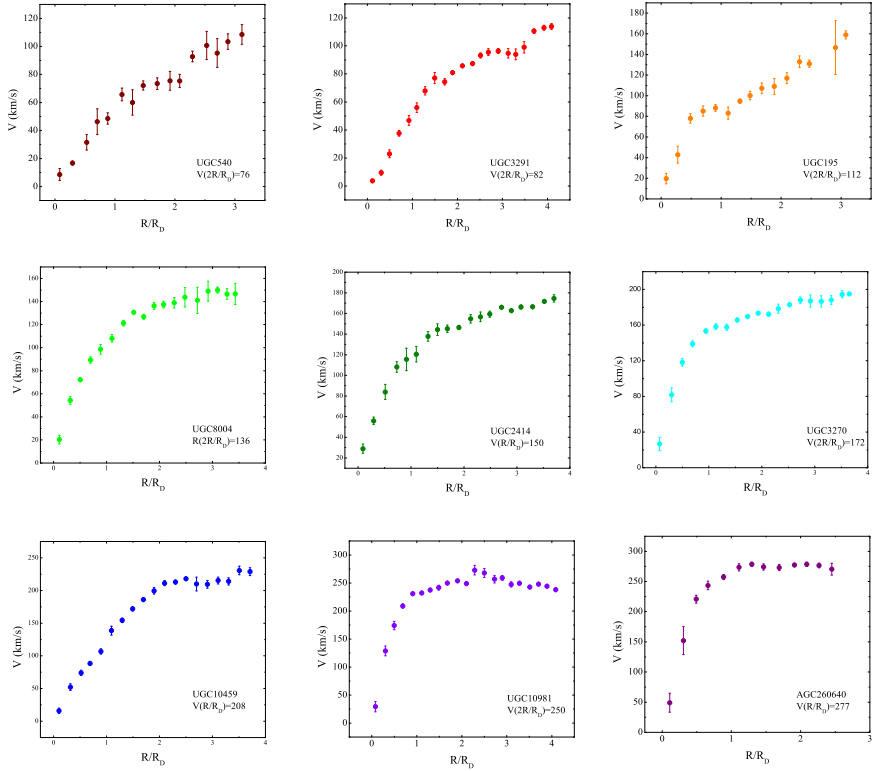}
\vskip -0.7cm
\caption{RCs of spiral galaxies of different luminosities.} 
\label{fig:non_flat}
\end{figure}

\section{Rotation Curves are not flat!}
The information about the distribution of luminous and DM in disk galaxies, as well as on the correlations among the main parameters that characterize both components, offers a fundamental clue to understand how galaxies form and evolve, what role DM plays in these processes, and what imprints DM leaves about its nature.
It is crucial to remark the observational fact that the RCs are not asymptotically flat (see a representative sample of RCs in Fig. \ref{fig:non_flat}), as it is assumed in a huge number of papers.
When in the late 1970s the phenomenon of DM  was discovered \cite{bosma81,rubin80} a few truly flat RCs were highlighted in order to rule out the claim that non Keplerian   velocity profiles originate from a  faint baryonic component distributed at large radii.  
\begin{figure} [h!]
\vskip -0cm
\hskip 1.5cm
\includegraphics[width=6.4cm,angle =-90]{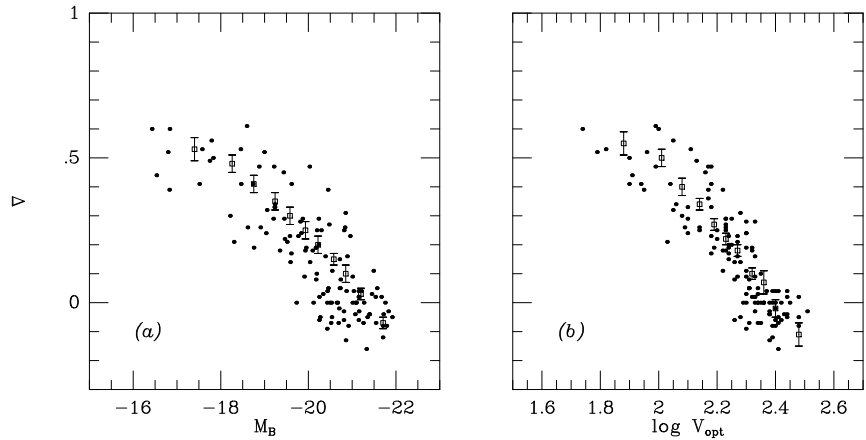}
\caption{Logarithmic gradient of the circular velocity $\nabla$ vs  $B$ absolute magnitude and vs $log \ V(R_{opt})$  \cite{salucci_gentile06}.}
\label{fig:slope}
\end{figure}
At that time a large part of the evidence for DM was provided by extended, low-resolution HI RCs of very luminous spirals (e.g. \cite{bosma81}) whose  velocity profile did show small radial variations.

The increase in the quality of the RCs soon leads to the conclusion that baryonic (dark) matter was not a plausible candidate for the cosmological DM and that the RCs did show variation with radius, even at large radii.
Later numerical simulations in the CDM scenario also predicted asymptotically declining RCs \cite{nfw96}.
The flat RC paradigm was hence dismissed in the 90's (e.g. \cite{PS88,ashman92,weinberg97}). Today,  the  structure of the DM  halos and their rotation speeds  is  thought to have   a central role  in Cosmology  and  a strong link to  Elementary Particles  via the Nature of their constituents (e.g. \cite{olive05}), and a  careful interpretation  of the spiral RCs is considered crucial.

It must be noticed that the circular velocity due to a  Freeman stellar disk   has a   flattish  profile between 2 and 3 disk scale-lengths which implies that a flat RC is not necessarily a proof for the existence of DM.
Its most solid evidence instead originates from the fact that even in very faint galaxies the RCs are often steeply rising already in their optical regions.

A quantitative analysis on the issue is shown in   \cite{salucci_gentile06}, where the concept of  RC logarithmic slope, defined as  $\nabla \equiv  \ (dlog \  V / dlog \  R)$,  is used.
By plotting the logarithmically slope at the optical radius for a huge sample of galaxies \cite{PSS96}, see Fig. \ref{fig:slope}, it is clear that:
 $$
-0.2 \leq \nabla \leq 1,
 $$
i.e. it covers most of the range that a circular velocity slope could take [-0.5 (Keplerian), 1 (solid body)].
Notice that a  flat RC means $\nabla =0$, while in the case of no DM the self-gravitating  Freeman disk lead to $\nabla =-0.27$ at 3 $R_D$.
It is also important to notice the strong correlation between the rotation shape ($\nabla$) and the galaxy luminosity \cite{PSS96,verheijen97,zavala03} (see Fig. \ref{fig:slope}).

The incredibly amount of theories that either imply or assume the existence of an observational scenario in which the RCs of spirals are asymptotically flat, is clearly in contradiction with observational evidences.

\section {The Universal Rotation Curve}       
The studies of spirals of type Sb-Im  in the '90, pioneered by  \cite{PS91} and further developed by \cite{PSS96}, led to the remarkable observation that the RCs of these objects present universal properties well correlated with other galactic properties like the disk mass or the virial mass. 
These works led to the construction of the so called ``Universal Rotation Curve''  $V_{URC}(R; P)$ \cite{urc2}, i.e. a function of the galactocentric distance $R$ tuned by the chosen parameter $P$ (e.g. the virial mass). 
Three different coordinate systems are normally used to measure the radius: the physical coordinate $R$, the radius expressed in terms of the scale length $R/R_D$ or in terms of the DM characteristic length $R/R_{vir}$. 
\begin{figure}[h!]
\centering
\vskip -0.5truecm
\includegraphics[width=9truecm]{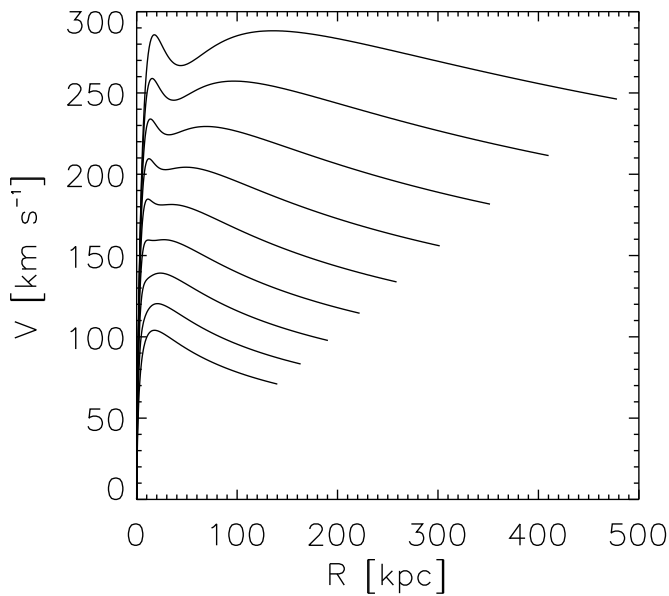}
\vskip -0.5truecm
\caption{The URC with the  radial coordinate in physical units \cite{urc2}. 
 Each curve corresponds to $M_{vir}=10^{11}  10^{n/5} \rm M_\odot $, with $n =
1 \ldots 9$ from the lowest to the highest curve. } 
\label{fig:urc1}
\end{figure}

In order to develop the URC, 11 synthetic curves $V_{coadd}(R/R_{opt}, M_I)$ were built by selecting 616 RCs of galaxies with negligible bulge or HI disk, subdividing them in 11 groups spanning in total the $I$-band luminosity range $-16.3 < {\rm M}_I< -23.4$.
\begin{figure} [h!]
\centering
\vskip -0.7truecm
\includegraphics[width=9truecm]{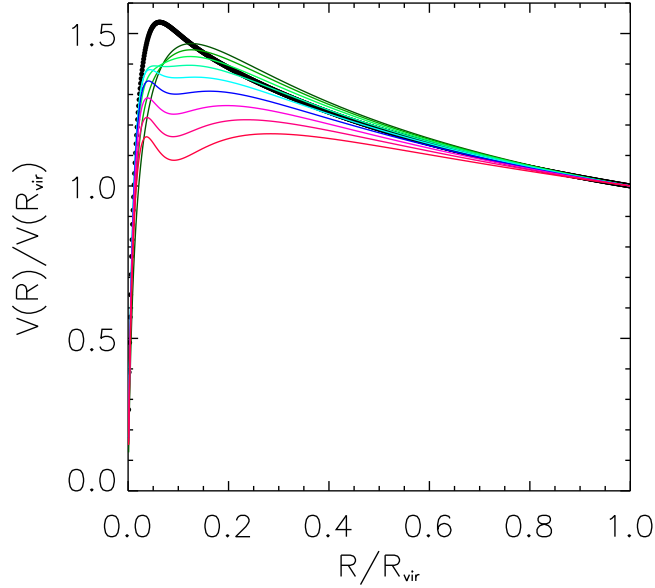}
\vskip-0.7truecm 
\caption{The URC, normalized at its
virial value   as a function of normalized 
radius $ R/R_{vir}$ \cite{urc2}. Each curve,  from the highest  to the
lowest, corresponds to  $M_{vir}$   as in Fig.\ref{fig:urc1}. The bold
line is the NFW velocity profile.} 
\label{fig:urc2}
\end{figure}
Each luminosity bin then contain about $1500$ velocity measurements (from different galaxies by with similar luminous properties) which are arranged in radial bins of size $0.3 \, R_D$ up to $\sim 4 R_D$ and then coadded.
The synthetic curves obtained are then free from most of the observational errors and non-axisymmetric disturbances present in individual RCs, smooth and with a very small intrinsic variance.
The properties of these curves are then found to strongly correlate with luminosity (see also \cite{catinella06}).
The additional data used in the URC are the empirical relationship between  RC slope at 2 $R_{opt} $ and  $log \  V_{opt}$ (see \cite{PSS96}) and  the halo virial velocity $V_{vir}\equiv (G M_{vir}/R_{vir})^{1/2}$, obtained from the disk mass vs virial mass relationship \cite{shankar06}.
  
The URC paradigm, which states that the halo or disk mass determines at any radii the circular velocity of any spiral by means of the URC {\it function}, is the  observational counterpart of the NFW velocity profile obtained by numerical simulations.

The URC function is modeled as the sum in quadrature of two terms:  $V^2_{URC} = V^2_{URCD} + V^2_{URCH}$, where  $V^2_{URCD}$ represent the disk contribution and  $V^2_{URCH}$ the DM halo.

\begin{figure}[h!]
\centering
\vskip -0.5cm
\includegraphics[width=13truecm]{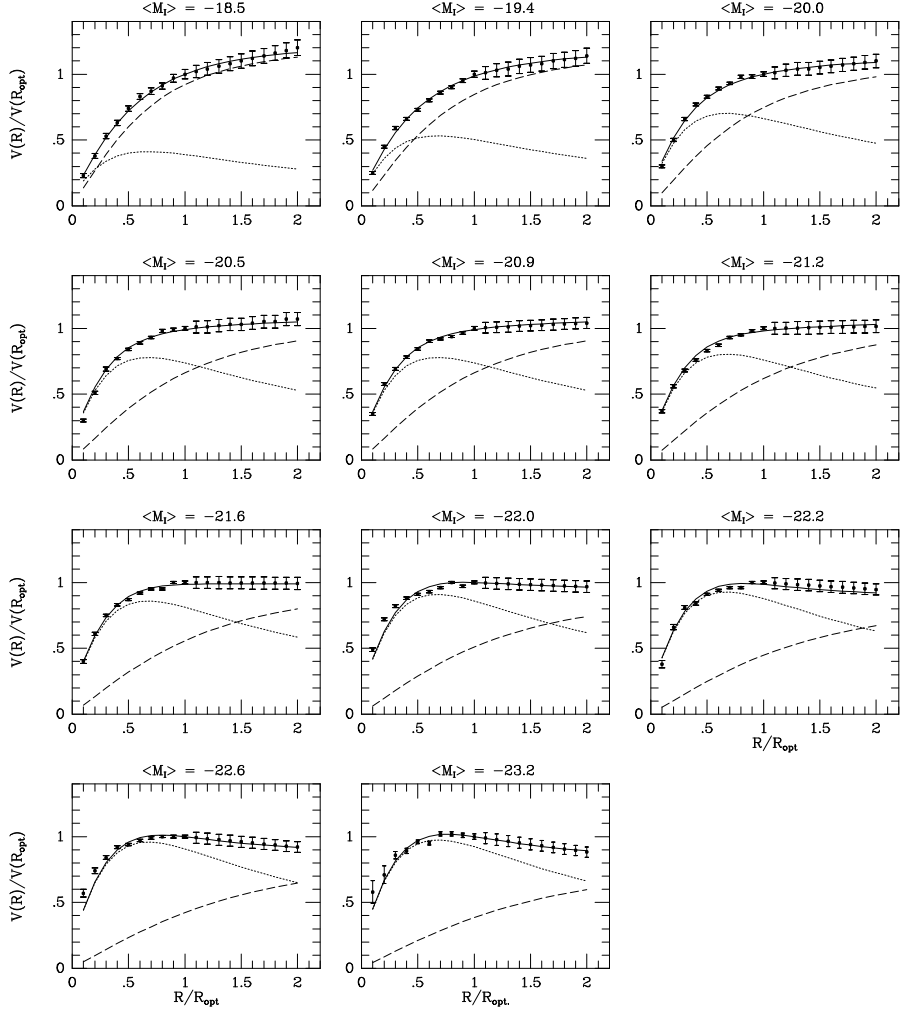}
\vskip -0.5cm
\caption {Best  disk-halo  fits to the URC (dotted/dashed line: disc/halo) \cite{urc2}.}
\vskip -0.1cm
\label{fig:urc1RC}
\end {figure}
The disk contribution is given by Eq. \ref{eq:3_V_freeman}.
For the  DM term it is assumed the empirically Burkert profile \cite{sb00}, a cored distribution that can  converges to a NFW one at  outer  radii:
\begin{equation}
\rho (R)={\rho_0\, r_0^3 \over (R+r_0)\,(R^2+r_0^2)}. \label{eq:burkert}
\end{equation}
$r_0$ is the core radius and $\rho_0$ the    central density
density.
Then:
\begin{equation}
V^2_{URCH}(R)= 6.4 \ G \ {\rho_0 r_0^3\over R} \Big\{ ln \Big( 1 +
\frac{R}{r_0} \Big) - \tan^{-1} \Big( \frac{R}{r_0} \Big) +{1\over
{2}} ln \Big[ 1 +\Big(\frac{R}{r_0} \Big)^2 \Big] \Big\}~.  
\end{equation}

The  URC  function then  has three free parameters   $\rho_0$,  $r_0$,  $M_D$  that are obtained from fitting   $V_{coadd}$ and the other data specified above.    
In Fig. \ref{fig:urc1} the URC function $V_{URC}(R; M_{vir})$ is shown expressing the radius  in physical units and identifying the objects  by the halo virial mass.
Each line refers to a given halo mass in the range $10^{11}\,  M_{\odot} -  10^{13}\,  M_{\odot}$.  
The halo mass determines both the amplitude and the shape of the curve.
Note however that the contribution of the baryonic component is negligible for small masses but becomes increasingly important in larger structures.
In Fig. \ref{fig:urc2} the URC $V_{URC}(R/R_{vir}; M_{vir})$ is shown as a function of the  radial dark coordinate $R/R_{vir}$  and  is  normalized   by $V_{vir} \propto M_{vir}^{1/3}$. 

The URC shows  that  the DM halos and stellar  disks  are both  self-similar, but  the whole system is not,
likely  due to the   baryons collapse that have broken it in   the innermost $30\%$ of the halo size.

RCs are {\it critically}  not flat:  their RC slopes     
take all of sort  of values from that of a solid-body  system (i.e. +1)  to   that  of  an  almost Newtonian point-mass (i.e. - 1/2).   
The maximum of the RC occurs at very different radii, for galaxies
of different mass,  viz. at $\simeq 2 R_D$ for the most massive objects and at $\sim 10 R_D$ for the least
massive ones. 
  
The existence of  systematical properties of the  mass distribution in  spirals  was first claimed by \cite{PS88} and then successively confirmed by independent works \cite{broeils92,PSS96,rhee97,swaters99}.
In order  to understand the whole process of cosmological galaxy formation we must take into account  the rich  scenario of the  dark-luminous  interplay occurred in galaxies.     

In detail, the mass distribution in  Spirals, as carefully obtained in \cite{urc2}, is obtained  by mass modeling   two very different and complementary  kinematical set of data {\it   a)} a large number of {\it individual} RCs of objects  of different luminosity and  {\it b)} the URC (see Fig. \ref{fig:urc1RC}).
The noticeably very similar results obtained from these two different sets of data strongly indicates their robustness and reliableness.
A clear  scenario   of  the mass distribution then emerges (see Fig. \ref{fig:scaling_relations}): 
\begin{figure}[h!]
\centering
\vskip -0.7cm
\includegraphics[width=12truecm]{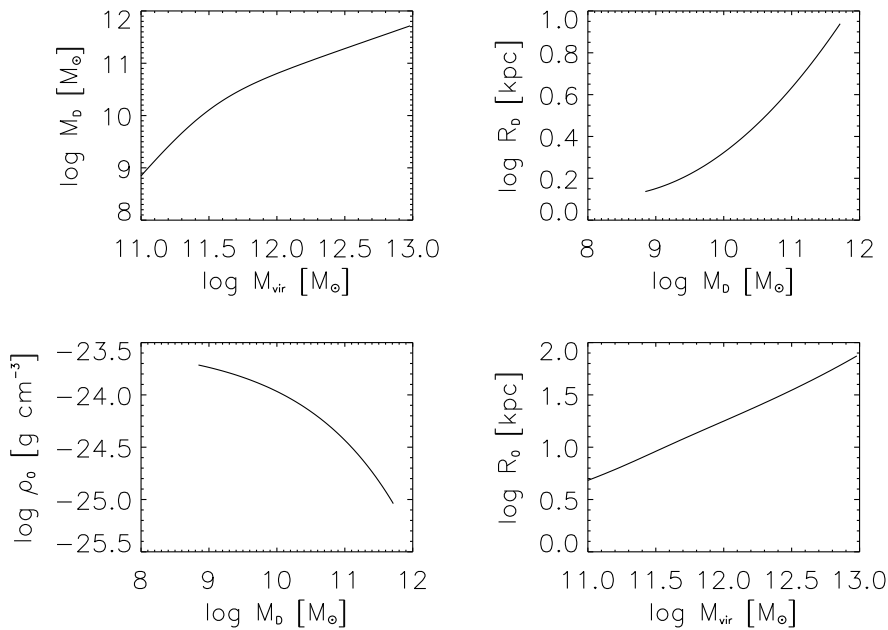}
\vskip -0.7truecm 
\caption{A summary of the empirical scaling relations between the structural parameters of the mass distribution \cite{urc2}; {\it top left:} stellar disk - hallo mass; {\it top right:} disk mass - scale-length; {\it bottom left:} disk mass - halo central density; {\it bottom right:} halo mass - core radius.} 
\label{fig:scaling_relations}
\end{figure}
\begin{itemize}  
\item The stellar disk  dominates  the galaxy's inner region  out to the  radius at which the DM halo contribution  starts to take over the stellar one.
This sets the properties of the  Radial Tully Fisher relation  and yields to  the paradigm of the Inner Baryon Dominance: the inner observed RC that can be accounted by the stellar matter alone are indeed saturated by this component. 
  
\item At any radii, galaxies with lower luminosities have progressively more proportion of DM i.e. a larger  dark-to-stellar mass ratio.
In detail, the disk mass is  $\propto M_{vir}^2$ at small halo virial  masses (e.g.  $M_{vir} = 10^{11}M_\odot$) and  $\propto M_{vir}$ at larger masses (e.g. $M_{vir}= 10^{13} M_\odot$).
The baryonic fraction is always much smaller than the cosmological value $\Omega_b/\Omega_{matter} \simeq 1/6  $, and it ranges between $7\times 10^{-3}$   to   $5\times 10^{-2}$ in line with  is the well-known evidence that SN  explosions    have removed (or made never condense)  a very large fraction of the original HI material.
  
\item Smaller spirals are denser, with the central density spanning 2 order of magnitudes over the mass sequence of spirals.

\item The  structural parameters of the mass distribution, $\rho_0$, $M_D$, $M_h$, $r_0$  are  remarkably all  related,  see Figs. 4 and 11 of \cite{PSS96}. 
  
\item The stellar mass-to-light ratio  is  found to lie between 0.5 and 4. The values of  disk masses  derived as above 
agree very well with those obtained by fitting their SED with spectro-photometric models \cite{SYD08}.  

\item  The HI component  is  almost always below  the kinematical detectably. However, in low mass  systems it cannot be neglected in the baryonic budget since it is more prominent than the stellar disk. 
\end{itemize}

\section{Dark halos from simulations}

In the standard picture of galaxy formation, DM halos provide the framework for the formation of luminous galaxies (e.g., \cite{WR78,BFPR84,WF91}).
The DM halos are assumed to form hierarchically bottom-up via gravitational amplification of initial density fluctuations.
The halos carry with them gas, which eventually cools and contracts to form luminous disk galaxies at the halo centres.
The halo profile has a direct dynamical role in determining the observable RC of the disc.
It  also affects gas cooling and in-fall and therefore the structural properties of the  resultant disc, such as size, luminosity and surface brightness.
In the 1970s numerical simulations were developed and used 
to understand the mechanisms of gravitational clustering, and the
evolving quality of the codes, together with the increasing resolution and
computational power, made them the preferred tool to study the formation of
Cold DM halos.
The success of numerical simulations in reproducing the observed
dynamical properties of galaxies and larger systems depends on the scale
investigated, and there is no agreement about the actual shape of DM halos and the mass distribution of substructures,
due to inconsistencies between the results of simulations and observations; however, simulations 
indeed reproduce well the mechanism of hierarchical clustering, and the latter
enjoys a much broader consensus in being the actual process responsible for structure
formation.

\begin{figure}[h!] 
\centering
\includegraphics[width=10.2cm]{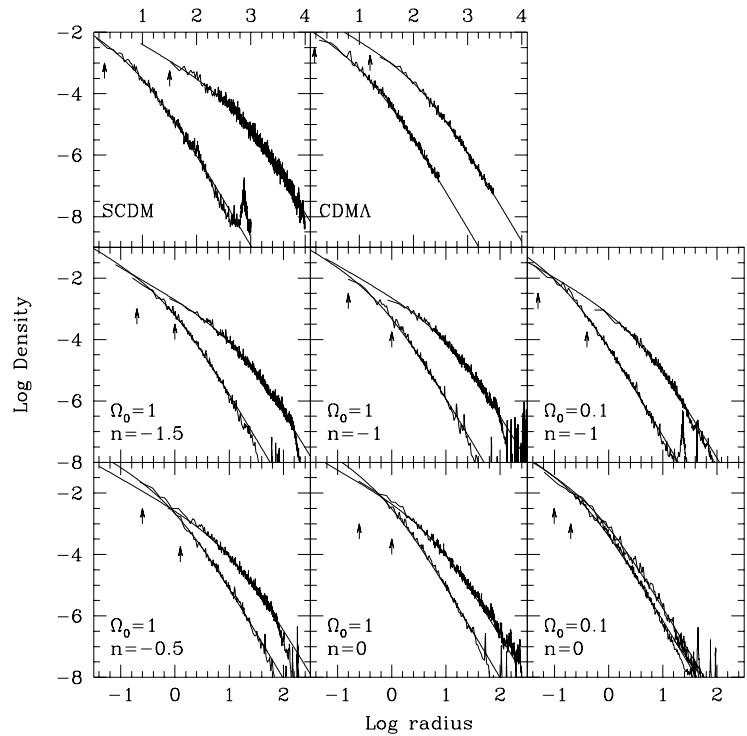}
\caption{Density profiles of simulated halos in different cosmologies
\cite{nfw97}. In each panel, the lower-mass halo is represented by the
leftmost curve; the {\it solid smooth curve} is the NFW fit.
{\it Left panels}: Standard CDM model ($\Lambda=0$).
{\it Right panels}: $\Lambda$CDM model. In each panel the varying cosmological parameters are specified.
Radii are in kiloparsecs ({\it scale at top}); 
the arrows indicate the softening length in each simulation.}
\label{fig:nfw}
\end{figure}
The most evident property of halos born through hierarchical clustering
is the self-similarity: no matter the
mass scale, they all belong to a one-parameter family of curves, known as Navarro, Frenk \& White \cite{nfw96,nfw97} profile
\begin{equation}
\rho_{NFW}(r) = \frac{\rho_s}{(r/r_s)\left(1+r/r_s\right)^2},
\label{eq:nfw}
\end{equation}
where $r_s$ is a characteristic inner radius, and $\rho_s$ the corresponding  inner density.
The outer, virial radius $R_{vir}$, of a halo of virial mass $M_{vir}$, is
defined as the radius within which the  mean density is $\Delta_{vir}$ times
the mean  universal   density $\rho_u$   at that  redshift:
\begin{equation}
M_{vir}  \equiv  \frac{4 \pi}{3} \Delta_{vir} \rho_u R_{vir}^3.
\label{eq:mvir}
\end{equation}
The associated virial  velocity is defined by $V_{vir}^2 \equiv G M_{vir}/R_{vir}$.
The one-to-one relations between the three virial  parameters are fully determined by the background cosmology.
The virial over-density $\Delta{vir}$
is provided by the  dissipationless  spherical top-hat collapse  model
\cite{peebles80,eke96}; it is a function of  the cosmological model, and it
may vary with time.  For the Einstein-deSitter cosmology, the familiar
value is  $\Delta_{vir} \simeq 178$  at all  times.
For  the family  of  flat
cosmologies ($\Omega_m+\Omega_{\Lambda}=1$), the value of  $\Delta_{vir}$ can be approximated by \cite{bryan98}   
$\Delta_{vir} \simeq   (18\pi^2     +  82x  - 39x^2)/\Omega(z)$, 
where $x\equiv \Omega(z) - 1$, and $\Omega(z)$ is the
ratio of mean matter density to critical density at redshift $z$.
In the $\Lambda$CDM cosmological model ($\Omega_m = 0.27$), the value is $\Delta_{vir}(z=0) \simeq 360$.

An associated useful characteristic  is the  concentration parameter,
$c_{vir}$, defined as the ratio between the virial and inner radii,
\begin{equation}
c_{vir} \equiv R_{vir} / r_s.
\label{eq:cvir}
\end{equation}

A third relation between the parameters of the NFW profile is
\begin{equation}
M_{vir} = 4\pi \rho_s r_s^3 A(c_{vir}), \quad
A(c_{vir}) \equiv \ln(1+c_{vir}) - \frac{c_{vir}}{1+c_{vir}}.
\label{eq:mcvir}
\end{equation}
The   three  relations    (Eqs.  \ref{eq:mvir},  \ref{eq:cvir}   and
\ref{eq:mcvir})  allow the  usage of any  pair  out of the parameters
defined so far as  the   two  independent  parameters that  fully
characterize the profile.
Finally the circular velocity curve for the halo is translated by
\begin{equation}
V_c^2(r) \equiv \frac{G M(r)}{r}=V_{vir}^2 \frac{c_{vir}}{A(c_{vir})}\frac {A(x)}{x},
\label{eq:vNFW}
\end{equation}
where $x \equiv r/r_s$.
The maximum velocity occurs at a radius $r_{max} \simeq 2.16 r_s$ and is given by $V_{max}^2/V_{vir}^2 \simeq 0.216 \:c_{vir}/ A(c_{vir})$.

Although in principle the NFW is a two-parameters family of curves, from statistical analysis of the simulated halos it turns out that there is an anti-correlation between the concentration and the halo mass \cite{bullock01b}.
Following \cite{duffy08} at $z=0$ one obtains:
\begin{equation}
c \simeq 8.8 \left( \frac{M_{vir}}{2 h^{-1}10^{12}M_{\odot}} \right)^{-0.09}, \label{eq:bullock}
\end{equation}
that leads to $r_{\rm s} \simeq 27 \left( \frac{M_{vir}}{10^{12}{\rm M}_{\odot}} \right)^{0.42} {\rm kpc}$.
The concentration $c_{vir}$ increases with the redshift of formation while decreasing with the halo mass, thus fulfilling the hierarchical clustering requirements. 

\begin{figure}[h!] 
\vskip -6cm
\hskip -3.5cm
\includegraphics[width=20cm]{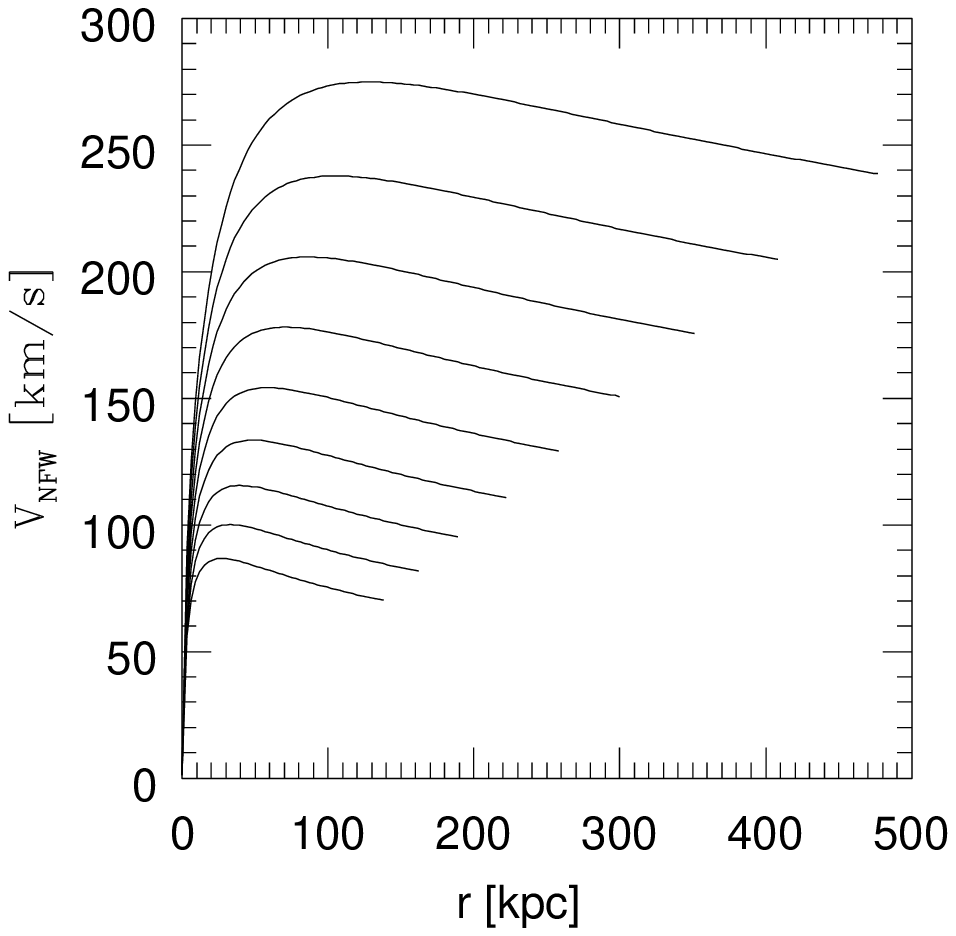}
\vskip -11.7cm
\hskip 5.2cm
\includegraphics[width=6.7cm]{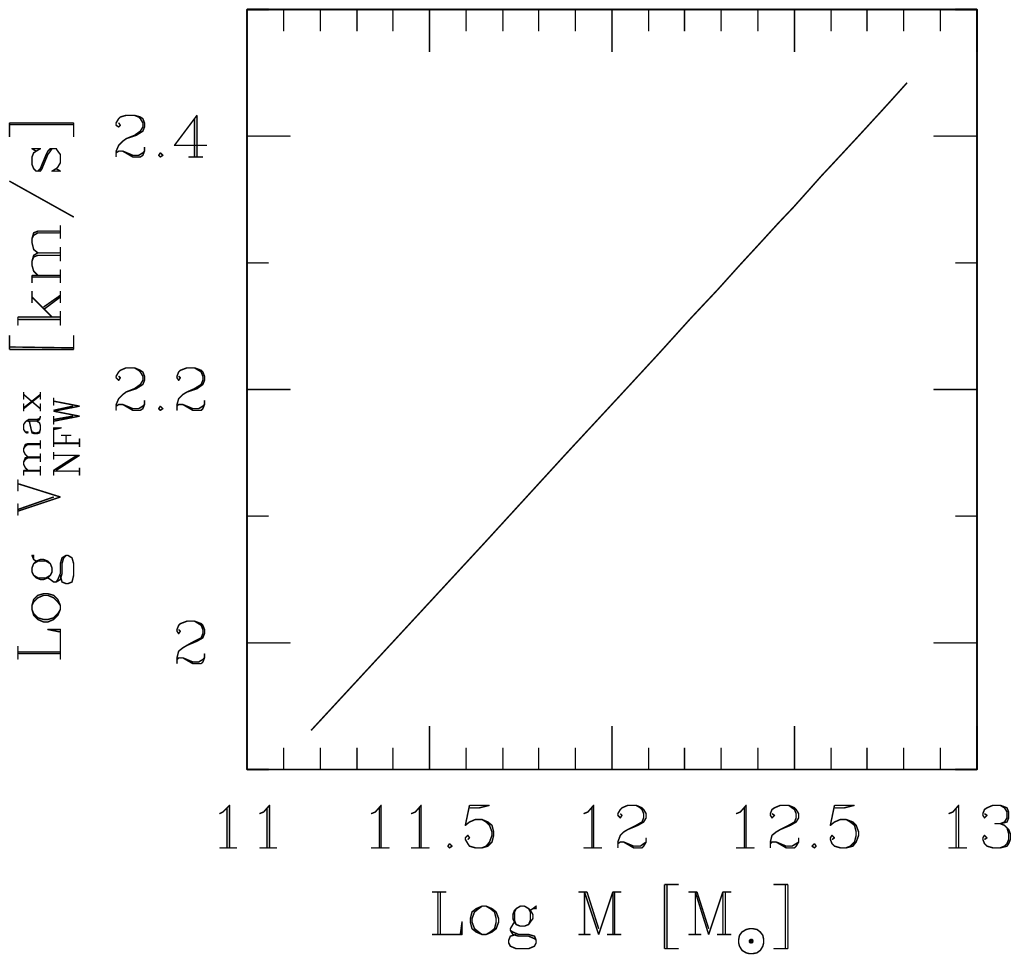}
\vskip 0.5cm
\caption{The NFW circular velocities with the radial coordinate in physical units.
Each curve corresponds to $M_{vir}=10^{11}  10^{n/5} \rm M_\odot $, with $n =
1 \ldots 9$ from the lowest to the highest curve.
Also shown the NFW maximum velocity dependence with virial mass.}
\label{fig:nfw_vel}
\end{figure}

In Fig. \ref{fig:nfw_vel} the NFW circular velocities from Eq. \ref{eq:vNFW} are shown, using the relation \ref{eq:bullock}, for different values of the virial mass.
Each curve corresponds to $M_{vir}=10^{11}  10^{n/5} \rm M_\odot $, with $n =
1 \ldots 9$ from the lowest to the highest curve.
It is also shown the NFW maximum velocity dependence with the same virial mass range.

The NFW result has been confirmed by a number of subsequent studies
(see e.g. \cite{cole96,huss99,fuku97,moore98,jing00}), although there is some disagreement regarding the innermost value of the logarithmic slope  $\gamma$.
NFW argued that a fitting formula where $\gamma=(1+3y)/(1+y)$ (where $y=r/r_s$ is the radial coordinate in units of a suitably defined scale-radius $r_s$) provides a very good fit to the density profiles of simulated halos over two decades in radius. 
Some authors (see \cite{moore98,ghigna00,fuku01}) have argued that
$\gamma$ converges to a value of $\sim -1.5$ near the center,
rather than $-1$ as expected from the NFW fit.
Others \cite{kravtsov98} initially obtained much shallower inner slopes ($\gamma \sim -0.7$) in their numerical simulations, but have now revised their conclusions; 
these authors now argue that CDM halos have steeply divergent density profiles but, depending on evolutionary details, the slope of a galaxy-sized halo at the innermost resolved radius may vary between $-1.0$ and $-1.5$.

\section{The cusp vs core issue}
Although the existence of DM has been inferred for several decades, it is only recently that we start to shed light on crucial aspects of the DM {\it distribution}.
Initially, the main focus was on the presence of a dark component \cite{rubin80}; 
this later shifted to investigating the ratio of dark to visible matter \cite{PS88,SAP91,SP97}.
Today, the focus is mainly on the actual density profile of dark halos (e.g. \cite{salucci01,salucci03}).

Any successful cosmological model must be able to reproduce both observed large and small scale structures, from galaxy clusters to galaxy halos.
A fundamental prediction of the cosmological CDM simulations is that virialized DM halos  have  an universal spherically averaged cuspy NFW density profile that disagrees with a number of observations.
Such cusps in the DM distribution would certainly have very interesting implication for particle DM searches.
For example, it could be possible to detect gamma rays from annihilations of very heavy DM particles in the centre of our Galaxy (e.g. \cite{flores94}), and present limits on radio and gamma-ray emission from the Galactic centre would then significantly constrain the mass of DM particles such as neutralinos.
This section is devoted to address the cusp vs core issue, that has stimulated a lot of discussions as it has the potential to provide interesting new insights into the nature of DM and its possible interactions with visible matter (for reviews, see \cite{primack04, ostriker03}).

\begin{figure}[h!]
\centering
\includegraphics[width=6.4cm]{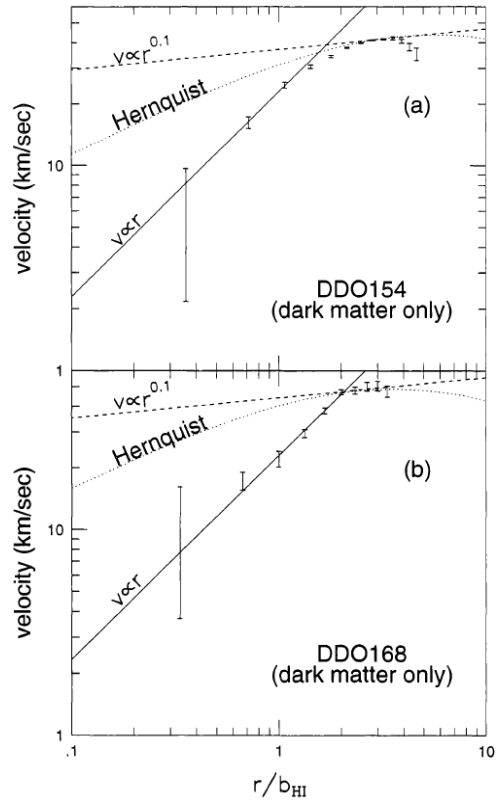}
\caption{DM contribution to the circular velocity of two dwarfs, as a function of distance from the center in units of the HI-disk scale length \cite{flores94}.
Lines show the radial behaviour assuming a DM Hernquist (dotted), $r^{-1.8}$ (dashed), and constant (solid) density profiles.}
\label{fig:flores}
\vskip -0cm
\end{figure} 
A cored distribution, i.e. a  density profile flat out to a radius that is a significant part of the disk size, has been often adopted (and represented by an isothermal profile, e.g. \cite{carignan_freeman85}), although the implications of this distribution appeared only 
after that cosmological $N$--body simulations found that CDM virialized halos achieve a cuspy density profile.
When the first simulations of CDM halos
became available (e.g. \cite{dubinski_carlberg91}), they had a central
density profile approximately $\rho(r) \propto r^{-1}$, which has come
to be known as the central ``cusp''.

\begin{figure}[h!] 
\centering
\includegraphics[width=11.3cm]{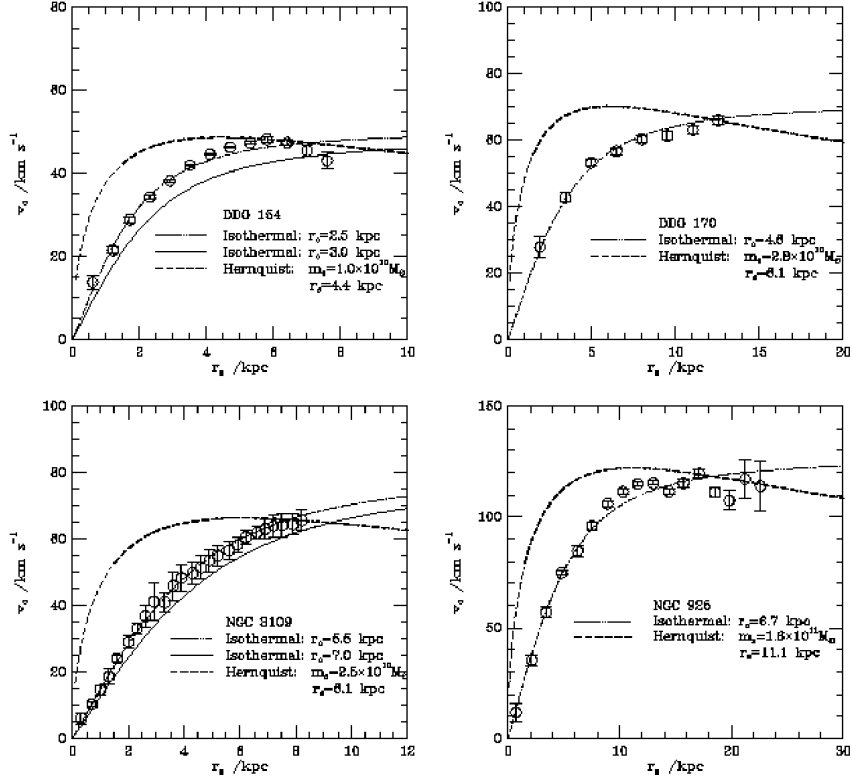}
\caption{Observed RCs as a function of galactocentric distance \cite{moore94}.
Lines show an approximately isothermal dark halo fits ($\rho(r) \propto 1/(r^2_c +r^2)$, where $r_c$ is the core radius) to the RCs before (dotted) and after (solid) including the luminous contributions.
Dashed lines are obtained with a Hernquist profile.}
\label{fig:moore94}
\end{figure}

The structure of the inner regions of galactic halos was soon investigated by \cite{flores94,moore94}, who used RCs measurements of some DM dominated dwarf galaxies (see Figs. \ref{fig:flores} and \ref{fig:moore94}).
It was pointed out a tension between the kinematical data and the predictions of simulations: DM halos seemed to prefer cored density distributions rather than cuspy ones.

To cope with this observational evidences, \cite{burkert95} proposed an empirical profile (see Eq. \ref{eq:burkert}) that successfully fitted the halo of those RCs, the so-called ``Burkert profile'' and since then has been mostly used to represent cored dark halos.

Meanwhile, theorists have done simulations with increasing resolution.
On the basis of simulations with tens of thousands of particles per DM halo, NFW \cite{nfw96,nfw97} showed that halos from galaxy to cluster scales have density profiles that are described fairly well by the fitting Eq. \ref{eq:nfw}.

An extensively 'galaxy by galaxy' comparison then started between the predicted NFW density distribution and those actually detected for the dark halos around disk galaxies highlighting a CDM crisis and becoming the main goal of several publications \cite{sb00,47salucci03,kravtsov98,salucci01,burkert_silk97,MdB98,dBMR01,blok_bosma02,trott02,binney01,blais02,bottema02,weldrake03,simon03}.

\begin{figure} [h!] 
\vskip -1.7cm
\hskip -1.9cm
\includegraphics[width=17.2cm]{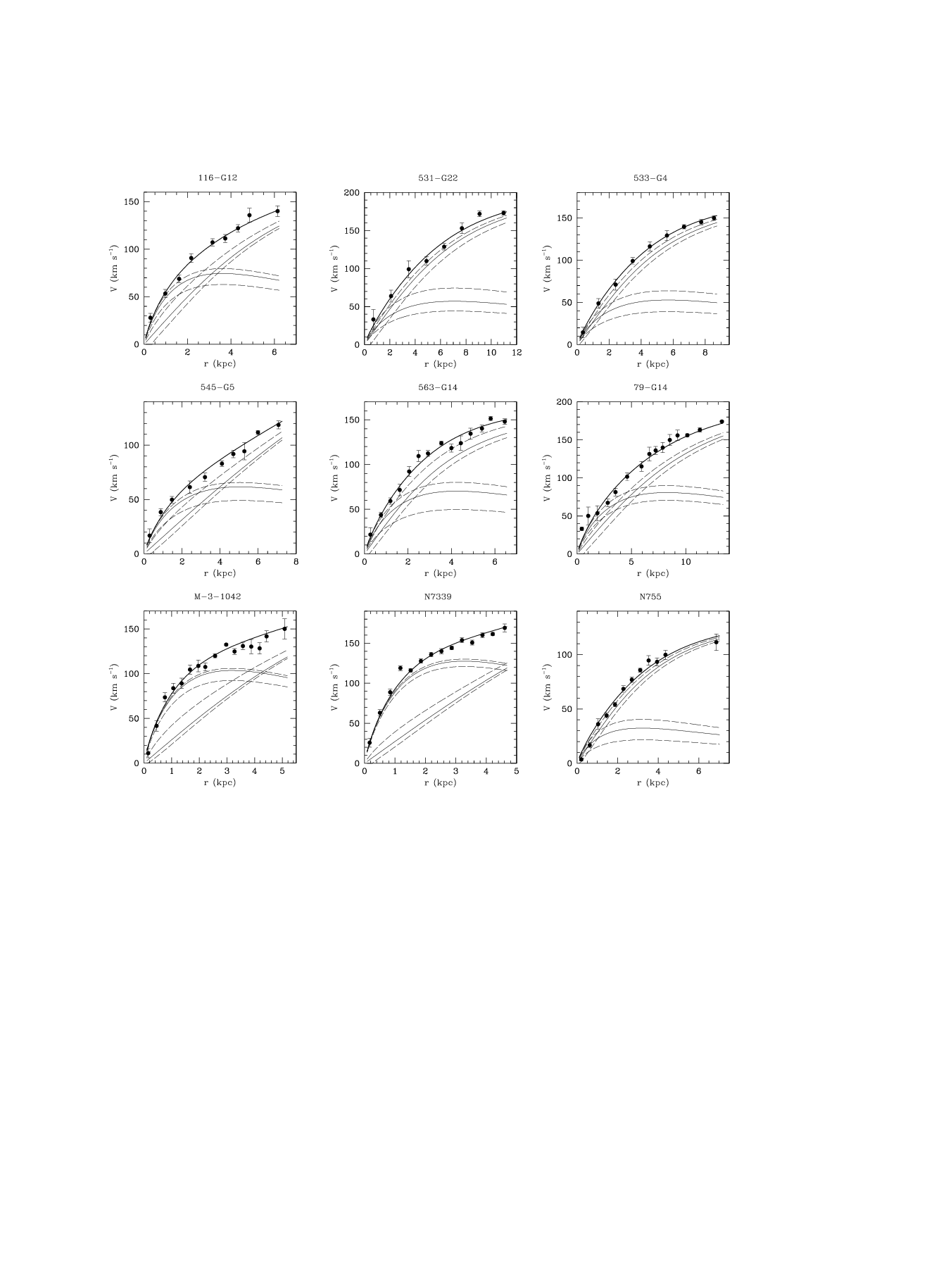}
\vskip -9.4cm
\caption{URC fits (thick solid line) to the RCs (points with errorbars) within the Constant Density Region \cite{salucci_borriello01b}.
Thin solid lines represent the disk and halo contributions.
The maximum disk and the minimum disk solutions are also plotted (dashed lines).}
\label{fig:Vplots}
\end{figure}

\begin{figure}[h!] 
\vskip -0.3cm
\hskip 2.8cm
\includegraphics[width=11cm]{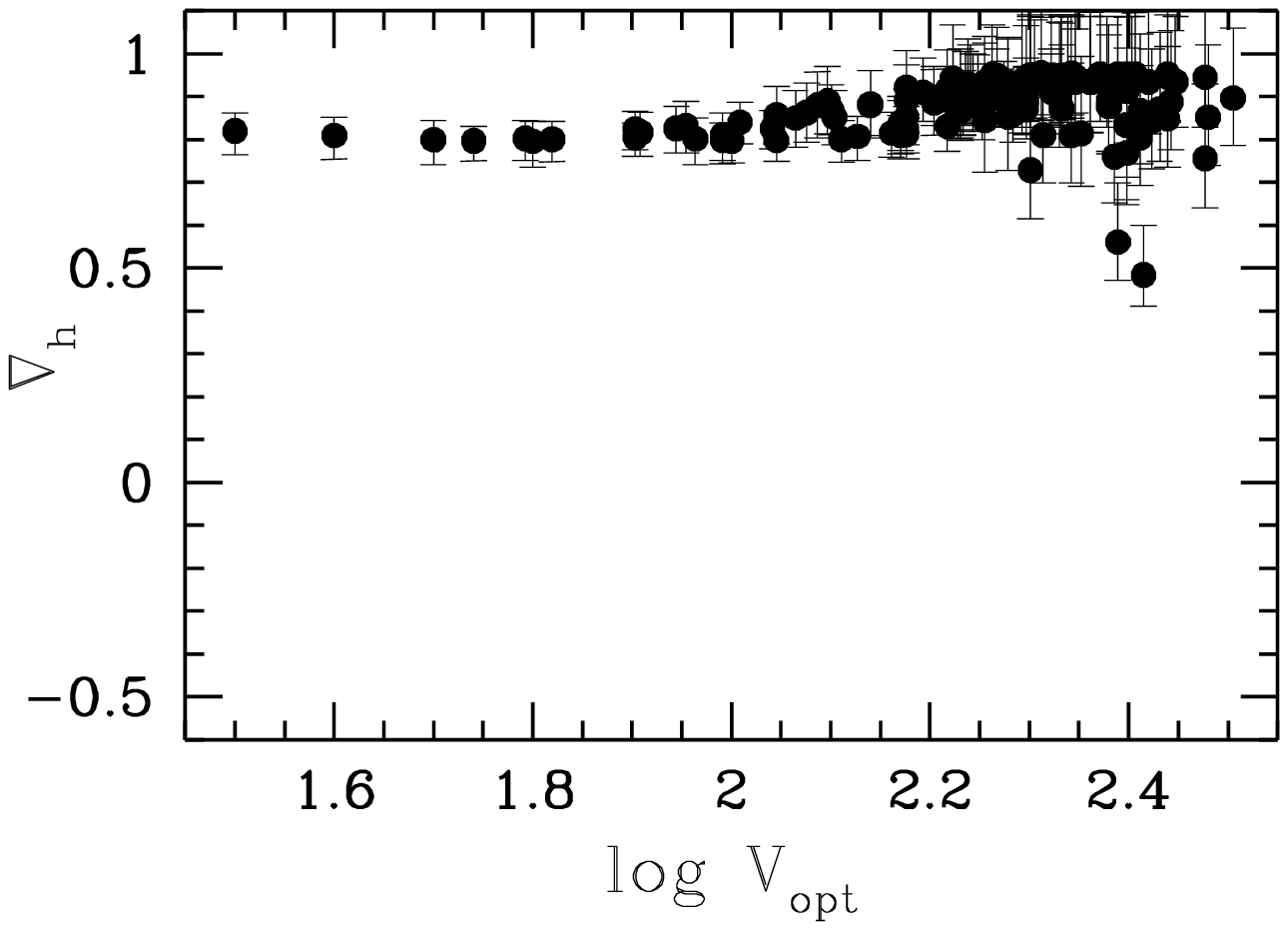}
\vskip -5.1cm
\caption{The dark halo slopes $\nabla_h$ as a function of $V_{opt}$ \cite{salucci01}.
As a comparison, in CDM $-0.1<\nabla_h \leq 0.5$.}
\label{fig:nablaH}
\end{figure} 
In the work of \cite{salucci_borriello01,salucci_borriello01b} an excellent sample of high-quality optical RCs, satisfying the following quality requirements were used to obtain the velocity profiles of the surrounding DM halos: 
i) data extend at least out to the optical radius,
ii) they are smooth and symmetric,
iii) they have small internal {\it rms},
iv) they have high spatial resolution and a homogeneous radial data coverage of 30-100 data points between the two arms,
v) each RC has 7-15 velocity points inside $R_{opt}$, each one being the average of 2-6 independent data,
vi) the RCs spatial resolution is better than 1/20 $R_{opt}$, the velocity {\it rms} is about 3\% and the RCs logarithimc derivative is generally known within 0.05.
It was found that they increase with linearly with radius at least out to the edge of the stellar disk, implying that, over the entire stellar region, the density of the dark halo is about constant.
The mass distribution was modeled as the sum of a stellar Freeman disk and a spherical halo ($V^2=V^2_D+V^2_H$), whose contribution to the circular velocity is given by \cite{PSS96,SP97}: $V^2_H(r)=V^2_{opt} (1-\beta) (1+a^2)x^2 /(x^2+a^2)$, where $x \equiv r/R_{opt}$, $a$ is the core radius measured in units of $R_{opt}$ and $\beta \equiv (V^2_D/V^2)_{R_{opt}}$.
It has been shown (e.g. \cite{PS90b,PS92}) that by taking  into 
account the logarithmic gradient of the circular velocity field defined as:
$\nabla(r)\equiv \frac{d \log V(r)}{d \log r}$, one can significantly increase the amount of information available from 
kinematics and stored in the  shape of the RC.
$\chi^2$ was   calculated on both velocities and logarithmic 
gradients: $\chi^2_V =\sum^{n_V}_{i=1}\frac{V_i-V_{model}(r_i; \beta,a)} 
{\delta V_i} $ and $\chi^2_{\nabla} = \sum^{n_{\nabla}}_{i=1}\frac{\nabla( 
r_i)-\nabla_{model}(r_i; \beta,a)}{\delta \nabla_i} $, 
and the parameters of the mass models derived by  minimizing  a total $\chi^2_{tot}$, defined as: $\chi^2_{tot} \equiv \chi^2_V+ \chi^2_{\nabla}$. 
The derived mass models are shown in  Fig. \ref{fig:Vplots}, alongside with the separate disk and halo contribution.
It is clear that the halo curve is steadily increasing out to the last data point.
Note also the uniqueness of the resulting halo velocity model: the maximum-disk and minimum-disk models almost coincide. 

This work is complementary to that of \cite{salucci01} who derived for 140 objects of different luminosity $\nabla_H$, the logarithmic gradient of the halo velocity at $R_{opt}$ (see blue points in Fig. \ref{fig:nablaH}; red points represent the results from \cite{sb00}).
The results are impressive:  the halo mass profiles at $R_{opt}$  turn out to 
be 
{\it i)}  independent of the galaxy  properties,     
{\it ii)}  Universal and  
{\it iii)}  essentially featureless in the sense that for  any spiral the stellar disk is embedded within a constant density sphere. 

The highest possible value for $\nabla_h^{CDM}$ is $0.5$, that is achieved on the $\sim 10$ kpc scale only for $c<5$ (see \cite{bullock01b,NS00}), i.e. for   low values of the  concentration parameter, a property of low-$\Omega$ universes.
This value is quite inconsistent with the average value found in spiral dark halos, especially if one considers that high resolution N-body simulations converge to a maximum value of $\nabla_h^{CDM}=1/4$ \cite{moore98}.

\begin{figure}[h!] 
\vskip -1.8cm
\hskip -2cm
\includegraphics[width=17.2cm]{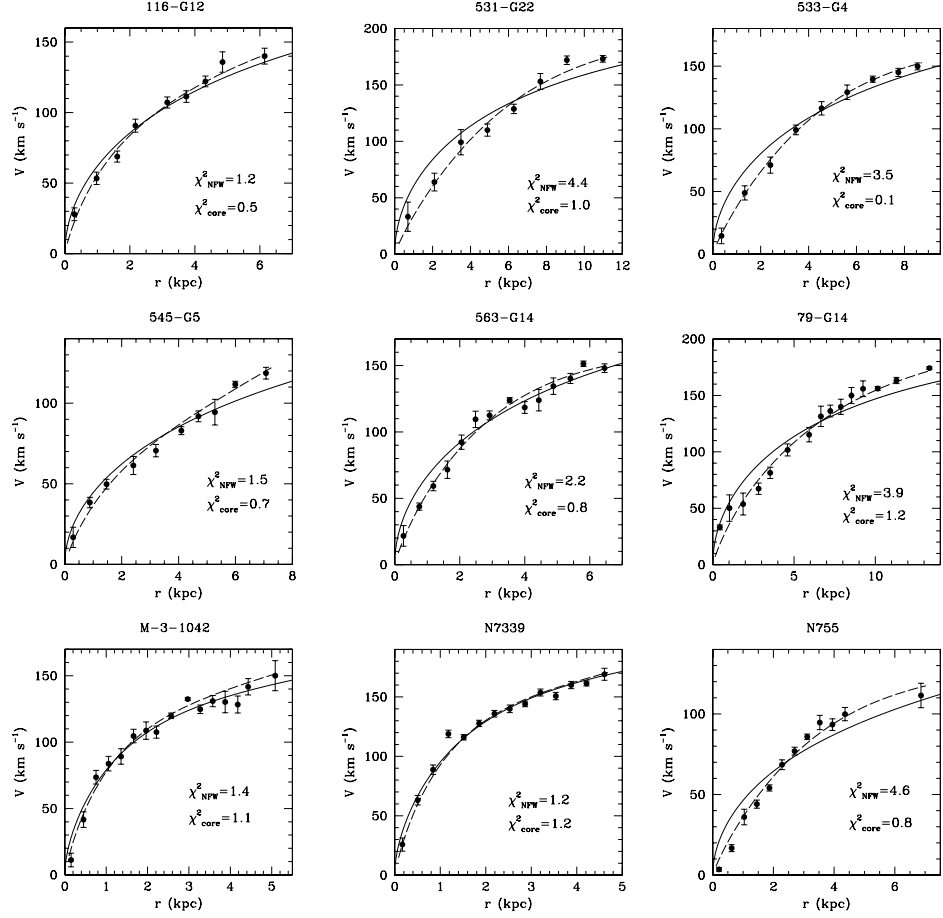}
\vskip -9.5cm
\caption{NFW best fits (solid lines) of the RCs (filled circles) compared with the CDR fits (dashed lines) \cite{salucci_borriello01b}.
The $\chi^2$ values are also indicated.}
\label{fig:Vtot}
\end{figure} 
Of crucial importance is also the absence of a significant scatter in the $\nabla_h$ vs. $ log V_{opt}$ relationship. 
In fact, the CDM  theory  predicts  that, in a very wide region
centered  at  $\sim 10$ kpc and including  $R_{opt}$ 
independently of its  relation with  the virial radius,   galactic    halos
with the same  mass    do not follow   a unique   velocity  curve   but    a 
family of  them.
These  can be  described by  a  set of  straight-lines with 
slopes varying between    $-0.1$ and $+0.5$ (e.g. see  Fig. 6 of  \cite{bullock01b}.
According to CDM the $\nabla_h-log V_{opt}$ plane should be filled
well beyond the tiny strip of Fig. \ref{fig:nablaH}.
Taken at its  face value,   the
observational constraint variance (${\nabla_h}<0.1 $) could  imply,   within
the   CDM scenario,  that protospiral halos  are coeval  and have  similar  
merging histories.
A second  possibility  may be that the disk length-scale
$R_{opt}$,  in units of virial radius,  is strongly coupled with 
the  structure of the DM halo (e.g., \cite{mo98,dal97,bosch00} but see also \cite{bullock01b}). 

\begin{figure}[h!] 
\centering
\vskip -1.4cm
\includegraphics[width=13.5cm]{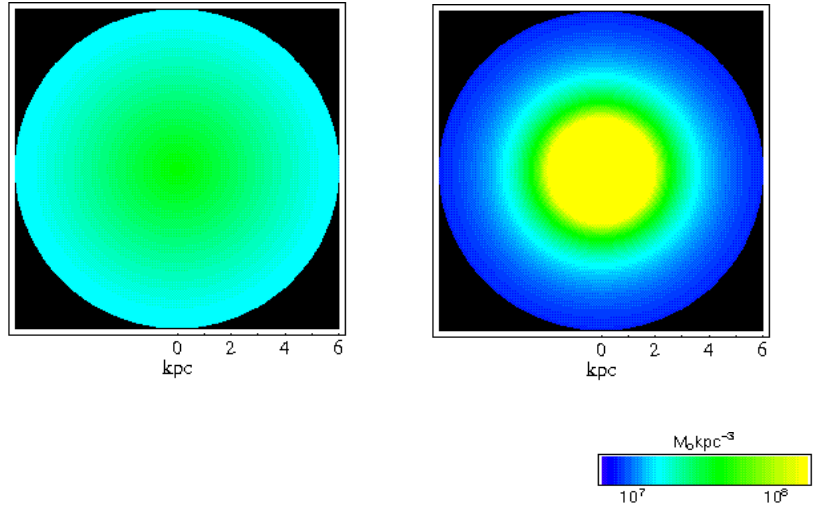}
\vskip -11cm
\caption{{\it Left}: ESO 116-G12 density dark halo \cite{salucci03}.
{\it Right}: CDM prediction.}
\label{fig:cuspVScore}
\end{figure}

Fig. \ref{fig:Vtot} shows the URC and NFW halo fits to the RCs, leaving $c$ and $r_s$ as free parameters, constraining a conservative halo mass upper limit of $2\times 10^{12} M_{\odot}$: for most objects the NFW models are unacceptably worse than the URC solutions, and the resulting CDM stellar mass-to-light ratios turn out to be in some cases unacceptable low. 
See the particular case of the ESO 116-G12 galaxy in Fig.\ref{fig:cuspVScore}.

Particular attention has been extensively given to RCs of Low Surface Brightness galaxies, where the implied cosmological parameters from NFW halos are inconsistent with $\Lambda$CDM picture, in what the observed concentrations of the NFW halos are too low \cite{McGaugh03,swaters03,deNaray08}.
Furthermore, much better fits to LSB observations are found when using cored halo models \cite{deNaray06,dBMR01,blok_bosma02,deNaray08,simon05,dBMBR01,dBBM03,marchesini02,bolatto02,blais01,cote00}.

\begin{figure}[h!] 
\vskip -0.4cm
\hskip -0.05cm
\includegraphics[width=13.2cm]{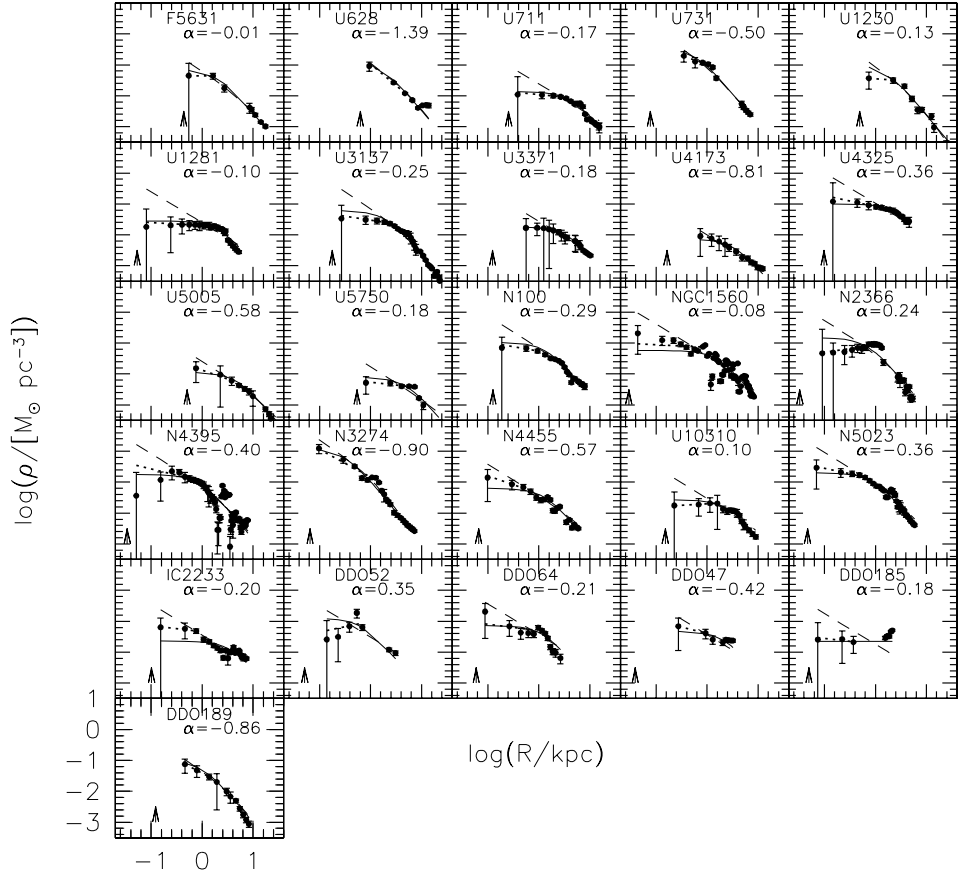}
\vskip -0.9cm
\caption{Mass profiles derived from HI high-resolution LSB RCs \cite{blok_bosma02}.
The profiles can be characterized by a steep $r^{-2}$ outer component, and a shallower inner cored component.
Lines represent the best-fitting minimum disk models:
pseudo-isothermal (full), the NFW (long-dashed) and a power-law fitted to the inner shallow part (thick short-dashed).
The slope $\alpha$ is given in the top-left corners of the panels.
The arrows indicate an angular size of 2$''$, the typical value of the seeing.}
\label{fig:dBB02}
\end{figure}

\begin{figure}[h!] 
\centering
\vskip -0.2cm
\includegraphics[width=8cm]{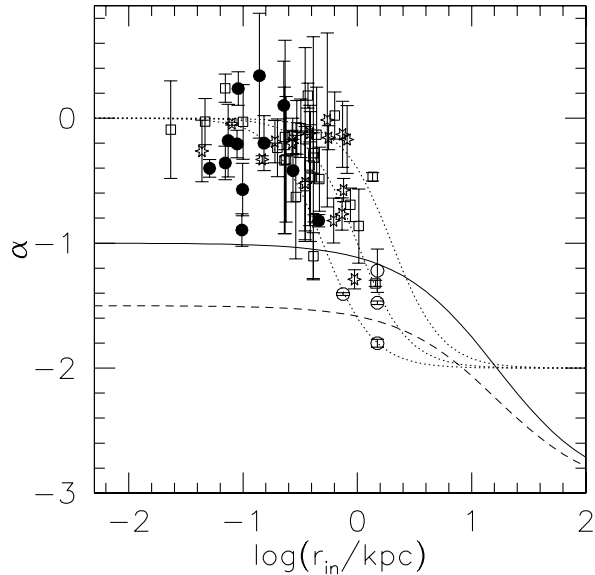}
\vskip -0.1cm
\caption{Value of the inner slope $\alpha$ of the LSB mass-density
profiles plotted against the radius of the innermost point \cite{blok_bosma02}.
Over-plotted are the theoretical slopes of a pseudo-isothermal halo model (dotted lines) with core radii of 0.5 (left-most), 1 (centre) and 2 (right-most) kpc.
The full line represents a NFW model, the dashed line a
CDM $r^{-1.5}$ model.}
\label{fig:dBB02slope}
\end{figure}

Fig. \ref{fig:dBB02} plots the derived mass profiles of the high-resolution LSB RCs sample of \cite{blok_bosma02}.
It is clear that most of the galaxies are 
characterized by an almost flat inner core with a radius of a few kpc,
in contrast with the steep inner $\alpha=-1.0$ power-law slope of the NFW
profile.
The values of the inner slope are plotted against the value of $r_{\rm in}$ in 
Fig. \ref{fig:dBB02slope}, showing that these galaxies are consistent with cored halos.

\begin{figure}[h!]
\centering
\vskip 0.4cm
\includegraphics[width=6.2cm]{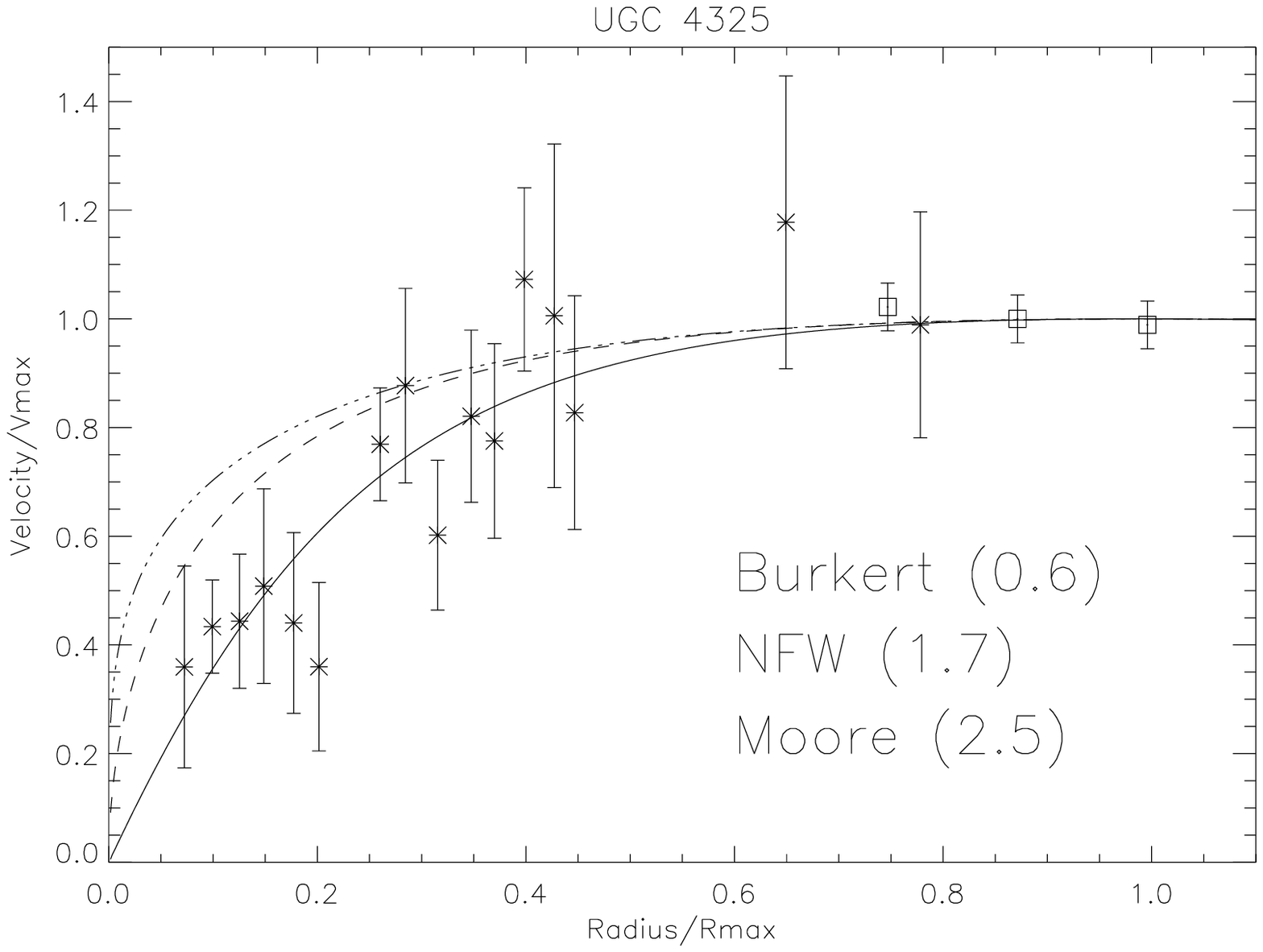}
\includegraphics[width=6.2cm]{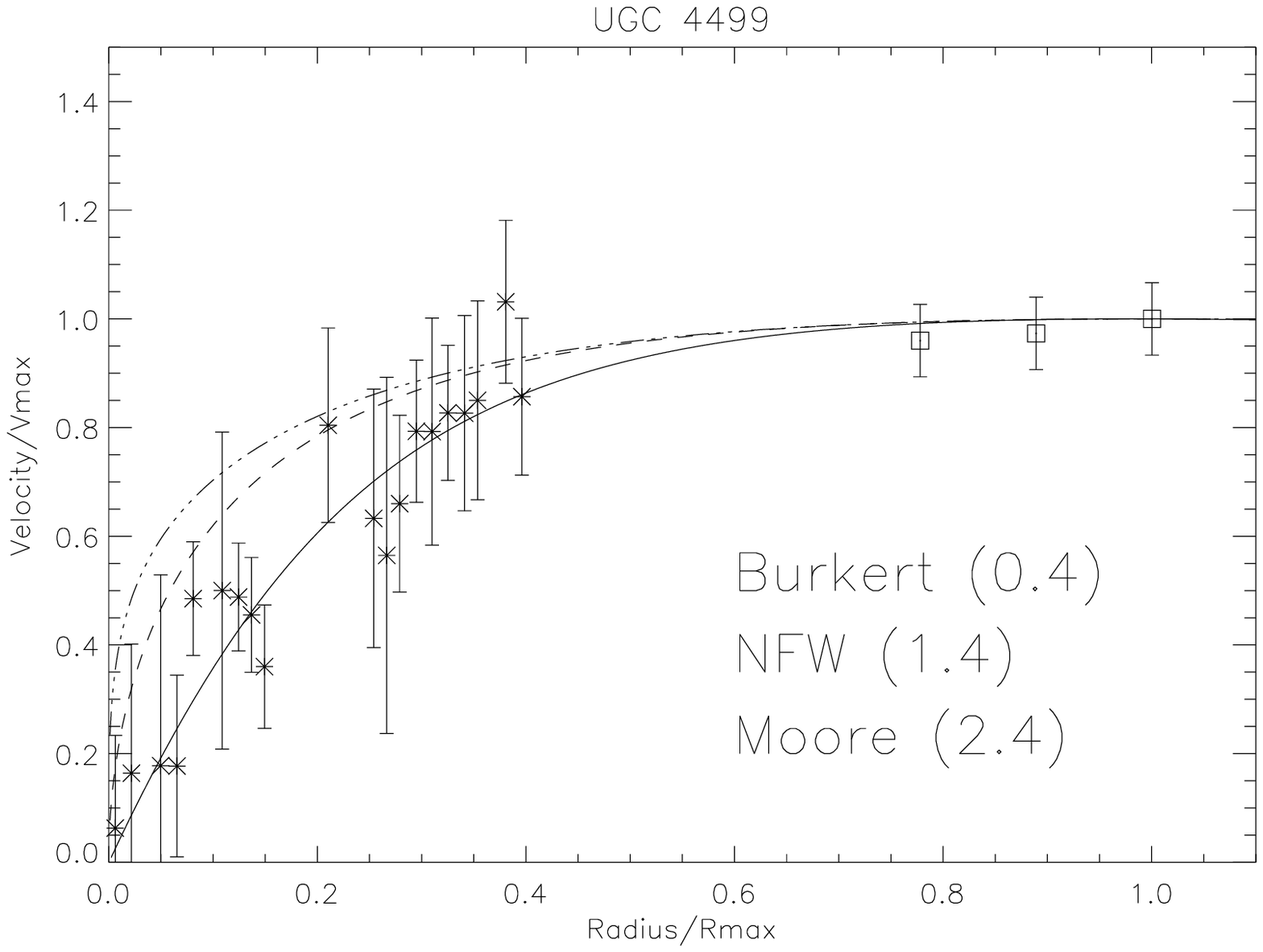}\\
\includegraphics[width=6.2cm]{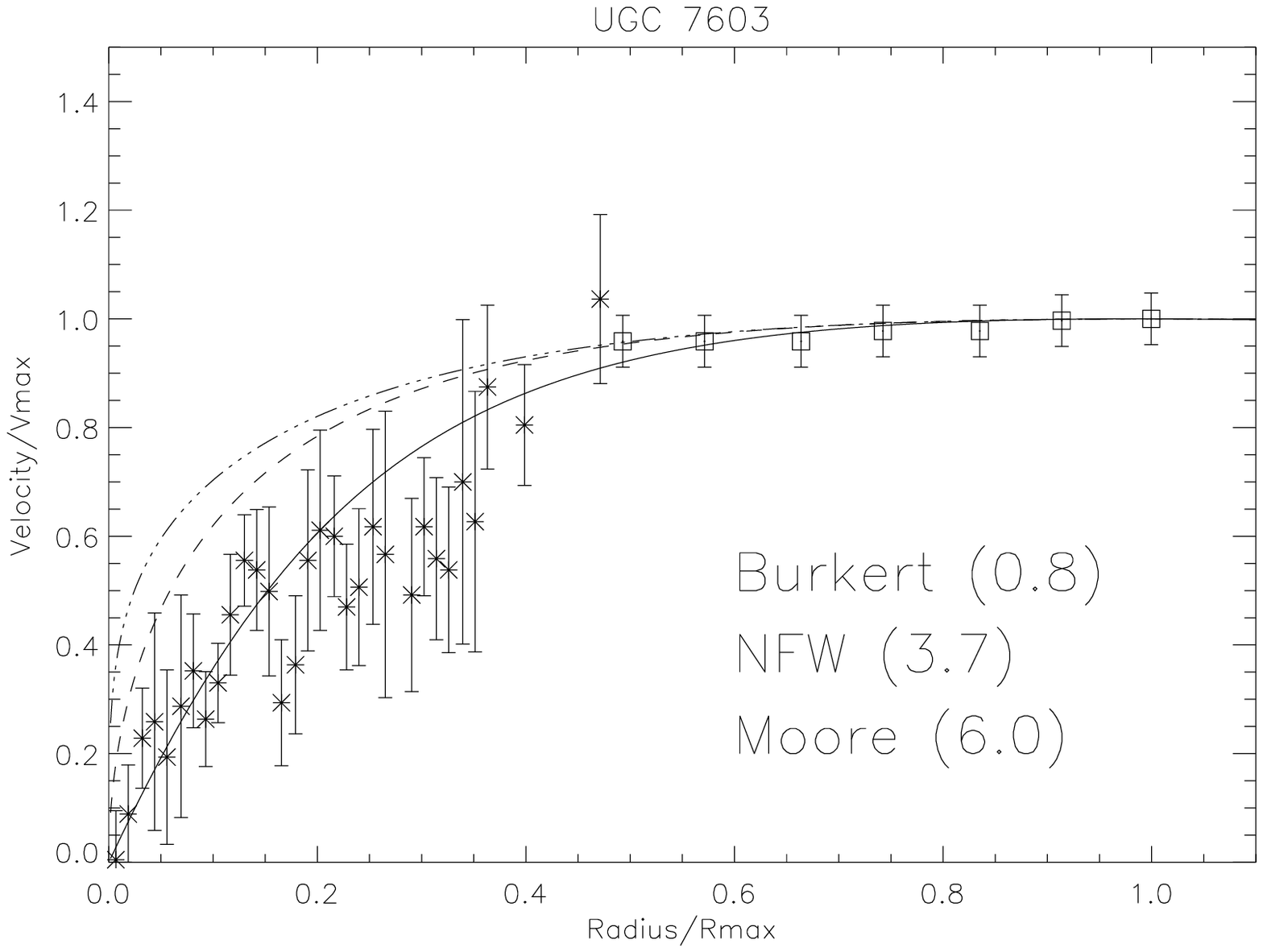}
\includegraphics[width=6.2cm]{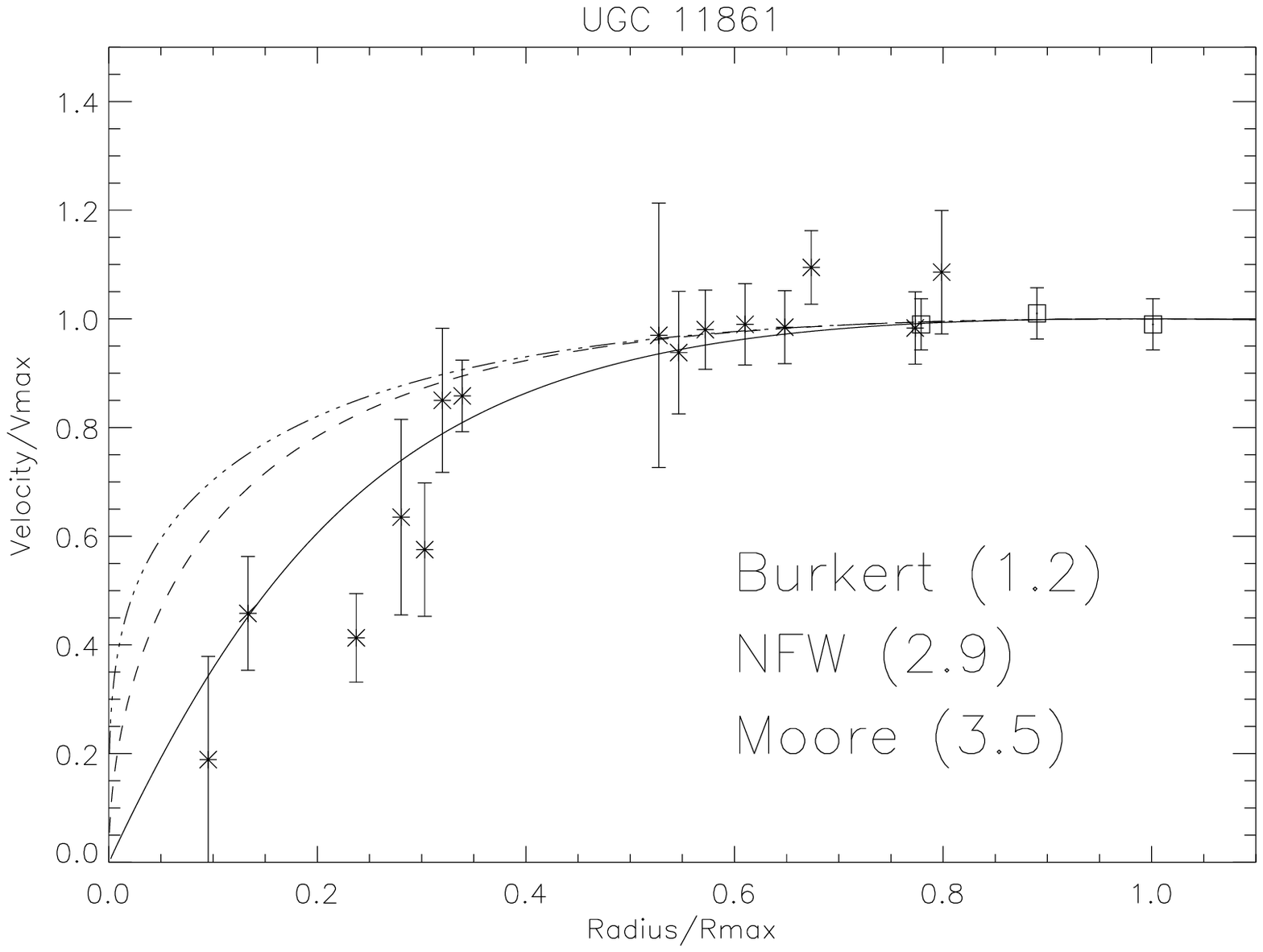}\\
\includegraphics[width=6.2cm]{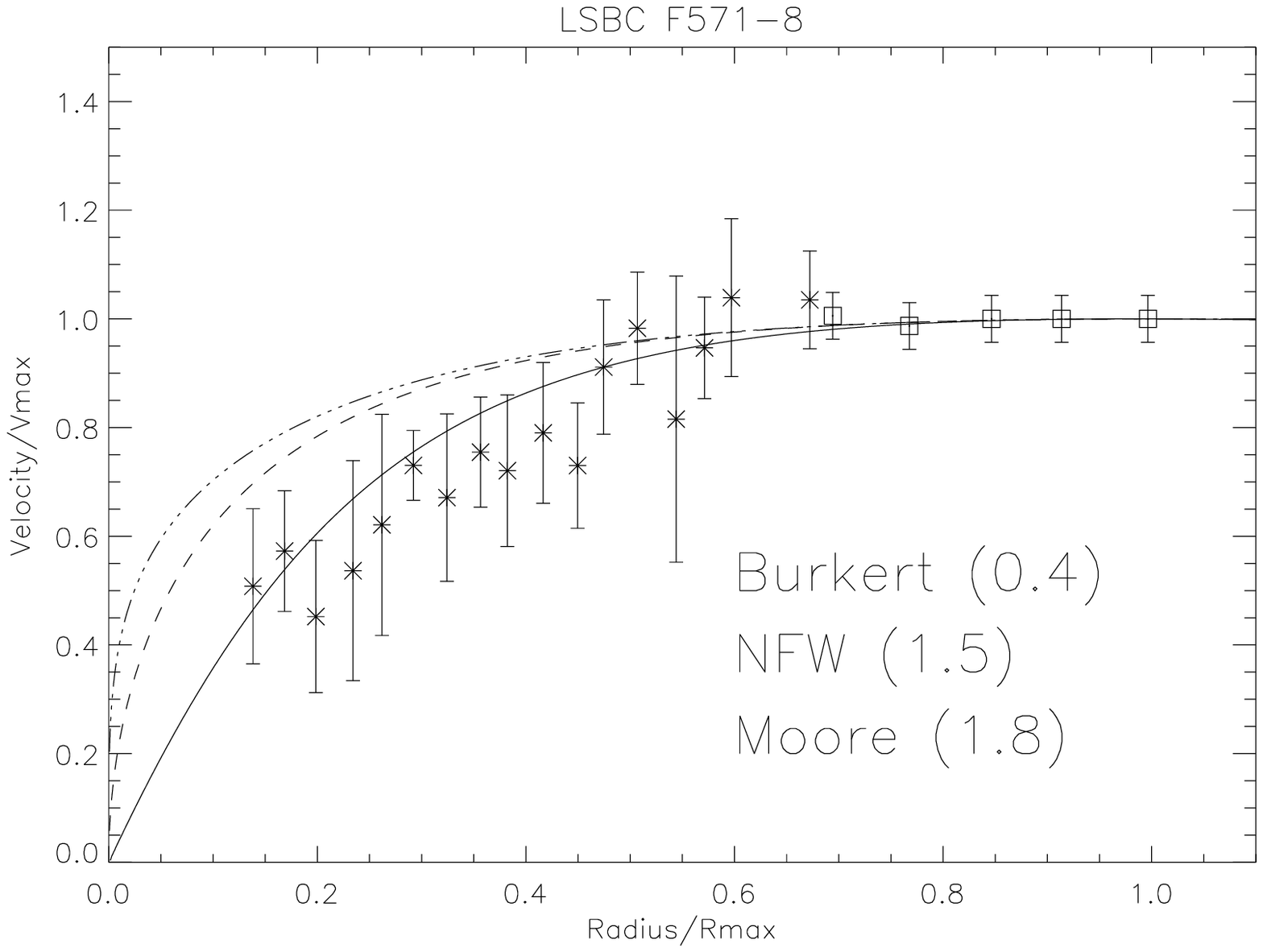}
\includegraphics[width=6.2cm]{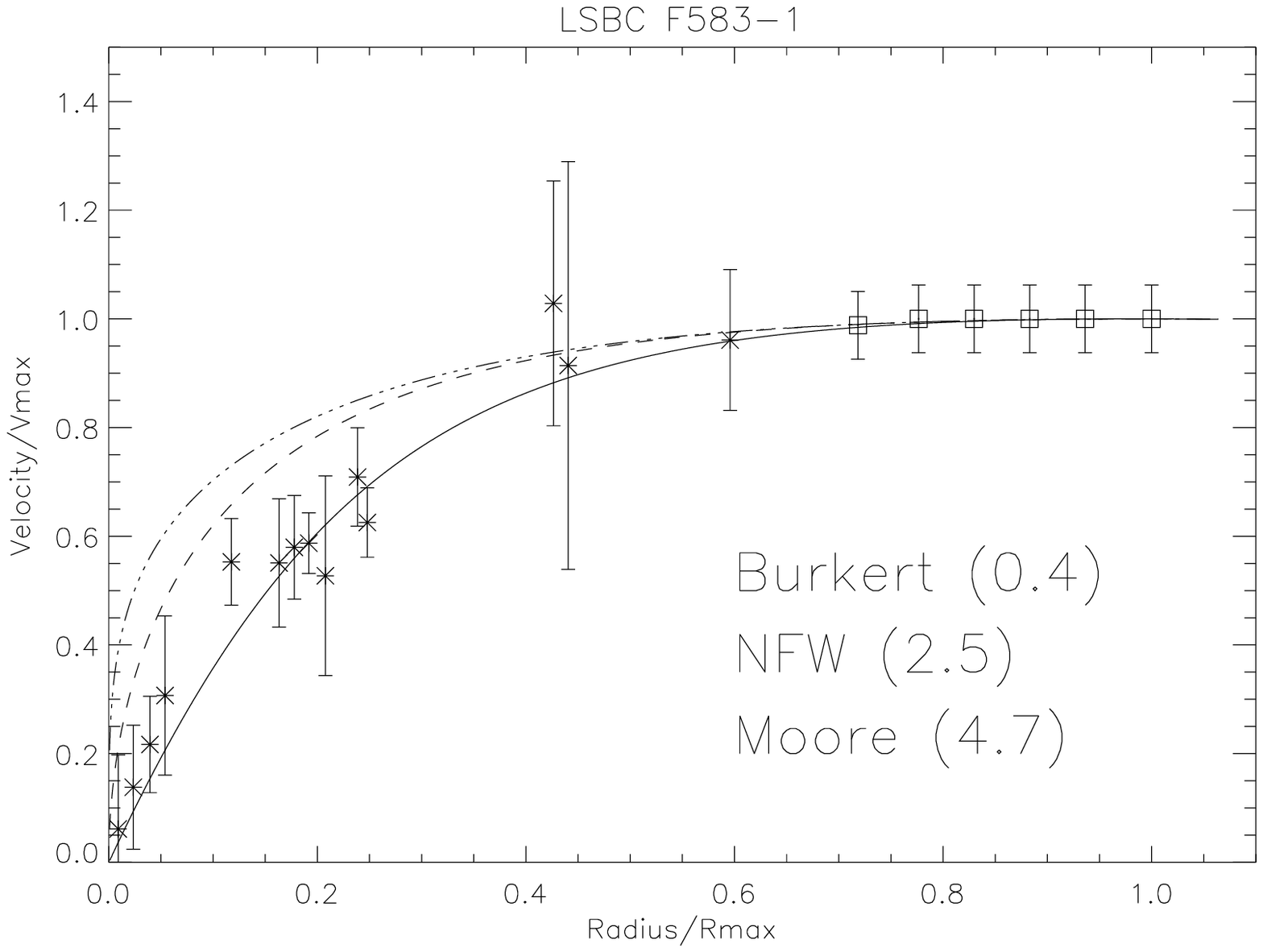}
\vskip -0cm
\caption{RCs of the LSB sample of \cite{marchesini02} compared to models in minimum disk hypothesis:
asterisks are the H$\alpha$ points, open squares are the HI data; the continuous line is the Burkert profile, the dashed line is the NFW profile, and the dotted-dashed line is the Moore profile; the numbers in parenthesis are the $\mathrm{\chi}_{\nu}^{2}$ for the three models}
\label{fig:marchesini02}
\vskip -0cm
\end{figure}

The analysis of \cite{marchesini02} on  high resolution H$\alpha$ and HI RCs of 4 late-type dwarf galaxies and 2 LSB galaxies, based on different halo models, is shown is Fig. \ref{fig:marchesini02}:
their findings are in favour of a Burkert profile.
NFW and the Moore profiles are inconsistent with the observed RCs in the inner regions in what they both predict a too fast rising RC because of the presence of the cuspy cores.

Owing to the many steps in the data analysis, however, there could be subtle systematics errors that could distort the results, or in any case render the results poorly constrained.
This has triggered the debate concerning the reliability of the data and the question of how well the mass models are constrained.

The earliest observations which indicated cores in LSB galaxies were 
two-dimensional 21 cm HI velocity fields \cite{flores94,moore94,blok_McGaugh_hulst96}.
Beam smearing (i.e., low spatial resolution) was
suggested to be a systematic effect that would erroneously indicate 
cores \cite{bosch00b,swaters00}.
This question was 
addressed by long-slit $H_{\alpha}$ observations which had an order of 
magnitude increase in spatial resolution (see, e.g., \cite{dBMR01,MRdB01}); 
yet cusps did not appear, showing that beam smearing had been 
of only minor importance in the HI observations.
Possible systematic
errors in the long-slit spectroscopy (e.g. \cite{simon03,rhee04,spekkens05}
like slit misplacement \cite{Rob} and non-circular motions
have since become the main concern.

An extensive modeling was then conducted  
in which the RCs of both cuspy and cored halos were
subjected to various effects and the conclusions were that no systematic effect
will entirely mask the presence of a cuspy halo for realistic
observing conditions (see \cite{dBBM03}, but also \cite{Rob}).

There are also claims that the observations could actually be
consistent with the DM density profiles predicted by
the CDM simulations, not only by considering the {H{\sc i}} data alone 
\cite{bosch00b,bosch01}, but also by combining H$\alpha$ and HI data \cite{Rob,primack02}.
This is the reason why particular care should be taken in choosing a suited
sample and in performing the data analysis.
Note that recent simulations (e.g. \cite{N03}) do not converge to a well-defined value of the inner slope down to the resolution limit,
even though the slope of the DM density profile 
(defined as $ -d{\rm ln}\rho/d{\rm ln}r$) at 1\% of the virial radius 
is still about $1.2$ for a typical galaxy.
Notice also that the observational results on
spiral galaxies show a discrepancy with the standard $\Lambda$CDM predictions
well beyond the resolution limit of the simulations.

In view of these discrepancies, together with the missing satellite problem (see e.g. \cite{moore99,kau93,kly99b}), many alternatives to the CDM  paradigm have been   proposed.
These include broken scale-invariance \cite{kam00,whi00}, warm DM \cite{som99,hog00}, scalar field DM \cite{peebles99,hu00,peebles00,matos00}, 
and various sorts of self-interacting or annihilating DM 
\cite{car92,spergel00,mohapatra00,firmani00,goodman00,kap00,bento00}.
Whereas  particle physics does not prefer CDM over these alternatives, it has the
advantage of having no free parameters.
Furthermore,  most of  these alternatives  seem  unable to solve  both
problems simultaneously \cite{hog00,moore99b,col00,dal00b},
and face their own problems \cite{burkert00,koc00,sellwood00,yos00,moore00}.
On the other hand, there are claims that the
sub-structure and core   problems might   be solved once    additional
baryonic    physics are taken   into   account.
Several studies  have
suggested  that processes such  as reionization and supernova feedback
can help to suppress star  formation and to decrease central densities
in  low-mass DM halos  (e.g., \cite{NEF96,gel99,bosch00b,BKW00,bin01}).

\begin{figure}[h!] 
\centering
\vskip -0cm
\includegraphics[width=9cm]{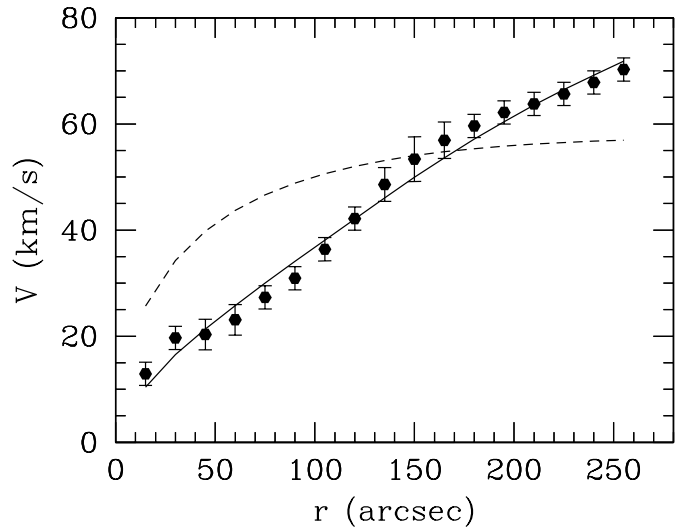}
\vskip -0cm
\caption{DDO 47 RC (filled circles) best-fitted by  Burkert halo + stellar disk (solid line) and by  NFW  halo + stellar disk (dashed line) mass models \cite{gentile05}.}
\label{fig:ddo47}
\end{figure} 

Whereas these processes  may  indeed   help to   solve the  problem   with the
over-abundance of  satellite  galaxies, the suggestion  that  feedback
processes can actually  destroy  steep  central cusps seems   somewhat
contrived  in light of  more   detailed simulations.
For instance,  as
shown by  \cite{NEF96}, the  effects are only  substantial if
large fractions of  baryonic mass are   expelled, which seems  hard to
reconcile  with the low  ejection  efficiencies found in more detailed
hydro-dynamical simulations (e.g., \cite{mac99,strickland00}).
In the recent work of \cite{tonini_erasing06}, it is proposed that angular momentum transfer from baryons to DM during the early stages of galaxy formation can actually flatten the halo inner density profile and modify the halo dynamics.
 
It is evident from the above discussion that the long-time popular CDM
paradigm is currently facing its biggest  challenge to date. 
The literature on the cusp vs core issue is vast and there is a clear consensus on that up to now there is not even one spiral that {\it requires} a NFW halo.

The importance of the issue, that concerns the very nature of DM, and the fact that these early results were questioned on several different aspects, has triggered new investigations characterized by the study of few proper test-cases with higher quality kinematical data, by means of properly devised  analysis \cite{gentile04}. 
These improvements were absolutely necessary in what to obtain reliable DM profiles requires extended, regular, homogeneous RCs reliable up to their second derivative and free from deviations from the axial symmetry.
Then, up to now, few tenths of objects have qualified to undergo such critics-free investigation (e.g. the list in \cite{gentile04,donato04,simon05,gentile05,3741}).
In all these cases data and simulations were found in plain disagreement on different aspects: the best-fit disk + NFW halo mass model 
\begin{itemize}
\item fits the RC  poorly and it implies  
\item an implausibly  low stellar mass-to-light ratio  and
\item an unphysical  high halo mass.    
\end{itemize}

\begin{figure}[h!]
\centering
\vskip -0cm
\includegraphics[width=3.9cm]{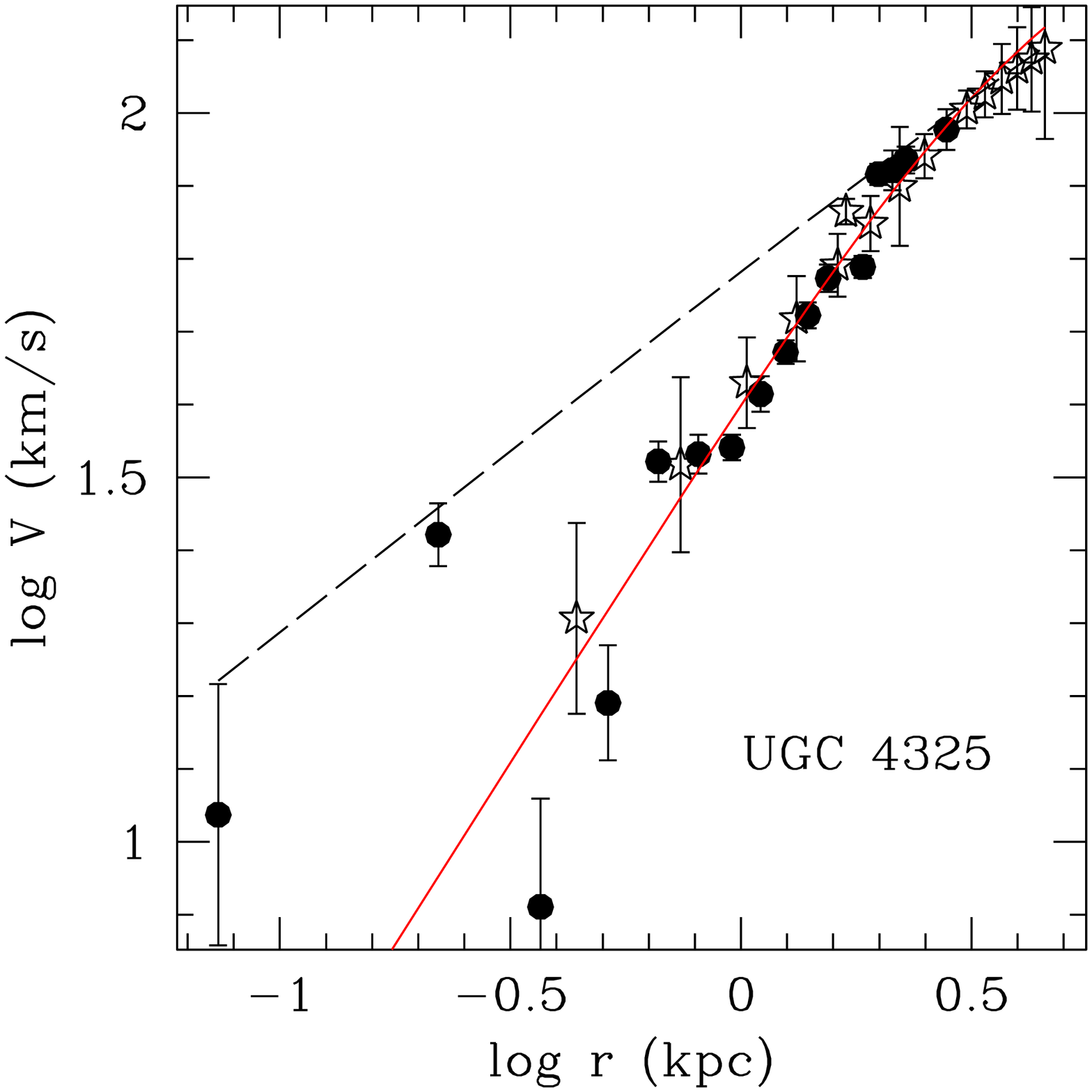}
\includegraphics[width=3.9cm]{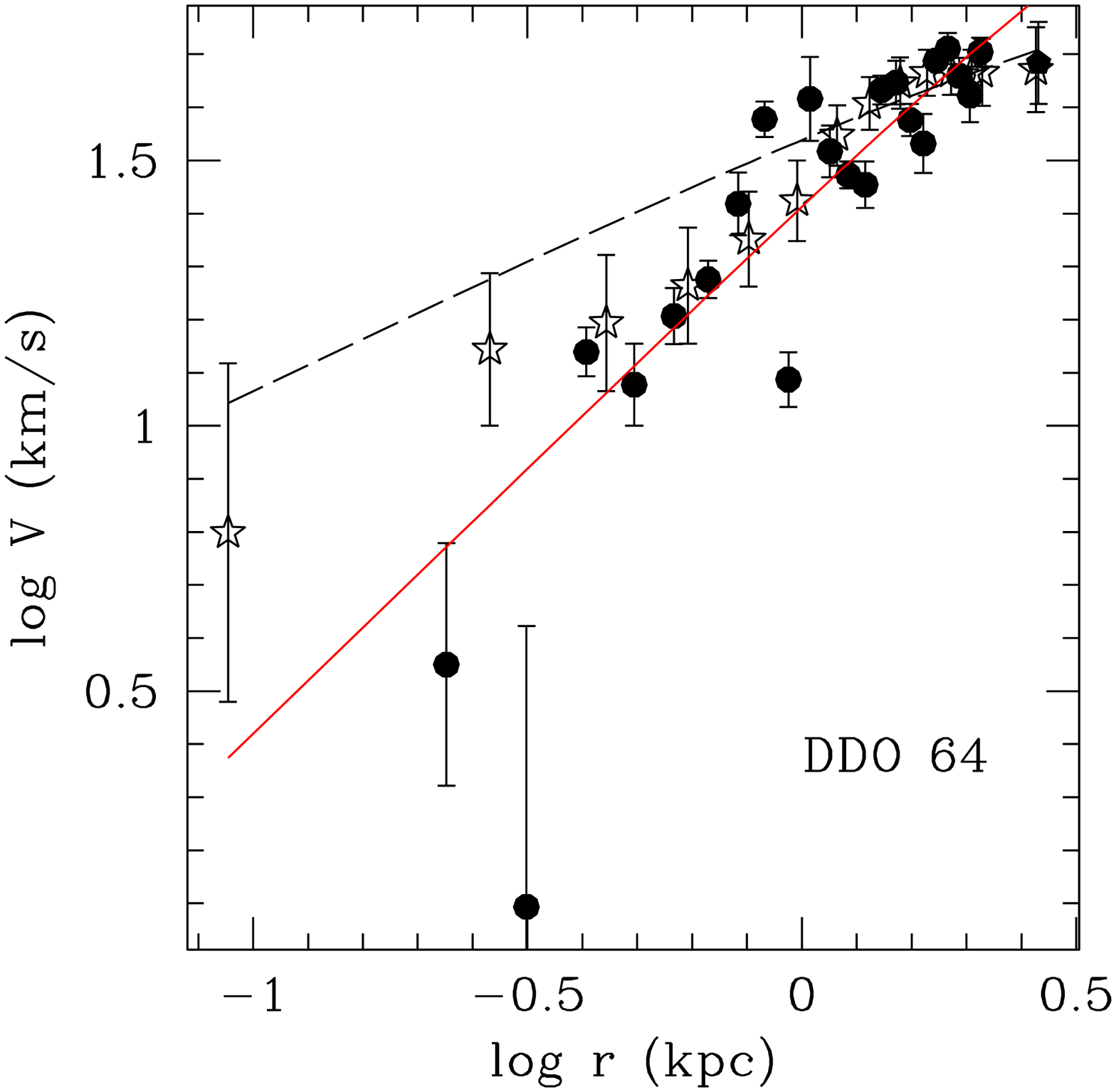}
\includegraphics[width=3.9cm]{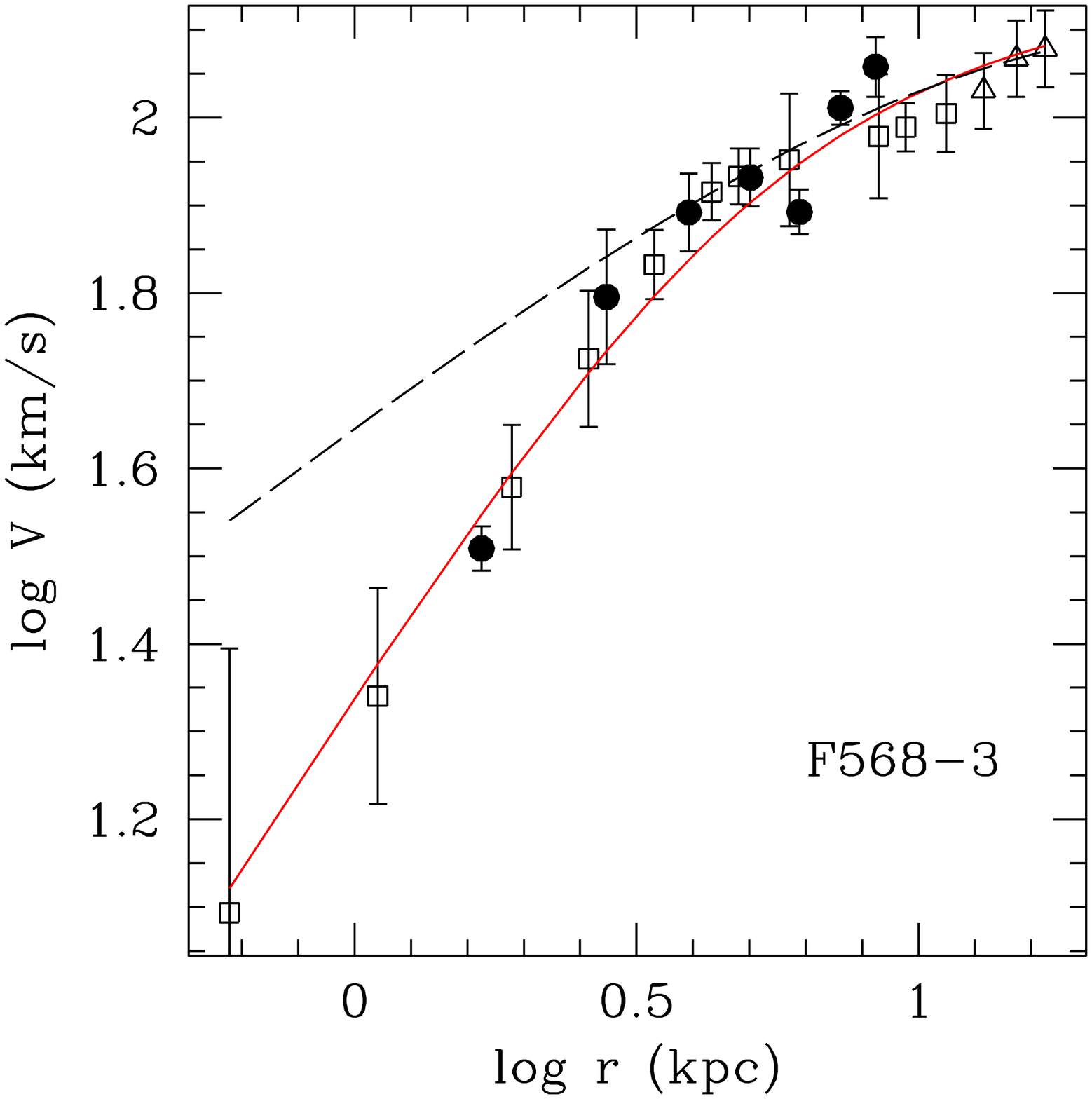}\\
\includegraphics[width=3.9cm]{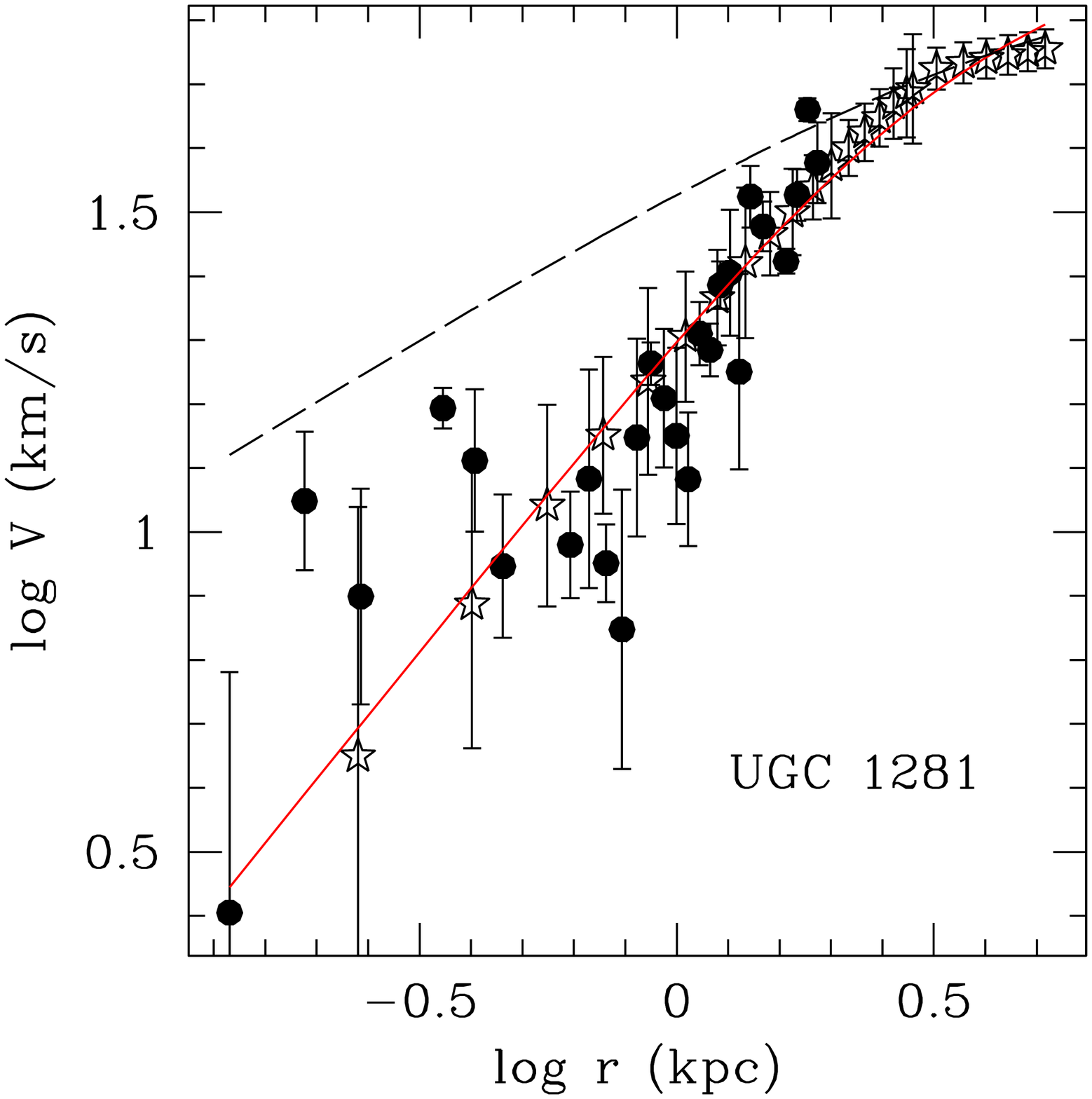}
\includegraphics[width=3.9cm]{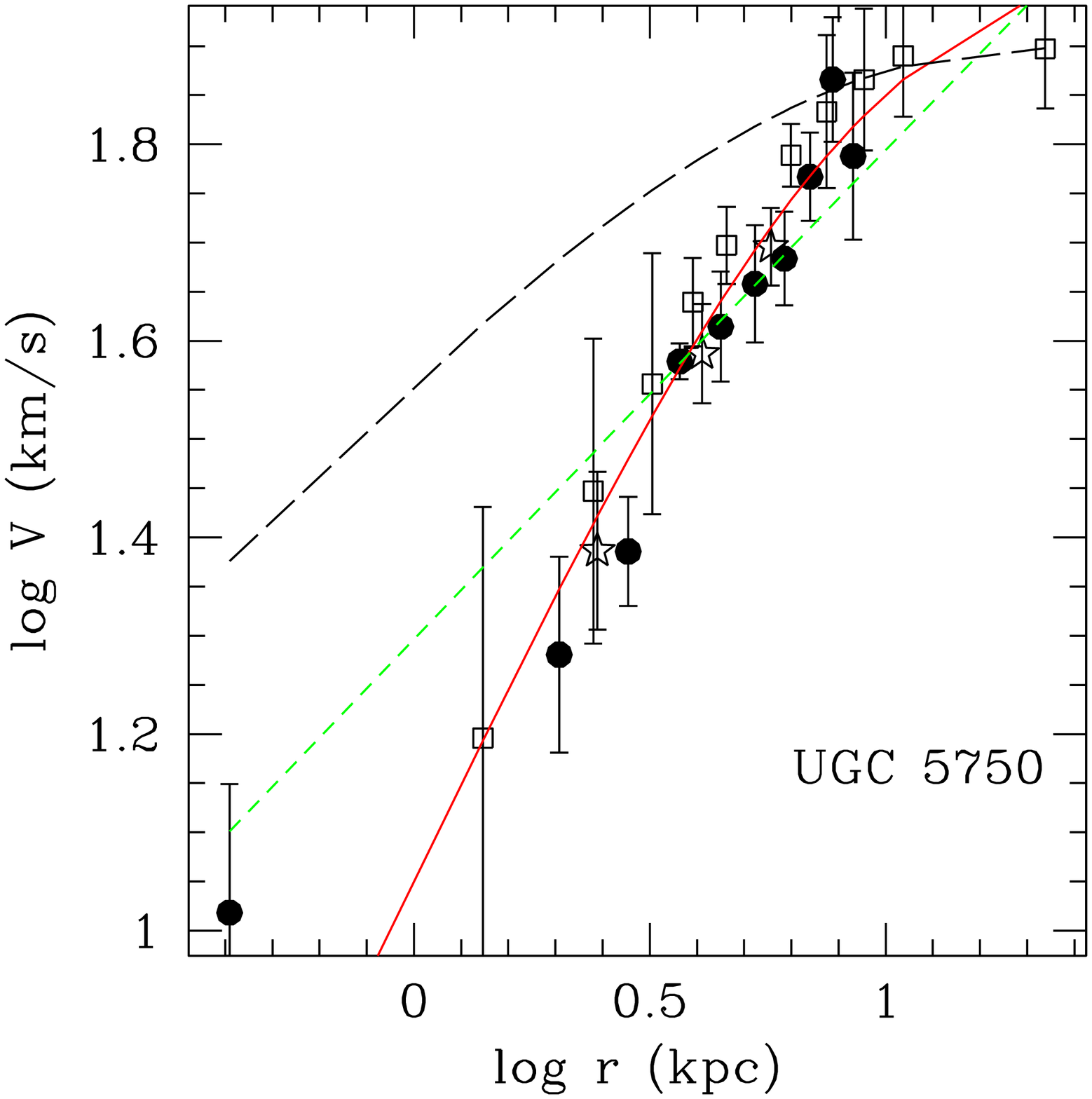}
\includegraphics[width=3.9cm]{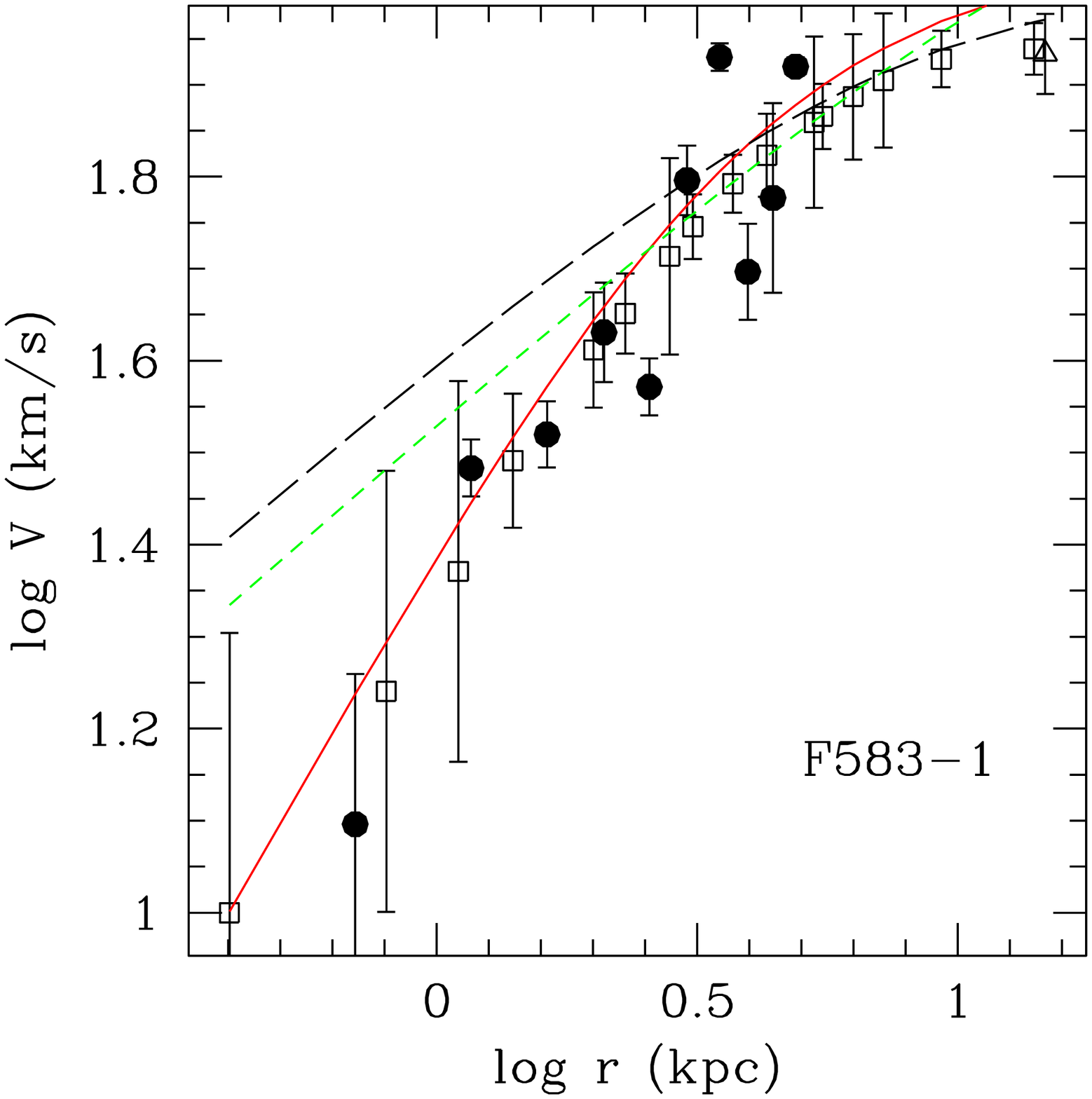}\\
\includegraphics[width=3.9cm]{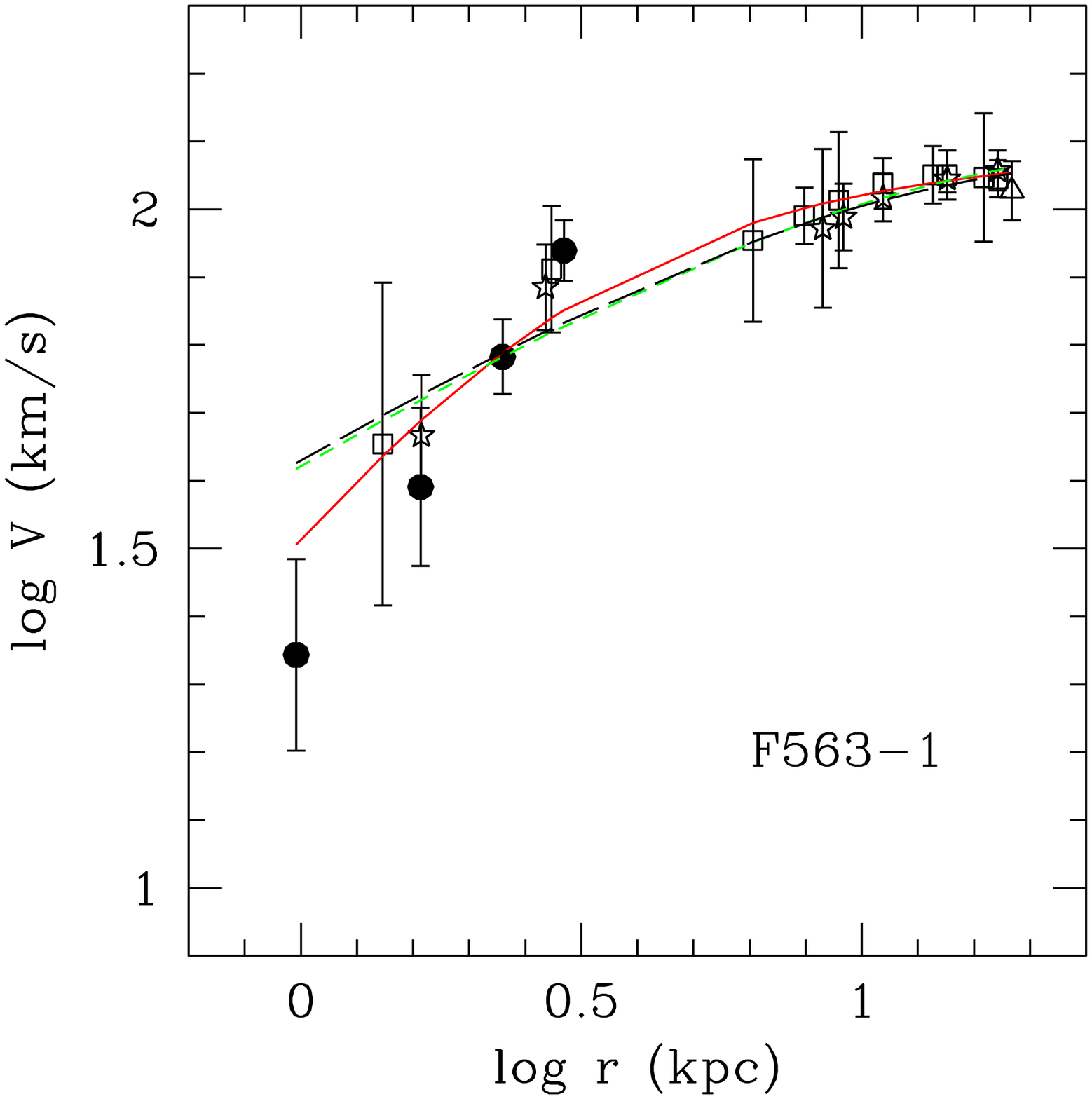}
\includegraphics[width=3.9cm]{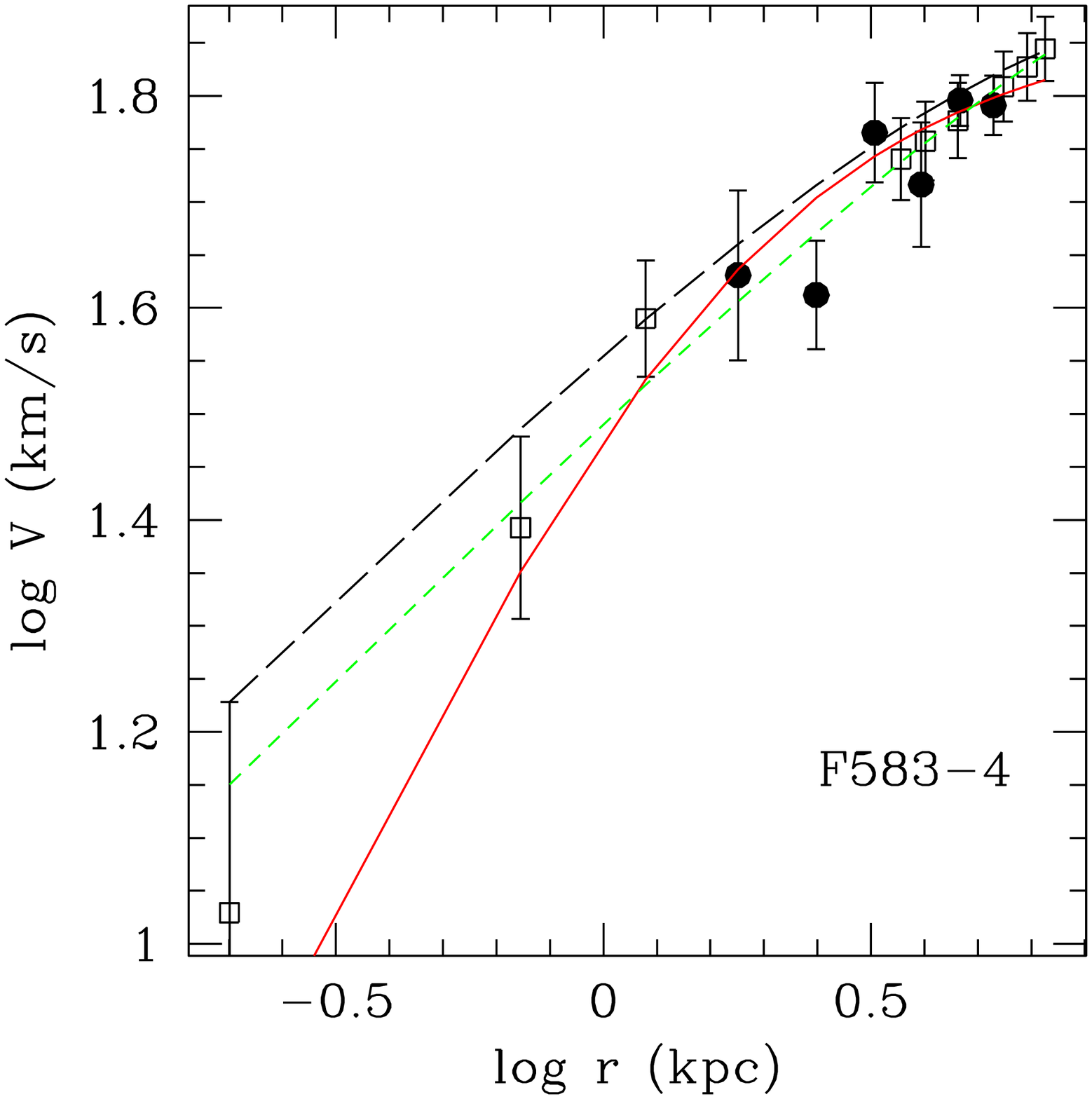}
\includegraphics[width=3.9cm]{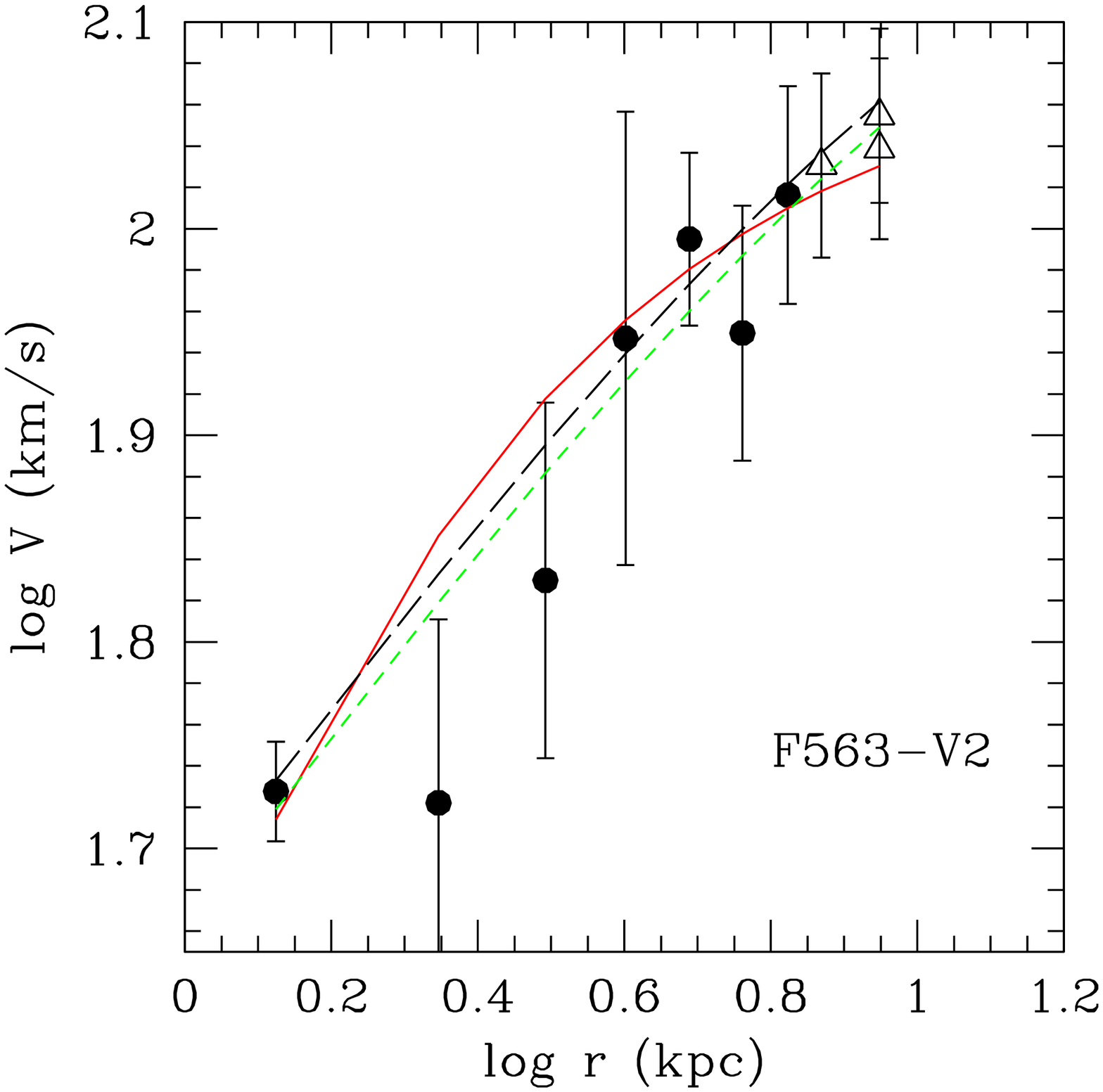}\\
\includegraphics[width=3.9cm]{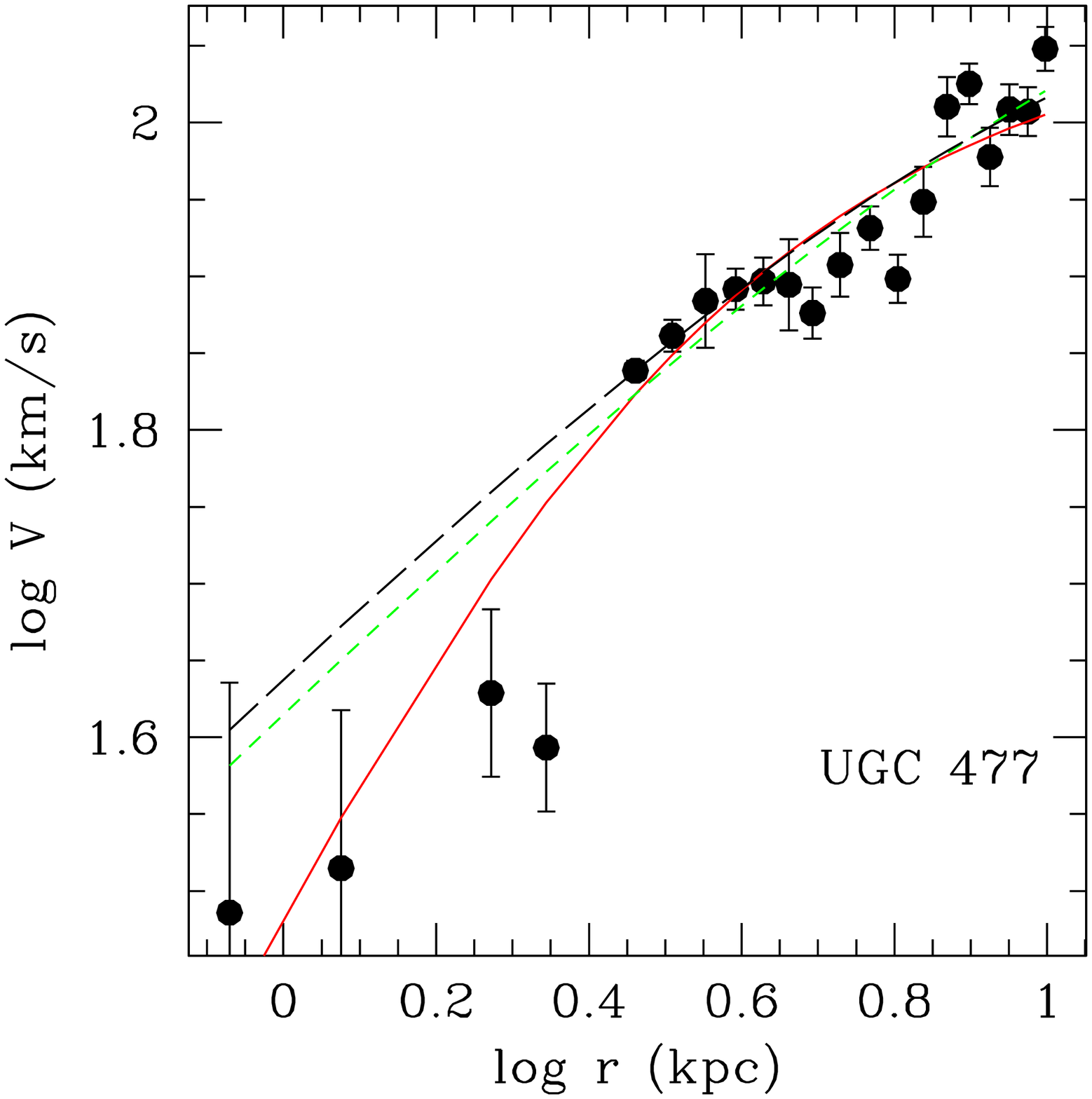}
\includegraphics[width=3.9cm]{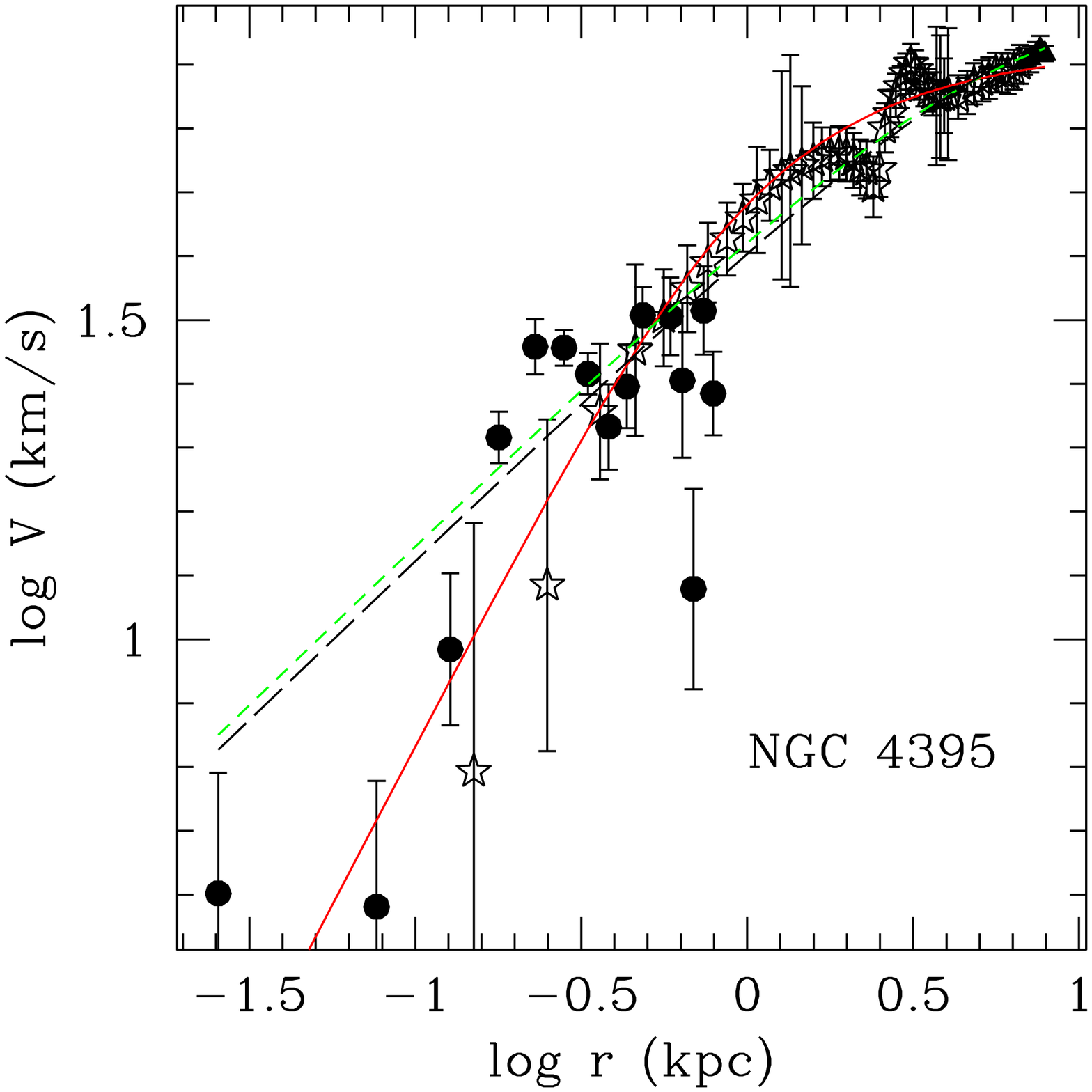}
\includegraphics[width=3.9cm]{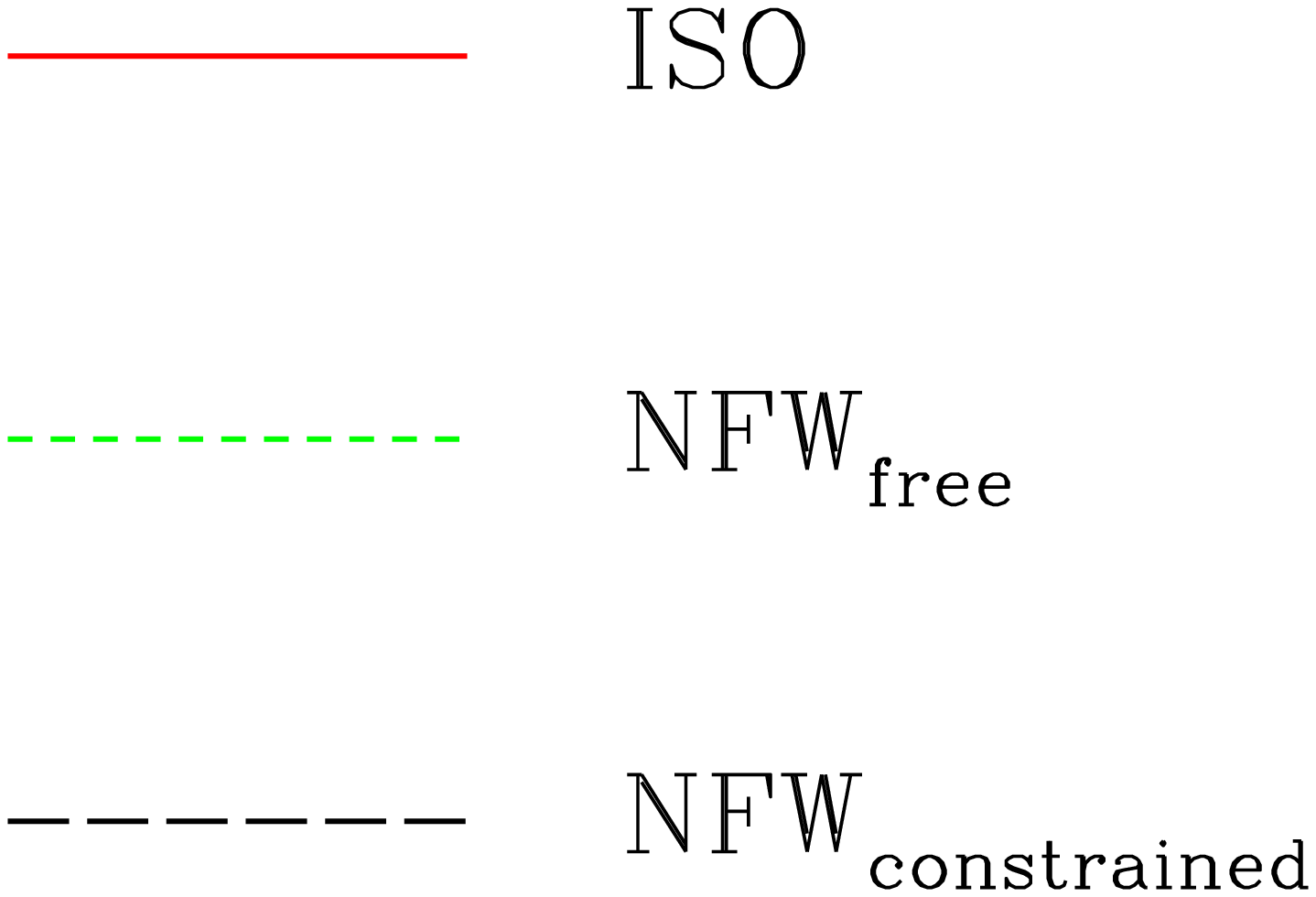}
\vskip -0cm
\caption{Halo pseudoisothermal (ISO) and NFW fits to the RCs sample of \cite{deNaray06} (circles), together with previous $H_{\alpha}$ (stars and squares) and HI data (triangles).
$NFW_{constrained}$ refers to fits that are constrained to matching the velocities at the outer radii while constraining reasonable values for the concentration parameter.}
\label{fig:deNaray}
\end{figure}

\begin{figure}[h!]
\centering
\vskip -0cm
\includegraphics[width=8cm]{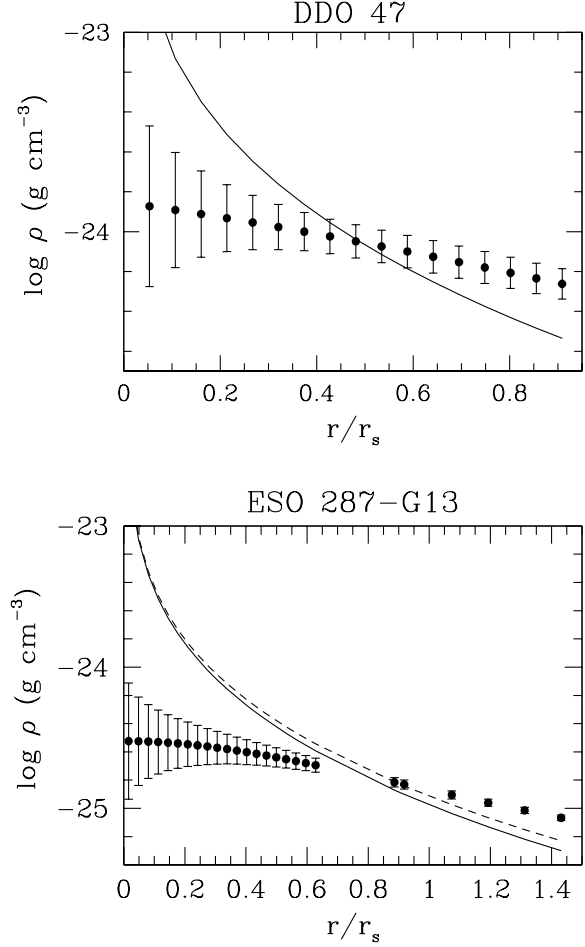}
\vskip -0cm
\caption{DM density profiles for DDO 47 and ESO 287-G13 ({\it dots}), as yielded by the best fits Burkert halo \cite{gentile_tonini07}.
{\it Solid lines}: NFW density profile such that the mass inside the last measured point is equal for the two profiles.
{\it Dashed line}: best-fit NFW for ESO 287-G13.}
\label{fig:rho_47_287}
\end{figure}

As an example, it is worth to discuss in detail the case of the nearby dwarf galaxy DDO 47 \cite{gentile05}.
The HI observations have adequate resolution and sensitivity, showing that the HI 2D kinematics is very regular, with a well-behaved velocity field.
The observed velocity
along the line of sight $V_{\rm los}$  has been decomposed  in terms of harmonic coefficients:  
$V_{\rm los} = c_0 + \sum_{j=1}^n [c_j {\rm cos}(j \psi) + s_j {\rm sin}(j \psi)]$ where $\psi$ is the azimuthal angle, $c_0 $ is the systemic velocity, $c_1$ is the rotation velocity; 
it is found that the coefficients $s_1$, $s_3$  $j_2$ have a small amplitude that excludes significant global elongation and lopsidedness of the potential and detects non-circular motions with amplitude and radial profile very different from that necessary to hide a cuspy density distribution in the observed RC.
The RC mass modeling, shown in Fig. \ref{fig:ddo47}, finds that the DDO 47 dark halo has a core radius of about 7 kpc and a central density $\rho_0 = 1.4 \times 10^{-24}$ g cm$^{-3}$, i.e. a {\it much} shallower distribution than that predicted by the NFW profile.
Is it possible to fit the data with Eq. \ref{eq:nfw} at all?
It gives the wrong shape and the clear prediction of CDM is simply not realized:
\begin{equation}
\rho_{NFW} \neq \rho_{obs}. 
\end{equation}

Fig. \ref{fig:deNaray} shows the best-fit pseudoisothermal ($\rho_{iso}(R)=\rho_0/[1+(R/R_c)^2]$) and NFW halos to the sample of \cite{deNaray06} of high-resolution 2D optical data combined with previous long-slit and HI RCs: 
the cored halo better represents the data in most cases.
Moreover, the NFW concentrations are mostly found to be too low.

An accurate mass modeling of the external regions of a couple of test-case spirals and a careful determination of the densities and enclosed masses of the DM halos at the farthest radii of 37 high quality RCs \cite{gentile_tonini07}, has brought to the discovery of a new problem for the NFW halos.
In addition to the well-known evidence for which in the inner 
regions of galaxies ($R< 2 R_{\rm D}$) the DM halos show a flattish density profile, with amplitudes up to  one order of magnitude lower than the $\Lambda$CDM predictions, at outer radii  ($R> 4  R_{\rm D}$)
the measured DM halo densities are found higher than the 
corresponding $\Lambda$CDM ones (see Fig. \ref{fig:rho_47_287}).
This implies an issue for $\Lambda$CDM that should be investigated
in future, when, due to improved observational techniques, the kinematic
information will be extended to the $\sim 100$ kpc scale \cite{3741}.
This new discrepancy provides additional information on the nature of the cusp vs core issue:
self-interacting or annihilating DM proposed as the cause 
for the inner discrepancy may be in difficulties in that it will cause a
rapid convergence to the NFW profile in the luminous parts of galaxies
and beyond once a critical density value is reached.

\section{Final remarks: intriguing evidences}
The distribution of luminous and DM in galaxies shows amazing properties and a  remarkable systematics  hat make it 
as one of the hottest cosmological issues.
There is no doubt that this emerging observational scenario   will be decisive 
in guiding  how  the $\Lambda$CDM-based theory of galaxy formation must evolve to meet the challenge that the observational data are  posing.     

\begin{figure}[h!]
\centering
\vskip -3.7cm
\includegraphics[width=12cm]{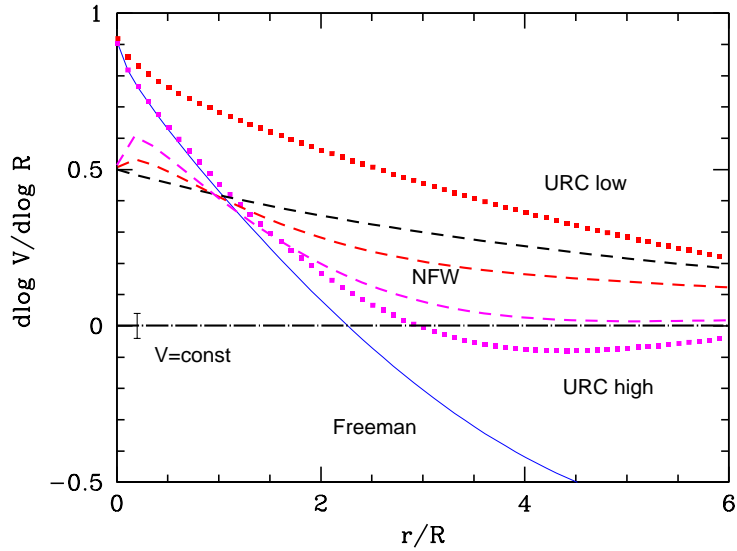}
\vskip -0.5cm
\caption{Logarithmic circular velocity slope as radius function in scale length units.
Lines represent: full blue, the Freeman disk for a typical massive spiral;
dashed (dotted) coloured, the NFW+disk (URC) model for a typical object with high (magenta) and low (red) luminosity;
dashed black, the NFW halo for a low luminosity object;
dot-dashed, a flat RC (V=const).}
\label{fig:n2}
\end{figure}

In all cases studied up to date   a  serious data-prediction  discrepancy  emerges, that becomes  definitive  when we  remind that the actual $\Lambda$CDM halo profiles  are  steeper than  the standard NFW ones  considered here  and that the   baryonic  adiabatic collapse  has  likely  contracted them further.       
As a final remark I present in Fig. \ref{fig:n2} a plot of the logarithmic circular velocity slope out to 6 scale lengths, of a stellar disk + halo model, defined as $\frac{d log \:V}{d log\:R}$.
The blue line represents the Freeman disk for a typical massive spiral;
Dashed coloured (dotted) lines represent the NFW+disk model (URC), for a typical object with high (magenta) and low (red) luminosity.
Dashed black line represent NFW halo for a low luminosity object.
Dot-dashed line represent a flat RC (V=const).
From these simple figure it clearly comes out that the ``flat rotation curve'' paradigm is not only a completely wrong assumption, but on top of it models having this starting point loose all crucial information of distinguishing one model to the other.
  
\chapter{Alternatives to Dark Matter}

Despite the important achievements of the CDM paradigm, as described in the previous chapters, this model requires however that the the overall dynamics of the Universe is dominated by two mysterious forms of matter and energy.

As far as the only evidence of dark matter or dark energy is of gravitational origin it is reasonable to imagine that what we observe is neither a new form matter nor energy but a deviation of the law of gravitation from General Relativity. 
It has to be noticed that deviations from general relativity are well motivated also from a pure theoretical point of view (string theory for example requires extra dimensions, possibly even ``large'', which may cause deviations from the Newtonian gravitational attraction in the sub-millimeter scale). 

In the this chapter two models of alternatives to dark matter are briefly discussed: MOND (MOdified Newtonian Dynamics) and $f(\mathcal{R})$ theories.

\section{MOND}

MOND (\cite{milgrom}, see \cite{mond_review} for a review) is certainly one of the most studied (and successful) model of modification of gravity.
In its original formulation it was a pure phenomenological description (without o proper ``theory'' behind) constructed to explain two observational systematics of spiral galaxies:
\begin{itemize}
\item the misleading paradigm of flat rotation curves
\item the existence of a relationship between rotational velocity and luminosity (the Tully-Fisher relation) which implies a mass-velocity relationship of the form $M\propto V^\alpha$, with $\alpha \sim 4$.  
\end{itemize}

The basic idea of MOND is to introduce a fundamental acceleration scale, $a_0$, below which deviations from the Newtonian dynamics appear; previous studies \cite{BBS} found that $a_0\sim  1.2 \times 10^{-8} \mbox{ cm s}^{-2}$ (notice that $a_0 \sim c H_0$).
Two formulations of the original model exist depending on whether the law of inertia or gravity is modified.
Notice that in both cases the model clearly does not respect the strong equivalence principle.

In the first case the modified law of inertia proposed by Milgrom is:
\begin{equation}
m \vec{a} \mu(\frac{\left|\vec{a}\right|}{a_0}) = \vec{F},
\end{equation}
where $\mu(x)$ is an interpolation function whose asymptotic values are $\mu(x)\simeq x$ for $x\ll 1$ (called Mondian regime) and $\mu(x)\simeq 1$ for $x\gg 1$ (called Newtonian regime).    

Formulated as a modification of gravity the model takes the following form:
\begin{equation}
\vec{g} \mu(\frac{\left|\vec{g}\right|}{a_0}) = \vec{g_N}, \label{eq:mondeq}
\end{equation}
where $\vec{g}$ is the effective acceleration and $\vec{g_N}$ is the standard Newtonian acceleration.

In the Mondian regime the effective gravitational acceleration takes the form $g=\sqrt{g_N a_0}$.
Assuming a gravitational field generated by a point source of mass $M$ and imposing the condition for circular orbits ($g=\frac{v^2}{r}$) the circular velocity can be calculated:
\begin{equation}
v^4=G M a_0 ,
\end{equation}
where $G$ is the gravitational constant.
Notice that the circular velocity does not depend on the radius in the Mondian regime (and hence the concept of flat RCs) and a Tully-Fisher relation is obtained respecting the original requirements of the model.

Even though in a general case a modified version of the Poisson equation
should be solved, Eq. \ref{eq:mondeq} can be shown to be a good approximation
for axisymmetric disks \cite{BM95}.
The interpolation function has been given usually the following functional form:\begin{equation}   
\mu_{\rm orig}(x)=\frac{x}{\sqrt{1+x^2}}.   
\label{eq:orig}   
\end{equation}

However, it is obvious that a whole family of functions are compatible with
the required asymptotic behaviours.
For instance, \cite{FB05} proposed that
\begin{equation}   
\mu_{\rm FB}(x)=\frac{x}{1+x}    
\label{eq:FB}   
\end{equation}   
could be a better choice in what it is compatible
with the relativistic theory of MOND put forward by Bekenstein \cite{teves}.
\cite{F07} showed that Eq. \ref{eq:FB} leads to a slightly
different value of $a_0$: $a_0 = 1.35 \times 10^{-8}$ cm s$^{-2}$.

The model as described above faces a fundamental conceptual difficulty in analyzing composite systems, in fact, at the microscopic level the characteristic acceleration of atoms and molecule is never in the Mondian regime.
However although micro-systems are not Mondian the composite system in certain circumstances is.
Recipes are then necessary  for the applicability of MOND on many body systems where the internal accelerations are above the MOND scale while the external acceleration is below that scale.
For this reason MOND as conceived in its original form can not be considered as a satisfactory theory.

It is clear that RCs are not asymptotically flat as originally assumed in construction of MOND.
An asymptotically gently decreasing RC however can be obtained whenever the disk surface density is of order $\Sigma_d \simeq a_0 / G$ or above.
In this case in fact the internal accelerations of the disk breaks the Mondian regime allowing for a quasi-Newtonian decline.
  
Although MOND is constructed to obtain ``flat'' RCs, it is able to fit a number of RCs and in many cases it correctly predicts general scaling relations linked to RCs \cite{mond_review,BBS,kent87,milgrom88,sanders96,blok_mcGaugh98,McGaugh04,McGaugh05}.

In the case that Eq. \ref{eq:orig} is used as the interpolation function, then within the MOND framework the observed circular velocity $V_{\rm obs}(r)$ can be expressed as a function of $a_0$ and the Newtonian baryonic contribution $V_{\rm bar}(r)$ to the RC:
\begin{equation}
V_{\rm obs}^2(r)=V_{\rm bar}^2(r)+V_{\rm bar}^2(r) 
\left(\sqrt{\frac{1+\sqrt{1+\left(\frac{2ra_0}{V_{\rm bar}^2(r)}\right)^2}}{2}}-1\right),
\label{eq:vorig}
\end{equation}
where $V^2_{\rm bar}(r)=V_{\rm stars}^2(r)+V_{\rm gas}^2(r)$ (ignoring the contribution of the bulge), $V_{\rm stars}(r)$ and $V_{\rm gas}(r)$ are the Newtonian
contributions to the RC of the stellar and gaseous disks, respectively (see \cite{milgrom}).
The amplitude of $V_{\rm stars}(r)$ can be scaled according to the chosen, or fitted, 
stellar mass-to-light ($M/L$) ratio. $V_{\rm gas}(r)$ is derived from
HI observations, when they are available.

If instead Eq. \ref{eq:FB} is used the equivalent of Eq. \ref{eq:vorig} becomes:
\begin{equation}
V_{\rm obs}^2(r)=V_{\rm bar}^2(r)+V_{\rm bar}^2(r) 
\left(\frac{\sqrt{1+\frac{4 a_0 r}{V_{\rm bar}^2(r)}}-1}{2}\right)
\label{eq:vFB}
\end{equation}
(see e.g. \cite{richtler08}). 
Note that the second term of the right-hand side of Eqs. \ref{eq:vorig} and  \ref{eq:vFB} acts as a ``pseudo-dark matter halo'' term and it is completely determined by the luminous matter. 
As expected, it vanishes in the limit $a_0 \rightarrow 0$.

\begin{figure}[h!]
\centering
\includegraphics[width=6cm]{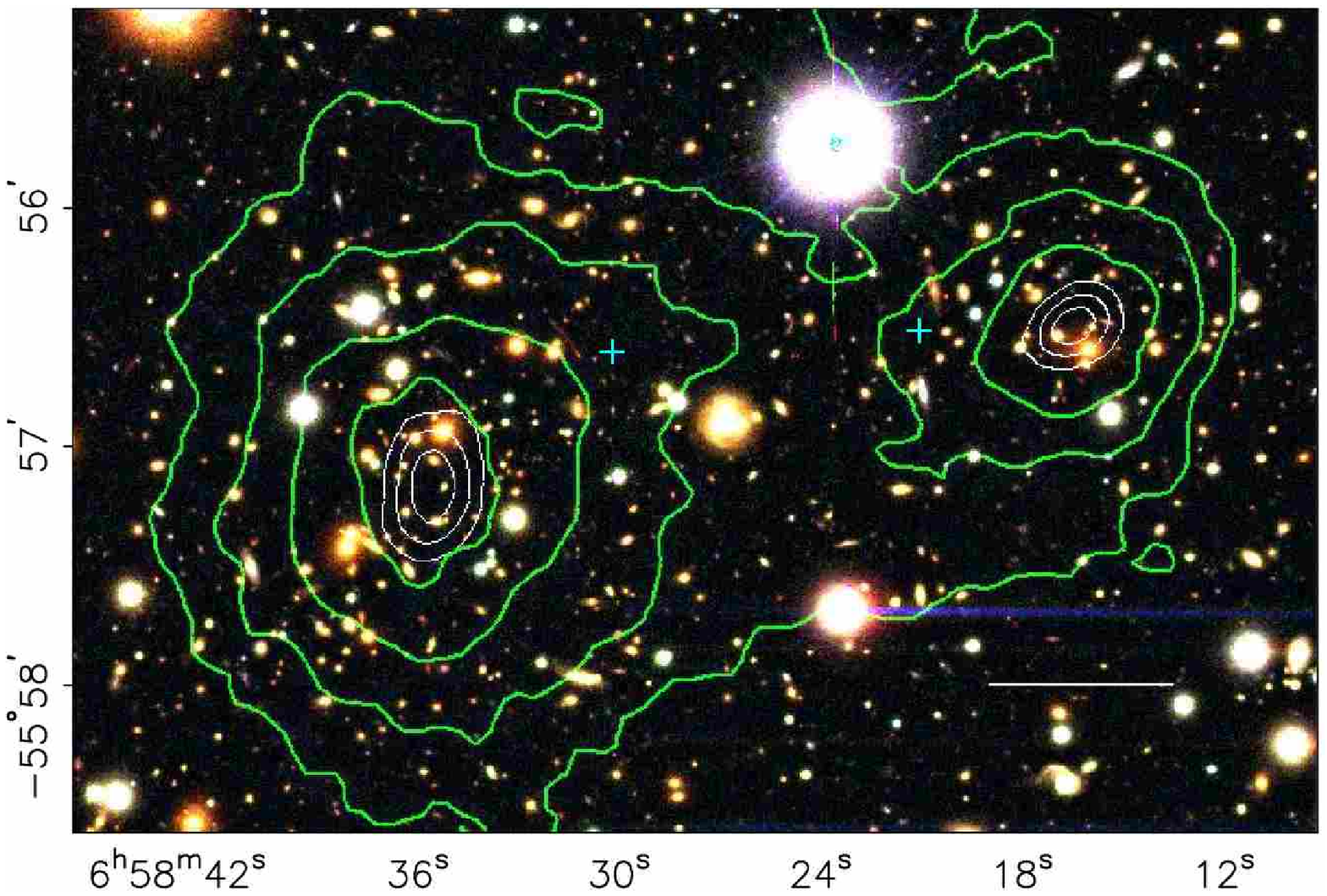}
\includegraphics[width=6cm]{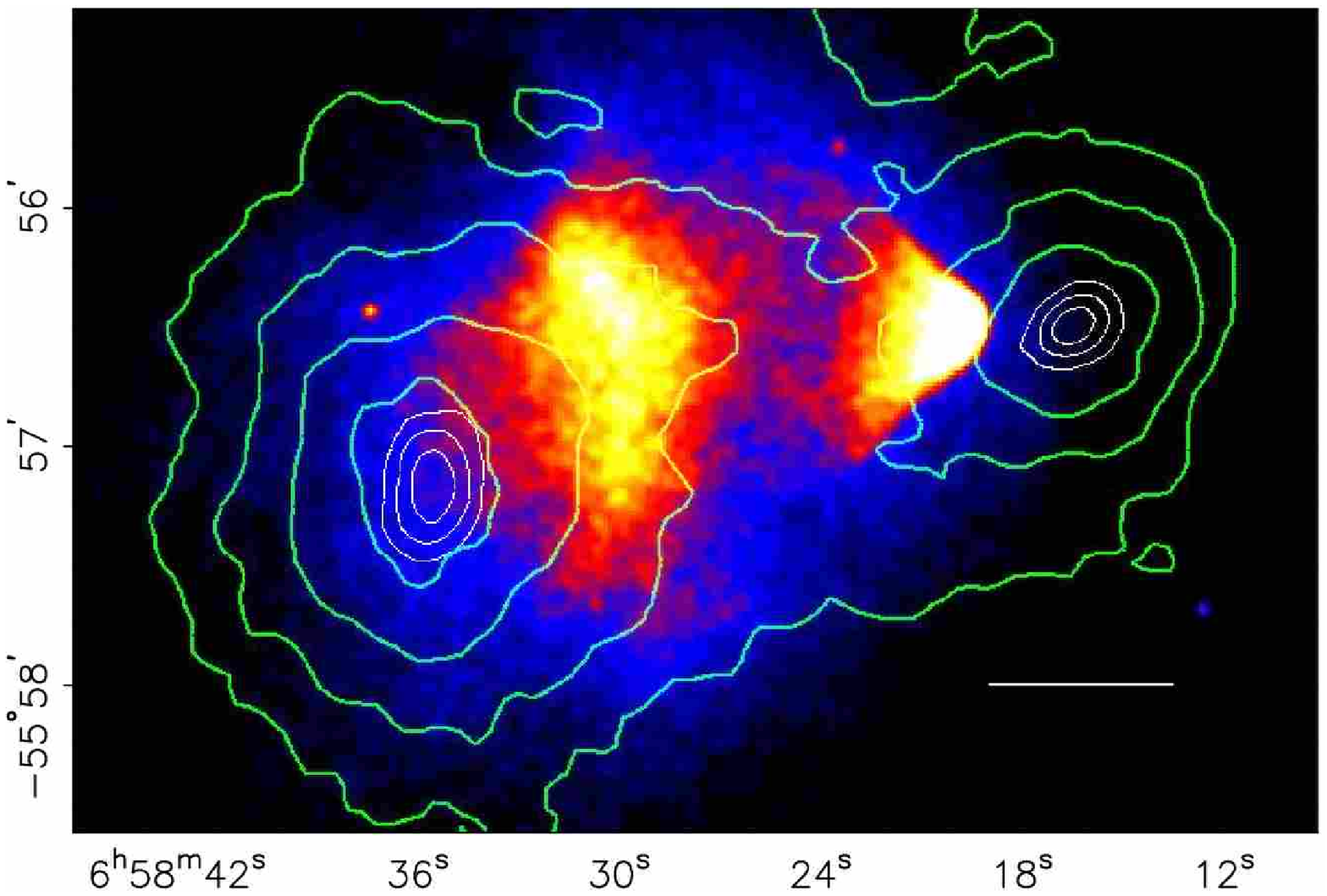}
\caption{
On the left is a color image of the merging cluster 1E0657$-$558, with the white bar indicating 200 kpc at the distance of the cluster \cite{bullet}.
On the right is an imagine of the same cluster from Chandra measurements.
Shown in green contours in both panels are the weak lensing reconstruction.
The blue $+$s show
the location of the centers used to measure the masses of the plasma
clouds.}
\label{fig:bullet}
\end{figure}

The MOND model can be applied also to pressure-supported systems.
Assuming an isotropic isothermal system of total mass $M$, it is possible to obtain a relation between the radial velocity dispersion $\sigma_r$ and the mass density distribution $\rho$:
\begin{equation}
\sigma_r^4=G M a_0 \left( \frac{d ln( \rho)}{d ln(r)} \right)^{-2},
\end{equation}
 which in turn implies:
\begin{equation}
\frac{M}{10^{11}M_\odot}\simeq\left(\frac{\sigma_r}{100 \mbox{km s}^{-1}}\right)^4,
\end{equation}
which is similar to the observed Faber-Jackson relation for elliptical galaxies.

Clearly the empirical formulation of MOND faces difficulties when compared with effects which originates from General Relativity (i.e. space-time curvature), in particular the gravitational lensing.
To properly study these effects a covariant MOND formulation is necessary.
After several attempts in 2004 (more than 20 years after the original MOND formulation!) it was proposed by  J. D. Bekenstein a Tensor-Vector-Scalar (TeVeS) field theory which correctly reproduces the main MOND features at small accelerations while preserving the Newtonian regime for higher accelerations \cite{teves}.
This theory allows not only the comparison of MOND with the lensing measurements but also the development of a theory for structure formation and a comparison with the modern cosmological measurements. 

Unfortunately the resulting theory is unable to fit the WMAP results in a pure baryonic framework and a form of DM is necessary \cite{McGaugh04_wmap}.
The kind of DM required by MOND however may well be represented by neutrinos with a mass of $2 \mbox{eV}$ (see e.g. \cite{angus07_neutrinos}), very close of the present experimental limits \cite{mass_neutrinos}.  

A serious challenge to the MOND-TeVeS theory came recently from the measurements of two merging cluster of galaxies (called ``Bullet Cluster'' \cite{bullet}).
In this cluster it has been shown by means of gravitational lensing  as well as x-rays measurements that the lensing source is misplaced with respect to the baryons (see Fig. \ref{fig:bullet}).
Also in this case however the presence of neutrinos with a mass of $2 \mbox{eV}$ would rescue MOND from a failure \cite{angus07_neutrinos}.
The on-going experiment KATRIN \cite{katrin} however will be able soon to probe the range of neutrino masses necessary for MOND.

Clearly the necessity of a form of DM also in MOND completely remove the beauty of the original proposal (although not excluding the model completely).

\section{$f(\mathcal{R})$ theories}

This class of theories of gravitation has been studied since the very beginning of the appearance of General Relativity \cite{weyl19,eddington23}.
General Relativity as formulated by Einstein suffers from well known problems (among them the problem of singularities and the lack of a full quantization).
The early attempts in modifying Einstein original theory went in the direction of solving the above mentioned problems \cite{utiyama62,stelle77,brandenberger93}.
More recently however the discovery of the dark components of the Universe modified the trend motivating theorists in finding a gravitational theory which correctly describes the observed space-time geometry without the hypothesis of unknown microphysics.

Considering  the Einstein equations:
\begin{equation}
 R_{\mu \nu} - \frac{1}{2}g_{\mu \nu}R = 8\pi G T_{\mu \nu},
\end{equation}
where $G$ is the Newton Constant, $R_{\mu \nu}$ and $R$ are the Ricci Tensor and Scalar and $T_{\mu \nu}$ is the stress-energy tensor, the dark component is defined by the difference between the observed stress-energy tensor and the measured geometrical quantities:
\begin{equation}
\left(T_{\mu \nu}\right)_{dark}  =  \frac{1}{8 \pi G} \left( R_{\mu \nu} - \frac{1}{2}g_{\mu \nu}R \right)_{meas.} -  \left(T_{\mu \nu}\right)_{vis.}.
\end{equation}
The idea behind the modern version of $f(R)$ theories is that $\left(T_{\mu \nu}\right)_{dark}$ originates from a modification of the fundamental Einstein equations rather than a new form of Dark Energy or DM.

The Einstein equations are obtained from the Einstein-Hilbert action which reads:
\begin{equation}
S_{EH}=\frac{1}{16 \pi G}\int d^4 x \sqrt{-g} R\label{eq:eh}.
\end{equation}
From this action however two variational  methods can be used to obtain the correct equations: the metric approach (where the variation is considered on the metric) or the Palatini approach (where variations of the metric and of the connection are assumed to be independent, for reviews see \cite{capoz_franca08,thomas_review}.

In $f(R)$ theories the starting action is a straightforward generalization of the Einstein-Hilbert one:  
\begin{equation}
S_{f}=\frac{1}{16 \pi G}\int d^4 x \sqrt{-g} \: f(R).
\end{equation}
In this case however the two variational approaches lead to different equations of motion.
Actually for these kinds of theories a third variational method exists called \emph{metric affine $f(R)$ gravity} \cite{sotiriou_liberati07} where also the matter action (not included in Eq. \ref{eq:eh}) is supposed to depend on the connection.
In what follows only the metric $f(R)$ theory will be considered.

In general the action will be the sum of the contribution from the theory of gravity and the contribution from the theory of matter: $S_{tot}=S_f+S_m$.
Applying the metric variation the following equations are obtained:
\begin{equation}
f'(R)R_{\mu \nu}-\frac{1}{2}f(R)g_{\mu \nu} - \left[ \nabla_\mu \nabla_\nu -g_{\mu \nu}\Box \right] f'(R) = k T_{\mu \nu},\label{eq:freq}
\end{equation}
where $T_{\mu \nu}$ is the usual stress-energy tensor of the matter action.
Clearly when $f(R)=R$ the usual Einstein equations are obtained.
The idea behind $f(R)$ theories is that the extra terms obtained in Eq. \ref{eq:freq} is responsible for the dark component of the Universe.
A remarkable feature of $f(R)$ gravity is that it is equivalent to the Jordan-Brans-Dicke scalar-tensor theory \cite{thomas_review}.

Taking as an example $f(R)\propto R^n$ and assuming a FLRW cosmology the extra terms of the equations give rise to a term which can be cast in the form of the contribution from a perfect fluid with state equation of the form $P_f=w \: \rho_f$  \cite{capozziello03} where:
\begin{equation}
w_f = -\frac{6 n^2 -7 n -1}{6 n^2 -9 n +3}.
\end{equation}
Assuming certain values of $n$ the Dark Energy value $w \simeq -1$ is obtained.
Important limits on the value of $n$ can be obtained both from cosmological analysis \cite{barrow_clifton06,clifton_barrow05,clifton_barrow06,zakharov06} and from the dynamics of the solar systems \cite{iorio_riggiero07a,iorio_riggiero07b}

The general treatment of $f(R)$ theories is rather difficult and most often important results are obtained for specific forms of the $f$ function.
In general however the viability of a specific form of $f(R)$ needs to fulfill the following criteria:
\begin{itemize}
\item the correct Newtonian or Post-Newtonian limit must be obtained in the weak-field approximation
\item the theory must be stable at the classical and semi-classical level
\item the theory must correctly describe the dynamics of the cosmological perturbations.
\end{itemize}

\chapter{Tests for dark matter mass models}

As described in the previous chapter, there are several proposals for the solution for the cusp vs core controversy.
In this Chapter I show how they can be validated or ruled out by a systematic comparison of their prediction with precision measurements of RCs.
As an example I develop a test of one of these suggestions, the Gravitational Suppression model (GraS), that can be easily extended to the other proposals in the literature.

\section{Introduction}
The gravitational suppression hypothesis \cite{PM} is a phenomenological model that addresses the complex understanding of the DM distribution on small, subgalactic scales.
High-resolution radio observations from spiral galaxies, along with their optical RCs, suggest that the DM is distributed in spherical halos with nearly constant density cores (see, e.g., \cite{gentile04,47salucci03,donato04} and references therein).
On the other hand, theoretical predictions from the well-known  N-body $\Lambda$CDM simulations (e.g., \cite{nfw96}) present a steep density distribution profile in the centre of the halos: 
\begin{equation}
\rho_{halo}(r)=\frac{\rho_{s}}{(r/r_{s})(1+r/r_{s})^{2}}.
\end{equation}
$r_{s}$ is a scale radius and $\rho _{s}$ its characteristic density, in principle independent, but found related within a reasonable scatter through the halo mass, by the Bullock \emph{et al.} \cite{bullock01b} equation: $c\equiv R_{vir}/r_{s}\sim18(\frac{M_{vir}}{10^{11}M_{\odot}})^{-0.13}$,
where $c$ is a concentration parameter and $R_{vir}$ and $M_{vir}$ are the virial radius and mass.
Mass models with a NFW density profile, given in Eq.(1), have two serious kinds of difficulty in reproducing the observed RCs: 
a) the fit is not satisfactory, i.e., $\chi^{2}_{red}\gg 1$ (see, e.g., \cite{gentile04} and references therein); b) the values of the parameters of the best-fit mass models are clearly unphysical.
In detail, the values for the halo mass result much higher than those we obtain from weak lensing halo models \cite{wlensing} and from the analysis of galaxy baryonic mass function \cite{shankar06}: $M_{halo}\approx 3\times 10^{12}M_{\odot}\: (L_{B}/10^{11}L_{\odot})^{1/2}$. 
In the same way the values of the disk mass-to-light ratio result much lower than those derived from colours of spirals \cite{shankar06,bruzual,luminosity}: $\log\: (M_{D}/M_{\odot})\approx -1.6+1.2 \log \:(L_{B}/L_{\odot}) $, i.e., $0.7< M_{D}/L_{B} < 4$.

Several solutions have been proposed for the above issue, most of them related either to a better comprehension of structure formation (e.g., \cite{chiara}) or to new fundamental physics (e.g., \cite{spergel00}).
Alternatively, the presence of noncircular motions in galaxies has been advocated to reconcile (up tp $70\%$ in the Low Surface Brightness of) the observed kinematics with the cuspy density profile (e.g., \cite{hayashi04,hayashi06}, but see also \cite{gentile05}). 

\section{The Gravitational Suppression model}
The original proposal by Piazza \& Marinoni (PM) GraS model, instead,  modifies the usual Newtonian potential of the DM felt by baryonic test particles in such a way that the NFW kinematics and the observed one  become in agreement.
According to PM, the NFW profile is used because GraS does not affect the DM dynamics, but only the dynamics in the mixed sector DM-baryons, so both primordial DM perturbations and halo formation are unaffected, and well-known N-body simulation results can be assumed.
The idea is adding a Yukawa contribution to the gravitational potential
\begin{equation}
\nabla^{2}\phi_{Newton}=4\pi G\: (\rho_{baryons}+\rho_{halo}),
\end{equation}
from a hypothetical short-range interaction just between dark and luminous matter
\begin{equation}
(\nabla^{2}-\lambda^{-2})\:\phi_{Yukawa}=4\pi G\:\rho_{halo},
\end{equation}  
where $\lambda$ is a \textit{scale range} parameter. The effect is damping the gravitational interaction on small scales.
The final potential is then
\begin{equation}
\phi_{halo}=\phi_{Newton}+\alpha \: \phi_{Yukawa}.
\end{equation}
$\alpha$ is a \textit{strength} parameter and taken to be $-1$ in order to have the maximum possible gravitational suppression \cite{veneziano}.
The circular velocity  is related to the potential by 
\begin{equation}
V_{halo}^2=V_{halo,\: Newton}^2+V_{halo,\: Yukawa}^2=r \: |d\phi_{halo}/dr|.
\end{equation}
In  PM model, for  a (small) sample of RCs  of  Low Surface Brightness galaxies GraS was able to eliminate the above core versus cusp discrepancy.
However, in order to allow a   simple analytic calculation, they have  taken   a number of assumptions and  approximations.  In detail,  the contribution to the gravitational potential from baryons (stars and HI disk)  was neglected and the DM distribution was modeled with the  simple form $\rho_{halo}(x)=\rho_{0}x^{-\beta}$, rather than by Eq. (1). 
Further support to GraS was given in \cite{piazza} where the dispersion velocity of two spheroidal dwarfs (Fornax and Draco) were studied in this scenario.
However, both large errors in the kinematic measurements and large geometric and orbital  uncertainties of the employed  mass model, limited the relevance of their findings. 

\section{Data and methodology of the test}
In the present analysis of GraS we abandon the above approximations and test a wider and fairer sample of spirals.
An in-depth review of the GraS model is beyond the scope of this work.
Our goal is to perform a check of GraS.
First, we assume the exact  NFW profile. Second, we consider the baryonic contribution, so that the total potential is
\begin{equation}
\phi_{model}=\phi_{halo}+\phi_{disk}+\phi_{gas},
\end{equation}
where the sum of the last two terms is $\phi_{baryons}$.  
This leads to 
\begin{equation}
V_{model}^2=V_{halo}^2+V^2_{disk}+V^2_{gas}.
\end{equation}
Finally, we use a sample of high-resolution RCs of Low and High Surface Brightness galaxies, in order to investigate the consistency and universality of the model.

Our sample represents the best available RCs to study the mass distribution of DM and it has been used in works concerning the core versus cusp discrepancy controversy \cite{gentile04, corbelli}.
The sample includes nearby Low and High Surface Brightness galaxies, all poorly fitted by mass models with NFW halos that also have unphysical values for their best-fit mass parameters: DDO 47 \cite{gentile05}; ESO 116-G12, ESO 79-G14 \cite{gentile04}; NGC 6822 \cite{weldrake03}; UGC 8017, UGC 10981, UGC 11455 \cite{vogt}; M 31 \cite{corbelli}.
Let us notice that in some cases H$_\alpha$ and HI RCs are  both available and they  agree well where they coexist.
Moreover the RCs we analyse are smooth, symmetric and extended to large radii.

We decompose the total circular velocity into stellar, gaseous and halo contributions, according to Eqs.($1$)-($5$), where the latter contains the additional DM-baryons interaction.
Available photometry shows that the stars in our sample of galaxies are distributed in a thin disk, with exponential surface density profile $\Sigma_{D}(r)=(M_{D}/2 \pi R_{D}^{2})\: e^{-r/R_{D}}$, where $M_{D}$ is the disk mass and $R_{D}$ is the scale length.
The circular velocity contribution is given by $V_{disk}^{2}(r)=(G M_{D}/2R_{D}) \:  x^{2}B(x/2)$, where $x\equiv r/R_{D}$ and $G$ is the gravitational constant.
The quantity $B=I_{0}K_{0}-I_{1}I_{1}$ is a combination of Bessel functions \cite{freeman}.
The contribution of the gaseous disk is directly derived from the HI surface density distribution.

In a first step, the RCs are $\chi^{2}$ best-fitted with the following free parameters: disk mass, NFW scale radius and characteristic density, and scale range of GraS. 
Then we redo the analysis fixing the GraS scale range parameter at the mean value found  of $\lambda=3.1$\:kpc.
Notice that the published mean value of PM for $\lambda$ is quite different from ours as an effect of their simplifications: $\lambda=1.1$\:kpc. 
Our value is the most favourable for the PM model: different values of $\lambda$ leads to worse performance.

\section{Results}
The test goes against the  GraS model.
For the RCs of our sample the NFW mass halo model fails to reproduce data according to the usual pattern explained in the introduction.
Data, not surprisingly, points to DM halos having inner density cores.
Applying a Yukawa potential to the cuspy NFW halo does not solve this discrepancy.
\begin{figure}[b!]
\centering
\vskip -1.4cm
\includegraphics[width=12cm]{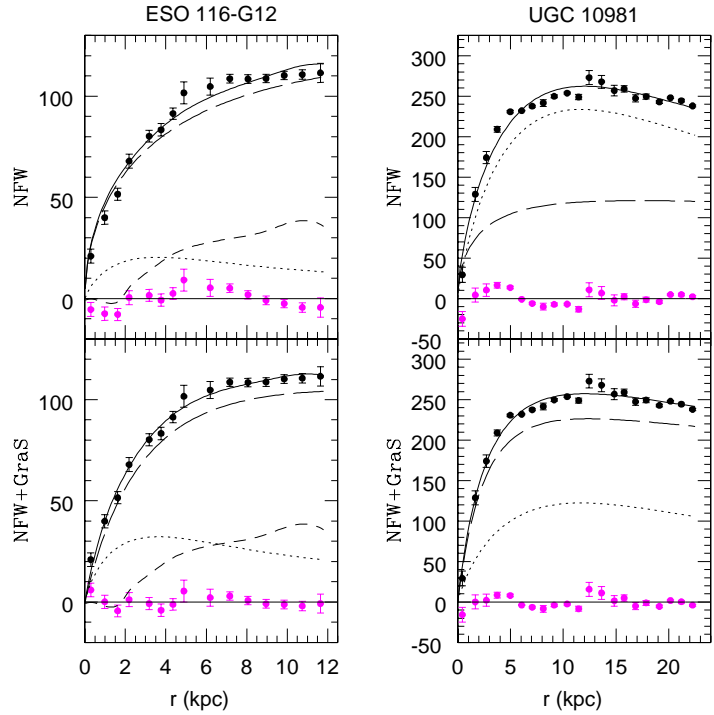}
\vskip  -1.2truecm
\caption{Galaxies in which GraS eliminates the core versus cusp discrepancy controversy. $Y$ axis is the velocity in km/s. The solid line represents the best-fit mass model, the long-dashed line is the contribution of the DM halo, and the dotted and short-dashed lines are those of the stellar and gaseous disks. Below the  RCs, we plot the residuals ($V_{obs}-V_{model}$).}
\label{fig:one}
\end{figure} 
The cusp is erased and RCs are fitted very well, but this success is illusory in that the corresponding values of the parameters of the best-fit mass model remain unphysical.
In table I we show the results of the test. We give: the values of the parameters of the mass model and global properties of the galaxies. 
$\chi^{2}_{red}$ is calculated with average typical velocity errors. In bold, unphysical values for halo mass and mass-to-light ratio, and $\chi^{2}_{red}>2.5$.The critical density of the Universe today is taken to be $\rho_{crit,\:0}=10^{-29}g/cm^{3}$.

In detail, in the cases of ESO 116-G12 and UGC 10891, we have that GraS fits sufficiently well the RCs unlike the NFW, confirming that this model could work in some objects (see Fig. $1$, table I).

\begin{figure}[b!]
\centering
\vskip -2.35cm
\includegraphics[width=11cm]{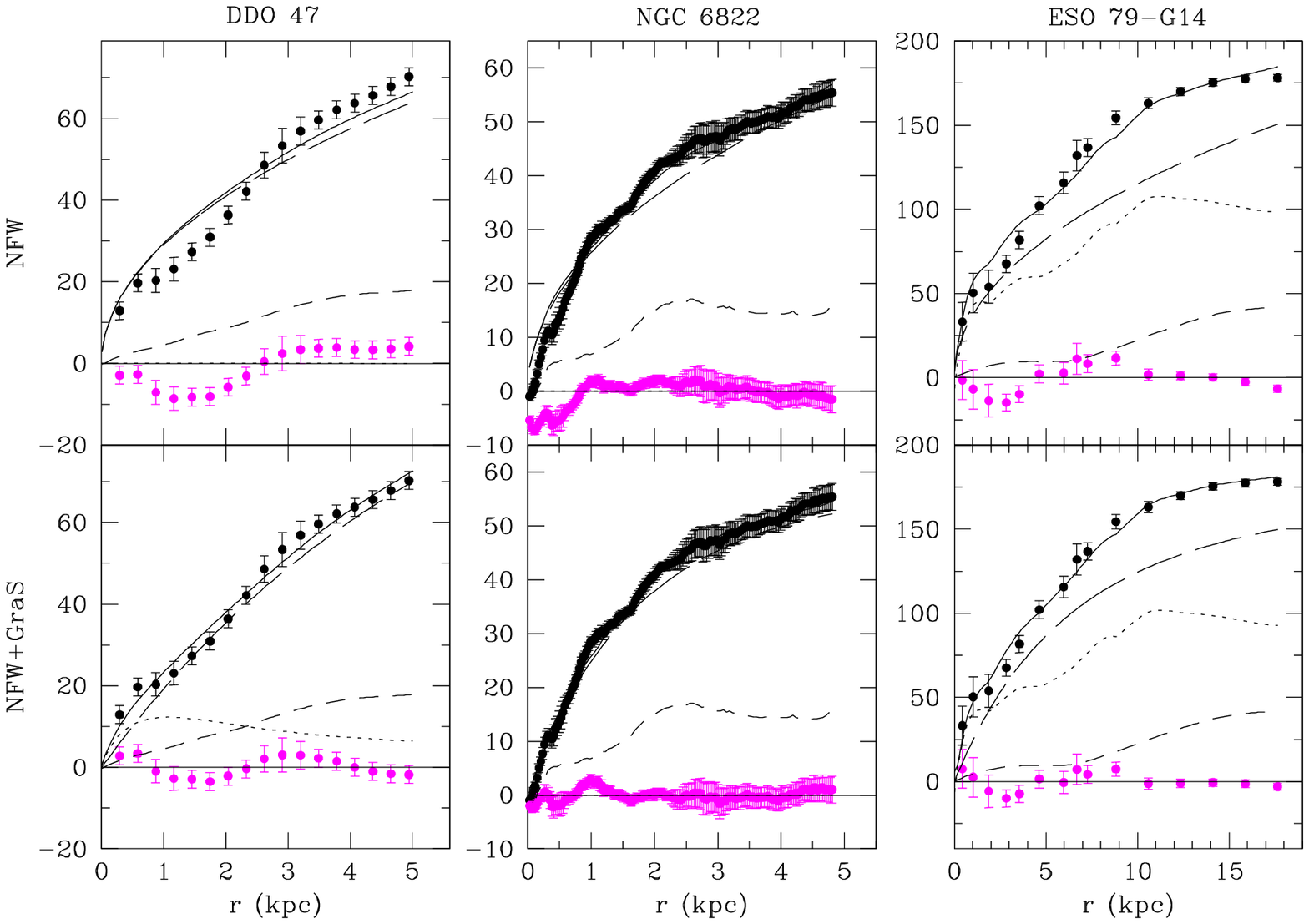}
\vskip -3.8cm
\includegraphics[width=11cm]{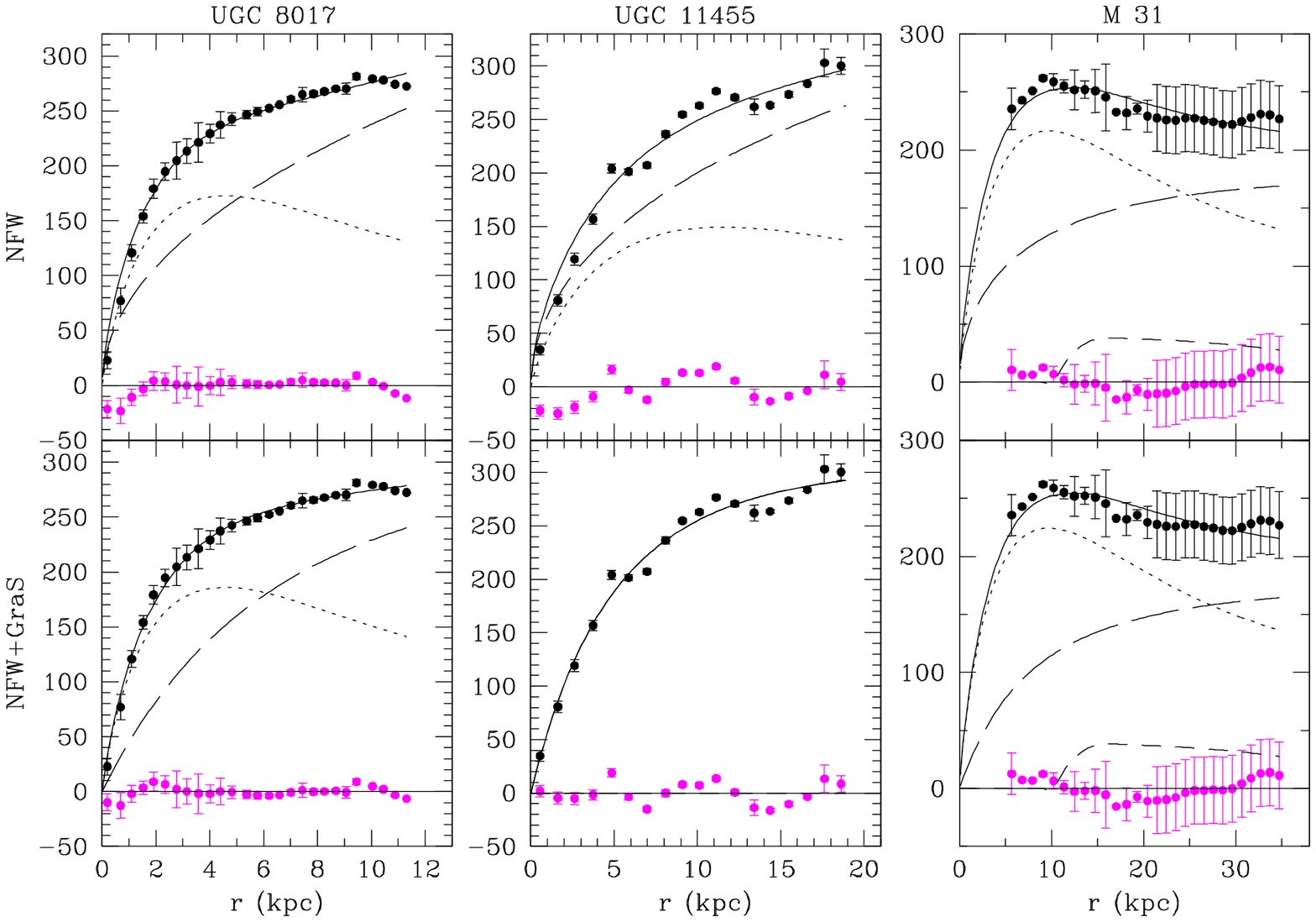}
\vskip  -2.1truecm
\caption{Galaxies in which GraS does not solve the core versus cusp discrepancy  controversy. The fitting values of the mass-to-light ratio (NGC 6822, ESO 79-G14, UGC 11455) and halo mass (DDO 47, UGC 8017) result unphysical. See Fig. $1$ and table I for details.}
\label{fig:two}
\end{figure}

However, in the other cases, although the fits are satisfactory, the best-fit values of the halo mass and mass-to-light ratio are unphysical.
In fact, we expect (see above) the mass-to-light ratios for NGC 6822, ESO 79-G14, UGC 11455, to be equal to (1, 2.6, 3.5), while we found much smaller best-fit values ($<$0.02, 0.3, $<$0.2).
In the same way, we expect halo masses for DDO 47 and UGC 8017 to be equal to ($9\times 10^{10}\: M_{\odot}$, $1.9\times 10^{12} \: M_{\odot}$), while we found much bigger best-fit values ($8.1\times 10^{11}\: M_{\odot}$, $1.5\times 10^{14} \: M_{\odot}$).
Furthermore, in M 31 the GraS modification is negligible and irrelevant (see Fig.\:$2$, table I).

Let us notice that by constraining the values  for  the mass  parameters within physically acceptable values, we obtain unacceptable fits for the GraS mass model, similar to those of the Newtonian NFW case.
As an example, in UGC 8017 with $M_{halo}=3\times 10^{12}M_{\odot}$ and $M_{D}/L_{B}=\: 3 M_{\odot}/L_{\odot}$, GraS shows an unacceptable fit to data (see Fig. 3).
More in general, we realize that for all six objects, all values of $\rho_{s}$ and $r_{s}$ within their  $1\sigma$ uncertainties imply unphysical halo masses and/or mass-to-light ratios. 

We now implement the Bullock \emph{et al.} concentration vs  halo mass  relation, that eliminates one parameter in the original NFW profile.
With this relation built in, GraS performs even worse than before. See in Fig. 3 the case for DDO 47. 

\begin{figure}[t!]
\vskip -3cm
\includegraphics[width=12cm]{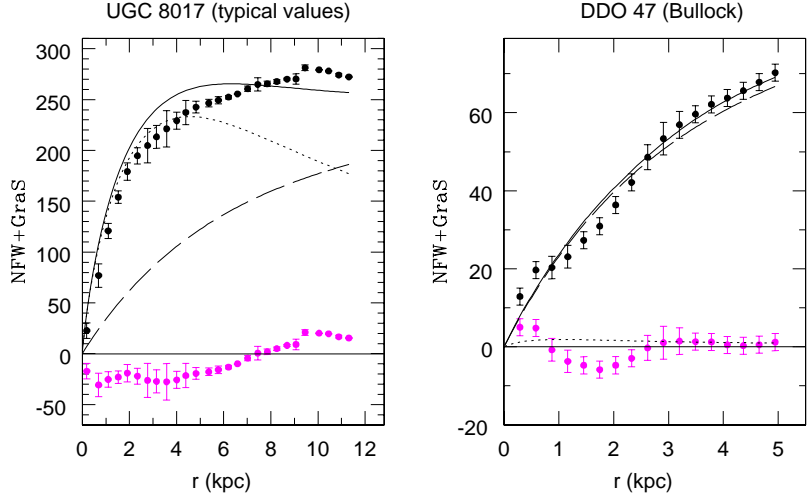}
\vskip -2.5cm 
\caption{NFW+GraS mass model. Left: with typical values for the halo mass and the mass-to-light ratio. Right: with the Bullock \emph{et al.} relation. See Figs. 1 and 2 and table I for details and comparison.}
\label{fig:three}
\end{figure}

\section{Conclusions}
In conclusion, the GraS-PM model fails to rescue the NFW profiles  in a number of  high quality well-suited RCs.
Moreover, let us point out that there is not  a pattern of  this inability, so that it is presently difficult to understand  how to  modify it in order to reach its original goal. 
Then the GraS model is a rather strong hypothesis that does not seem solve the core versus cusp discrepancy problem of the mass distribution of the center of DM halos.

Finally, let us remark that also in this work it has emerged that the available kinematics of galaxies is very constraining for non-Newtonian theories of gravity.

\begin{sidewaystable}[b!]\small
\caption{Parameters of the mass models.}
\begin{tabular}{lcccccccrc}\hline\hline
Galaxy&$L_{B}\:(L_{\odot}$)&Mass model&$M_{D}/L_{B}$&$M_ {halo}\:(M_{\odot})$&$\chi^{2}_{red}$&$r_{s}\:(kpc)$&$\rho_{s}\:(10^{4}\rho_{crit,\:0})$&$M_{D}\:(M_{\odot})$&\emph{c}\\\hline

\multicolumn{5}{l}{Positive results}\\

\multirow{2}{*}{ESO 116-G12}&\multirow{2}{*}{$4.6\times 10^{9}$}&NFW&\textbf{0.1}&3.8$\times 10^{11}$&\textbf{2.8}&$14.5\pm14$&$4.0\pm6.6$&$(4.2\pm27)\times 10^{8}$&13\\
&&NFW+GraS&0.3&1.5$\times 10^{11}$&1&$5.1\pm2.3$&$26\pm25$&$(1\pm1.7)\times 10^{9}$&26\\

\multirow{2}{*}{UGC 10981}& \multirow{2}{*}{$1.2\times 10^{11}$}&NFW&1.5&2.6$\times 10^{11}$&\textbf{4.2}&$8\pm2.9$&$13\pm9$&$(1.8\pm0.3)\times 10^{11}$&21\\
&&NFW+GraS&0.4&7.7$\times 10^{11}$&2.5&$4.2\pm0.3$&$180\pm40$&$(4.9\pm4.4)\times 10^{10}$&55\\

\multicolumn{5}{l}{Negative results: unphysical parameters}\\
\multirow{2}{*}{DDO 47}&\multirow{2}{*}{$10^{8}$}&NFW&$<0.2$&$\mathbf{7.4\times 10^{12}}$&1.9&$176\pm10$&$0.12\pm0.1$&$<2.3\times10^{7}$&2.8\\
&&NFW+GraS&0.5&$\mathbf{8.1\times 10^{11}}$&0.4&$26\pm18$&$1.8\pm1.4$&$(4.5\pm2.2)\times 10^{7}$&9.2\\

\multirow{2}{*}{NGC 6822}&\multirow{2}{*}{$1.6\times 10^{8}$}&NFW&$\mathbf{<0.04}$&1.7$\times 10^{12}$&2.3&$87\pm49$&$0.19\pm0.12$&$<6.7\times10^{6}$&3.5\\
&&NFW+GraS&$\mathbf{<0.02}$&2.5$\times 10^{10}$&0.5&2.9$\pm0.1$&24$\pm0.7$&$<2.9\times10^{6}$&26\\

\multirow{2}{*}{ESO 79-G14}&\multirow{2}{*}{$2\times 10^{10}$}&NFW&\textbf{0.3}&$\mathbf{3.9\times 10^{13}}$&\textbf{5}&$330\pm1400$&$0.1\pm0.49$&$(6.4\pm1.9)\times 10^{9}$&2.6\\
&&NFW+GraS&$\mathbf{0.3}$&$1.1\times 10^{12}$&2&$22.9\pm6$&$3.2\pm1.4$&$(6\pm0.9)\times 10^{9}$&11.2\\

\multirow{2}{*}{UGC 8017}&\multirow{2}{*}{$4\times 10^{10}$}&NFW&1&$\mathbf{4.4\times10^{17}}$&\textbf{4}&$379\pm3600$&$150\pm60$&$(3.8\pm0.8)\times 10^{10}$&51\\
&&NFW+GraS&1.1&$\mathbf{1.5\times10^{14}}$&1.6&$22\pm9$&$250\pm50$&$(4.4\pm0.3)\times 10^{10}$&62\\

\multirow{2}{*}{UGC 11455}&\multirow{2}{*}{$4.5\times 10^{10}$}&NFW&1.4&3.6$\mathbf{\times 10^{13}}$&\textbf{7.2}&$121\pm13$&$0.9\pm0.1$&$(7\pm2)\times 10^{10}$&7\\
&&NFW+GraS&$\mathbf{<0.2}$&3.2$\times 10^{12}$&3.9&$13.7\pm0.5$&$28\pm2.6$&$<10^{10}$&27\\

\multicolumn{5}{l}{Negative result: no change}\\
\multirow{2}{*}{M 31}&\multirow{2}{*}{$2\times 10^{10}$}&NFW&6.5&1.4$\times 10^{12}$&2&$28.5\pm1$&$2.2\pm0.1$&$(1.3\pm0.1)\times 10^{11}$&10\\
&&NFW+GraS&7&1.4$\times 10^{12}$&2.2&$31\pm1.1$&$1.8\pm0.1$&$(1.4\pm0.1)\times 10^{11}$&9.2\\\hline\hline
\end{tabular}
\end{sidewaystable}

\chapter[Rotation Curves \& $R^n$ gravity]{Analysis of Rotation Curves in the framework of $R^n$ gravity}
 
As seen in Chapter 4, modifications of the law of gravity are appealing alternatives to the yet undetected DM particles.
In this chapter I develop a test of $R^n$ gravity in galactic scales that represents a step forward on the issue in what for the first time a complete analysis of a devised sample of Rotation Curves has been performed.

\section{Introduction}
It is well-known that the RCs of spiral galaxies show a non-Keplerian circular velocity profile which cannot be explained by considering a Newtonian gravitational potential generated by the baryonic matter \cite{PSS96}.
Current possible explanation of this controversy includes, among others, the postulate of a new yet not detected state of matter, the DM \cite{rubin83}, a phenomenological modification of the Newtonian dynamics \cite{milgrom,brownstein,mond_review,bekenstein}, and higher order gravitational theories (originally devoted to solve the dark energy issue, see e.g., \cite{carroll,capozziello04}).

The recent theory proposed by Capozziello, Cardone $\&$ Troisi 2007 (hereafter CCT, \cite{CCT}), modifies the usual Newtonian potential generated by baryonic matter in such a way that the predicted galaxy kinematics and the observed one have a much better agreement.
They consider power-law fourth order\footnote{The term comes from the fact that the generalized Einstein equations contain fourth order derivatives of the metric.} theories of gravity obtained by replacing in the gravity action the Ricci scalar $R$ with a function $f(R)\propto R^n$, where $n$ is a slope parameter. 
The idea is that the Newtonian potential generated by a point-like source gets modified in to
\begin{equation}
\phi(r) = -\frac{G m}{r} \{1+\frac{1}{2}[(r/r_c)^\beta-1]\} \label{eq: phi},
\end{equation}
where $\beta $ is a function of the slope $n$, and $r_c$ is a scale length parameter.
It turns out that in this theory $\beta$ is a universal constant while $r_c$ depends on the particular gravitating system being studied.
In a virialized system the circular velocity is related to the derivative of the potential through $V^2=r \:d\phi(r)/dr$.
It is clear that (\ref{eq: phi}) may help in the explanation of the circular velocity observed in spirals.

We remark that any proposed solution to the galaxy RC phenomenon must not only fit well the kinematics but, equally important, also have best-fit values of the mass model parameters that are consistent with well studied global properties of galaxies.

For a sample of 15 Low Surface Brightness galaxies the model described in CCT  was fairly able to fit the RCs. 
However, in our view, the relevance of their finding is limited by the following considerations: 
\begin{itemize}
\item the sample contains several objects whose RCs are not smooth, symmetric and extended to large radii
\item the sample contains only Low Surface Brightness galaxies while a wider sample is desirable 
\item the universal parameter $n$ is not estimated by the analysis itself but it is taken from other observations. 
\end{itemize}  

In the present work we generalize the results of CCT and test a wider and fairer sample of spirals, improving the analysis methodology.
Our goal is to perform a check of their model on galactic scales in order to investigate its consistency and universality.

The plan of this work is the following: in Sect.2 we briefly summarize the main theoretical results described in CCT relevant for the analysis of our sample.
In Sect.3 we present our sample and methodology of analysis.
In Sect.4 the results are presented and finally the conclusions in Sect.5. 

\section{Newtonian limit of $f(R)$ gravity}

The theory proposed by CCT is an example of $f(R)$ theory of gravity \cite{nojiri,carloni}.
In these theories the gravitational action is defined to be:
\begin{equation}
{\cal S}=\int d^4 x \: \sqrt{-g} \:[f(R)+{\cal L}_m] \label{eq: action}
\end{equation}
where $g$ is the metric determinant, $R$ is the Ricci scalar  and ${\cal L}_m$ is the matter Lagrangian.
They consider:
\begin{equation}
f(R)=f_0 R^n \label{eq: f}
\end{equation}
where $f_0$ is a constant to give correct dimensions to the action and $n$ is the slope parameter.
The modified Einstein equation is obtained by varying the action with respect to the metric components.

Solving the vacuum field equations for a Schwarzschild-like metric in the Newtonian limit of weak gravitational fields and low velocities, the modified gravitational potential for the case of a point-like source of mass $m$, is given by (\ref{eq: phi}), where the relation between the slope parameter $n$ and $\beta$ (see detailed calculation in CCT) is given by:
\begin{equation}
\beta = \frac{12 n^2 -7 n - 1 - \sqrt{36 n^4 + 12 n^3 - 83 n^2 + 50 n + 1}}{6 n^2 -4 n + 2}. \label{eq: beta}
\end{equation}
Note that for $n=1$ the usual Newtonian potential is recovered.  
The large and small scale behavior of the total potential constrain the parameter $\beta$ to be $0 < \beta < 1$.

The solution (\ref{eq: phi}) can be generalized to extended systems with a given density distribution $\rho(r)$ by simply writing:
\begin{eqnarray}
\phi(r) & = & -G \int d^{3}r' \:\frac{\rho(\textbf{r'})}{|\textbf{r}-\textbf{r'}|}\:\:\{1+\frac{1}{2}[\frac{|\textbf{r}-\textbf{r'}|^\beta}{r_{c} ^\beta}-1]\}\nonumber\\
~ & = &\phi_{N}(r)+\phi_{C}(r) \label{eq:tot},
\end{eqnarray}
where $\phi_{N}(r)$ represents the usual Newtonian potential and $\phi_{C}(r)$ the additional correction.
In this way, the Newtonian potential can be recuperated when $\beta=0.$
The solution for the specific density distribution relevant for spiral galaxies is described in the following paragraph.

\section{Data and Methodology of the test}
We selected two samples of galaxies: a first with 15 galaxies, called \emph{Sample A}, that represents the best available RCs to study the mass distribution of luminous and/or DM, and it has been used in works concerning modifications of gravity and the  cusp vs core controversy \cite{gentile04,corbelli,frigerioPRL}.

This sample includes nearby galaxies of different Surface Brightness: DDO 47 \cite{gentile05}; ESO 116-G12, ESO 287-G13, NGC 7339, NGC 1090 \cite{gentile04}; UGC 8017, UGC 10981, UGC 11455 \cite{vogt}; M 31, M 33 \cite{corbelli}; IC 2574 \cite{2574}, NGC 5585 \cite{5585}, NGC 6503 \cite{6503}, NGC 2403 \cite{2403}, NGC 55 \cite{55}.
This sample is the most suitable for a fair test of theories like the one of CCT:
\begin{itemize}
\item The RCs are smooth, symmetric and extended to large radii.
\item The galaxies present a very small bulge so that it can be neglected in the mass model to a good approximation.
\item The luminosity profile is well measured and presents a smooth behavior
\item The data are uniform in quality up to the maximal radii of each galaxy. 
\end{itemize}
Let us notice that in some of these galaxies H$_\alpha$ and HI RCs are  both available and in these cases they  agree well where they coexist.

We also considered a second sample called \emph{Sample B} consisting of 15 selected objects from Sanders \& McGaugh 2002 that has been used to test MOND.
This sample consists of the following galaxies: UGC 6399, UGC 6983, UGC 6917, NGC 3972, NGC 4085, NGC 4183, NGC 3917, NGC 3949, NGC 4217, NGC 3877, NGC 4157, NGC 3953, NGC 4100 \cite{umajorPhotometry,umajorHI}; NGC 300 \cite{300}; UGC 128 \cite{128}. 
Although these galaxies do not fulfill all the requirements of \emph{Sample A} we have analyzed them for completeness sake.
The properties of the galaxies of the two samples are listed in table 1.
Notice that the theory of CCT requires an analysis with a sample of high quality galaxies, as described above, where each luminous profile plays an important role, whereas this is not the case in MOND. 

We decompose the total circular velocity into stellar and gaseous contributions.
Available photometry and radio observations show that the stars and the gas in our sample of galaxies are distributed in an infinitesimal thin and circular symmetric disk.
While the HI surface luminosity density distribution $\Sigma_{gas}(r)$ gives a direct measurement of the gas mass, optical observations show that the stars have an exponential distribution:
\begin{equation}
\Sigma_{D}(r)=(M_{D}/2 \pi R_{D}^{2})\: e^{-r/R_{D}}\label{eq:sigma},
\end{equation}
where $M_{D}$ is the disk mass and $R_{D}$ is the scale length, the latter being measured directly from the optical observations, while $M_D$ is kept as a free parameter of our analysis.

The distribution of the luminous matter in spiral galaxies has to a good extend cylindrical symmetry, hence using cylindrical coordinates, the potential (\ref{eq:tot}) reads
\begin{equation}
\phi(r)=-G\int^{\infty}_{0}dr'\:r'\Sigma(r')\int^{2\pi}_{0} \frac{d\theta}{|\textbf{r}-\textbf{r'}|}\{1+\frac{1}{2}[\underbrace{\frac{|\textbf{r}-\textbf{r'}|^\beta}{r_{c} ^\beta}}-1]\}.\label{eq:cyl}
\end{equation}
$\Sigma(r')$ is the surface density distribution of the stars, given by (\ref{eq:sigma}) , or of the gas, given by an interpolation of the HI data points up to the last measured point.
$\beta$ and $r_c$ are free parameters of the theory, with the latter galaxy dependent.
We neglected the gas contribution to the mass density for radii larger than the last measured point, however we checked the goodness of this approximation by extending the distribution with a different kind of decreasing smooth curves and realized that error made in the truncated approximation is small enough to be neglected.  

Defining $k^{2}\equiv \frac{4r\:r^{'}}{(r+r^{'})^2}$, we can express the distance between two points  in cylindrical coordinates as $|\textbf{r}-\textbf{r'}|=(r+r)^2 (1-k^2cos^2(\theta/2))$.
The derivation of the circular velocity due to the marked term of equation (\ref{eq:cyl}), that we call $\phi_{\beta}(r)$, is now direct:
\begin{equation}
r\:\frac{d}{dr}\:\phi_{\beta}(r) =  -2^{\beta -3} r^{-\beta}_{c}\: \pi \: \alpha \: (\beta -1)\: G\:I(r) \label{eq: vel}, 
\end{equation}
where the integral is defined as
\begin{equation}
{\cal{I}}(r) \equiv \int^{\infty}_{0} dr' r' \frac{\beta -1}{2}k^{3-\beta}\: \Sigma(r')\:{\cal{F}}(r) \label{eq: int},
\end{equation}
with ${\cal{F}}(r)$ written in terms of confluent hyper-geometric function: ${\cal{F}}(r) \equiv 2(r+r')\:_{2} F_{1}[{\frac{1}{2},\frac{1-\beta}{2}},{1},k^{2}]+[(k^{2}-2)r'+k^2 r]\:_{2} F_{1}[{\frac{3}{2},\frac{3-\beta}{2}},{2},k^{2}]$.

The total circular velocity is the sum of each squared contribution:  
\begin{equation}
V_{CCT}^{2}(r)=\:V^{2}_{N,stars}+V^{2}_{N,gas}+V^{2}_{C,stars}+V^{2}_{C,gas}\label{eq: vtot}
\end{equation}
where the $stars$ and $gas$ subscripts refer to the different contributions of luminous matter to the total potential (\ref{eq:tot}). The \emph{N} and \emph{C} subscripts refer to the Newtonian and the additional correction potentials.

Let us recall that we can write
\begin{equation}
V_stars^{2}(r)=(G M_{D}/2R_{D}) \:  x^{2}B(x/2)\label{eq: vN},
\end{equation}
where $x\equiv r/R_{D}$, $G$ is the gravitational constant and the quantity $B=I_{0}K_{0}-I_{1}K_{1}$ is a combination of Bessel functions \cite{freeman}.

Galaxies UGC 8017, M 31, UGC 11455 and UGC 10981 presents a very small amount of gas and for this reason it has been neglected in the analysis.
Notice that the correction to the Newtonian potential in equation (\ref{eq: phi}) may be negative and this would lead to a negative value of $V^2_C$.
In Figs. 1 and 2 however the velocities $V_C$ are shown only in the ranges of $r$  where their square are positive.

In a first step, the RCs are $\chi^{2}$ best-fitted with the following free parameters: the slope ($\beta$) and the scale length ($r_{c}$) of the theory, and the gas mass fraction ($f_{gas}$) related to the disk mass simply by $M_{D}=M_{gas}(1-f_{gas})/f_{gas}$.
The errors for the best fit values of the free parameters are calculated at one standard deviation with the $\chi^2_{red}+1$ rule.
From the results of these fits we get a mean value of $\beta=0.7 \pm 0.25$ ($n\simeq2.2$). 
In the second step we redo the best-fit fixing the slope parameter at $\beta =0.7$ keeping as free parameters only $r_c$ and $f_{gas}$.
Notice that in a previous paper \cite{capozziello06}, a mean value of  $\beta=0.58 \pm 0.15$ ($n\simeq1.7$)  has been obtained, perfectly compatible with our result.
This parameter however, is well constrained from SNeIa observations to be $\beta=0.87$ ($n\simeq3.5$), also compatible with our measurements. 
In our analysis the value $\beta=0.7$ is the most favorable for explaining the RCs: different values of $\beta$ from the one we adopt here lead to worse performance.

\begin{figure} [t!]
\hskip -1.3cm
\includegraphics[width=16.2cm]{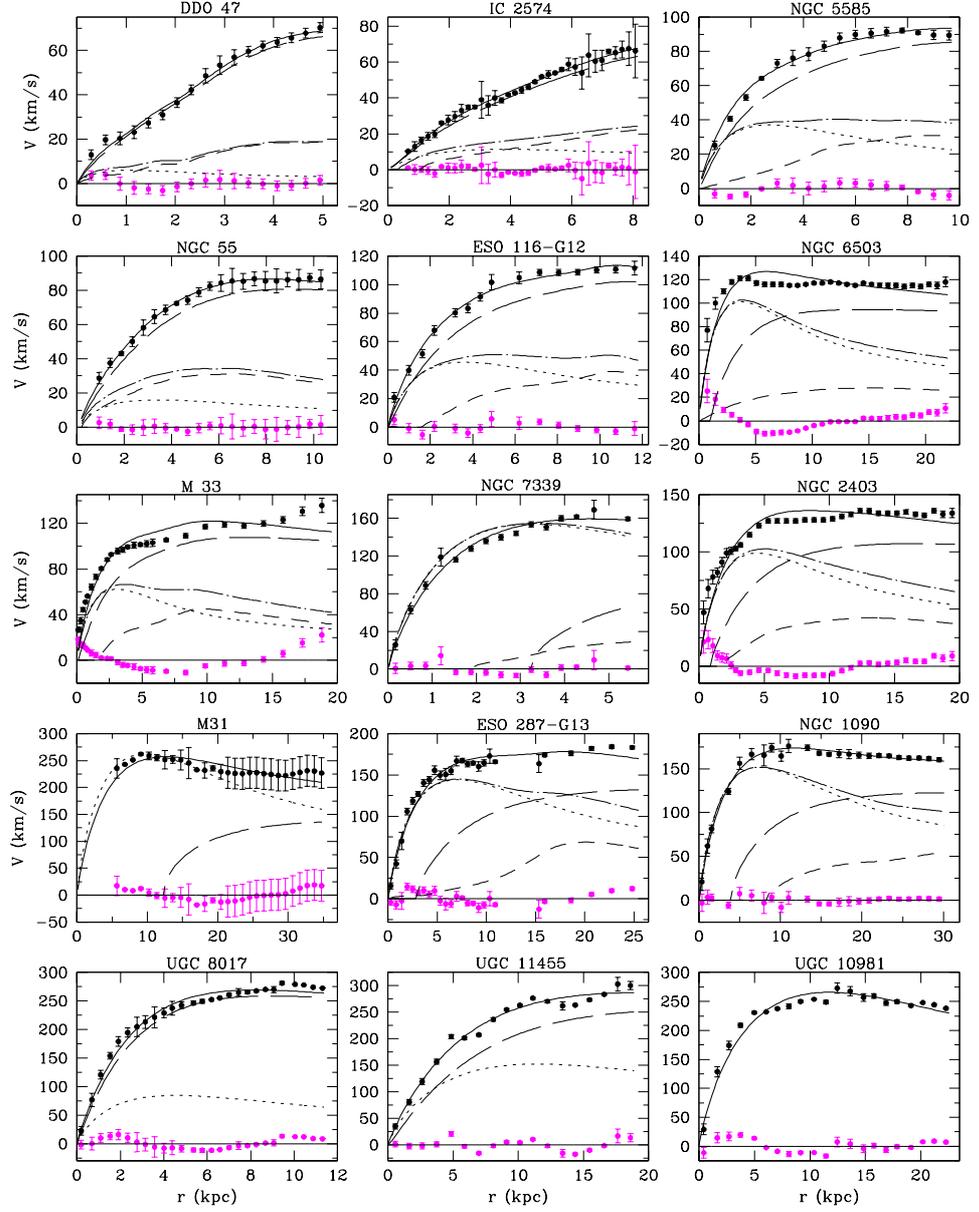}
\caption{\emph{Sample A}: The solid line represents the best-fit total circular velocity $V_{CCT}$.
The dashed and dotted lines are the Newtonian contributions from the gas and the stars, while the dot-dashed represents their sum.
The long-dashed line is the non-Newtonian contribution of the gas and the stars to the model.
Below the  RCs, we plot the residuals ($V_{obs}-V_{CCT}$).
See table 1 for details.}
\label{fig:bf}
\end{figure}

\begin{figure} [t!]
\hskip -1.3cm
\includegraphics[width=16.2cm]{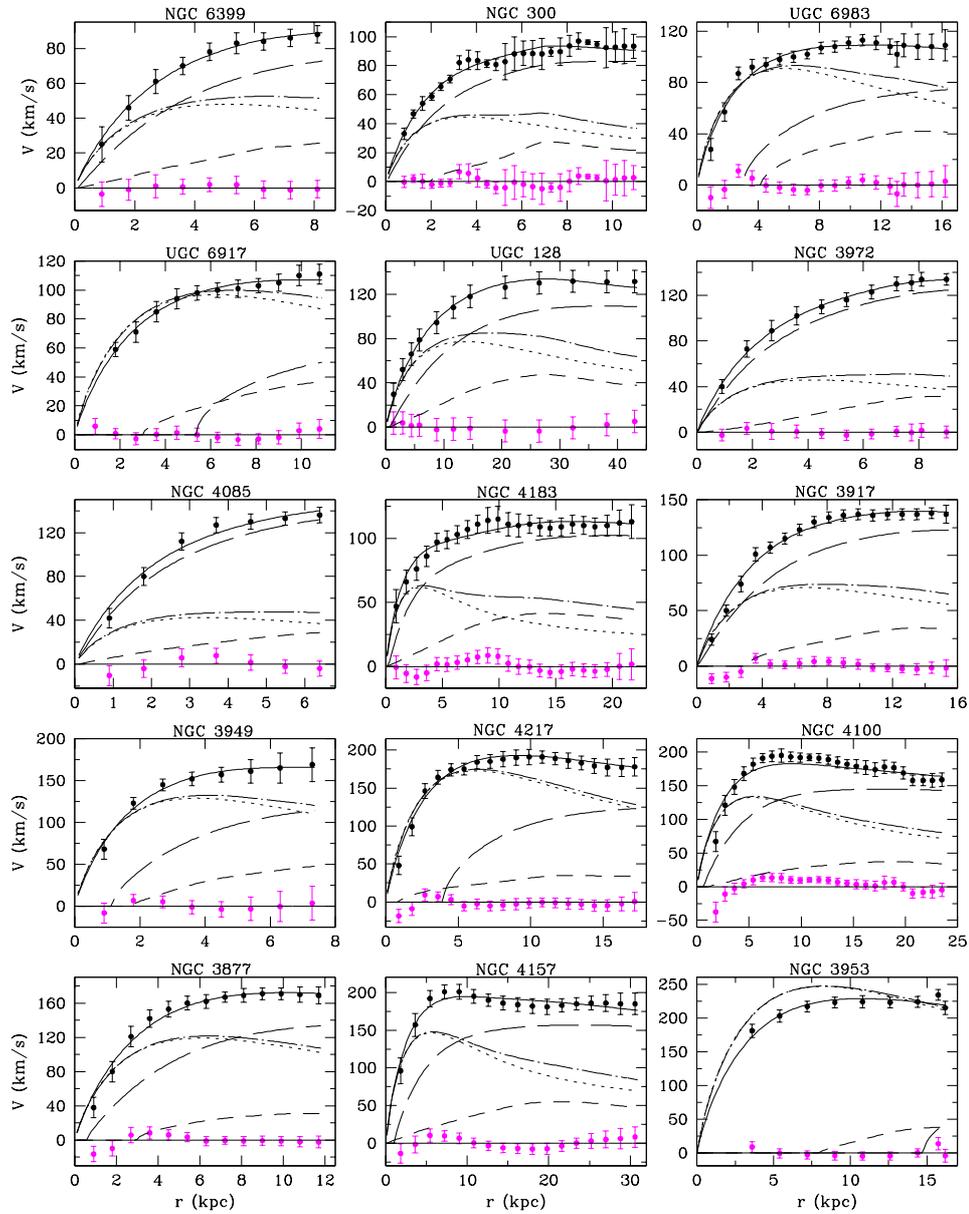} 
\caption{\emph{Sample B}: Best-fit curves superimposed to the data from
  selected objects from Sanders \& McGaugh 2002. See Fig. 1 for details.}
\end{figure}

\section{Results}

We summarize the results of our analysis in Figs. 1 and 2 and table 1\footnote{Numerical codes and data used to obtain these results can be found at the address http://people.sissa.it/$\sim$martins/home.html}.
In general we find for all galaxies:
\begin{itemize}
\item the velocity model $V_{CCT}$ well fitting the RCs 
\item acceptable values for the stellar mass-to-light ratio
\item too vast range for values of the gas fraction (0$\% < f_g  < 100\%$)
\item not clear comprehension for the big variation of values for the scale length parameter (0.005 kpc$<r_c<$1.53 kpc).
\end{itemize}

The residuals of the measurements with respect to the best-fit mass model are in most of the cases compatible with the error-bars, see Figs. 1 and 2, though three galaxies show significant deviations: NGC 6503, NGC 2403 and M 33.
\begin{table} [b!] \tiny
\caption{Properties and parameters of the mass model of the analyzed Samples ($\beta=0.7$).
From left to right, the columns read: name of the galaxy, Hubble type as reported in the NED database, adopted distance in $Mpc$, B-band luminosity in $10^{9}L_{B \odot}$, disk scale length in $kpc$, gas mass in $10^{9} M_{\odot}$ until last measured point, gas fraction in $\%$, disk mass in $10^{9} M_{\odot}$, scale length CCT parameter in $10{-2}\: kpc$, mass-to-light ratio in $\Upsilon_{\odot}^{B}$, and $\chi^{2}_{red}$.
The galaxies are ordered from top to bottom with increasing luminosity.}
\begin{center}
\begin{tabular}{l|c|c|c|c|c|c|c|c|c|c}\hline
\emph{Galaxy}&\emph{Type}&\emph{D}&\emph{$L_{B}$}&\emph{$R_{D}$}&\emph{$M_{gas}$}&{$f_{gas}$}&\emph{$M_D$}&\emph{$r_c$}&\emph{$\Upsilon_{\star}^{B}$}&\emph{$\chi_{red}^2$}\\ \hline
\multicolumn{11}{c}{\emph{Sample A}}\\\hline
DDO 47&IB&4&0.1&0.5&2.2&96$\pm$1&0.01&0.5&0.1&0.5\\
IC 2574&SABm&3&0.8&1.78&0.5&79$\pm$12&0.14&1.7$\pm$0.3&0.2&0.8\\
NGC 5585&SABc&6.2&1.5&1.26&1.5&58$\pm$3&1&3.8$\pm$0.4&0.7&1.4\\
NGC 55&SBm&1.6&4&1.6&1.3&84$\pm$7&0.24&2.4$\pm$0.4&0.06&0.1\\
ESO 116-G12&SBcd&15.3&4.6&1.7&21&50&2.1&5$\pm$1&0.5&1.2\\
NGC 6503&Sc&6&5&1.74&2.3&18$\pm$0.7&10.6&21$\pm$1.4&2.1&18\\
M 33&Sc&0.84&5.7&1.4&3.7&53$\pm$2&3.3&7.5$\pm$0.4&0.6&25\\
NGC 7339&SABb&17.8&7.3&1.5&6.2&2.8$\pm$0.2&22&41$\pm$7&3&2.3\\
NGC 2403&Sc&3.25&8&2.08&4.5&27$\pm$0.9&12.1&21$\pm$1.5&1.5&19\\
M 31&Sb&0.78&20&4.5&-&-&180$\pm$70&153$\pm$19&9&3.4\\
ESO 287-G13&Sbc&35.6&30&3.3&14&25$\pm$1&41&48$\pm$5&1.4&3.2\\
NGC 1090&Sbc&36.4&38&3.4&100&18$\pm$1&47&59$\pm$4&1.2&0.9\\
UGC 8017&Sab&102.7&40&2.1&-&-&9.1$\pm$0.3&1$\pm$1&0.2&5.2\\
UGC 11455&Sc&75.4&45&5.3&-&-&74$\pm$3&14$\pm$1&1.6&5\\
UGC 10981&Sbc&155&120&5.4&-&-&460$\pm$200&$\sim 10^{11}$&3.8&4.9\\\hline
\multicolumn{11}{c}{\emph{Sample B}}\\\hline
UGC 6399&Sm&18.6&1.6&2.4&1&23$\pm$3&3.3&10$\pm$3&2&0.1\\
NGC 300&Scd&1.9&2.3&1.7&1.3&39$\pm$4&2&5.2$\pm$1&0.9&0.4\\
UGC 6983&SBcd&18.6&4.2&2.7&4.1&24$\pm$2&13&4.6$\pm$10&3.1&0.9\\
UGC 6917&SBd&18.6&4.4&2.9&2.6&14$\pm$1&16&71$\pm$17&3.6&0.5\\
UGC 128&Sd&60&5.2&6.4&10.7&32$\pm$5&23&39$\pm$11&4.4&0.1\\
NGC 3972&Sbc&18.6&6.7&2&1.5&39$\pm$3&2.5&2.5$\pm$0.4&0.4&0.1\\
NGC 4085&Sc&18.6&6.9&1.6&1.3&44$\pm$4&1.7&1.4$\pm$0.3&0.3&1\\
NGC 4183&Scd&18.6&9.5&1.4&4.9&60$\pm$6&3.2&9$\pm$2.3&0.3&0.3\\
NGC 3917&Scd&18.6&11&3.1&2.6&22$\pm$1.5&9.2$\pm$0.9&9.8$\pm$1.4&0.8&1\\
NGC 3949&Sbc&18.6&19&1.7&4.1&19$\pm$2.2&17&22$\pm$6&0.9&0.3\\
NGC 4217&Sb&18.6&21&2.9&3.3&6.1$\pm$0.7&52&55$\pm$15&2.5&0.4\\
NGC 4100&Sbc&18.6&25&2.5&4.4&13$\pm$1.5&28&20$\pm$3&1.1&1.5\\
NGC 3877&Sc&18.6&27&2.8&1.9&7.3$\pm$0.8&24&20$\pm$4&0.9&0.8\\
NGC 4157&Sb&18.6&30&2.6&12&26$\pm$2.6&33&25$\pm$4&1.1&0.5\\
NGC 3953&SBbc&18.6&41&3.8&4&2.8$\pm$0.18&140&190$\pm$50&3.4&0.8\\\hline
\end{tabular}
\end{center}
\end{table}

We also find acceptable values for the B-band mass-to-light ratio parameter for most of the galaxies, for which we should have approximately $0.5<\Upsilon_{\star}^{B}<6$ and a positive correlation between B-luminosity ($\Upsilon_{\star}^B \equiv M_{D}/L_{B}$; $M_D$ is the disk mass and $L_{B}$ is the B-band galaxy luminosity) and $\Upsilon_{\star}^{B}$ \cite{luminosity}:
\begin{equation}
M_D(L_B)\simeq 3.7 \times 10^{10} \times [(\frac{L_B}{L_{10}})^{1.23} \: g(L_B)+0.095(\frac{L_B}{L_{10}})^{0.98}]M_{\odot}\label{eq: salucci99},
\end{equation}
where $L_{10}\equiv 10^{10} L_{B \odot}$ and $g(L_B)=exp[-0.87 \times (log \frac{L_B}{L_{10}}-0.64)^2]$.
In detail we find discrepancies for NGC 55, UGC 8017, NGC 3972, NGC 4085 and NGC 4183.
Values for the scale length parameter ($r_c$) are in general smaller for less massive galaxies and bigger for more massive ones.
We obtained a Newtonian fit for UGC 10981, as shown by the exceedingly large value for $r_c$, see Fig. 1.

The model analyzed in this work yields better results on galactic scales than CDM models, where in the latter these galaxies have serious problems like marginal fits and unreasonable values for the stellar mass-to-light ratio, see e.g., \cite{gentile04,frigerioPRL}.

\section{Conclusions}

We have investigated the possibility of fitting the RCs of spirals with a power-low fourth order theory of gravity of CCT, without the need of DM.
We remark the relevance of our sample that contains objects in a large range of luminosity and with very accurate and proper kinematic.
We find in general a reasonable agreement, with some discrepancies, between the RCs and the CCT circular velocity model, encouraging further investigations from the theoretical point of view.

\chapter[dSphs \& spirals scaling laws]{Universal scaling relations in the luminous and dark mass distributions of spirals and dwarfs spheroidals}

I now turn the discussion to a more in depth analysis of DM halos in galaxies on a wide range of galaxy luminosity. 
Kinematic surveys of the dSph satellites of the
Milky Way are revealing tantalizing hints about the structure of DM halos at the low end of the galaxy luminosity
function. In brighter galaxies, observations and modeling of spiral
galaxies suggest that their dark halo parameters follow a number of
scaling relations. In this work, we investigate whether the
extrapolation of these relations to the dSph regime is consistent with
the observed internal kinematics of dSphs. The negligible fraction
($\sim10^{-2}-10^{-3}$) of baryonic matter inside the optical regions
of dSphs is consistent with the declining trend of baryon fraction
with baryonic (and DM) mass seen in spirals. The dSph data do not
currently discriminate between cored and cusped halos, due to our
lack of knowledge about the anisotropy of the stellar velocity
distribution and the limited spatial extent of the stellar tracers
relative to the DM. Nevertheless, although the DM densities in dSphs
are typically almost two orders of magnitude higher than those found
in (larger) disk systems, we find that the dSph kinematics are
consistent with their occupancy of (cored) Burkert DM halos whose
core radii and central densities lie on the extrapolation of the
scaling laws seen in spiral galaxies. We discuss the potential
implications of this scaling relation, if confirmed by future
observations, for understanding the nature of DM.

\section{Introduction}
DM provides the gravitational potential wells in which
galaxies form and evolve. Over the past decades, observations have
provided detailed information about the distribution of DM within
those regions of spiral galaxies where the baryons reside (\cite{urc2} and references therein, \cite{sofue01,PSS96,ashman92}).
Similar information is now also
becoming available for Low Surface Brightness galaxies~ \cite{blok05,deNaray06}.
In these disk systems,
the ordered rotational motions and known geometry of the tracers has
facilitated this achievement and an intriguing phenomenological
picture has emerged.  Spiral galaxies are composed of a disk
surrounded by a dark halo. Inside the optical regions ($R<R_{opt}$),
the disk is almost self-gravitating in the most luminous objects but
contributes a negligible amount to the gravitational potential at the
lower end of the luminosity function.  Mass modeling of both
individual and co-added RCs shows: (1) cored DM
 halos generally provide a better fit to the observed data than cusped
halos; more specifically, the Burkert density profile

\begin{equation}
\rho (r)={\rho_0\, r_0^3 \over (r+r_0)\,(r^2+r_0^2)}~,
\end{equation}
that contains two free parameters, the core radius $r_0$ and the
central halo density $\rho_0$, reproduces the available kinematical
data \cite{sb00,gentile04,3741,PSS96,dBMR01,blok_bosma02,marchesini02,gentile05}.
(2) When the data for
spiral galaxies are modeled assuming a Burkert distribution for the
DM and a Freeman disk for the luminous matter, the
parameters (DM central densities, core radii, disk masses and length
scales) are all related by a series of scaling laws \cite{PSS96,urc2,shankar06}.

In contrast to the results gathered for disk-dominated systems, our
knowledge of the mass distribution in pressure-supported systems like
elliptical galaxies is still limited (see \cite{Napolitano07} for a recent summary of the state of art).
However, on-going observations of
Local Group dwarf spheroidal galaxies (dSph), which occupy the faint
end of the luminosity function of pressure-supported systems, are
currently yielding crucial information about the properties of the
dark and luminous components in these objects and, in turn, on the
underlying physical properties of DM
halos (e.g. \cite{Gilmore07,Tolstoy04,Walker07}).
A number of important questions remain unanswered. These include: 
\begin{itemize}
\item Is the distribution of DM on galactic (i.e. kpc) scales universal?
\item Why do the dark and luminous mass distributions appear to be related, even though baryons dominate, at most, only the inner regions of galaxies?
\end{itemize}

\begin{figure}[h!]
\centering
\vskip -2.3cm
\includegraphics[width=12cm]{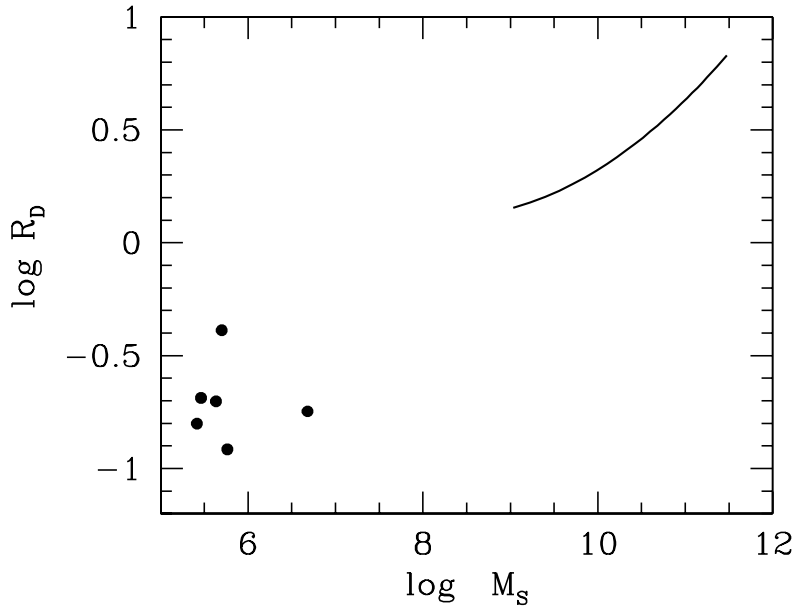}   
\vskip -0.9cm
\caption{Comparison of the distribution of characteristic baryonic scale 
$R_{D}$ versus stellar mass $M_{\rm s}$ for dSphs (points) with the
corresponding relation in Spirals (see \cite{PSS96}). For the dSphs, the
stellar mass is estimated from the $V$ band galaxy luminosity,
assuming a stellar mass-to-light ratio of unity (in solar units).}
\label{fig:r83}
\end{figure}

Work on spiral galaxies performed over the past 20 years has suggested
some answers to these questions for this Hubble type.  In contrast,
observations of the internal kinematics of dSphs have only recently
begun to provide hints of the distribution of DM in these
low-luminosity systems. The dSphs are indispensable for building up an
observational picture of the process of galaxy formation as they
extend the exploration of the dark and luminous mass distribution in
galaxies over a much wider range of Hubble type and luminosity. An
indication of this is given in Fig.~\ref{fig:r83} where we show the
relationship between a characteristic baryonic length scale (see below
for definitions) and the stellar mass in spirals and dSphs. The figure
illustrates the very different ranges of baryonic mass and size scale
in these two classes of stellar system.

The dSphs are typically at least two orders of magnitude less luminous
than the faintest spirals, and show evidence of being DM dominated at
all radii.  They are a primary laboratory for the bottom-up theory of
galaxy formation.  Moreover, being predominantly old,
pressure-supported, spheroidal systems, their evolutionary histories
are significantly different from those of spirals, especially in the
baryonic components. There is some
evidence of universality in the global properties of the mass
distribution of dSphs. \cite{Mateo98} found that the variation of the
mass to light ratios of dSphs with total luminosity was consistent
with the hypothesis that all dSphs contain similar masses of dark
matter interior to their stellar distributions that implies a larger  {\it proportion } of DM in the less luminous objects,  a  main global  characteristic of the  spiral mass distribution,  well-known since \cite{PS88}. 

More recent
analysis~\cite{Gilmore07,Koch07a}, based on extended velocity
dispersion profiles rather than central velocity dispersions, have
generally supported this conclusion. In dSphs, we note that due to the
limited spatial extent of the stellar distributions, the radial limit
of kinematic observations may be only a small fraction of the actual
DM halo size.

Obviously, the existence of common features, or scaling laws, relating
the \textit{structural parameters} of the mass distributions of dSphs
with those of very different stellar systems (e.g. spirals) would be
of potentially great significance and indicative of a Grand Picture
drawn by the fundamental physical processes in the formation and
evolution of galaxies. A number of recent papers have studied various
scaling relations between the properties of hot stellar
systems~\cite{Zaritsky2006,Dabringhausen2008,Forbes2008}. A common
conclusion is that the dSphs are outliers from other spheroidal
systems in terms of many of their properties (mass-to-light ratios,
sizes, etc.). In this paper, therefore, we examine whether the
properties of the dSphs are consistent with another class of stellar
system, namely luminous spiral galaxies. Additionally, while previous
works have studied the properties of the stellar distributions, or
global mass-to-light ratios, in this paper we make a tentative first
attempt to compare the DM halo parameters of different
systems. In particular, we will extrapolate the picture emerging in
spirals to the region of parameter space occupied by the dSphs, thus
comparing systems across a broad span of galaxy global properties and
morphologies. 

The outline of the paper is as follows. In Section~\ref{sec:data}, we
summarize the observational data used in our study and describe in
detail the analysis of the dSph data.  Section~\ref{sec:dmprop}
compares the properties of the dark halos of spiral and dSph
galaxies, while Section~\ref{sec:DMBaryons} discusses the relations
between the baryonic and DM properties of these
systems. Section~\ref{sec:conc} summarizes our findings and speculates
on the implications for the nature of DM.

\section{Data}
\label{sec:data}
\subsection{Spiral Galaxies}

Our aim in this paper is to test the consistency of the dSph data with
the scaling relations seen in spiral galaxies. When the mass
distribution in spirals is modeled using a Burkert DM halo (with
parameters $\rho_0$ and $r_0$) and a Freeman stellar disk, a tight
relation between $\rho_0$ and $r_0$ emerges \cite{PSS96}.
Noticeably , as it can be seen in Fig. 6, we find 
similar  $\rho_0$ vs $r_0$  relationships   {\em independently} of whether the mass profiles are obtained from kinematics (i. e. from  RCs) or from  gravitational lensing data or   from the analysis  individual or coadded  objects.
 
More in detail here, we make use  and show in Fig. 6 of the values of these parameters obtained  in  galaxies in which the profiles have been determined via one of
(1) the Universal Rotation Curve (see Fig.~\ref{fig:rho0r0} and \cite{shankar06});
(2) the analysis of weak lensing signals around spirals (see Fig.~\ref{fig:rho0r0});
(3) the mass modeling of individual RCs \cite{gentile04,donato04,spano08}. 

\subsection{dSph galaxies}

The study of the internal kinematics of the Milky Way dSphs has been
revolutionized by the availability of multi-object spectrographs on 4m
and 8m-class telescopes. Large data sets comprising several hundred
individual stellar velocities per galaxy have now been acquired for
all the luminous dSphs surrounding the Milky Way~\cite{Walker07,Koch07a,Wilkinson04,Kleyna04,Munoz05,Munoz06,Koch07b,Battaglia08}.
The volume of the currently available data is sufficient to place
reliable constraints on the dynamical masses interior to the stellar
distributions of the dSphs. However, the mass profiles are less
well-determined, and the velocity dispersion profiles alone cannot
distinguish between cored and cusped halos due to the degeneracy
between mass and velocity anisotropy (see, e.g. \cite{Koch07a,Battaglia08}).
However,
\cite{Gilmore07} recently showed that the kinematic data in six of
the well-studied dSphs are consistent with their occupying cored DM
halos, under the assumptions of spherical symmetry and velocity
isotropy. Further, \cite{Gilmore07} note that two dSphs exhibit
additional features which suggest that their halos are not cusped.

Before comparing the properties of dSphs with those of spiral
galaxies, we first re-visit the DM density profiles derived in
\cite{Gilmore07} for six Milky Way dSphs. In particular, we
investigate whether the Burkert DM profile which, let us recall,
generally reproduces the RCs data for spiral galaxies
(\cite{sb00}, see also \cite{gentile04}), is also consistent with the observed data for dSphs.
We note that for spiral galaxies, it has been shown that this choice of halo model is not prejudicial. In the region probed by the data,  for appropriate values of the halo parameters, actually  very different from those  that we actually find, the (cored) Burkert profile could have  mimicked, to a very good approximation,  a (cusped) NFW \cite{nfw96}.
To proceed, we would ideally require estimates of the Burkert
parameters $\rho_0$ and $r_0$, as well as their associated errors, for
our six dSphs. However, as we discuss below, an unambiguous
determination of whether dSph halos are cored and, if so, the sizes
of their core radii, is beyond the scope of the present paper. Instead
we will investigate the consistency of Burkert halos with the
velocity dispersion profiles of the dSphs, as published in
~\cite{Gilmore07}.

\begin{table}[h!]\tiny
\begin{center}
\begin{tabular}{l|c|c|c|c|c|c|c}\hline
Name & $R_{\rm b}$ & $\sigma_0$ & $R_{\rm s}$ &   $\rho_0 
$              & $r_0$   & $M_{\rm b}$  & $M(R_{83}/2)$ \\
      & (kpc)     &   (km/s)    &  (kpc)      & ($10^8 M_\odot$ kpc$^{-3} 
$)  & (kpc) & ($10^5 M_\odot$)        & ($10^7 M_\odot$) \\\hline
LeoI$^1$ & $0.28\pm0.01$ & $10.4\pm1.0$ & $1.9$ & $5.3 
\pm1.3 $  & $0.27\pm0.02$  & $48$  & $3.1\pm0.6$ \\
LeoII$^1$ & $0.19\pm0.01$ & $7.5\pm0.6$ & $0.9$ & $6.1 
\pm1.8 $ & $0.18\pm0.02$ & $5.8 $  &$1.1\pm0.2 $\\
Carina$^2$ & $0.31\pm0.01$ & $7.5\pm0.4$ & $1.2$ & $2.1 
\pm0.3 $ & $0.32\pm0.02$ & $4.3 $ &$1.8\pm0.2 $\\
Sextans$^1$ & $0.64\pm0.04$ & $6.3\pm1.0$ & $1.9$ & $3.5 
\pm1.5 $  & $0.65\pm0.06$  & $5.0 $  &$2.6\pm0.8 $\\
Draco$^3$ & $0.247\pm0.002$ & $10.5\pm0.8$ & $1.5$ & $6.9 
\pm1.2 $ & $0.24\pm0.01$ & $2.6 $ &$2.8\pm0.4 $\\
Ursa Minor$^1$ & $0.321\pm0.014$ & $12.8\pm1.2$ & $1.1$ &  
$6.6\pm1.6 $ & $0.28\pm0.02$ & $2.9 $  &$5.2 \pm0.9 $\\
\end{tabular}
\caption{Parameters obtained from mass modeling of six Milky Way dSphs.
Columns: (1) name of dSph and reference for surface brightness profile used. 
1: \cite{IH95};
2: \cite{Majewski00};
3: \cite{Wilkinson04};
(2) scale-length $R_{\rm b}$ of Plummer fit
to light distribution; (3),(4) central velocity dispersion $\sigma_0$
and $3\sigma$ lower limit on the scale-length $R_{\rm s}$ of Plummer
function fit (equation~\protect\ref{eq:plummer_velocity}) to velocity
dispersion profile; (5),(6) central density $\rho_0$ and scale-length
$r_0$ of median Burkert fit to density profile from Jeans equations;
(7) total stellar mass $M_{\rm b}$; (8) total mass inside $R_{83}/2$
($R_{83}$ is the three dimensional radius enclosing $83\%$ of the
light).
Quoted errors indicate $1\sigma$ ranges of parameters obtained from
1000 random realizations of the observed data, but do not include
modeling uncertainties. The large range of $R_{\rm s}$ values in each
dSph indicates that both flat ($R_{\rm s} > 10^7$kpc) and
falling ($R_{\rm s}\sim1$kpc) dispersion profiles are compatible with
the observations, although in all cases except Ursa Minor, the median
profile is flat.  See \cite{Gilmore07} for sources of
velocity data.}
\label{tab:dSphs}
\end{center}
\end{table}

\begin{figure}[h!]
\centering
\includegraphics[height=8.5cm]{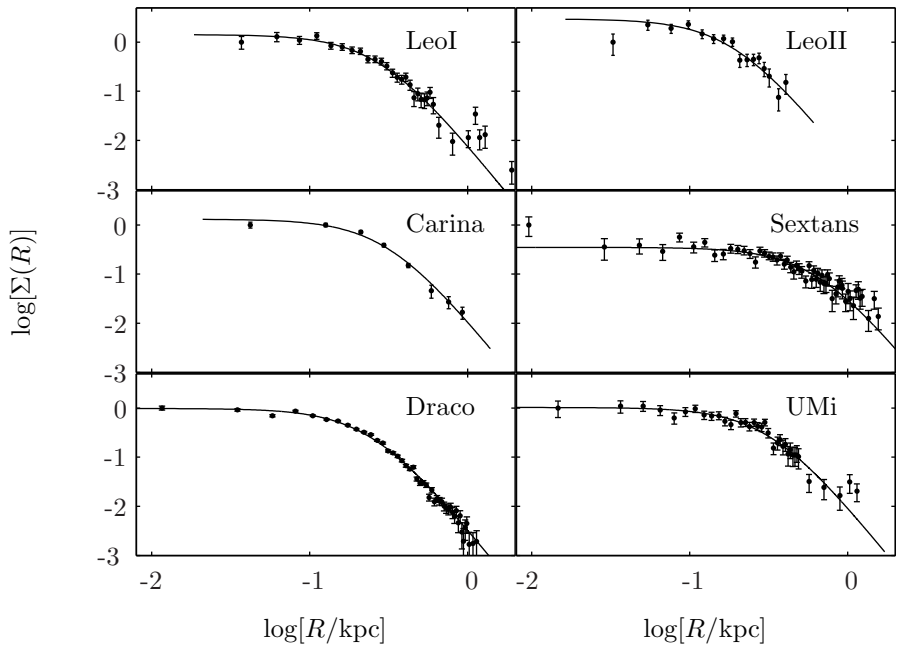}
\caption{Surface brightness profiles of six Milky Way dSphs.
Observed profiles are shown as points with error bars. Best-fit
Plummer profiles are shown as solid curves. In all cases, the Plummer
model is a good match to the light distribution.}
\label{fig:dSphs_SBP}
\end{figure}

For each object, we generate $1000$ random realizations of the surface
brightness profile and velocity dispersion profile by drawing values
within the observed error bars. We fit each surface brightness profile
with a Plummer~\cite{Plummer1915} distribution
\begin{equation}
\label{eq:plummer}
\Sigma(R) = \frac{\Sigma_0}{\left( 1 + \left(\frac{R}{R_{\rm b}}  
\right)^2\right)^2} ,
\end{equation}
where $\Sigma_0$ is the central surface density and $R_{\rm b}$ is the
scale-length.  Fig.~\ref{fig:dSphs_SBP} shows the observed surface
brightness profiles obtained from the literature (see column 9 of
table~\ref{tab:dSphs} for references) and the best-fitting Plummer
distributions for each dSph in our sample. The median values for
$R_{\rm b}$ obtained from the random realizations are given in
table~\ref{tab:dSphs}.

\begin{figure}[h!]
\includegraphics[height=8.5cm]{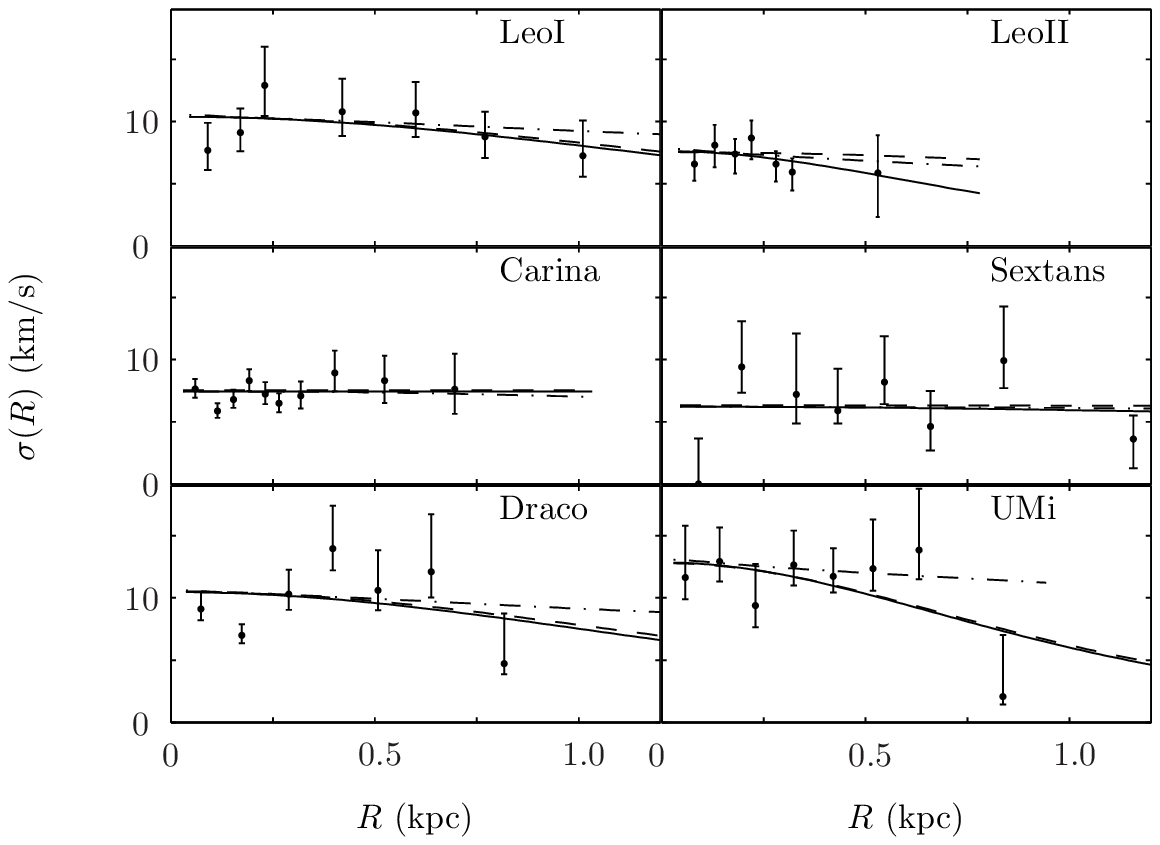}
\caption{Line-of-sight velocity dispersion profiles of six Milky Way
dSphs. Observed profiles are shown as data points with error bars. The
solid curves show the best-fit Plummer functions
(equation~\protect\ref{eq:plummer_velocity}) to the observed dispersion
profiles. The dashed curves show the median Plummer function based on
the Monte Carlo realizations of the observed data. The dot-dashed line
shows the dispersion profile obtained using the best-fit Plummer
profile to the light distribution and the best-fit Burkert model to
the halo mass distribution.}
\label{fig:dSphs_sigma}
\end{figure}

We fit each line of sight dispersion profile with a function of the
form
\begin{equation}
\label{eq:plummer_velocity}
\sigma(R) = \frac{\sigma_0}{\left( 1 + \left(\frac{R}{R_{\rm s}}  
\right)^2\right)^2} ,
\end{equation}
where $\sigma_0$ is the central velocity dispersion and $R_{\rm s}$ is
the scale length of the dispersion profile. Table~\ref{tab:dSphs} also
gives all the relevant parameters for these fits. Notice that, as in
the analysis presented in ~\cite{Gilmore07}, the fits to the
dispersion profiles are merely functional fits to smooth the data and
reproduce the general shape of the observed dispersion profiles. We
observe generally flat velocity dispersion profiles, the scale radii
$R_s$ being much larger than $R_b$ (see column 4 of
table~\ref{tab:dSphs}).  The observation that our measured dispersion
profiles for these six dSphs are consistent with being flat, and that
deviations from this are not statistically significant, is in
agreement with the more recent (and more extensive) data
of~\cite{Walker07} for five of these systems (Ursa Minor was not
included in their sample) in which the dispersion profiles are found
to remain flat to very large projected radii. Our conclusions in this
paper would thus be the same if we had used the~\cite{Walker07}
data. Fig.~\ref{fig:dSphs_sigma} shows the observed velocity
dispersion profiles, the best-fitting Plummer functions to the raw
dispersion data and the median Plummer function for each dSph.

\begin{figure}[h!]
\centering
\includegraphics[height=8.5cm]{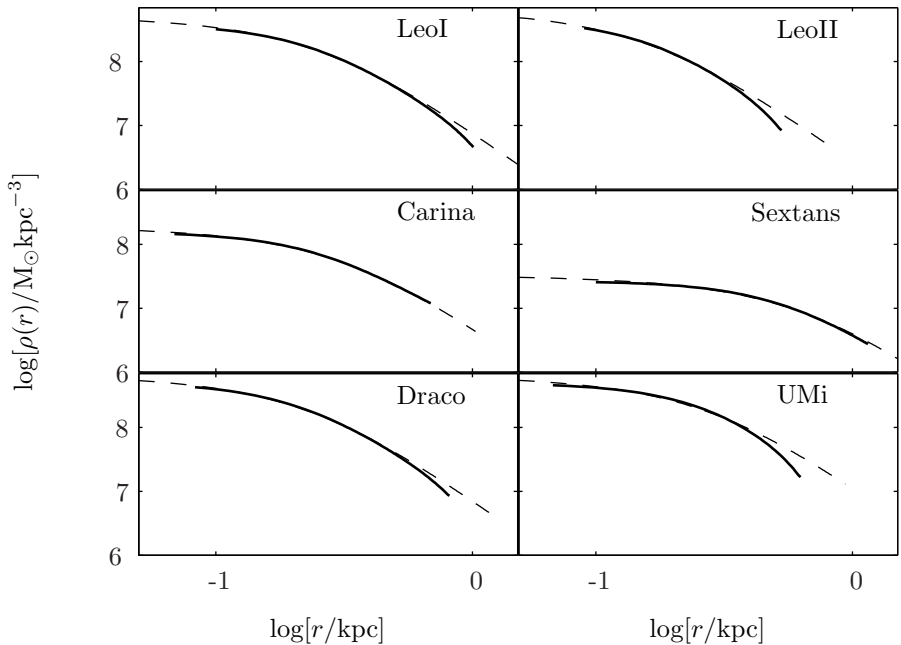}
\caption{Three-dimensional mass density profiles (solid curves) for our
dSphs obtained using the Jeans equation, assuming spherical symmetry
and velocity isotropy. The dashed curves show the best-fit Burkert
profiles. }
\label{fig:dSphs_rho}
\end{figure}

We use the Jeans equations to determine the three-dimensional mass
profile corresponding to each realization of the light distribution
and velocity dispersion profile, under the assumptions of spherical
symmetry and velocity isotropy and fit a Burkert profile to the three
dimensional density profiles thus obtained. The median values and
$1\sigma$ ranges of the Burkert parameters are presented in
table~\ref{tab:dSphs}. Fig.~\ref{fig:dSphs_rho} presents the density
profiles obtained from the best-fit surface-brightness and velocity
dispersion profile as well as the corresponding best-fit Burkert
profile. We note that the Burkert profile obtained from the median
values of $\rho_0$ and $r_0$ is very similar to the best-fit profile
shown. Moreover, as a sanity check, in Fig.~\ref{fig:dSphs_sigma} we
overplot the observed dispersion profile for each dSph with the
profile obtained from the the best-fit Burkert halo and the
best-fitting Plummer light distribution. The figure shows that the
observed stellar data in each dSph can be reproduced by a Plummer
distribution of stars embedded in a Burkert halo.

The aim of the present paper is to investigate whether the scaling
laws found by \cite{PSS96} for the luminous and DM mass distributions in
Spirals are compatible with the available dSph kinematics and
photometry.  To facilitate some of this comparison, we must define a
stellar length scale for the dSphs which plays the same role as the
disk scale length $R_D$ in spirals.  One way to do this is to identify
the location of the peak of the Plummer dSph stellar spheroid
``rotation curve'', occurring at $1.4R_{\rm b}$, with the peak, at
$2.2.\ R_D$, of the stellar Freeman disk RC. 
Thus, in the
dSphs, we associate the Spiral length scale $R_D$ with the radius
$0.64R_{\rm b}$. However, we note that most of our conclusions in this
paper do not make use of this length scale.

\subsubsection{dSph core radii and halo central densities}

Under the assumptions of spherical symmetry and velocity isotropy, the
observation of a flat velocity dispersion profile implies that the DM
mass profile inferred via the Jeans equations (e.g. \cite{BT87}) is
dictated by the distribution of the luminous matter: $M(r)
\propto - r \: d\log \rho_b(r)/d\log r$. If the stellar density  
distribution
is cored, e.g. it is represented by a Plummer distribution, the DM
distribution obtained is also cored with the two core radii being
proportional. It turns out that the scale length of a Burkert halo
fitted to the profile obtained in this case is equal to the Plummer
radius of the light distribution, as can be seen by comparing columns
2 and 6 of table~\ref{tab:dSphs}.

The accurate determination of the size of the DM core radii in dSphs
requires the construction of dynamical models which include velocity
anisotropy and which can be compared to the full velocity distribution
rather than just the velocity dispersion as in the Jeans
equations. This analysis is beyond the scope of the current paper, and
will be presented elsewhere in connection with a larger velocity data
set (Wilkinson {\em et al.}, in prep.). In the absence of such
constraints, it is important to consider whether the $r_0$ value we
use are physically meaningful. It is possible that the stellar core
radii of dSphs may have evolved from their original values since their
formation due to various processes, both internal (e.g. supernovae)
and external (e.g. tidal disturbance). Since our analysis is based on
a scale which is essentially the present-day core radius of the
stellar distribution, we must be cautious in drawing conclusions from
this about the underlying DM distribution.
In what follows,
therefore, we restrict ourselves to the possible compatibility  between the values of the 
the halo parameters as  extrapolated at low luminosity  from  spiral
galaxy scaling laws and  the observed  kinematics 
of dSphs.  We defer a  more robust demonstration of the
physical nature of these parameters in dSphs (and in particular of
their relation to actual DM halo parameters) to future work. 

In contrast to the halo core radius $r_0$ which is determined  from the kinematics and  the light distribution,  via an assumption on the  anisotropy of stellar motions,  the normalization of the halo density
$\rho_0$ is constrained by the amplitude of the velocity dispersion
profile. In our models, $\rho_0$ corresponds to the mean mass density
inside one core radius of the light distribution, and is therefore
likely to be accurate to better than a factor of three, allowing for
uncertainty in the velocity anisotropy.

The assumptions of velocity isotropy and spherical symmetry that have
allowed us to solve the Jeans equation are supported by additional
arguments in a two particular dSphs (Ursa Minor and Fornax; see \cite{Gilmore07}).
Although models with larger cores (and
appropriate velocity anisotropy profiles) might also reproduce the
observations, our goal in this paper is to explore whether the gross
properties of the DM halos around dSphs are {\it consistent} with the
better-determined relations that characterize the $\sim10^4$ times
more massive halos around spirals. In this context, it is thus
interesting to investigate first the consistency of the simplest
models.

\section{Dark matter properties}
\label{sec:dmprop}

\begin{figure}[h!]
\centering
\vskip -2.5cm
\includegraphics[width=10cm]{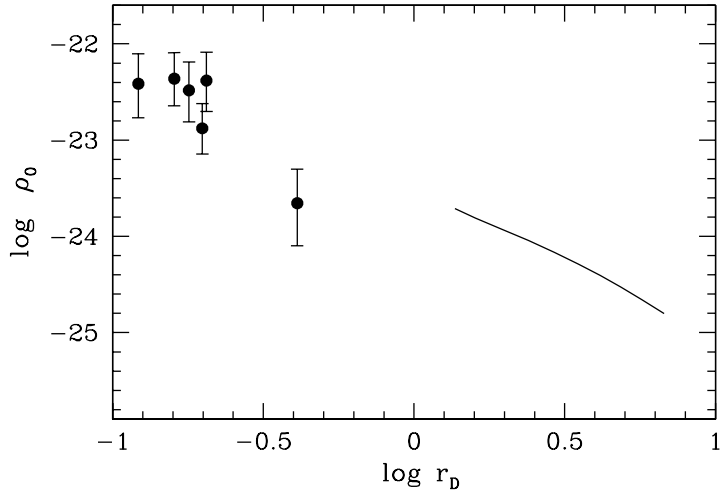}
\vskip -0.5cm
\caption{Halo central density $\rho_0$  versus stellar length  
scale $R_{\rm D}$ for spirals (solid curve) and dSphs (points)}
\vskip -0.0cm
\label{fig:rho0RD}
\end{figure}

\begin{figure}[h!]
\centering
\vskip -3cm
\includegraphics[width=11cm]{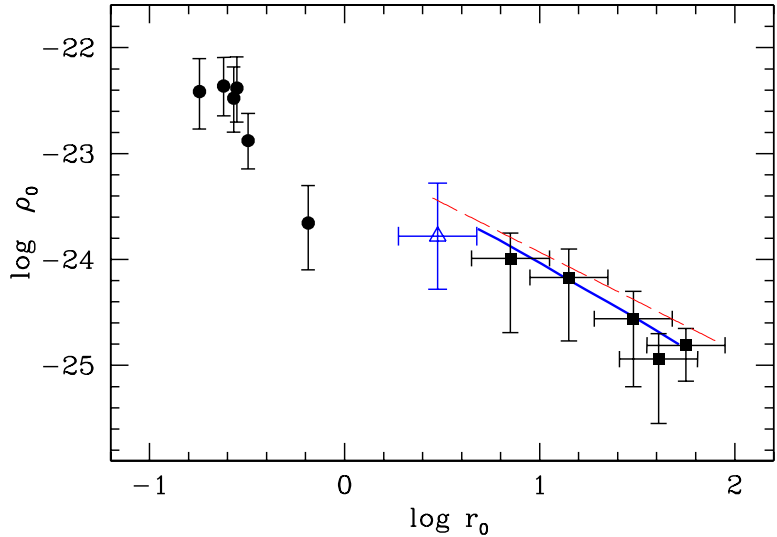}   
\caption{Structural halo parameters derived in i) spirals, by means  
of the URC (solid line), and the 
weak lensing shear (squares), ii) NGC 3741 (triangle) the darkest
spiral in the Local Universe by means of its kinematics, iii) Milky
Way dSph satellites, by means of their internal stellar
kinematics. The \cite{sapno08} relation is shown as a dashed
line.
All these data can be reproduced by $log \rho_0  \simeq  \alpha log r_0 +cost$  with $0.9 <\alpha<1.1$}.
\label{fig:rho0r0}
\end{figure}

In spiral galaxies, \cite{PSS96} have shown that the DM distribution is closely
related to that of the luminous matter. Their structural parameters
are all correlated: the mass and the length-scale of the luminous
matter correlate with similar quantities of the DM (\cite{PSS96}).
We start to frame the DM properties 
in galaxies of different luminosity and Hubble Types by analyzing for Spirals and dSphs the 
 $\rho_0$ vs $R_D$ relationship  in Fig.~\ref{fig:rho0RD} which is not  affected by 
an  anisotropy assumption in a way  relevant relevant for our scopes.  The data are taken from \cite{urc2} and are in good agreement with those in \cite{donato04,spano08}.
The  "central"  densities of  DM halos regularly  increase as  the size of 
the stellar component decreases.
In detail we obtain the intriguing result that although dSph
halos are much denser, they are found to lie on the extrapolation of
the spiral relationship. Although the observational
evidence for this relation is relatively strong, we stress that  its physical
interpretation is presently unknown.

We  continue  our comparison between
spirals and dSphs by testing the dSphs for consistency with  the internal halo  
relationship.
We next consider how the parameters $\rho_0$ and $r_0$ for the dSphs,
which we discussed in the previous section, compare to those of spiral
galaxies. In Fig.~\ref{fig:rho0r0}, we plot $\rho_0$ versus $r_0$,
recalling that $r_0$ in dSph halos is an assumption-dependent
quantity and its errorbar does not include the significant uncertainty
which arises from our lack of knowledge about the velocity anisotropy.
Interestingly, the figure shows that the extrapolation to higher
central densities of the $\rho_0-r_0$ relation for spirals would
predict halos halos for the dSphs which, as we have seen, are
consistent with the observed kinematics.

Although the observed data we are using for the dSphs neither require
cored halos, nor constrain their values in a model-independent way,
the ease with which a family of dSph halos can be obtained by simple
re-scaling of larger spiral galaxy halos is intriguing. If confirmed
by future data, the existence of such a scaling law, spanning three
orders of magnitude in each halo parameter, would indicate that the
physical processes of galaxy formation tend to produce DM cores of
sizes roughly equal to the stellar cores, in all galaxies. This would
potentially require a significant revision of our picture of galaxy
formation: it is difficult to explain the origin of such a scaling law
in that it relates quantities which do not exist in the standard
galaxy formation theory (i.e. a core radius and finite central density
in the DM distribution).

\section{The baryonic-dark matter interplay}
\label{sec:DMBaryons}
In this section, we investigate the coupling between the distributions
of dark and luminous matter at the level of the global mass
properties.  This is easier to investigate than the relationship among
the structural mass parameters as done in the previous section: in
both Hubble Types, the global properties are less dependent on the
modeling assumptions and are less strongly affected by observational
uncertainties.

\cite{PSS96} found that the dark and stellar mass inside a reference radius are
very closely related in Spirals. We now investigate whether this
general behaviour is also seen in dSphs.  We plot the ratio of the
stellar mass to the halo mass at a radius of $R_{83}/2$ corresponding
to the region inside which, in spirals and ellipticals, the baryonic
matter is always a major component of the dynamical mass budget.
Moreover, in both spirals and dSphs, the baryonic matter inside this
radius roughly coincides with the total stellar content, the HI
content being negligible inside this radius in dwarf spirals (see
e.g. Figs. 4.13 in \cite{rhee97}).
Finally, a convenient coincidence is
that this radius is approximately the farthest one for which we have
kinematic data for all objects (dSphs and spirals).

\begin{figure}[h!]
\centering
\vskip -4.3cm
\includegraphics[width=12cm]{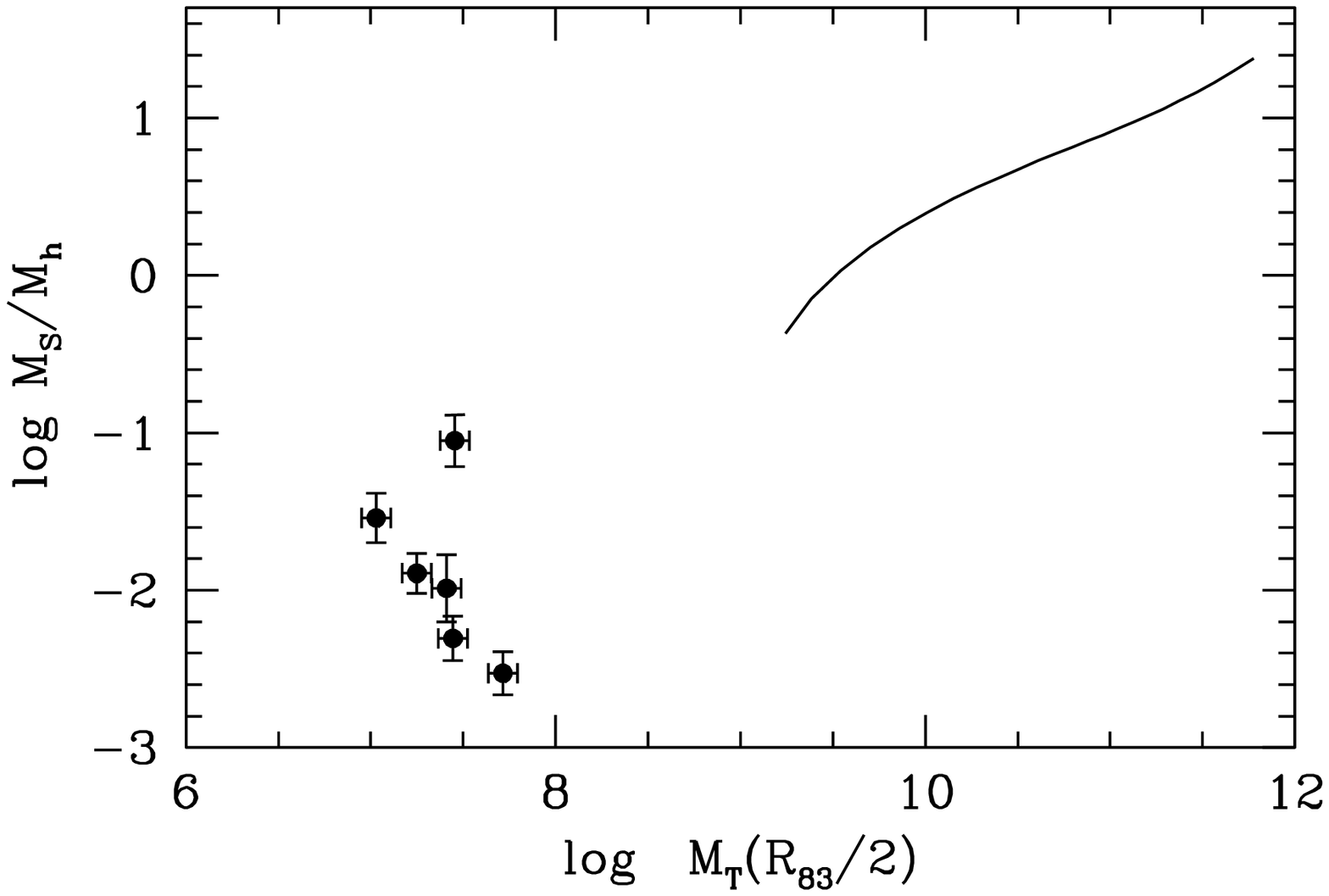}   
\vskip -0.3cm
\caption{Ratio of the stellar mass to the halo  mass at  $R_{83}/2$   
versus the
total mass {\em inside} $R_{83}/2$
for the dSphs (points).  The solid curve represents the relation
obtained for spires.}
\label{fig:MR83}
\vskip -0.4truecm
\end{figure}

In Fig.~\ref{fig:MR83} we show the well-established result that, in
contrast to galaxies of other Hubble type (and of much larger stellar
mass), dSphs are always dominated by DM even in their inner
regions. The fraction $M_{\rm s}/M_{\rm h}$ sets an important physical
quantity, namely the percentage of baryonic mass residing inside the
luminous part of a galaxy and (for dSph also the the percentage of
baryonic mass tout court, given the general absence of an ``external''
HI component). Bearing in mind that all galaxies are thought to have
formed with the same initial baryon fraction of roughly $17$ per
cent~\cite{wmap}, the data in Fig.~\ref{fig:MR83} imply that
star formation was very inefficient in processing gas into stars in
galaxies with stellar masses smaller than $10^{10}$M$_\odot$. In
particular, in dSphs we find values for the baryon fraction smaller
than $10^{-2}$ already at the optical radius, that imply even smaller
global values (i.e. at their virial radii).
Fig.~\ref{fig:MR83}
supports the view that dwarf systems, i.e.  objects less massive than
$10^{10}$M$\odot$ (irrespective of their Hubble type) must have
experienced  massive supernova feedback that has strongly limited
their star formation efficiency (see \cite{shankar06} for a
discussion).

An individuality of the dSphs as compared to other Hubble types~ is
that the baryonic fraction at any radius exhibits considerable object
to object variations, of magnitude about $1$ dex. This is several
times larger than those seen in ellipticals and disk systems~ \cite{PSS96,RTF07,deNaray08,Gerhard}, and cannot be explained by merely the
uncertainties in the determination of this quantity from the
observations. A number of authors have noted that the increase of mass
to light ratio with decreasing total luminosity seen in the dSphs is
consistent with a common halo mass scale (interior to $\sim 0.6$ kpc)
but a systematically varying baryon
fraction~\cite{Gilmore07,Mateo98,Koch07a}.
It is possible that this
may arise from environmental effects, perhaps related to their varied
orbits about the Milky Way, and in particular to their minimum
perigalactic distances~\cite{Piatek03,Piatek05,Piatek06,Piatek07}.

\section{Conclusions}
\label{sec:conc}
Dwarf spheroidal galaxies are the lowest luminosity stellar systems
which show evidence of dynamically significant DM. Moreover, (i) their
typical stellar masses lie in the range $3\times 10^5M_\odot$ to
$2\times 10^7M_\odot$, although the luminous masses of some recently
discovered objects are as low as $10^3M_\odot$~\cite{Martin2008};
(ii) the central densities of their DM halos reach almost $10^7$
times the critical density of the universe; (iii) their stellar length
scales are of order $0.3$ kpc; (iv) the DM in these systems typically
outweighs the baryonic matter by a large factor (from a few tens, up
to several hundred).

Let us stress that all the above quantities are about two
orders of magnitude different from those observed for spiral and
elliptical galaxies. Therefore, due to these extreme structural
properties, an understanding of the formation of dSphs is crucial for
the development of a complete picture of galaxy formation.

The main result of this paper is the finding that these galaxies,
despite being very separate in their physical properties from spirals
and ellipticals and having a large individual scatter in their
baryonic properties, exhibit kinematics which could be consistent with
the presence of DM halos which are essentially scaled-down versions
of those found in galaxies of much higher mass and different Hubble
type. We have shown that a Burkert halo density profile can reproduce
the available kinematic data for the dSphs. We find that the derived
central densities and the stellar core radii are consistent with the
extrapolation of the relationship between these quantities seen in
spiral galaxies. In addition, we have shown that if we extrapolate the
relation between halo central density and DM core radius previously
found in ellipticals and spirals, the halo parameters expected for the
dSphs would be consistent with their observed kinematics.

This potential consistency is intriguing, and could point to a common
physical process responsible for the formation of cores in galactic
halos of all sizes, or to a strong coupling between the DM and
luminous matter in dSphs.  If confirmed, this would suggest a Grand
Picture for galaxy formation in which in galaxies of all Hubble Types,
the DM is "aware" of the length scale of the luminous matter and vice
versa. It is worth noting that a potential connection between spiral
galaxies and dSphs does not appear as natural as one between dSphs and
other hot, spheroidal systems. For example, while the sizes of spiral
galaxies are presumably fixed by the angular momentum of the gas from
which they form, most of the present-day dSphs show no signs of
rotation~ (\cite{Battaglia08} have recently found evidence of rotation in the
Sculptor dSph). However, \cite{Mayer2001} have proposed a
formation scenario for dSphs in which they are initially low-mass disk
galaxies that are subsequently transformed into spheroids by tidal
interaction with the Milky Way. More recently, such models have been
shown to provide reasonable models for the properties of the
Fornax~\cite{Klimentowski2007} and LeoI~\cite{Lokas2008} dSphs.  If
the halos of dSphs do indeed follow the scaling laws defined by more
massive disk galaxies, this could lend indirect support to
evolutionary histories of this kind.

We also find evidence that the depletion of primordial gas through
supernova feedback has proceeded in a similar manner across all Hubble
types, with the resulting luminous to DM ratio depending mostly on the
depth of the gravitational potential. In the dSph potentials, which
correspond to a virial temperature of order $10^4$K, we find a
depletion by a factor of one hundred at $R_{83}$, and a factor which
may reach and exceed $10^3$ at the virial radius.

As we have emphasized throughout this work, further dynamical analysis
is needed in the dSphs to show directly that they possess DM cores
and, if so, to constrain their core radii. Nevertheless, it
interesting to speculate on the possible implications of these scaling
laws for our understanding of DM. Warm DM has been invoked as
a potential solution to the over-prediction of substructure by
$\Lambda$CDM simulations, and to the cusp-core
issue (e.g. \cite{moore99b}).
However, the existence of scaling
relations between the central density and core radius over three
orders of magnitude in both quantities would rule out this
explanation, unless the warm DM spectrum is extremely
fine-tuned. Further, such DM relations cannot arise due to either
self-annihilation or decay of DM which would predict a narrow range in
$\rho_0$ and no clear correlation of the latter with the core radius.

~\cite{dal00b} argued that the phase-space densities of DM
halos suggested that warm DM (either collisional or collisionless)
could not be the cause of cores in galaxy halos on all scales. These
authors suggested a dynamical origin for the cores of larger
galaxies. A universal scaling relation suggesting that any core
formation process has to proceed with approximately comparable
efficiency across three orders of magnitude in scale, would render
dynamical core formation scenarios (e.g. angular momentum transfer
from the baryons to the halo, expulsion of baryonic matter by
supernovae, or spiralling binary black holes at the centre of the
galaxy, etc.) more difficult to envisage.  We can speculate that a
physical property of DM which has the potential to explain the origin
of the observed trends among the structural DM parameters in
primordial NFW halos would be a self-interaction with an appropriate
velocity-dependent cross-section. Alternatively, some currently
unknown interaction between DM particles and baryonic matter or
photons may be required to explain core formation at the galactic
scale.

Clearly, direct kinematic evidence for or against the presence of
cores in dSph halos is now required to resolve the situation. If
cores are detected and are found to have parameters consistent with
those discussed in this paper, this will provide important information
about the properties of the DM of which they are composed. On the
other hand, if it turns out that cored halos are not a general
feature of dSphs (restricted perhaps to the cases of the Ursa Minor
and Fornax dSphs which require cores to allow survival of their
internal substructure), the similarity of the apparent interplay
between dark luminous matter in dSphs and spirals, as suggested by
Fig.~\ref{fig:rho0RD}, would remain an intriguing observation.

\chapter[Constant DM halo surface density]{A constant Dark Matter  Halo Surface density in Galaxies}

In the same line of the previous chapter I investigate further the DM halo properties. 
In particular I discuss our work where we confirm and extend an earlier claim by Spano {\it  et al.} 2008 \cite{spano08} that the
central surface density $\mu_{0D} $ of galaxy DM halos is
nearly constant, independent of galaxy luminosity. Based on the
co-added RCs of $\sim1000$ spiral galaxies, mass models of
individual dwarf irregular and spiral galaxies with high-quality
RCs, and the galaxy-galaxy weak lensing signals from a
sample of spiral and elliptical galaxies, we find that $ \log \mu_{0D}
= 2.05 \pm 0.15$, in units of M$_{\odot}$ pc$^{-2}$.  We also show
that the observed kinematics of Local Group dwarf spheroidal galaxies
are consistent with this value. Our results are obtained for galactic
systems spanning a wide range in magnitude, belonging to different
Hubble Types, and whose mass profiles have been determined by
independent modeling methods. The constancy of $\mu_{0D}$ is in sharp
contrast to the variation, by several orders of magnitude, of the halo
density and stellar surface density in the same objects.

\section{Introduction}
\label{sec:introduction}
 
It has been known for several decades that the kinematics of disk
galaxies exhibit a mass discrepancy: in their outermost optical
regions the circular velocity profile cannot be explained by the
ordinary stellar or gaseous matter. This is usually solved by adding
an extra mass component, the DM halo.
RCs have been used to assess the existence, the amount and the
distribution of this dark component (e.g. \cite{rubin80,PSS96}).
Recent debate in
the literature has focused on the "cuspiness" of the DM
density profile in the centres of galaxy halos that emerges in CDM simulations of structure formation \cite{nfw96,N03,moore99b,Neto07} but is not seen in observed data (e.g. \cite{gentile04,3741,dBMR01,blok_bosma02,marchesini02,gentile05}), as well as on the various systematics of
the DM distribution (see \cite{urc2}).
A significant
contribution to this debate was recently made by Spano {\it  et al.} 2008 \cite{spano08},
who fitted the RCs of 36 spiral galaxies using a mass model involving
a cored dark sphere of density
\begin{equation}
\rho(r) = \frac{\rho_0}{\left(1+\left(\frac{r}{r_0}\right)^2\right)^{3/2}}, 
\label{rho}
\end{equation}
where $\rho_0$ is the central density and $r_0$ is the core
radius. The authors found that the quantity $\mu_{0D}\equiv \rho_0
r_0$, proportional to the central halo surface density $ \Sigma(R) = 2
\int_{0}^{\infty}\rho(R,z)dz$, is independent of the galaxy blue
magnitude:
\begin{equation}
{\rm log} (\mu_{0D}/{\rm M_{\odot}  pc^{-2}}) =  2.2 \pm 0.25 \,\,\,\,\, {\rm or}  \,\,\,\,\, 
 \mu_{0D}= 150^{+100}_{-70} \,\,  {\rm M_{\odot}  pc^{-2}}.
\label{spano}
\end{equation}
For the sake of completeness, we note that a constant $\mu_{0D}$ of
about $100 \, \rm M_{\odot} pc^{-2}$, but with a much larger r.m.s
(0.4 dex), was found in the earlier work of \cite{KF04}, for a sample of 50 spiral and dwarf galaxies.

In this work, we will investigate the constancy of $\mu_{0D}$ found
in \cite{spano08} for objects whose central densities and core
radii vary by 1-2 orders of magnitude. We aim to confirm or rule out
this property by investigating independent samples of galaxies that
include a large number of objects of different Hubble Type and
magnitude and whose halo properties have been estimated using
different and independent methods of mass modeling. Given the
wide-ranging nature of the data and models we include, a positive
result arising in this study would be difficult to dismiss as a
coincidence.  

In this work, we make use of data from galactic systems spanning wide
ranges in luminosity and Hubble Type. Moreover, their mass
distributions are modeled by means of different techniques.  In
particular, our results are obtained from mass models of: (a) a large
sample of Spiral galaxies, analyzed by means of their URC; (b) the darkest Spiral in the local Universe,
studied through its kinematics; (c) a large sample of Spiral and
Elliptical galaxies, for which weak-lensing shear measurements are
available. We also compare the value of $\mu_{0D}$ obtained from these
luminous galaxies with the halo parameters consistent with the
kinematics of six dwarf spheroidal satellite galaxies of the Milky Way
for which extensive stellar kinematic data sets are available. We note
that with the exception of the weak lensing results which are
presented in this work, the values of $\rho_0$ and $r_0$ (and their
relative uncertainties) that we use to compute $\mu_{0D}$ and then to
investigate Eq. \ref{spano} are obtained and discussed in previous
works. We will refer interested readers to those publications for
details of the data and models.

In Section~2, we compute the quantity $\mu_{0D}$ for different
families of galaxies and compare it with the \cite{spano08} result.
A discussion of our result is given in Section 3.

\section {The $\rho_0 r_0$ vs magnitude  relationship}

In this work, we assume that the DM halo in each galaxy
follows the Burkert profile \cite{burkert95}:
\begin{equation}
\rho (r)={\rho_0\, r_0^3 \over (r+r_0)\,(r^2+r_0^2)} .
\label{rho_bur}
\end{equation}
This profile, when combined with the appropriate baryonic gaseous and
stellar components, has been found to generally reproduce very well
the available kinematics of disk systems out to 6 $R_D$ (\cite{sb00,47salucci03,gentile04}; see \cite{3741} for the case of the most extended RC).

The possible existence of a constant central surface density of DM for all galaxies does not depend on which specific (cored)
density distribution we assume for the DM, whether we adopt
any of the following: Spano {\it et al.} (2008; labeled as S hereafter),
Donato {\it et al.} (2004; D) \cite{donato04} or the present one (B).
Since different {\it
cored} mass models provide equally good fits to the same kinematical
data sets (e.g. \cite{gentile04}), with all of them (presumably)
describing the true, underlying halo mass profile $M_h$, the relations
$ M_h(r, B)= M_h(r, S)= M_h(r, D)=M_h(r, true)$ must hold, to within
observational uncertainties. This enables us to derive proportionality
factors between the corresponding parameters of the different cored
profiles. These can easily be computed: $\log \mu_{0D}(D)=
\log \mu_{0D}(B) +0.1 = \log \mu_{0D}(S) +0.3$ showing  that the
correction terms needed to compare different profiles are quite
negligible for each specific profile, relative to the observed
object-to-object variance of $\mu_{0D}$ at a fixed magnitude. 

Let us consider the case in which the  halos around galaxies  are cusped rather than
cored, as predicted by cosmological simulations of structure formation
(e.g. \cite{nfw96,Dehnen2005}), then does the use of the Burkert profile  introduce a bias into the results we obtain?
We first remind that  this  that possibility is  unlikely in view of the   many cases in which the NFW  profile  fails to
fit  the observed spiral kinematics.  However,   in any case,  in the range $0.2 R_D- R_{vir}$,
  the Burkert profile (with a   small  value for the
core radius and  a appropriate value 
for  the "central density"),  can mimic quite well    the velocity profile  of a  NFW halo   with a  standard value of the concentration 
parameter.   The Burkert profile is  therefore  an empirical one able to "measure" the  level of cuspiness of the underlying DM  density distribution.    As general result,   with the
same number of free parameters (i.e. a length scale and a density
scale) the Burkert profile  is able to fit all  available  kinematical data  within the observational uncertainties;  moreover,  differently from NFW mass modeling the present one  is able  to estimate very properly the disk mass,  which turns out  in
agreement with  the expectations from stellar population synthesis 
models (e.g. \cite{gentile04,spano08,SYD08}, see also \cite{frigerioPRL}). 
 Thus, the halo parameters we use   in this work  are suitable and  unbiased measures of  the spirals  physical properties.

Immediate, though indirect, support for the \cite{spano08} claim
comes from the results of \cite{donato04}.
In Fig.~1 we plot
$\mu_{0D}$ as a function of the stellar exponential scale-length $R_D$
for the sample of 25 disk systems (Spirals and LSB) analyzed by \cite{donato04}.
We see that the derived values for $\mu_{0D}$ are
almost constant, although $R_D$ varies by more than one order of
magnitude. In addition, there is no obvious difference between the
results from High Surface Brightness (HSB) galaxies and Low Surface
Brightness (LSB) galaxies.  This result is good agreement with Eq.
\ref{spano}. However, it is important to note that the two samples 
are similar, with five objects in common, and the analysis employed is
essentially the same.

\begin{figure}[h!]   
\centering
\vskip -2.5cm
\includegraphics[width=13cm]{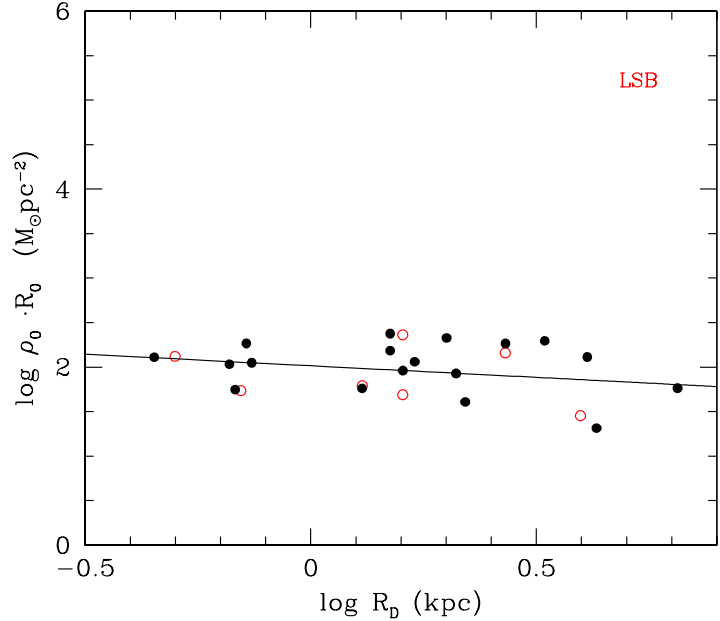}   
\vskip -1.3cm
\caption{The central halo surface density  
$\rho_0 r_0 $ as a function of disk scale-length $R_D$ for the Donato
et al. (2004) sample of galaxies. Open and filled circles refer to LSB
and HSB galaxies, respectively. The solid line is our best fit to the
data. }
\end{figure}   

\begin{figure} [h!]  
\centering
\vskip -4.5cm
\includegraphics[width=13cm]{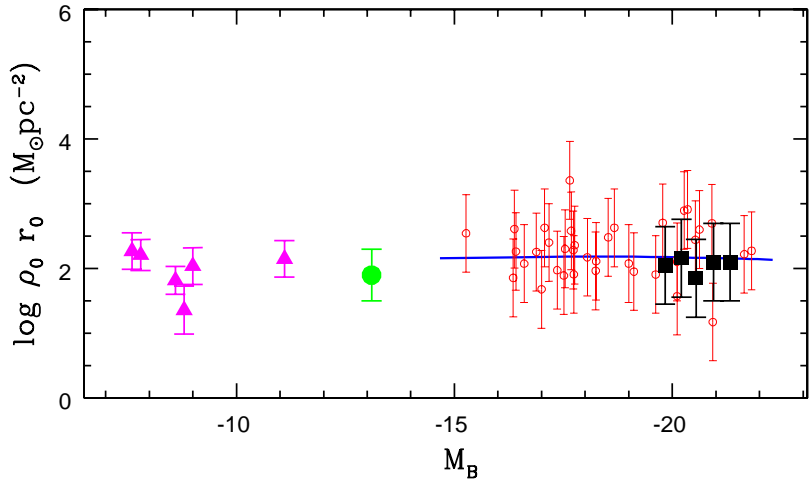}
\vskip -1cm
\caption{ $\rho_0 r_0$ in units of $M_\odot$pc$^{-2}$ as a function of
galaxy magnitude for different galaxies and Hubble Types.  The
original Spano et. al. (2008) data (empty small red circles) are
plotted alongside those for Spirals obtained by the URC (solid line).
The full (green) circle corresponds to the the dwarf galaxy N3741,
full (black) squares to Spirals and Ellipticals obtained by weak
lensing and the (pink) triangles to the dSphs obtained by their
kinematics.}
\end{figure}   

We now calculate the central surface density $\mu_{0D}$ for the family of Spirals by means
of their URC. 
This curve, {\it on average},
reproduces well \cite{PSS96,urc2} the RCs of individual
objects out to their virial radii $R_{vir}$ (the radius at which the
halo mass is 100 times the background mass).
The URC is built from
(a) the co-added kinematical data of a large number of Spirals (\cite{PSS96}; see also \cite{catinella06}) and
(b) the disk mass versus halo virial
mass relationship found by \cite{shankar06} and it leads, for
objects of given luminosity (or disk mass), to specific values of
$\rho_0$ and $ r_0 $ (see equations 6a, 7 and 10 of \cite{urc2} for details).
The solid line in Fig.~2 shows the resulting
$\mu_{0D}$ as a function of galaxy magnitude. Because the URC is
derived from co-added RCs, the particularities of
individual galaxy curves (e.g. observational errors or
non-axisymmetric motions due to bars or spiral structure) are averaged
out.  The URC therefore allows us to trace the general form of the
gravitational potential of Spirals over their full luminosity range.
A natural concern is that the values of the halo parameters we obtain are biased by  the smoothing process itself.
However, the  values of $\mu_{0D}$ obtained from detailed mass modeling of 36 RC of  spirals by \cite{spano08} shown in Fig. 2 as open circles (the $\mu_{0D}$'s for the 25 mass models in  \cite{donato04} not reported here are in very good agreement with the latter)  are consistent with those obtained from the  the URC, all  suggesting that these various  mass  modeling  it is returning physically meaningful values of physically meaningful mass  parameters.
More in detail,  the URC provides, for Spirals of a given luminosity, a reliable estimate of their average value of $\mu_{0D}$, although not of their cosmic variance around it.
In the estimation of the latter quantity, the detailed studies of
individual objects such as those of \cite{spano08} and \cite{donato04} are indispensable to  provide us with the needed quantity, that results negligible for the present aim.

\begin{figure} [h!]
\centering
\vskip -1cm
\includegraphics[width=13cm]{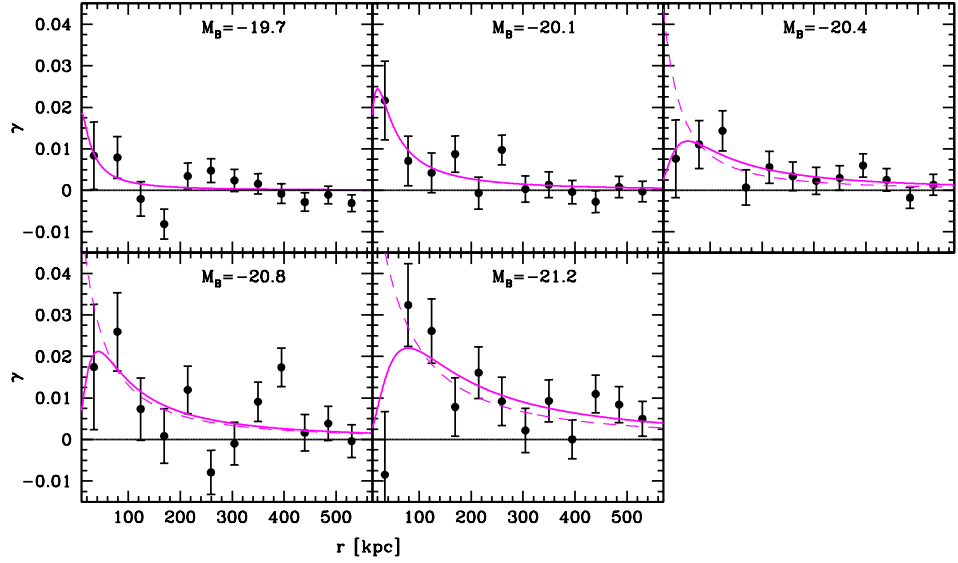}
\vskip -3.5cm
\caption{Tangential shear measurements from \cite{hoek}
as a function of projected distance from the lens in five B-band
luminosity bins. In this sample, the lenses are at a mean redshift
z$\sim$0.32 and the background sources are, in practice, at
$z=\infty$.  The solid (dashed) magenta line indicates the Burkert
(NFW) model fit to the data.}
\label{fig:weak_lensing}
\end{figure}
In this work,  we estimate the values of
$\mu_{0D}$ from the DM structural parameters obtained in a third different way:  from analyzing  the
galaxy-galaxy weak-lensing signals for a large sample of Spiral and
Elliptical galaxies. The details are presented as follows.

Recent developments in weak gravitational lensing have made it
possible to probe the ensemble-averaged mass distribution around
galaxies out to large projected distances.  These new data provide
crucial information, complementary to that obtained from kinematics.
The tidal gravitational field of the DM halos generates weak-lensing
signals, by introducing small coherent distortions in the images of
distant background galaxies, which can be detected in current large
imaging surveys. We can measure, from the centre of the lenses out to
large distances (much greater than the distances probed by the
kinematic measurements), the azimuthal-averaged tangential shear
$\gamma_{\rm t}$
\begin{equation}
<\gamma_{\rm t}> \equiv
\frac{\overline{\Sigma}(R) - \Sigma(R)}{\Sigma_{\rm c}},
\end{equation}
where $\Sigma(R) = 2 \int_{0}^{\infty}\rho(R,z)dz$ is the projected
mass density of the object distorting the galaxy image, at projected
radius $R$ and $\overline{\Sigma}(R)= \frac{2}{R^2}\int_0^R x\Sigma(x) dx$
is the mean projected mass density interior to the radius $R$. The
critical density $\Sigma_{\rm c}$ is given by 
\begin{equation}
\Sigma_c \equiv \frac{c^2}{4\pi G} \frac{D_{\rm s}}{D_{\rm l} D_{\rm ls}}, 
\end{equation}
where 
$D_{\rm s}$ and $D_{\rm l}$ are the distances from the observer to the
source and lens, respectively, and $D_{\rm ls}$ is the source-lens
distance.  The above relations directly relate observed signals with
the underlying DM halo density. For our analysis we use the weak
lensing measurements from \cite{hoek} available out to a
projected source-lens distance of $530$ kpc.
The sample, which
contains about $10^5$ isolated objects and spans the whole luminosity
range of Spirals, is split into 5 luminosity bins of magnitudes given
in table~\ref{tab:weak_lensing}. The most luminous bin is likely dominated by the
biggest Ellipticals.
By adopting a density profile, we
model $\gamma_{\rm t}$ (see Fig.~\ref{fig:weak_lensing}) and obtain
the structural free parameters $\rho_0$ and $r_0$ by means of standard
best-fitting techniques.  The Burkert profile given by
equation~\ref{burkert} provides an excellent fit to the tangential
shear (see Fig.~\ref{fig:weak_lensing} and
table~\ref{tab:weak_lensing}). The NFW density profile provides a less
satisfactory fit to the gravitational shear around the most luminous
objects (Fig.~\ref{fig:weak_lensing}; see also Fig.~6 of \cite{hoek}.
Notice that at fainter luminosities ($M_B> -20.1$) the
signal-to-noise is too low to discriminate between mass models, so
that while the Burkert profile remains a working assumption, NFW
profiles cannot be excluded.
Assuming the Burkert halo profile we plot the resulting $\mu_{0D}$ values in
Fig. 2 as solid squares. 

\begin{table}[h!]
\begin{center}
\begin{tabular}{c|c|c|c}\hline
\emph{$M_{B}$}&\emph{$r_0\:$(kpc)}&\emph{$\rho_0 \:(10^{6}
M_{\odot}/kpc^3$)}&$\chi^2_{red}$\\ \hline
-19.7&$7^{+3}_{-6}$&$15^{+15}_{-15}$&1.6\\

-20.1&$14^{+6}_{-10}$&$10^{+10}_{-10}$&1\\

-20.4&$40.4^{+20}_{-20}$&$1.7^{+1.5}_{-1.5}$&0.7\\

-20.8&$30^{+10}_{-20}$&$4.1^{+4}_{-4}$&2.2\\

-21.1&$56^{+20}_{-20}$&$2.3^{+1.2}_{-1.2}$&1.1\\\hline
\end{tabular}
\caption{Structural parameters  and goodness of fit for a Burkert profile  to the  weak lensing signal of \cite{hoek}.}
\label{tab:weak_lensing}
\end{center}
\end{table}

The nearby dwarf galaxy NGC 3741 ($M_B=-13.1 $) is a very interesting
case: it represents the very numerous dwarf disk objects which are
DM dominated down to one disk length-scale or less and in
which the HI gaseous disk is the main baryonic component.  In
addition, this specific galaxy has an extremely extended and very
symmetric HI disk, which allowed \cite{3741} to carefully
trace the RC and therefore its gravitational potential out to
unprecedented distances relative to the extent of the optical
disk. The data probe to radii of 7 kpc (equivalent to 42 B-band
exponential scale lengths), and have several independent points within
the estimated halo core radius. The RC was decomposed into its
stellar, gaseous and dark (Burkert) halo components, yielding a very
good fit \cite{gentile_tonini07}: the corresponding $\mu_{0D}$ is
plotted in Fig. 2 as a filled circle.  The relatively large error-bar
is due to uncertainties in the distance.

At the level of 0.2 dex,  no large differences 
emerges between  the values of $\mu_{0D}$ estimated in different way or referring to a
Spiral or an  Elliptical population. It thus appears that the central
surface density of DM halos assumes a nearly constant value with
respect to galaxy luminosity, over a range of at least nine magnitudes.

The Milky Way satellite dwarf spheroidal (dSph) galaxies are the
smallest and most DM dominated systems known in the universe
(see e.g. \cite{Gilmore07,Mateo98} and references therein).
Their low HI gas content is another property that sets them
apart as a galaxy class (e.g. \cite{Grebel2003}).
In a recent study of six dSphs, \cite{Gilmore07} showed that,
assuming spherical symmetry and velocity isotropy, the stellar
kinematics and photometry of dSphs are consistent with their occupying
cored DM halos. \cite{salucci2008} subsequently showed that, for
the same simplifying assumptions, cored Burkert profiles are able to
reproduce the dSph kinematic observations. Our current lack of
knowledge about the anisotropy of the velocity distribution means that
the density profiles of dSphs are not uniquely constrained by the
data, and both cored and cusped models can reproduce the data in most
dSphs \cite{Gilmore07,Koch07a,Battaglia08}.
Bearing this caveat in mind, it is nevertheless
interesting to compare the value of $\mu_{0D}$ from our spiral and
elliptical galaxy samples with the values obtained from the halo
parameters which \cite{salucci2008} showed to be consistent with
the dSph kinematics. These are plotted in Fig.~2 as triangles.  Note
that the errorbars shown reflect only the statistical errors in the
estimation of the parameters from the observed data, and do not
account for any modeling uncertainties.  We emphases that the
relatively small range of both halo density and core radius found for
the dSphs means that the current data in these galaxies would be
consistent with the approximate constancy of any product of $\rho_0$
and $r_0$. In particular, it has been noted that all the dSph data are
consistent with their occupying halos which contain roughly equal
masses interior to about 0.6-1.0\,kpc (i.e. $\rho_0 r_0^3 \approx$
constant: \cite{Gilmore07,MateoLeo98}. Nevertheless, the consistency of the dSph data with the value
of $\mu_{0D}$ suggests that the relation $\rho_0 r_0 \approx$ constant
may extend to fainter systems, and thus be valid over a range of
fourteen magnitudes in luminosity.

\section{Discussion and Conclusions}

We have compiled data on the DM halo mass distribution in many
galactic systems of different Hubble Type (including Dwarfs disk
galaxies, Spirals, Ellipticals) spanning a luminous range of about $-8
< M_B < -22$ and a gaseous-to-stellar mass fraction of many orders of
magnitude. The mass modeling of such objects has been carried out
using different and independent methods.  The halos are all well
reproduced by a cored profile with two structural parameters: a
central halo density $\rho_0$ and a core radius $r_0$, whose
respective values range over several orders of magnitude: $ 6\times
10^{-23} {\rm g/cm^3}
\leq \rho_0 \leq 10^{-25} {\rm g/cm^3}$ and $ 0.3$ kpc $\leq r_0 \leq
30$ kpc.  In spite of dealing with galaxies with such different
physical properties, we have found that their central DM surface
density $\mu_{0D}\equiv \rho_0 r_0$ remains almost constant:
\begin{equation}
\mu_{0D}= 110^{+50}_{-30} \ {\rm M_{\odot}  pc^{-2}}
\label{concl}
\end{equation}
independent of galaxy luminosity. In addition, we have compared this
value of $\mu_{0D}$ with the kinematics of dSphs and found that these
are also be consistent.

Our results support the pioneering analysis by \cite{spano08},
based on a sample of 36 spiral galaxies, in which they found a nearly
constant halo surface density around 150 ${\rm M_{\odot} pc^{-2}}$
independent of galaxy luminosity. In addition to investigating many
more objects across more Hubble-types and a much wider luminosity
range, we have obtained the halo surface density $\mu_{0D}$ both from
{\it individual} galaxy kinematics and from {\it co-added}
kinematical/shear measurements.  The approximate constancy of
$\mu_{0D}$ is in stark contrast to the stellar central surface density
in galaxies of different Hubble Type and magnitudes which shows large
variations (see the relevant works cited above for details). In
Spirals, it ranges between $ 800 \, {\rm M_\odot pc^{-2}}$ at about
$M_B= -22.5$ to $\sim 50 \, {\rm M_\odot pc^{-2}}$ at $M_B=-17$, in
dSph it probably does not reach $ 1 \, {\rm M_\odot pc^{-2}}$, while
in Ellipticals it easily exceeds $ 10000 \, {\rm M_\odot pc^{-2}}$,
with large variations with luminosity and object-to-object.

It is important to consider how the approximate constancy of
$\mu_{0D}$ with $M_B$ is related to the correlation between $r_0$ and
$\rho_0$,
\begin{equation}
\log r_0 = A  \log \rho_0 +C
\label{burkert}
\end{equation}  
which has been claimed in Spiral galaxies \cite{burkert95}.
First,  the former relationship (Eq. \ref{spano}) 
links two very different physical properties of galaxies (i.e. the
central DM surface density and galaxy magnitude), while the latter
(Eq. \ref{burkert}) relates two "internal" DM halo structural
parameters. Further, let us stress that   $A $ could be near,  but maybe relevantly not coinciding with   -1, 
 see  \cite{sb00,gentile04,donato04,urc2,burkert95} and even   show  some non (log) linearities  (see  \cite{gentile04}), but still the quantity $r_0 \ \rho_0$ could be found   constant, within a factor 2, over several orders of magnitudes. The study of Eq. \ref{concl}  and Eq. \ref{burkert}  must therefore  proceed separately. 
 
The evidence that the DM halo central surface density $\rho_0 r_0$,
over at least nine (and possibly up to fourteen) galaxy magnitudes and
across several Hubble types, remains constant to within less than a
factor of two, suggests that $\mu_{0D}$ may be an important physical
quantity in the DM distribution of galaxies. This is a surprising
finding, as it is difficult to envisage how such a relation can be
maintained across galaxies which range from DM-dominated to
baryon-dominated in the inner regions. In addition, these galaxies
have experienced significantly different evolutionary histories
(e.g. numbers of mergers, significance of baryon cooling, stellar
feedback, etc.). Further investigation is clearly required in order to
verify and interpret this relation.

\chapter{Conclusions}

Although CDM model is able to correctly describe observations made on the largest cosmological scales down to roughly those of galactic scales, and from the early Universe to the present epoch, on subgalactic scales it predicts that there should be more DM than is detected gravitationally.

There are several suggestions that could account for the lack of the cusps, both from fundamental physics and through astrophysical processes.
These suggestions make definite predictions of other observables that could
be used to test the variant properties of DM.
In this Thesis I show how these tests can be performed with the analysis of the RCs with a proper sample of spirals.

As gravity is by far the dominant interaction at cosmological scales and the force governing the evolution of the universe, another perspective to the current picture of the evolution and the matter content of the Universe arises: 
the description of the gravitational interaction at the relevant scales may be not sufficiently adequate and a modification of gravity could answer the cosmological and astrophysical riddles.
Conceivable alternatives are numerous and each of them produces distinctive modifications on small scales that can be tested through improved astronomical observations and numerical simulations.
In my Thesis I explore how such modifications may well account for the phenomenon of the RCs.
Of course such a solution pays the price of renouncing the  great success of the actual theory of structure formation and  evolution  envisaging a pure baryonic scenario.

In my Thesis I also investigate the extension of the well-known scaling relations of DM halo properties on a large range in galaxy luminosity, including the latest observations of the Milky Way satellites.
Within some assumption I show that the halos of the faintest objects have properties scaled down with respect of the bigger ones.
This  possible consistency could well point to a common physical process on the formation of galactic halo cores of all sizes.
I also find a surprising evidence of a DM constant halo central surface density over a large range in galaxy magnitude, suggesting that it could be an important physical quantity in the DM distribution of galaxies, even though these galaxies have different evolutionary histories.
Further investigations are necessary to better interpret these results.

I conclude that properties of dark matter are written on the kinematical features of the luminous matter and that their detailed study can give the right glasses to decipher its nature.

\backmatter


\begin{thebibliography}{250}

\addcontentsline{toc}{chapter}{Bibliography}

\bibitem{wmap}
D.~N. Spergel, R. Bean, O. Dor{\'e}, M.~R. Nolta, C.~L. Bennett, J. Dunkley, G. Hinshaw, N. Jarosik, E. Komatsu, L. Page, H.~V. Peiris, L. Verde, M. Halpern, R.~S. Hill, A. Kogut, M. Limon, S.~S Meyer, N. Odegard, G.~S. Tucker, J.~L. Weiland, E. Wollack and E.~L. Wright,
{\it Astrophys. J. S.} {\bf 170}, 377 (2007).

\bibitem{ostriker}
J.P. Ostriker,
{\it Ann. Rev. Astron. Astrophys.} {\bf 31}, 689-716 (1993).

\bibitem{nfw96}
J.~F. Navarro, C.~S. Frenk and S.~D.~M. White,
{\it Astrophys. J} {\bf 462}, 563 (1996).

\bibitem{sb00}
P. Salucci and A. Burkert,
{\it Astrophys. J} {\bf 537}, L9 (2000).

\bibitem{47salucci03}
P. Salucci, F. Walter and A. Borriello,
{\it Astron. Astrophys.} {\bf 409}, 53 (2003).

\bibitem{gentile04}
G. Gentile, P. Salucci, U. Klein, D. Vergani and P. Kalberla,
{\it Mon. Not. Roy. Astron. Soc.} {\bf 351}, 903 (2004).

\bibitem{donato04}
F. Donato, G. Gentile and P. Salucci, 
{\it Mon. Not. Roy. Astron. Soc.} {\bf 353}, L17 (2004).

\bibitem{blok05}
W., J., G. de Blok,
{\it Astrophys. J.} {\bf 634}, 227 (2005).

\bibitem{deNaray06}
R. Kuzio de Naray, S.S. McGaugh, W.~J.~G.de Blok and A. Bosma,
{\it Astrophys.J.S} {\bf 165}, 461 (2006).

\bibitem{3741}
G. Gentile, P. Salucci, U. Klein and G.L. Granato, 
{\it Mon. Not. Roy. Astron. Soc.} {\bf 375}, 199 (2007).

\bibitem{spano08}
M. Spano, M. Marcelin, P. Amram, C. Carignan, B. Epinat and O. Hernandez, 	
{\it Mon. Not. Roy. Astron. Soc.} {\bf 383}, 297 (2008).

\bibitem{moore99}            
B. Moore, S. Ghigna, F. Governato, G. Lake, T. Quinn, J. Stadel and P. Tozzi, 
{\it Astrophys. J.} {\bf 524}, L19 (1999). 


\bibitem{olive}
K.~A. Olive, G. Steigman and T.~P. Walker, 
{\it Physics Reports} {\bf 333}, 389 (2000). 

\bibitem{pdg}
C. Amsler {\it et al.} (Particle Data Group),
{\it Physics Letters B} {\bf 667}, 1 (2008).

\bibitem{wmap5}
G. Hinshaw, {\it et al.},
{\it Astrophys.J.Suppl.} {\bf 180}, 225 (2009).

\bibitem{tegmark06} 
M. Tegmark, {\it et al.},
{\it Phys. Rev. D} {\bf 74}, 123507 (2006). 

\bibitem{peebles70}
P.J.E. Peebles,
{\it Astron. J.} {\bf 75}, 13 (1970).

\bibitem{kolb}
E.~W. Kolb and M.~S. Turner, 
{\it eaun.book} (1990). 

\bibitem{bardeen}
J.~M. Bardeen, J.~R. Bond, N. Kaiser and A.~S. Szalay, 
{\it Astrophys. J.} {\bf 304}, 15 (1986). 

\bibitem{eisenstein_hu99}
D.~J. Eisenstein, W. Hu,
{\it Astrophys. J.} {\bf 511}, 5 (1999). 

\bibitem{jenkins}
A. Jenkins, C.~S. Frenk, S.~D.~M. White, J.~M. Colberg, S. Cole, A.~E. Evrard, H.~M.~P. Couchman and N. Yoshida, 
{\it Mon. Not. R. Astron. Soc.} {\bf 321}, 372 (2001). 

\bibitem{masiero}
M. Taoso, G. Bertone and A. Masiero, 
{\it JCAP} {\bf 3}, 22 (2008). 




\bibitem{cdms}
$http://cdms.berkeley.edu/$

\bibitem{edelweiss}
$http://edelweiss.in2p3.fr/$

\bibitem{warp}
$http://warp.lngs.infn.it/$

\bibitem{xenon}
$http://xenon.astro.columbia.edu/$

\bibitem{cresst}
$http://www.cresst.de/darkmatter.php$

\bibitem{dmtpc}
G. Sciolla, {\it et al.}, arXiv:astro-ph/0805.2431

\bibitem{dama}
$http://people.roma2.infn.it/~dama/$

\bibitem{dama_exp}
R. Bernabei, {\it et al.}, 
{\it Eur.Phys.J.C} {\bf 56}, 333 (2008).

\bibitem{xmm}
$http://xmm.esac.esa.int/$

\bibitem{chandra}
$http://chandra.harvard.edu/$

\bibitem{integral}
$http://www.esa.int/esaMI/Integral/$

\bibitem{cgro}
$http://heasarc.gsfc.nasa.gov/docs/cgro/index.html$

\bibitem{agile}
$http://agile.rm.iasf.cnr.it/$

\bibitem{glast}
$http://www.nasa.gov/mission_pages/GLAST/main/index.html$

\bibitem{cangaroo}
$http://ptp.ipap.jp/link?PTPS/151/85/$

\bibitem{hess}
$http://www.mpi-hd.mpg.de/hfm/HESS/$

\bibitem{magic}
$http://wwwmagic.mppmu.mpg.de/$

\bibitem{veritas}
$http://veritas.sao.arizona.edu/$

\bibitem{amanda}
$http://amanda.uci.edu/$

\bibitem{icecube}
$http://icecube.wisc.edu/$

\bibitem{antares}
$http://antares.in2p3.fr/$

\bibitem{pamela}
$http://pamela.roma2.infn.it/index.php$

\bibitem{ams}
$http://ams.cern.ch/AMS/ams_homepage.html$

\bibitem{hooper07}
D. Hooper,
{\it eprint} arXiv:hep-ph/0710.2062 

\bibitem{olive08}
K.~A. Olive, 
{\it Eur.Phys.J.C} {\bf 59}, 269 (2009).

\bibitem{raffelt07}
G.~G. Raffelt, 
{\it J. Phys. A} {\bf 40}, 6607 (2007). 

\bibitem{faber79}
S. M. Faber and J.S. Gallagher, 
{\it Annual Rev. Astron. Astrophys.} {\bf 17}, 135 (1979).

\bibitem{bosma81}
A. Bosma, 
{\it Astron. J} {\bf 86}, 1825 (1981).

\bibitem{rubin80}
V.C. Rubin, W.K. Ford, Jr. and N. Thonnard, 
{\it  Astrophys. J.} {\bf 238}, 471 (1980).

\bibitem{rubin85}
V.C. Rubin, D. Burstein, W.K. Ford, Jr. and N. Thonnard, 
{\it  Astrophys.J.} {\bf 289}, 81 (1985).

\bibitem{PS88}
M. Persic and P. Salucci, 
{\it Mon. Not. R. Astron. Soc.} {\bf 234}, 131 (1988).

\bibitem{broeils92}
A.H. Broeils, 
{\it Astron. Astrophys.} {\bf 256}, 19 (1992).

\bibitem{sofue01}
Y. Sofue and V. Rubin, 
{\it Annual Rev. Astron. Astrophys.} {\bf 39}, 137 (2001).

\bibitem{slipher14}
V.M.Slipher, 
{\it Lowell Obs. Bull.} {\bf 62}, 11 (1914).

\bibitem{wolf14}
M. Wolf, 
{\it Vierteljahresschr Astron. Ges.} {\bf 14}, 162 (1914).

\bibitem{babcock39}
H.W. Babcock, 
{\it Lick Obs. Bull.} {\bf 19}, 41 (1939).

\bibitem{oort40}
J.H. Oort, 
{\it Astrophys. J.} {\bf 91}, 273 (1940).

\bibitem{zwicky37}
F. Zwicky, 
{\it Astrophys. J.} {\bf 86} ,3 (1937).

\bibitem{whitehurst72}
R.N. Whitehurst and M.S. Roberts, 
{\it Astrophys. J.} {\bf 175}, 347 (1972).

\bibitem{PSS96}
Persic, M., Salucci, P. and Stel, F., 
{\it Mon. Not. R. Astron. Soc.} {\bf 281}, 27 (1996).

\bibitem{freeman}
K. C. Freeman, 
{\it Astrophys. J.} {\bf 160}, 811 (1970).

\bibitem{aceves06}
H.Aceves, H.Ve\'{a}zquez and F. Cruz, 
{\it Mon. Not. R. Astron. Soc.} {\bf 373}, 632 (2006).

\bibitem{kent}
S.M. Kent, 
{\it Astron. J.} {\bf 91}, 6 (1986).

\bibitem{TF77}
R.B. Tully and J.R. Fisher, 
{\it Astron. Astrophys.} {\bf 54}, 661 (1977).

\bibitem{RTF07} 
I.A. Yegorova and P. Salucci, 
{\it Mon. Not. R. Astron. Soc.} {\bf 377}, 507 (2007).

\bibitem{ashman92} 
K.M. Ashman, 
{\it Publ. Astron. Soc. Pac.} {\bf 104}, 1109 (1992).

\bibitem{weinberg97} 
D. Weinberg, 
{\it ASPC} {\bf 117}, 578 (1997).

\bibitem{olive05}
K.A. Olive, 
{\it preprint} arXiv: astro-ph/0503065 

\bibitem{salucci_gentile06} 
P. Salucci and G. Gentile,  
{\it Phys. Rev.D} {\bf 73}, 128501 (2006).

\bibitem{verheijen97}
M.A.W. Verheijen, 
PhD. Thesis, Groningen University (1997).

\bibitem{zavala03} 
J. Zavala, V. Avila-Reese, H. Hern{\'a}ndez-Toledo and C. Firmani, 
{\it Astron. Astrophys.} {\bf 412}, 633 (2003).


\bibitem{PS91}
M. Persic, P. Salucci, 
{\it Astrophys. J.} {\bf 368}, 60 (1991).

\bibitem{urc2}
P. Salucci, A. Lapi, C. Tonini, G. Gentile, I. Yegorova and U. Klein, 
{\it Mon. Not. Roy. Astron. Soc.} {\bf 378}, 41 (2007).

\bibitem{catinella06}
B. Catinella, R. Giovanelli and M.P. Haynes, 
{\it Astrophys. J.} {\bf 640}, 751 (2006).
 
\bibitem{shankar06}
F. Shankar, A. Lapi, P. Salucci, G. De Zotti and L. Danese,
{\it Astrophys. J.} {\bf 643}, 14 (2006).

\bibitem{rhee97}
M-H Rhee, 
PhD Thesis, University of Groningen (1996)

\bibitem{swaters99}
R.A. Swaters, 
Ph.D. Thesis, Groningen University (1999).

\bibitem{SYD08}
P. Salucci, I.A. Yegorova and N. Drory, 
{\it Mon. Not. Roy. Astron. Soc.} {\bf 720} (2008).



\bibitem{WR78}
S.D.M. White and M.J. Rees, 
{\it Mon. Not. R. Astron. Soc.} {\bf 183}, 341 (1978).

\bibitem{BFPR84}
G. R. Blumenthal, S. M. Faber, J.R. Primack, and M.J. Rees, 
{\it Nature} {\bf 311}, 517 (1984).

\bibitem{WF91}
S.D.M. White and C.S. Frenk, 
{\it Astrophys. J.} {\bf 379}, 52 (1991).

\bibitem{nfw97}
J.F. Navarro, C.S. Frenk and S.D.M. White, 
{\it Astrophys. J.} {\bf 490}, 493 (1997).

\bibitem{peebles80}
P.J.E. Peebles, 
\textit{Physical cosmology} (1980 Princeton university press)

\bibitem{eke96}
V.R. Eke, S. Cole and C.S. Frenk, 
{\it Mon. Not. Roy. Astron. Soc.} {\bf 282}, 263 (1996).

\bibitem{bryan98}
G.L. Bryan and M-L. Norman, 
{\it Astrophys. J.} {\bf 495}, 80 (1998).

\bibitem{bullock01b}
J.S. Bullock, T.S. Kolatt, Y. Sigad, R.S. Somerville, A.V. Kravtsov,
A.A.  Klypin, J.R. Primack and A. Dekel, 
{\it Mon. Not. Roy. Astron. Soc.} {\bf 321}, 559 (2001).

\bibitem{duffy08}
A. R. Duffy, J. Schaye, S. T. Kay and C. Dalla Vecchia,
{\it Mon.Not.Roy.Astron.Soc.} {\bf 390}, L64 (2008).

\bibitem{cole96}
S. Cole and C. Lacey, 
{\it Mon. Not. Roy. Astron. Soc.} {\bf 281}, 716 (1996).

\bibitem{huss99}
A. Huss, B. Jain and M. Steinmetz, 
{\it Astrophys. J.} {\bf 517}, 64 (1999).

\bibitem{fuku97}
T. Fukushige and J. Makino, 
{\it Astrophys. J. L} {\bf 477}, L9 (1997).

\bibitem{moore98}
M. Moore, F. Governato, T. Quinn, J. Stadel and G. Lake, 
{\it Astrophys. J.} {\bf 499}, L5 (1998).

\bibitem{jing00}
Y.P. Jing and Y. Suto, 
{\it Astrophys. J. L} {\bf 529}, L69 (2000).

\bibitem{ghigna00}
S. Ghigna, B. Moore, F. Governato, G. Lake, T. Quinn and J. Stadel, 
{\it Astrophys. J.} {\bf 544}, 616 (2000).

\bibitem{fuku01}
T. Fukushige and  J. Makino, 
{\it Astrophys. J.} {\bf 557}, 533 (2001).

\bibitem{kravtsov98}
A.V. Kravtsov, A.A. Klypin, J.S. Bullock and J.R. Primack, 
{\it Astrophys. J.} {\bf 502}, 48 (1998).



\bibitem{SAP91}
P. Salucci,  K.M. Ashman and M. Persic, 
{\it Astrophys. J.} {\bf 379}, 89 (1991).

\bibitem{SP97}
P. Salucci and M. Persic, in {\it Dark and visible matter in
Galaxies}, 
{\it ASP. Conf. Ser.} {\bf 117}, Ed. Persic and Salucci (1997).

\bibitem{salucci01}
P. Salucci, 
{\it Mon. Not. Roy. Astron. Soc.} {\bf 320}, 1 (2001).

\bibitem{salucci03} 
P. Salucci, 
{\it preprint} arXiv:astro-ph/0310376 (2003).

\bibitem{flores94}
R.A. Flores and J.R. Primack, 
{\it Astrophys. J.} {\bf 427}, L1 (1994).

\bibitem{primack04}
J.~R. Primack, 
{\it IAU Symposium} {\bf 220}, 53 (2004).

\bibitem{ostriker03}
J.~P Ostriker  and P. Steinhardt, 
{\it Science} {\bf 300}, 1909 (2003).

\bibitem{carignan_freeman85}
C. Carignan and K.C. Freeman, 
{\it Astrophys. J.} {\bf 294}, 494 (1985).

\bibitem{dubinski_carlberg91}
J. Dubinski and R.G. Carlberg, 
{\it Astrophys. J.} {\bf 378}, 496 (1991).

\bibitem{moore94}
B. Moore, 
{\it Nature} {\bf 370}, 629 (1994).

\bibitem{burkert95}
A. Burkert, 
{\it Astrophys. J.} {\bf 447}, L25 (1995).

\bibitem{burkert_silk97}
A. Burkert and J. Silk, 
{\it Astrophys. J.} {\bf 488}, L55 (1997).

\bibitem{MdB98}
S.S. McGaugh and W.J.G. de Blok, 
{\it Astrophys. J.} {\bf 449}, 41 (1998).

\bibitem{dBMR01}
W.J.G. de Blok and V.C. Rubin, 
{\it Astron. J.} {\bf 122}, 2396 (2000).

\bibitem{blok_bosma02}
W.J.G. de Blok and A. Bosma, 
{\it Astron. Astrophys.} {\bf 385}, 816 (2002).

\bibitem{trott02}
C.M. Trott and R.L. Webster, 
{\it Mon. Not. Roy. Astron. Soc.} {\bf 334}, 621 (2002).

\bibitem{binney01}
J.J. Binney and N.W. Evans,
{\it Mon. Not. Roy. Astron. Soc.} {\bf 327}, L27 (2001).

\bibitem{blais02}
S. Blais-Ouellette, C. Carignan and P. Amram, 
{\it ASPC} {\bf 282}, 129 (2002).

\bibitem{bottema02}
R. Bottema, 
{\it Astron. Astrophys.} {\bf 388},809 (2002).

\bibitem{weldrake03}
D.T.F. Weldrake, W.J.G. de Blok and F. Walter, 
{\it Mon. Not. Roy. Astron. Soc.} {\bf 340}, 12 (2003).

\bibitem{simon03}
J.D. Simon, A.D. Bolatto, A. Leroy and L. Blitz,  
{\it Astrophys. J.} {\bf 596}, 957 (2003).

\bibitem{salucci_borriello01}
P. Salucci and A. Borriello, 
{\it dmap.conf.} {\bf 12} (2001).

\bibitem{salucci_borriello01b}
A. Borriello and P. Salucci, 
{\it Mon. Not. Roy. Astron. Soc.} {\bf 323}, 285 (2001).

\bibitem{PS90b}
M. Persic and P. Salucci, 
{\it Mon. Not. Roy. Astron. Soc.} {\bf 245}, 577 (1990).

\bibitem{PS92}
M. Persic. and P. Salucci, 
{\it Astro. Lett. and Communications} {\bf 28}, 307 (1992).

\bibitem{NS00}
J.F. Navarro and M. Steinmetz, 
{\it Astrophys. J.} {\bf 528}, 607 (2000).

\bibitem{mo98}
H. Mo, S. Mao and S.D.M. White, 
{\it MNRA} {\bf 295}, 319 (1998).

\bibitem{dal97}
J. Dalcanton, F. Summers and D. Spergel, 
{\it Astrophys. J.} {\bf 482}, 659 (1997).

\bibitem{bosch00}
F.C. van den Bosch, 
{\it Astrophys. J.} {\bf 530}, 177 (2000).

\bibitem{McGaugh03}
S.S. McGaugh, M.K. Barker and W.J.G. de Blok, 
{\it Astrophys. J.} {\bf 584}, 566 (2003).

\bibitem{swaters03}
R.A. Swaters, M.A.W. Verheijen, M.A. Bershady and D.R. Andersen, 
{\it Astrophys. J.} {\bf 587}, 19 (2003).

\bibitem{deNaray08}
R .Kuzio de Naray, S.S. McGaugh and W.J.G. de Blok, 
{\it Astrophys. J.} {\bf 676}, 920 (2008).

\bibitem{simon05}
J.D. Simon, A.D. Bolatto, A. Leroy, L. Blitz and E. Gates, 
{\it Astrophys. J.} {\bf 621}, 757 (2005).

\bibitem{dBMBR01}
W.J.G. de Blok, S.S. McGaugh, A. Bosma and V.C.  Rubin, 
{\it Astrophys. J.} {\bf 552}, L23 (2001).

\bibitem{dBBM03}
W.J.G. de Blok, A. Bosma and S.S. McGaugh, 
{\it Mon. Not. Roy. Astron. Soc.} {\bf 340}, 657 (2003).

\bibitem{marchesini02}
D.M. Marchesini, E. D'Onghia, G. Chincarini, C. Firmani, P. Conconi, E. Molinari and A. Zacchei, 
{\it Astrophys. J.} {\bf 575}, 801 (2002).

\bibitem{bolatto02}
A.D. Bolatto, J.D. Simon, A. Leroy and L. Blitz, 
{\it Astrophys. J.} {\bf 565}, 238 (2002).

\bibitem{blais01}
S. Blais-Ouellette, P. Amram and C. Carignan, 
{\it Astron. J.} {\bf 121}, 1952 (2001).

\bibitem{cote00}
S. C\^{o}t\'{e}, C. Carignan and K.C. Freeman, 
{\it Astron. J.} {\bf 120}, 3027(2000).

\bibitem{blok_McGaugh_hulst96}
W.J.G., McGaugh, S.S. and van der Hulst, J.M., 
{\it Mon. Not. Roy. Astron. Soc.} {\bf 283}, 18 (1996).

\bibitem{bosch00b}
F.C. van den Bosch, B.E. Robertson, J.J. Dalcanton and W.J.G. de Blok, 
{\it Astron. J.} {\bf 119}, 1579 (2000).

\bibitem{swaters00}
R.A. Swaters, B.F. Madore and M. Trewhella, 
{\it Astrophys. J.} {\bf 531}, L107 (2000).

\bibitem{MRdB01}
S.S. McGaugh, V.C. Rubin and W.J.G. de Blok, 
{\it Astron. J.} {\bf 122}, 2381 (2001).

\bibitem{rhee04}
G. Rhee, O. Valenzuela, A. Klypin, J. Holtzman and B, Moorthy, 
{\it Astrophys. J.} {\bf 617}, 1059 (2004).

\bibitem{spekkens05}
K. Spekkens, R. Giovanelli and M.P. Haynes, 
{\it Astron. J.} {\bf 129}, 2119 (2005).

\bibitem{Rob}
R.A. Swaters, B.F. Madore, F.C. van den Bosch and M. Balcells, 
{\it Astrophys. J.} {\bf 583}, 732 (2003).

\bibitem{bosch01} 
F.C. van den Bosch and R.A. Swaters, 
{\it Mon. Not. Roy. Astron. Soc.} {\bf 325}, 1017 (2001).

\bibitem{primack02} 
J.R. Primack, 
{\it NuPhS} {\bf 124}, 3 (2003).

\bibitem{N03}
J. F. Navarro, E. Hayashi, C. Power, A. Jenkins, C.S. Frenk, S.D.M. White, V. 
Springel, J. Stadel and T.R. Quinn, 
{\it Mon. Not. Roy. Astron. Soc.} {\bf 349} 1039 (2004).

\bibitem{kau93}
G. Kauffmann, S.D.M. White and B. Guiderdoni, 
{\it Mon. Not. Roy. Astron. Soc.} {\bf 264}, 201 (1993).

\bibitem{kly99b}
A.A. Klypin, A.V. Kravtsov, O. Valenzuela and F. Prada, 
{\it Astrophys. J.} {\bf 522}, 82 (1999).

\bibitem{kam00}
M. Kamionkowski and A.R. Liddle, 
{\it Phys. Rev. Let.} {\bf 84}, 4525 (2000).

\bibitem{whi00}
M. White and R.A. Croft, 
{\it Astrophys. J.} {\bf 539}, 497 (2000).

\bibitem{som99}
J. Sommer-Larsen and A. Dolgov, 
{\it Astrophys. J.} {\bf 551}, 608 (2001).

\bibitem{hog00}
C.J. Hogan and J.J. Dalcanton, 
{\it Phys.Rev. D} {\bf 62}, 063511 (2000).

\bibitem{peebles99}
P.J.E. Peebles and A. Vilenkin, 
{\it Phys. Rev. D} {\bf 60}, 103506 (1999).

\bibitem{hu00}
W. Hu and P.J.E. Peebles, 
{\it Astrophys. J.} {\bf 528}, L61 (2000).

\bibitem{peebles00}
P.J.E. Peebles, 
{\it Astrophys. J.} {\bf 534}, L127 (2000).

\bibitem{matos00}
T. Matos, G. Siddhartha Guzmán and L.A. Urena-L\'opez, 
{\it Clas. Quantum Grav.} {\bf 17}, 1707 (2000).

\bibitem{car92}
E.D. Carlson, M.E. Machacek and L.J. Hall, 
{\it Astrophys. J.} {\bf 398}, 43 (1992).

\bibitem{spergel00}
D. N. Spergel and P. J. Steinhardt, 
{\it Phys. Rev. Lett.} {\bf 84}, 17 (2000).

\bibitem{mohapatra00}
R.N. Mohapatra and V.L. Teplitz, 
{\it Phys.Rev. D} {\bf 62}, 063506 (2000).

\bibitem{firmani00} 
C. Firmani, E. D'Onghia, V. Avila-Reese, G. Chincarini and X. Hern\'andez, 
{\it Mon. Not. Roy. Astron. Soc.} {\bf 315}, L29 (2000).

\bibitem{goodman00}
J. Goodman, 
{\it New Astron.} {\bf 5}, 103 (2000).

\bibitem{kap00}
M. Kaplinghat, L. Knox and M.S. Turner, 
{\it Phys. Rev. Let.} {\bf 85}, 3335 (2000).

\bibitem{bento00}
M.C. Bento, O. Bertolami, R. Rosenfeld and L. Teodoro, 
{\it Phys. Rev. D} {\bf 62}, 041302 (2000).

\bibitem{moore99b} 
B. Moore, T. Quinn, F. Governato, J. Stadel and G. Lake, 
{\it Mon. Not. Roy. Astron. Soc.} {\bf 310}, 1147 (1999).

\bibitem{col00}
P. Col\'{i}n, V. Avila-Reese and O. Valenzuela, 
{\it Astrophys. J.} {\bf 542}, 622 (2000).

\bibitem{dal00b} 
J.J. Dalcanton and C.J. Hogan, 
{\it Astrophys. J.} {\bf 561}, 35, (2001).

\bibitem{burkert00}
A. Burkert, 
{\it Astrophys. J.} {\bf 534}, L143 (2000).

\bibitem{koc00}
C.S.Kochanek and M. White, 
{\it Astrophys. J.} {\bf 543}, 514 (2000).

\bibitem{sellwood00}
J.A. Sellwood, 
{\it Astrophys. J.} {\bf 540}, L1 (2000).

\bibitem{yos00}
N. Yoshida, V. Springel, S.D.M. White and G. Tormen, 
{\it Astrophys. J.} {\bf 535}, L103 (2000).

\bibitem{moore00}
B. Moore, S. Gelato, A. Jenkins,  F.R. Pearce and V. Quilis, 
{\it Astrophys. J.} {\bf 535}, L21 (2000).

\bibitem{NEF96}
J.F. Navarro, V.R. Eke and C.S. Frenk, 
{\it Mon. Not. Roy. Astron. Soc.} {\bf 283}, 72 (1996).

\bibitem{gel99}
S. Gelato and J. Sommer-Larson, 
{\it Mon. Not. Roy. Astron. Soc.} {\bf 303}, 321 (1999).

\bibitem{BKW00}
J.S. Bullock, A.V. Kravtsov and D.H. Weinberg, 
{\it Astrophys. J.} {\bf 539}, 517 (2000).

\bibitem{bin01} 
J.J. Binney, O.E. Gerhard and J. Silk, 
{\it Mon. Not. Roy. Astron. Soc.} {\bf 321}, 471 (2001).

\bibitem{mac99}
M.-M. Mac Low and A. Ferrrara, 
{\it Astrophys. J.} {\bf 513}, 142 (1999).

\bibitem{strickland00}
D.K. Strickland and I.R. Stevens, 
{\it Mon. Not. Roy. Astron. Soc.} {\bf 314}, 511 (2000).

\bibitem{tonini_erasing06}
C. Tonini, A. Lapi and P. Salucci, 
{\it Astrophys. J.} {\bf 649}, 591 (2006).

\bibitem{gentile05}
G. Gentile, A. Burkert, P. Salucci, U. Klein and F. Walter, 
{\it Astrophys. J.} {\bf 634}, L145 (2005).
  
\bibitem{gentile_tonini07}
G. Gentile, C. Tonini and P. Salucci, 
{\it Astron. Astrophys.} {\bf 467}, 925 (2007).





\bibitem{milgrom} 
M. Milgrom, 
{\it Astrophys. J.} {\bf 270}, 365 (1983).

\bibitem{mond_review}
R.~H. Sanders and S.~S. McGaugh, 
{\it Annual Rev. Astron. Astrophys.} {\bf 40}, 263 (2002).

\bibitem{BBS}
K.~G. Begeman, A.~H. Broeils and R.~H. Sanders, 
{\it Mon. Not. Roy. Astron. Soc.} {\bf 249}, 523 (1991).

\bibitem{BM95}
R. Brada and M. Milgrom, 
{\it Mon. Not. Roy. Astron. Soc.} {\bf 276}, 453 (1995).

\bibitem{FB05}
B. Famaey and J. Binney,
{\it Mon. Not. Roy. Astron. Soc.} {\bf 363}, 603 (2005).

\bibitem{teves} 
J.~D. Bekenstein, 
{\it Phys. Rev. D} {\bf 70}, 083509 (2004).

\bibitem{F07}
B. Famaey, G. Gentile , J.-P. Bruneton and H. S. Zhao,
{\it Phys. Rev. D} {\bf 75}, 063002 (2007).

\bibitem{kent87}
S.~M. Kent, 
{\it Astron. J.} {\bf 93}, 816 (1987).

\bibitem{milgrom88}
M. Milgrom,
{\it Astrophys. J.} {\bf 333}, 689 (1988).

\bibitem{sanders96}
R.~H. Sanders,
{\it Astrophys. J.} {\bf 473}, 117 (1996).

\bibitem{blok_mcGaugh98}
W.~J.~G. de Blok and S.~S. McGaugh
{\it Astrophys. J.} {\bf 508}, 132 (1998).

\bibitem{McGaugh04}
S. S. McGaugh,
{\it Astrophys. J.} {\bf 609}, 652 (2004).

\bibitem{McGaugh05}
S. S. McGaugh,
{\it Astrophys. J.} {\bf 632}, 859 (2005).

\bibitem{richtler08}
T. Richtler, Y. Schuberth, M. Hilker, B. Dirsch, L. Bassino and A.~J. Romanowsky,
{\it Astron. Astrophys.} {\bf 478}, L23 (2008).

\bibitem{McGaugh04_wmap} 
S.~S. McGaugh, 
{\it Astrophys. J.} {\bf 611}, 26 (2004). 

\bibitem{angus07_neutrinos} 
G.~W. Angus, H.~Y. Shan, H.~S. Zhao and B. Famaey,
{\it Astrophys. J.} {\bf 654}, L13 (2007). 

\bibitem{mass_neutrinos}
Ch. Kraus, B. Bornschein, L. Bornschein, J. Bonn, B. Flatt, A. Kovalik, B. Ostrick, E.W. Otten, J.P. Schall, Th. Thümmler and Ch. Weinheimer,
{\it Eur.Phys.J. C} {\bf 40} 447 (2005).

\bibitem{bullet}
D. Clowe, M. Brada{\v c}, A.~H. Gonzalez, M. Markevitch, S.~W. Randall, 
C. Jones and D. Zaritsky, 
{\it Astrophys. J.} {\bf 648}, L109 (2006).

\bibitem{katrin}
KATRIN collaboration,
{\it preprint} arXiv:hep-ex/0109033




\bibitem{weyl19}
H. Weyl, 
{\it Ann.~Phys.} {\bf 59}, 101 (1919).

\bibitem{eddington23}
A.~S. Eddington,
{\it The Mathematical Theory of Relativity} (Cambridge University Press, Cambridge) (1923).

\bibitem{utiyama62}
R. Utiyama and B.~S. DeWitt, 
{\it J. Math. Phys.} {\bf 3}, 608 (1962).

\bibitem{stelle77}
K.~S. Stelle,
{\it Phys. Rev. D} {\bf 16}, 953 (1977).

\bibitem{brandenberger93}
R.~H. Brandenberger, V.~F. Mukhanov and A. Sornborger,
{\it Phys. Rev. D} {\bf 48}, 1629 (1993).

\bibitem{capoz_franca08}
S. Capozziello and M. Francaviglia,
{\it Gen. Rel. Grav.} {\bf 40}, 357 (2008).

\bibitem{thomas_review}
T. P. Sotiriou and V. Faraoni,
{\it preprint} arXiv: gr-qc/0805.1726

\bibitem{sotiriou_liberati07}
T.~P. Sotiriou and S. Liberati,
{\it Annals Phys.} {\bf 322}, 935 (2007).

\bibitem{capozziello03}
S. Capozziello, S. Carloni and A. Troisi,
{\it Recent Res. Dev. Astron. Astrophys.} {\bf 1}, 625 (2003).

\bibitem{barrow_clifton06}
J.~D. Barrow and T. Clifton,
{\it Class. Quant. Grav.} {\bf 23}, L1 (2006).

\bibitem{clifton_barrow05}
T. Clifton and J.~D. Barrow,
{\it Phys. Rev. D} {\bf 72}, 103005 (2005).

\bibitem{clifton_barrow06}
T. Clifton and J.~D. Barrow,
{\it Class. Quant. Grav.} {\bf 23}, 2951 (2006).

\bibitem{zakharov06}
A.~F. Zakharov, A.~A. Nucita, F. De~Paolis and G. Ingrosso,
{\it Phys. Rev. D} {\bf 74}, 107101 (2006).

\bibitem{iorio_riggiero07a}
L. Iorio and M.~L. Ruggiero,
{\it Int. J. Mod. Phys.} {\bf A22}, 5379 (2007).

\bibitem{iorio_riggiero07b}
L. Iorio and M.~L. Ruggiero, 
{\it eprint} arXiv: gr-qc/0711.0256.

\bibitem{PM}
F. Piazza and C. Marinoni, 
{\it Phys. Rev. Lett.} {\bf 91}, 141301 (2003).

\bibitem{wlensing}
H. Hoekstra, 
{\it EAS Pub. Ser.} {\bf 20}, 153 (2006).

\bibitem{bruzual}
G. Bruzual and S. Charlot, 
{\it Mon. Not. R. Astron. Soc.} {\bf 344}, 1000 (2003).

\bibitem{luminosity}
P. Salucci and M. Persic, 
{\it Mon. Not. R. Astron. Soc.} {\bf 309}, 923 (1999).

\bibitem{chiara}
C. Tonini, A. Lapi, F. Shankar and P. Salucci,
{\it Astrophys. J.} {\bf 638}, L13 (2006).

\bibitem{hayashi04}
E. Hayashi, J.F. Navarro, C. Power, A. Jenkins, C.S. Frenk, S.D.M. White, V. Springel, J. Stadel and T.R. Quinn,
{\it Mon. Not. R. Astron. Soc.} {\bf 355}, 794 (2004).

\bibitem{hayashi06}
E. Hayashi and J. F. Navarro, 
{\it Mon. Not. R. Astron. Soc.} {\bf 373}, 1117 (2006).

\bibitem{veneziano}
T. Damour, F. Piazza and G. Veneziano, 
{\it Phys. Rev. D} {\bf 66}, 46007 (2002).

\bibitem{piazza}
C. Marinoni and F. Piazza, 
{\it preprint} arXiv: astro-ph/0312001

\bibitem{corbelli}
E. Corbelli and P. Salucci, 
{\it Mon. Not. R. Astron. Soc.} {\bf 347}, 1051 (2007).

\bibitem{vogt}
N.P. Vogt, M.P. Haynes, T. Herter and R. Giovanelli,
{\it Astron. J.} {\bf 127}, 3273 (2004).




\bibitem{rubin83}
V.C. Rubin,  
{\it Science (ISSN 0036-8075)} {\bf 220}, 1339 (1983).

\bibitem{brownstein}
J.R. Brownstein and J. W. Moffat, 
{\it Astrophys. J.} {\bf 636}, 721 (2006).

\bibitem{bekenstein} 
J. Bekenstein, 
{\it ConPh} {\bf 47}, 387 (2006).

\bibitem{carroll}
S.M. Carroll, V. Duvvuri, M.  Trodden and M.S. Turner,
{\it Phys. Rev.D} {\bf 70}, 4, 043528 (2004). 

\bibitem{capozziello04}
S. Capozziello, V.F. Cardone, S.  Carloni and A. Troisi,
{\it Phys. Lett. A} {\bf 326}, 5-6, 292 (2004).

\bibitem{CCT}
S. Capozziello, V.F.  Cardone and A. Troisi, 
{\it Mon. Not. R. Astron. Soc.} {\bf 375}, 4 (2007).   

\bibitem{nojiri}
S. Nojiri and S.D.  Odintsov, 
{\it Int. J. Geom. Meth. Mod. Phys.} {\bf 4}, 115 (2007).

\bibitem{carloni}
S. Carloni, P. K. S. Dunsby, S. Capozziello and A. Troisi,
{\it Class. Quant. Grav.} {\bf 22}, 22, 4839 (2005).

\bibitem{frigerioPRL}
C. Frigerio Martins and P.  Salucci, 
{\it  Phys. Rev. Lett.} {\bf 98}, 151301 (2007).

\bibitem{2574}
N. Martimbeau, C. Carignan and J.  Roy, 
{\it Astron. J.} {\bf 107(2)}, 543 (1994).

\bibitem{5585}
S. C\^{o}t\'{e}, C.  Carignan and R. Sancisi, 
{\it Astron. J..} {\bf 102}, 3, 904 (1991).

\bibitem{6503}
B.M.H.R. Wevers, P.C. van der Kruit and R.J. Allen, 
{\it Astron. Astrophys.} {\bf 4}, 86 (1986).

\bibitem{2403}
F. Fraternali, G.  van Moorsel, R. Sancisi and T. Oosterloo,
{\it Astron. J.} {\bf 123}, 3124 (2002).

\bibitem{55}
D. Puche, C. Carignan and R.J. Wainscoat, 
{\it Astron. J.} {\bf 101}, 2, 447 (1991).

\bibitem{umajorPhotometry}
R. B. Tully, M.A.W.  Verheijen, M.J.  Pierce, J-S. Huang and R.J. Wainscoat, 
{\it Astron. J.} {\bf 112}, 2471 (1996).

\bibitem{umajorHI}
M.A.W. Verheijen and R. Sancisi, 
{\it Astron. Astrophys.} {\bf 370}, 765 (2001).

\bibitem{300}
D. Puche, C. Carignan and A. Bosma, 
{\it Astron. J.} {\bf 100}, 5, 1468 (1990).

\bibitem{128}
J.M. van der Hulst, E.D. Skillman, T.R.  Smith, G.D.  Bothun, S.S. McGaugh and W. J. G. de Blok,
{\it Astron. J.} {\bf 106}, 2, 548 (1993).

\bibitem{capozziello06}
S. Capozziello, V.F. Cardone and A. Troisi,  
{\it JCAP} {\bf 8}, 001 (2006).




\bibitem{Napolitano07}
N.~R. Napolitano,
{\it IAU Symposium} {\bf 244}, 289 (2007).

\bibitem{Gilmore07}
G. Gilmore, M.~I. Wilkinson, R.~F.~G. Wyse, J.~T. Kleyna,  
A. Koch, N.~W. Evans and E.~K. Grebel, 
{\it Astrophys. J.} {\bf 663}, 948 (2007).

\bibitem{Tolstoy04}
E. Tolstoy, {\it et al.}, 
{\it Astrophys. J.} {\bf 617}, L119 (2004).

\bibitem{Walker07}
M.~G. Walker, M. Mateo, E.~W. Olszewski, O.~Y. Gnedin, X. Wang, B. Sen and
M. Woodroofe, 
{\it Astrophys. J.} {\bf 667}, L53 (2007).

\bibitem{Mateo98}
M.L. Mateo,
{\it Annual Rev. Astron. Astrophys.} {\bf 36}, 435 (1998).

\bibitem{Koch07a}
A. Koch, M.~I. Wilkinson, J.~T. Kleyna, G.~F. Gilmore, E.~K. Grebel, A.~D. Mackey, N.~W. Evans and R.~F.~G. Wyse,
{\it Astrophys. J.} {\bf 657}, 241 (2007).

\bibitem{Zaritsky2006}
D. Zaritsky, A.~H. Gonzalez and A.~I. Zabludoff,  
{\it Astrophys. J.} {\bf 638},  725 (2006).

\bibitem{Dabringhausen2008}
J. Dabringhausen, M. Hilker and P. Kroupa,  
{\it Mon. Not. R. Astron. Soc.} {\bf 386},  864 (2008).

\bibitem{Forbes2008}
D. Forbes, P. Lasky, A. Graham and L. Spitler, 
{\it eprint} arXiv:astro-ph/0806.1090.

\bibitem{Wilkinson04}
M.~I. Wilkinson, J.~T. Kleyna, N.~W. Evans, G.~F. Gilmore, M.~J. Irwin and E.~K. Grebel,
{\it Astrophys. J.} {\bf  611}, L21 (2004)

\bibitem{Kleyna04}
J.~T. Kleyna, M.~I. Wilkinson, N.~W. Evans and G. Gilmore
{\it Mon. Not. R. Astron. Soc.} {\bf 354}, L66 (2004).

\bibitem{Munoz05}
R.~R. Mu{\~n}oz, {\it et al.},
{\it Astrophys. J.} {\bf 631}, L137 (2005).

\bibitem{Munoz06}
R.~R. Mu{\~n}oz, {\it et al.},
{\it Astrophys. J.} {\bf 649}, 201 (2006).

\bibitem{Koch07b}
A. Koch, J.~T. Kleyna, M.~I Wilkinson, E.~K. Grebel, G.~F. Gilmore, N.~W. Evans, R.~F.~G. Wyse and D.~R Harbeck
{\it Astron. J.} {\bf 134}, 566 (2007).

\bibitem{Battaglia08}
G. Battaglia, A. Helmi, E. Tolstoy, M. Irwin, V. Hill and P. Jablonka,
{\it eprint} arXiv:astrp-ph/0802.4220.

\bibitem{IH95}
M. Irwin and D. Hatzidimitriou
{\it Mon. Not. R. Astron. Soc.} {\bf 277}, 1354 (1995).

\bibitem{Majewski00}
S.~R. Majewski, J.~C. Ostheimer, R.~J. Patterson, W.~E. Kunkel, K.~V.Johnston and D. Geisler,
{\it Astron. J.} {\bf 119}, 760 (2000)

\bibitem{Plummer1915}
H.~C. Plummer,
{\it Mon. Not. R. Astron. Soc.} {\bf 76}, 107 (1915).

\bibitem{BT87}
J. Binney and S. Tremaine,  
{\it Galactic Dynamics}, Princeton University Press, Princeton (1987).

\bibitem{Gerhard}
O. Gerhard, A. Kronawitter, R.~P. Saglia and R. Bender, 
{\it Astron. J.} {\bf 121}, 1936 (2001).

\bibitem{Piatek03}
S. Piatek, C. Pryor, E.~W. Olszewski, H.~C. Harris, M. Mateo, D. Minniti and C.~G. Tinney, 
{\it Astron. J.} {\bf 126}, 2346 (2003).

\bibitem{Piatek05}
S. Piatek, C. Pryor, P. Bristow, E.~W. Olszewski, H.~C. Harris, M. Mateo, D. Minniti and C.~G. Tinney,
{\it Astron. J.} {\bf 130}, 95 (2005).

\bibitem{Piatek06}
S. Piatek, C. Pryor, P. Bristow, E.~W. Olszewski, H.~C. Harris, M. Mateo, . inniti and C.~G. Tinney,
{\it Astron. J.} {\bf 131}, 1445 (2006).

\bibitem{Piatek07}
S. Piatek, C. Pryor, P. Bristow, E.~W. Olszewski, H.~C. Harris, M. Mateo, D. Minniti, C.~G. Tinney,
{\it Astron. J.} {\bf 133}, 818 (2007).

\bibitem{Martin2008}
N. F. Martin, J. T. A. de Jong and H.-W Rix
{\it eprint} arXiv:astro-ph/0805.2945v2.

\bibitem{Mayer2001}
L. Mayer, F. Governato, M. Colpi, B. Moore, T. Quinn, J. Wadsley, J. Stadel, G.  Lake,
{\it Astrophys. J.} {\bf 559},  754 (2001).

\bibitem{Klimentowski2007}
J. Klimentowski, E.~L. {\L}okas, S. Kazantzidis, F. Prada, L. Mayer, G.~A. Mamon
MNRAS,  
378,  353 (2007).

\bibitem{Lokas2008}
E.~L. {\L}okas, J. Klimentowski, S. Kazantzidis and L. Mayer,
{\it Mon.Not.Roy.Astron.Soc.} {\bf 390}, 625 (2008)



\bibitem{Neto07}
Neto, A.~F., {\it et al.},
{\it Mon. Not. R. Astron. Soc.} {\bf 381}, 1450 (2007).

\bibitem{KF04}
J. Kormendy and K. C. Freeman,
IAU Symposium, Sydney, Astronomical Society of the Pacific., {\bf 220}, 377 (2004).

\bibitem{Dehnen2005}
W. Dehnen and D.~E. McLaughlin,  
{\it Mon. Not. R. Astron. Soc.} {\bf 363},  1057 (2005).

\bibitem{hoek}
H. Hoekstra {\it et al.}
{\it Astrophys. J.} {\bf 635}, 73 (2005).

\bibitem{Grebel2003} 
E.~K. Grebel, J.~S Gallagher, III and D. Harbeck,
{\it Astron. J.} {\bf 125}, 1926 (2003).

\bibitem{salucci2008}
Salucci, P., {\it et al.}, 2008, MNRAS, in submition

\bibitem{MateoLeo98}
M. Mateo, E.~W. Olszewski, S.~S. Vogt and M.~J. Keane
{\it Astron. J.} {\bf 116},  2315 (1998).


\end{thebibliography}
\end{document}